\documentclass[10pt,
onecolumn,
amsmath,
paper=a4,
bibtotocnumbered,	  
liststotocnumbered,  
DIV=calc,		  
tablecaptionabove,	  
headinclude,
]{revtex4-1}
\usepackage{amssymb}
\usepackage{graphicx}
\usepackage{dcolumn}
\usepackage{array}
\usepackage{bm}
\usepackage{fancyheadings}
\usepackage{longtable}
\usepackage{float}
\usepackage{morefloats}
\usepackage{afterpage}  
\usepackage{color}
\setlongtables
\usepackage[breaklinks=true,linkbordercolor={1 1 1}]{hyperref}
\begin{document}
\setcounter{page}{1}
\title{
     \qquad \\ \qquad \\ 
     \qquad \\  \qquad \\  
     \qquad \\ \qquad \\ 
Extensive study of the quality of fission yields \\ from experiment, evaluation and
GEF \\ for antineutrino studies and applications
}

\author{K.-H.~Schmidt}
\email[Corresponding author: ]{schmidt-erzhausen@t-online.de \\ URL: http://www.khschmidts-nuclear-web.eu\\ORCID iD
https://orcid.org/0000-0002-3467-9213}
\author{M.~Estienne}
\email[Corresponding author: ]{magali.estienne@subatech.in2p3.fr}
\author{M.~Fallot}
\author{S.~Cormon}
\author{A.~Cucoanes}
\author{T.~Shiba}
\affiliation{Subatech, CNRS/IN2P3-Universit\'e de Nantes-IMTA\\ 
4 rue Alfred Kastler, F-44307 Nantes, France}
\author{B.~Jurado}
\affiliation{CENBG, CNRS/IN2P3, Chemin du Solarium B.P. 120, F-33175 Gradignan, France }
\author{K.~Kern}
\affiliation{Prokerno Corp.,
Apartado Postal 0823-04789,
Panam\'a - Bella Vista,
Republic of Panama} 
\author{Ch.~Schmitt}
\affiliation{IPHC, CNRS/IN2P3,
23 rue du Loess, B.P. 28, F-67037 Strasbourg, France}


\begin{abstract}
{\bf Abstract:} The understanding of the antineutrino production in fission and the theoretical calculation of the antineutrino energy spectra in different, also future, types of fission reactors rely on the application of the summation method, where the individual contributions from the different radioactive nuclides that undergo a beta decay are estimated and summed up. The most accurate estimation of the independent fission-product yields is essential to this calculation. This is a complex task, because the yields depend on the fissioning nucleus and on the energy spectrum of the incident neutrons. 

In the present contribution, the quality of different sources of information on the fission yields is investigated, and the benefit of a combined analysis is demonstrated. 
The influence on antineutrino predictions is discussed.

In a systematic comparison, the qualilty of fission-product yields emerging from different experimental techniques is analyzed. 
The traditional radiochemical method, which is almost exclusively used for evaluations, provides an unambiguous identification in $Z$ and $A$, but it is restricted to a limited number of suitable targets, is slow, and the accuracy suffers from uncertainties in the spectroscopic nuclear properties. 
Experiments with powerful spectrometers, for example at LOHENGRIN, provide very accurate mass yields and a $Z$ resolution for light fission products from thermal-neutron-induced fission of a few suitable target nuclei. 

On the theoretical side, the general fission model GEF has been developed. It combines a few general theorems, rules and ideas with empirical knowledge. GEF covers almost all fission observables and is able to reproduce measured data with high accuracy while having remarkable predictive power by establishing and exploiting unexpected systematics and hidden regularities in the fission observables. In this article, we have coupled for the first time the GEF predictions for the fission yields to fission-product beta-decay data in a summation calculation of reactor antineutrino energy spectra. The first comparisons performed between the spectra from GEF and those obtained with the evaluated nuclear databases exhibited large discrepancies that highlighted the exigency of the modelisation of the antineutrino spectra and showing their usefulness in the evaluation of nuclear data. Additional constraints for the GEF model were thus needed in order to reach the level of accuracy required by the antineutrino energy spectra. The combination of a careful study of the independent isotopic yields and the adjunction of the LOHENGRIN fission-yield data as additional constraints led to a substantially improved agreement between the antineutrino spectra computed with GEF and with the evaluated data. 
The comparison of inverse beta-decay yields computed with GEF with those measured by the Daya Bay experiment shows the excellent level of predictiveness of the GEF model for the fundamental or applied antineutrino physics.

The main results of this study are:

- an improved agreement between the antineutrino energy spectra obtained with the newly tuned GEF model and the JEFF-3.1.1 and JEFF-3.3 fission yields for the four main contributors to fission in standard power reactors;

- indications for shortcomings of mass yields for $^{\text{241}}$Pu(n$_{\text{th}}$,f) and other systems in current evaluations;

- a demonstration of the benefit from cross-checking the results of different experimental 
  approaches and GEF for improving the quality of nuclear data;

- an analysis of the sources of uncertainties and erroneous results from different experimental
  approaches;

- the capacity of GEF for predicting the fission yields (and other observables) in cases
  (in terms of fissioning systems and excitation energies) which are presently not accessible 
  to experiment;

- predictions of antineutrino energy spectra that aim to assess the prospects for reactor monitoring, based on the GEF fission yields associated with the beta-decay data of the most recent summation model.



\end{abstract}

\maketitle


\newpage
\newpage

\tableofcontents{}

\newpage

\section{Introduction} \label{S1}
When a heavy nucleus breaks apart, the two fragments, even after prompt-neutron emission, are usually situated on the neutron-rich side of the nuclear chart.
Thus, most of them undergo a sequence of several beta-minus decays, until the beta-stability line is reached. In each beta decay, an antineutrino is produced. Each beta emitter is characterized by a specific antineutrino spectrum, which is determined by the beta Q value and the relative population of ground and excited states in the respective daughter nucleus. 
Fission reactors form particularly strong antineutrino sources \cite{Roskovec18}, which can be used for 
particle physics studies~\cite{ReinesCowan,DC,DB,RENO} or for technical purposes. 
The total spectrum of all these contributions from all the fissioning species in a fission reactor is characteristic for the operation method of the reactor and was proposed to be exploited for reactor monitoring~\cite{Mikaelian}. 

Until recently, integral measurements of the beta spectra~\cite{SchreckU5-1,SchreckU5-2,SchreckU5Pu9,Hahn} of the main fission sources of a power reactor, $^{\text{235}}$U, $^{\text{239}}$Pu, $^{\text{241}}$Pu and $^{\text{238}}$U~\cite{Haag}, were used to obtain the antineutrino emission by the reactor-neutrino experiments. In 2011, these converted spectra were computed again, and the comparison between the newly obtained predictions and reactor antineutrino experiment results showed a 6\% discrepancy~\cite{Mueller,Huber} called the "reactor anomaly"~\cite{Mention}. A little later, a shape discrepancy between 5 and 7 MeV in antineutrino energy was evidenced between measured antineutrino spectra and the same predictions, called the shape anomaly~\cite{ShapeAnomaly}. These unexplained discrepancies triggered numerous studies in several directions: search for sterile neutrinos at reactors~\cite{Mention,WhitePaper} ; exploration of potential biases of the conversion model~\cite{Hayes,FangBrown,Hayen} ; development of an alternative model based on nuclear data i.e. the summation method~\cite{Mueller,Fallot,Zak,Sonzogni,Estienne}. 
An important pre-requisite of a
summation calculation of these antineutrino spectra is an accurate estimation of the independent fission-product yields, that means the yields before beta decay. 
The crucial importance of this point is demonstrated by the considerably diverging antineutrino spectra obtained by using different evaluations \cite{Hayes15,Xubo18,Dwyer,Sonzo2015}. 
In particular, drastic discrepancies were found in the antineutrino spectrum, which amount to more than 30\% around 5 to 6 MeV for $^{\text{235}}$U(n$_{\text{th}}$,f) when using fission yields from ENDF/B-VII.1, JEFF-3.1.1 and JENDL-4.0, respectively. 
Also products with small yields can have strong influence on the antineutrino spectrum, because only few beta emitters may contribute to certain regions in the antineutrino spectrum.

In the present contribution, we investigate how a combined analysis of experiment, evaluation and theory can lead to an improved quality of fission-product yield estimations. In particular, we add the antineutrino observable to the ones already used such as decay heat, delayed neutron fractions or prompt neutron multiplicities, and we
demonstrate the benefit of including a theoretical model in this process.
The GEF model \cite{Schmidt16} seems to us best suited for this purpose.
The calculation of antineutrino energy spectra with fission yields resulting from different sets of parameters of the GEF model allows tuning these parameters to better reproduce those computed with the JEFF fission yields with the constraint to keep the consistency of the parameters among the various fissioning systems. 

Antineutrino detection for reactor monitoring is another motivation for improving the quality  of the fission yields stored in the evaluated databases for fission products and of the beta-decay properties.
The property of antineutrinos of crossing large quantities of matter without interaction makes them a naturally temper-proof probe. 
The detection of antineutrinos close to reactors presents several advantages: it could be performed remotely, it reflects the fuel content and the thermal power of the reactor. 
The monitoring of on-load reactors as well as some of the future reactor designs is challenging for conventional safeguard techniques. 
In the case of on-load reactors, such as CANDU reactors or Pebble Bed Modular Reactors or Molten Salt Reactors, it is not necessary to stop the operation of the core to refuel it. 
For these reactors, an antineutrino detector placed outside the containment walls at a moderate distance could offer an instrument for bulk accountancy of the fuel content of the core. 
To infer to which extent antineutrinos could provide a diversion signature, the characterization of the antineutrino source associated to different contemporary or future reactor designs and fuels is mandatory. This is to be the first step of a feasibility study and necessitates the development of simulation tools \cite{CormonND}.
The summation method is the only predictive method that could allow such calculations. 
Potential applications of antineutrino detectors at reactors were listed if this novel technology is approved~\cite{IAEAReport}. 
These designs imply fission induced by thermal and fast neutrons for various fuels. The  fission-yield data are still scarce for fuels deviating from the most standard ones in use in today's power plants, and the GEF model can provide a means to get reliable predictions with uncertainties.

In the first part of this article, after a presentation of the GEF model, we present comparisons between antineutrino energy spectra built with the JEFF and the GEF fission yields. We explain how it led to improvements of the model through the adjunction of experimental constraints such as the LOHENGRIN sets of fission yields. We then show the level of agreement reached between JEFF and GEF on the antineutrino energy spectra from a standard power reactor fuel. In the second part of this article, we review the experimental methods available to bring additional experimental constraints to the evaluated fission yields. We then give a general view on a large variety of fissioning systems with the aim to test the validity of the postulated regularities of the GEF model, which are crucial for its predictive power.
The comparison is made for all systems, for which empirical fission-product yields from evaluations 
or from selected highly accurate kinematic experiments on thermal-neutron-induced fission are available.
(In this work, evaluated data are considered as empirical information, because they are essentially based on measured data.)
In addition, the fast-neutron-induced fission of $^{238}$U is included due to its contribution to the antineutrino production in a reactor.  
This wide overview that includes also many systems, which do not contribute to the antineutrino production in currently operated reactors, allows us to obtain a complete picture of the deviations between GEF and the available empirical data and to locate their origin. 
It is also useful for estimating the antineutrino production in future fission reactors with different kinds of fuel. In the last section of this article, we provide predictions of antineutrino energy spectra for the corresponding fissioning systems.


\section{The GEF model} \label{S2}
The fully theoretical (microscopic) description of the complete fission process has not yet attained the accuracy that makes it suitable for technical applications.
Only the description of pre-saddle and post-scission phenomena, in particular the fission cross section and the de-excitation of the fission fragments, is well mastered by highly developed and rather sophisticated optical-model and by dedicated statistical de-excitation codes, respectively, while the dynamical evolution of the system between saddle and scission, which is decisive for the fission yields, poses still a severe challenge to theory, see \cite{Schmidt18}.

Therefore, we focus in this contribution on a semi-empirical approach, the general fission model GEF \cite{Schmidt16}, which is based on a number of concepts and laws of general validity. GEF has shown to reproduce measured data remarkably well, and, thus, it is reasonable to expect its predictive power to be most reliable. GEF covers the whole fission process, starting with the formation of an excited system and ending after the radioactive decay of the fission products towards the beta-stable end products. This model has a set of empirical parameters, which are adjusted to the available empirical information. 
The GEF model with a set of well adjusted parameters is able to predict the fission quantities of other systems with an accuracy comparable with the uncertainties of the experimental data used for the parameter fit \cite{Schmitt18}.


\subsection{Concept} \label{S2.1}

A detailed description of the GEF model can be found in Ref.~\cite{Schmidt16}. 
Here, we only give a succinct and somewhat simplified description of the main ideas that are specific to the GEF model. The calculations in this work were performed with the version GEF-Y2019/V1.2 \cite{GEF}.

\begin{figure}[h]
\centering
\includegraphics[width=0.5\textwidth]{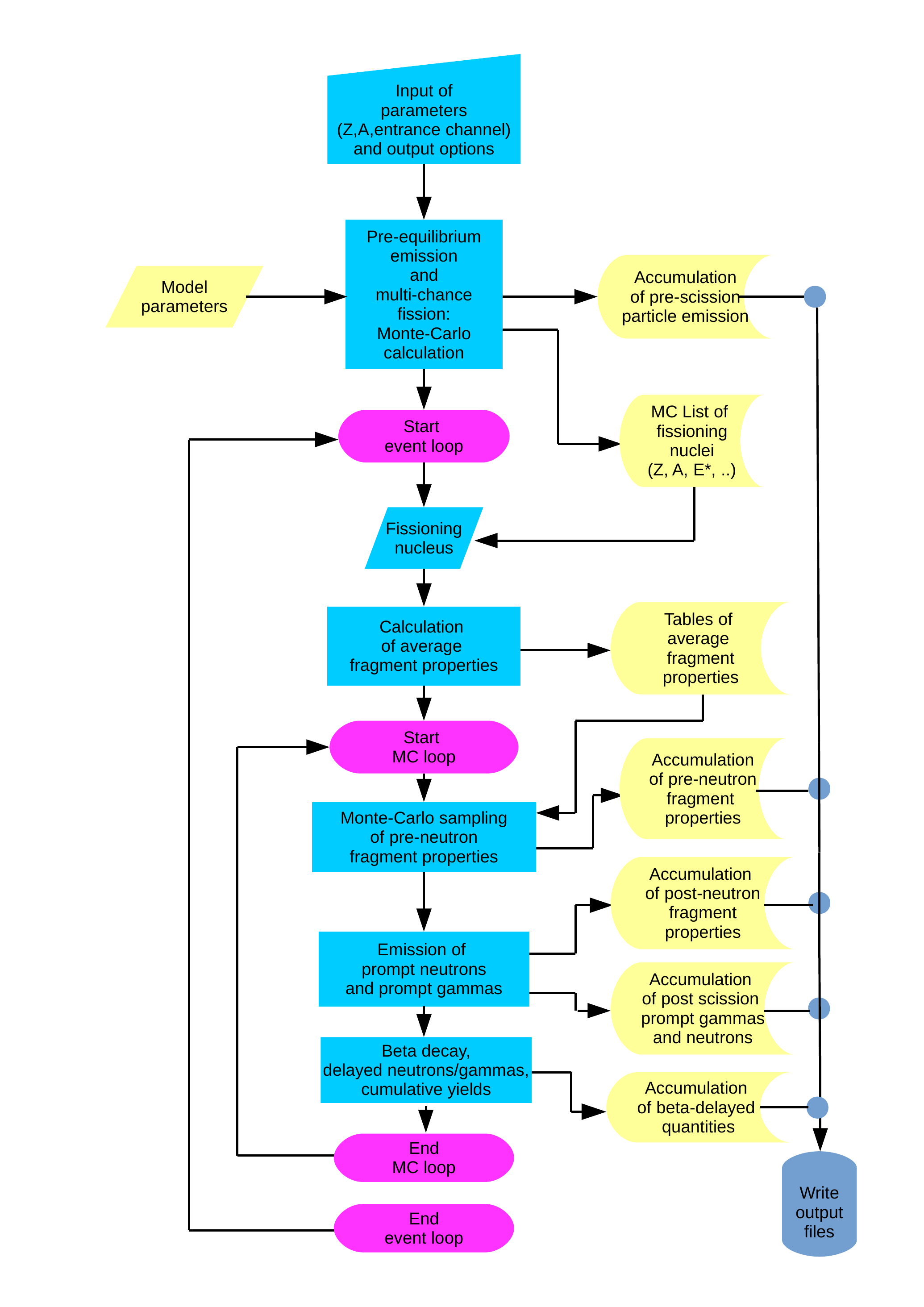}
\caption{Flow chart of the GEF code.} 
\label{Flowchart}       
\end{figure}

Fig. \ref{Flowchart} shows a flow diagram of the GEF code, which documents the treatment of the different steps of the fission process.

- The GEF code uses the Monte-Carlo approach to generate event-by-event information of nearly all observables.

- Each event starts from a specific, possibly excited and rotating, nucleus as given by the user (spontaneous fission from the nuclear ground state or by specifying the reaction, e.g. neutron or proton bombardment, or by indicating the compound nucleus and its excitation energy and angular momentum directly). GEF calculates the decay of the system by fission in competition with the emission of neutrons, protons and photons. Pre-equilibrium emission is also included whenever suited.

-In case the system is committed to fission, the distributions of the fragment properties at scission ($A$, $Z$, kinetic energies, excitation energies and deformation) are calculated. Then, the de-excitation of the primary fragments is calculated by a competition between neutron, proton and gamma emission, till the cold secondary products reach the ground state or an isomeric state. For those products which are radioactive, GEF can also compute their decay by beta emission, delayed neutrons etc.

The main ingredients of the GEF code entering the modeling of the fission probability and fragment properties, and which are often specific to GEF, are shortly discussed below. The modeling of particle evaporation and gamma emission, in an extended Weisskopf theory with explicit consideration of angular-momentum-dependent nuclear properties, is more standard and is not mentioned further. Details can be found in \cite{Schmidt16}.

\paragraph{Fission barriers:}
The most important physical property for the modeling of the fission probability is the fission barrier. 
The fission barriers are calculated by use of the topographic theorem \cite{Myers96} as the sum of the macroscopic barrier and the additional binding energy by the ground-state shell correction. 
This approach avoids the uncertainties of the theoretical shell-correction energies.

\paragraph{Fission channels:}
Fission-fragment yields are given by the sum of the yields associated to different fission channels.
The fission channels are related to the statistical population of quantum oscillators in the mass-asymmetry degree of freedom that form the fission valleys in the multidimensional potential-energy landscape. 
The three parameters of the oscillators (position, depth, and curvature) are traced back to the macroscopic potential (symmetric, 'super-long' fission channel SL) and to shells in the proton and neutron subsystems of both fragments ('standard' fission channels S1 and S2), which are assumed to be effective already at or little behind the outer saddle \cite{Mosel71}. 
The description of the S2 fission channel requires two additional parameters, because its shape is parametrized as a rectangular distribution convoluted with two Gaussian distributions at the inner and the outer side, respectively.

These shells are assumed to be essentially the same for all fissioning systems. 
Only the superposition of different shells and the interaction with the macroscopic potential cause the different mass distributions found for different systems \cite{Schmidt08}. 
These shells also determine the shapes (mainly the quadrupole deformation) of the nascent fragments at scission. 
According to Strutinsky-type calculations, the fragment shapes are found to be characterized by a linearly increasing quadrupole deformation as a function of the number of protons, respectively neutrons, in regions between closed spherical shells \cite{Wilkins76}. 
Also the charge polarization (deviation of the $N/Z$ degree of freedom at scission - mean value and fluctuations - from the $N/Z$ value of the fissioning nucleus) is treated by the corresponding quantum oscillator \cite{Nifenecker80}.

\paragraph{Energy sorting:}
The excitation energy of the fragments at scission is essential to determine the de-excitation of the fragments via prompt neutron and gamma emission after scission. To infer the excitation energy of the individual fragments at scission, it is necessary to model how the total available intrinsic excitation energy at scission is shared between the two fragments. In GEF, this is ruled by the so called energy-sorting process.
By the influence of pairing correlations, the nuclear temperature below the critical pairing energy is assumed to be constant \cite{Schmidt12}. 
Therefore, the di-nuclear system between saddle and scission consists of two coupled microscopic thermostates \cite{Schmidt10}. 
This leads to a sorting process of the available intrinsic energy before scission \cite{Schmidt11a,Schmidt11b}, where most of the excitation energy available at scission goes to the heavy fragment. The energy sorting has an important influence on the odd-even effect in the fragment $Z$ distribution \cite{Jurado15} and on the fragment-mass-dependent prompt-neutron multiplicity \cite{Schmidt10}. 

\subsection{Strengths and weaknesses} \label{S2.2}

The GEF model combines a well defined theoretical framework of basic concepts and laws of general validity with the ability to closely reproduce measured fission observables by adjusting the values of the model parameters in a rather direct and flexible way. Thus, it goes well beyond the purely empirical description of systematics without the necessity of a complete and accurate quantitative understanding of the physics in an ab-initio approach. The concept of the GEF model combines the strength of empirical systematics with the strength of a rather far-reaching understanding of the physics. This leads to a good reproduction of measurements and a good predictive power \cite{Schmidt16,Schmitt18,Schmidt18}. 
On the other hand, a good description of a specific feature is only possible, if the relevant experimental data for determing the corresponding GEF parameters are available.


\section{GEF improvements using reactor antineutrinos and application to antineutrino production} \label{3}

Antineutrino energy spectra of individual fission products obtained from nuclear databases have been used to refine the GEF code in order to improve its potential of predictiveness for reactor antineutrinos. In the present section, reactor antineutrino energy spectra have been computed using summation calculations with decay data taken from nuclear databases and fission yields taken respectively from JEFF-3.1.1, JEFF-3.3 and GEF. The direct comparison of the three calculations allowed us extensively improving the predictions of GEF for antineutrinos by acting on a few well identified parameters, depending on the fission channel concerned.

\subsection{Summation calculations for antineutrinos} \label{Summation}

The summation method is based on the use of nuclear data combined in a sum of all the individual contributions of the beta branches of the fission products, weighted by the amount of the latter nuclei. Two types of datasets are thus involved in the calculation: fission yields and fission-product decay data. This method was originally developed by~\cite{King} followed by~\cite{Avignonne} and then by~\cite{Vogel81,Tengblad}. The $\beta$/$\bar{\nu}$ spectrum per fission of a fissionable nuclide $S_k(E)$ can be broken-up into the sum of all fission products $\beta$/$\bar{\nu}$ spectra weighted by their activity $A_{\text{fp}}$
\begin{equation}
S_k(E)=\sum_{fp=1}^{N_{fp}}{A_{fp}\times S_{fp}(E)}
\label{Sk}
\end{equation} 

Eventually, the $\beta$/$\bar{\nu}$ spectrum of one fission product is the sum over $N_\text{B}$ beta branches (b) of all beta-decay spectra (or associated antineutrino spectra), $S_{fp}^{b}$ (in eq~\ref{Sfp}), of the parent nucleus to the daughter nucleus weighted by their respective branching ratios ($BR_{fp}^{b}$) as
\begin{equation}
S_{fp}(E)=\sum_{b=1}^{N_{\text{B}}}{BR_{fp}^{b}\times S_{fp}^{b}(Z_{fp},A_{fp},E_{0 fp}^{b},E)}
\label{Sfp}
\end{equation}

$E_{0 fp}^{b}$ being the endpoint of the $b^{th}$ branch of a given fission product.

In the summation spectra presented in this article, the beta-decay properties of the fission products have been selected following the prescription of~\cite{Estienne} and include the most recent Total Absorption Gamma-ray Spectroscopy (TAGS) data which are free from the Pandemonium effect~\cite{Hardy77}. The Pandemonium effect is the main bias of the antineutrino energy spectra computed with the summation method, its impact being larger than other nuclear effects such as forbidden non-unique shape factors or the weak magnetism correction. It arises from the use of germanium detectors to detect the beta branches of beta decays with large Q-value. In some cases, the lack of efficiency of these detectors to high energy or multiple gamma-rays induce the misdetection of beta branches towards high energy states in the daughter nucleus. This leads to the distortion of the beta and antineutrino spectra with an overestimate of the high energy part. The measurement of beta-decay properties with the TAGS technique~\cite{RubioGelletly} allows circumventing the problem, and experimental campaigns focused on nuclei contributing importantly to the reactor antineutrino spectra have been performed in Jyv\"askyl\"a since 2009~\cite{Fallot,Zak,PRLAlgora,Valencia,Rice,Guadilla2018}, leading to an impressive improvement of the agreement between the summation method predictions and the Daya Bay experimental results~\cite{Estienne}.

\subsection{Sensitivity of antineutrino spectra to the fission yields} \label{S3.2}

\begin{figure}[h]
\centering
\includegraphics[width=0.5\textwidth] 
{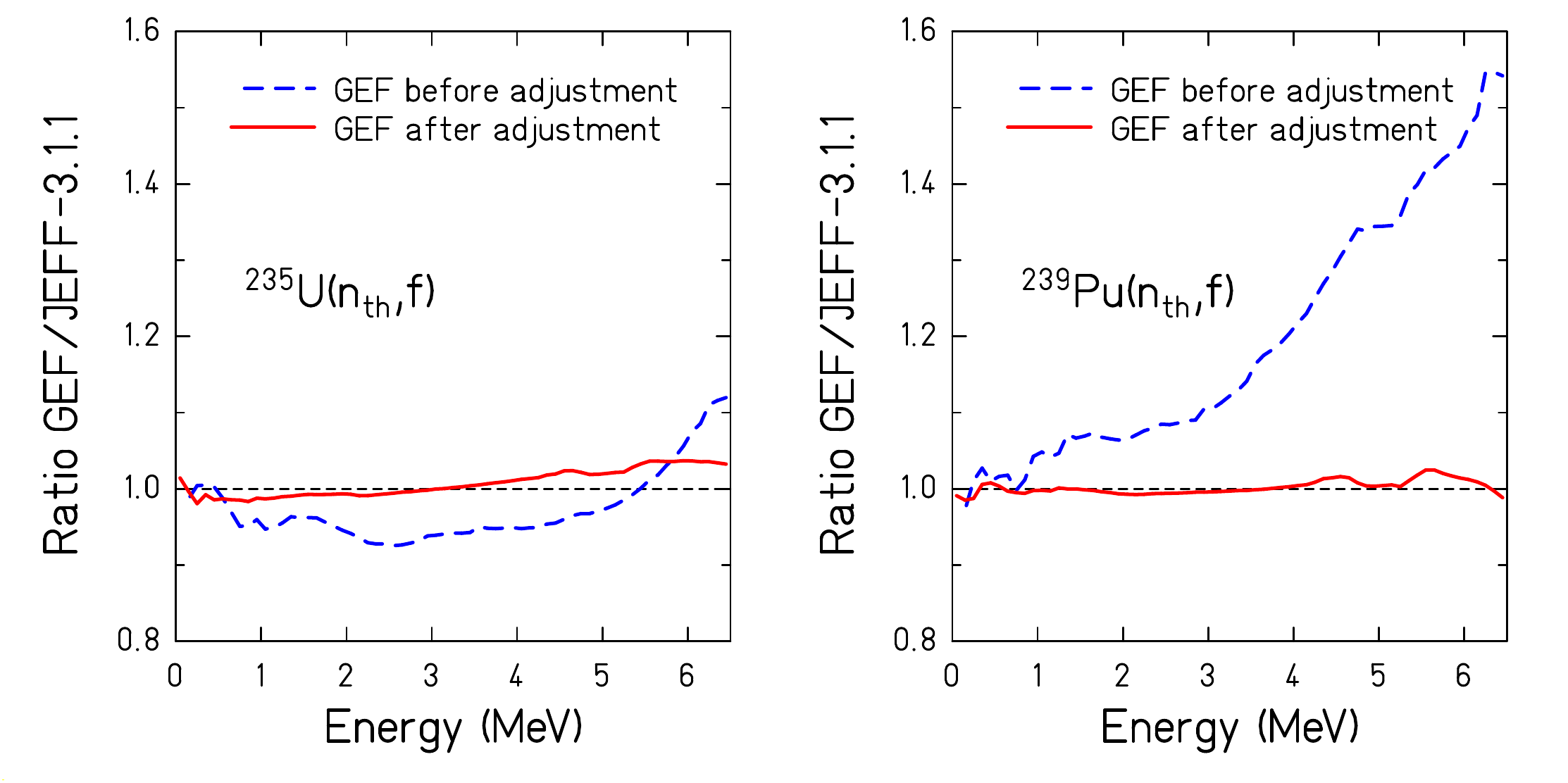}
\caption{Ratio of the antineutrino spectra GEF/JEFF-3.1.1 before the tuning (blue line) using antineutrino energy spectra and additional fission-yield data and after (red line).} 
\label{ratio_NOTtuned}       
\end{figure}

Figure~\ref{ratio_NOTtuned} (dashed line) shows the level of agreement that was reached between the antineutrino spectra of the GEF predictions and those obtained with the JEFF-3.1.1 fission yields with a previous version of GEF (GEF-Y2017/V1.2) that was in good agreement with the integral data of the decay heat after fission pulses of various fissioning systems \cite{Schmidt18}. The antineutrino energy spectra of $^{235}$U and $^{239}$Pu computed with the two sets of fission yields were in agreement only at the 10-30\% level even in a restricted energy range up to 6~MeV. Though this previous version of GEF already provided very good reproduction of decay heat after thermal fission pulses for $^{235}$U and $^{239}$Pu, it was not satisfactory to meet the additional exigence of the dependence in energy of the antineutrino spectra. The adjustment of the GEF model documented in~\cite{Schmidt16} had been performed with a general fit to all mass yields from ENDF/B-VII.0, because this evaluation provided the widest coverage of fissioning nuclei \cite{ENDF}. 

This way, data of very different quality, including faulty data, which spoiled the quality of the result, were included in the fit on the same footing, see section \ref{S6.2}. This may explain why the antineutrino spectra obtained with the fission yields of the previous version of GEF deviate so strongly from those obtained with the yields from the more recent JEFF-3.1.1 evaluation.

The extraction of the list of nuclei contributing importantly to the antineutrino energy spectra (lists commented in section~\ref{S7}) has allowed to evidence the causes of the  discrepancies between GEF and experimental fission yields for these nuclei. 
The antineutrino spectra are particularly sensitive to the yields of specific nuclides, especially at the higher antineutrino energies. In addition, the relatively large uncertainties of JEFF-3.1.1 and JEFF-3.3 fission yields suggested a good reproduction by the GEF model with rather large deviations. Deviations inside the error bars of the evaluations lead to substantial variations in the antineutrino spectrum. These remaining discrepancies had only little impact on other observables such as the decay heat after fission pulses but showed to impact a lot the antineutrino spectra. Additional experimental constraints on fission yields were needed, and this conclusion triggered the use of the LOHENGRIN data which eventually allowed to improve the predictiveness of the model because they are much more accurate, as shown in the sections~\ref{S5} and~\ref{S6} of this article.
The reactor antineutrino observable is thus a stringent additional constraint for the evaluation of nuclear data, and its combination with the GEF model allows to tackle the source of remaining inconcistencies in the data.
It is important to underline here that the GEF parameters have been tuned globally so that these results constrain also all the other predictions for different fissioning systems.

The solid lines in Fig.~\ref{ratio_NOTtuned} illustrate the improvement achieved after the tuning with GEF-Y2019/V1.2 using the reactor antineutrino observable as explained above. It is more
extensively quantified in section~\ref{S7} where useful results associated to reactor antineutrino of interest for the fission process and for reactor surveillance are also presented. The level of accuracy reached for the mass yields between GEF and the evaluated data and the experiments is presented and discussed in the section~\ref{S6}.
Comparisons of independent yields from JEFF-3.3 and GEF are shown in Appendix I.

\subsection{Parameter values} \label{Parameter_values}


The study performed on the fission yields led to the introduction of a few modifications of the GEF parameters 
to better represent the empirical fission yields of the JEFF-3.1.1 and the JEFF-3.3 evaluations as well as the LOHENGRIN experiments (comparison provided in section~\ref{S6}), which were not considered before using antineutrino summation calculations as explained above. In particular, the very accurate mass yields of the LOHENGRIN experiments required an individual adjustment of some GEF parameters, depending on the $Z$ value of the fissioning nucleus. Still, all isotopes of a given element are described with the same parameter set. The parameter values of GEF-Y2019/V1.2 are very close to the ones documented in Ref.~\cite{Schmitt18}. The modified parameter values (all related to the modeling of the fission channels) are documented in Table \ref{Parameters}.

The strength of the shell effect for symmetric fission has to be determined for individual fissioning systems, because the nuclides formed in symmetric fission depend on the composition of the fissioning nucleus. However, this was only possible in a limited number of cases, where the required experimental information is available. 
These values are listed in Table \ref{Delta_S0}. They vary by a few 100 keV. Such a small difference underlines the high sensitivity of the fragment yield at symmetry to this critical parameter.

\begin{table}[h]
\begin{center}
\caption{List of locally adjusted parameter values.}
\label{Parameters}
~\\
\begin{tabular}{|c|c|c|c|} \hline \hline
                    & Global values  & \multicolumn{2}{c|}{ Locally adjusted values }       \\ \hline \hline
  Parameter         &                &  $Z = 90$   & $Z = 93, 94, 95$ \\ \hline \hline
   \verb+P_DZ_Mean_S1+  &   0         &   (0)        &  0.25       \\ \hline
   \verb+P_DZ_Mean_S2+  &   0         &   0.6        &  -0.3       \\ \hline
   \verb+P_Shell_S2+    &   -4.4 MeV  &   -4.8 MeV   &  (-4.4 MeV) \\ \hline
   \verb+P_Z_Curv_S2+   &   0.098     &   0.25       &  0.08       \\ \hline
   \verb+S2leftmod+     &   0.75      &   (0.75)     &  0.65       \\ \hline
   \verb+P_A_Width_S2+  &   11.5      &   12.5       &  (11.5)     \\ \hline
\end{tabular}
\small{~\\~\\Note:
Local values that are identical with the global ones are given in parentheses.
Global parameter values are used for all elements with $Z$ different 
from 90, 93, 94, and 95. 
}
\end{center}
\end{table}

The parameter names are defined as:
 
\begin{itemize}  
\item          
\verb+P_DZ_Mean_S1+:     Shift of the S1 fission channel in $Z$ with respect to the global value.
\item
\verb+P_DZ_Mean_S2+:     Shift of the S2 fission channel in $Z$.
\item
\verb+P_Shell_S2+:       Strength of the shell behind the S2 fission channel.
\item
\verb+P_Z_Curv_S2+:     This parameter determines the smoothing of the inner side of 
                      the potential pocket of the S2 fission channel.
\item
\verb+S2leftmod+:          This parameter determines the smoothing of the outer side of
                      the potential pocket of the S2 fission channel.
\item
\verb+P_A_width+:        Flat part of the S2 potential pocket.
\end{itemize} 

For a detailed explanation of these parameters, see also \cite{Schmidt16,Schmitt18}.
 
\begin{table}[h]
\begin{center}
\caption{Adapted values of the strength of the shell effect for symmetric fission.}
\label{Delta_S0}
~\\
\begin{tabular}{l l l l l l l l } \hline \hline
$Z = 89$ &                 &      &     &     &     &     &    \\ 
       & $A =$             & 226  &     &     &     &     &      \\  
       & \verb+Delta_S0+/MeV = & -0.3 &     &     &     &     &      \\ \hline 
$Z = 90$ &                 &      &     &     &     &     &    \\  
       & $A =$             & 228  & 229 & 230 & 231 & 232 & 233  \\  
       & \verb+Delta_S0+/MeV = & 0.2  & 0.4 & 0.7 & 0.8 & 0.9 & 0.9  \\ \hline 
$Z = 92$ &                 &      &     &     &     &     &    \\  
       & $A =$             & 233  & 234 & all  &  &  &   \\  
       &                 &      &     & other &  &  &   \\  
       & \verb+Delta_S0+/MeV = & 0.4  & 0.4 & 0.2 &  &  &   \\ \hline 
$Z \ge 93$ &                 &      &     &     &     &     &    \\ 
       & $A =$             & all       &  &  &  &  &   \\  
       & \verb+Delta_S0+/MeV = & -0.3          &  &  &  &  &   \\ \hline 
       
\end{tabular}
\small{~\\~\\Note:
The values are adjusted to the relative yield of the symmetric channel
in measured mass distributions. Data from JEFF-3.3 (n$_{\text{th}}$,f) and 
refs. \cite{Specht74a,Specht74b} (transfer reactions) were used. For all other cases: \verb+Delta_S0+ = 0. See refs. \cite{Schmidt16,Schmitt18} for the exact meaning of \verb+Delta_S0+.  
}
\end{center}
\end{table}

In the next section we will detail the existing experimental approaches used to constitute the pool of available fission-yield data and provide their strengths and weaknesses.
Then we illustrate how the yields of the tuned version of GEF eventually compare to experiments.


\section{Experimental approaches} \label{S4}
The fission process ends up in two fission products, which populate about a thousand different nuclides. (We do not consider ternary fission here, where a third light particle is formed, in addition, with low probability.) Several experimental approaches have been developed for measuring the yields of the different fission products formed in the fission of a specific nucleus at a certain excitation energy and angular momentum. We will consider some of those, which are most often used. See Ref.~\cite{Andreyev18} for an extensive overview on presently used experimental approaches in fission.

\subsection{Radiochemistry} \label{S4.1}

\subsubsection{The method} \label{S4.1.1}
The traditional method for measuring fission-product yields consists of exposing samples to a flux of neutrons. After irradiation, the samples are investigated by gamma spectroscopy \cite{Laurec10}. The fission products are identified unambiguously in $Z$ and $A$ by measuring the gammas emitted directly or in their radioactive decay chain, and their yields are deduced from the intensities of the gamma lines. Chemical separation is often applied in order to purify the gamma spectra by reducing the background radiation. 

\subsubsection{Independent and cumulative yields} \label{S4.1.2}
The primary fission fragments, as they are formed at scission, normally carry some excitation energy that gives rise to a cascade of prompt neutrons and prompt gammas, until the ground state or a longer-lived isomeric state is reached. (Processes are called to be prompt, if they occur inside a certain time window that is much shorter than typical beta-decay half-lifes, which are in the milli-second range or longer.) 
The yields of the fission products formed right after the prompt processes are called independent yields.


Only the gamma radiation emitted in a time range starting a few seconds (or longer) after fission, which is needed for the extraction of the target and, possibly, the chemical separation, can be measured by radiochemistry. Therefore, the yields of the most neutron-rich fission products, which are especially short-lived, cannot be determined directly.

The yields including also the products of the consecutive radioactive decay are called cumulative yields. Because beta-delayed neutron emission, which changes the nuclear mass number, is a rare process, the last cumulative yields near the beta stability are a rather good measure of the mass yields. 
However, the application of the summation method for estimating the antineutrino production is based on the independent yields, which requires the additional knowledge of the fission-product atomic number before beta decay.


\subsubsection{Yields of short-lived products} \label{S4.1.3}
Methods have been developed to determine even the independent yields of short-lived radioactive fission products, fully identified in $Z$ and $A$, by requiring consistency between the neutron-deficient wing of the nuclide distribution in the light fission product and the neutron-rich wing of the nuclide distribution in the heavy product (and vice versa) with the mass-dependent multiplicity of prompt neutrons \cite{Wahl80}. The application of this method on the basis of incomplete or even fragmentary experimental data requires a good knowledge of the behavior of fission-fragment nuclide distributions and prompt-neutron multiplicities. One of the most popular systematics used for this purpose was developed by Wahl \cite{Wahl02}.  

\subsubsection{Strengths and weaknesses} \label{S4.1.4}
The main strength of the radiochemical method is the unambiguous identification of the fission products in $Z$ and $A$. Also the sensitivity down to very low yields (10$^{-8}$\% or lower \cite{Mills95}) is a strength of this method.

However, there are several weaknesses of this method: Due to the time delay between irradiation and measurement, this method is slower than the lifetimes of many fission products, in particular of the most neutron-rich ones. Therefore, the independent yields of short-lived fission products cannot directly be measured, and their indirect determination (see above) depends on certain assumptions. 

Another weakness is the uncertainty introduced by the spectroscopic information that is used to infer the number of fission products from the intensities of the gamma lines
. Misidentification of a gamma line can also lead to erroneous results. Moreover, target impurities may be an issue, regardless of the measurement method.
 
The application of this method is limited to suitable targets and available neutron sources with suitable energies. Most of the available data were obtained with thermal neutrons, "fast" neutrons with energies around 1 MeV that are produced in the evaporation process, possibly partly moderated, and with 14-MeV neutrons. 

\subsection{Experiments with particle detectors in direct kinematics} \label{S4.2}

\subsubsection{The method} \label{S4.2.1}
Instead of exploiting the radioactivity of the fission products with the radiochemical method, their high kinetic energies have been used to detect and identify the fission products by their ionization signals in different kind of detectors. This way, the energy loss in thin detectors, the total energy in thick detectors, and/or the time-of-flight between two detectors were measured, eventually those of both fission products simultaneously. However, in most cases the additional measurement of the deflection in the electric and/or magnetic field in powerful spectrometers was used to determine the yields of individual nuclides with sufficient resolution in $Z$ and $A$. We consider here the LOHENGRIN spectrometer \cite{Moll75}, where the full mass distribution and the $Z$ distribution in the lighter fission product were measured for the thermal-neutron-induced fission of a number of systems.   

\subsubsection{Strengths and weaknesses} \label{S4.2.2}
A great advantage of kinematic measurements at the LOHENGRIN spectrometer is the rather direct determination of fission-product yields by ion counting with 100\% detection efficiency for ions that reach the detector. Nevertheless, a few corrections must be applied in order to account for the burn-up of the target material and the deterioration of the target quality by diffusion of the target material into the backing \cite{Koester10}. Furthermore, the fission products appear with a distribution of ionic charge states. These distributions have to be measured separately, and the associated yields have to be added up. A peculiar difficulty consists in the shift of the ionic charge-state distribution due to internal conversion and a consecutive Auger cascade for specific nuclides \cite{Wohlfarth78}. These cannot be calculated with sufficient accuracy and must be determined experimentally by a scan over the charge-state distribution of all fission products.  
Therefore, a good quality of the data requires a very careful analysis and correction of these disturbing effects.

In addition to the limitation to thermal-neutron-induced fission of a few suitable target nuclei, the kinetic-energy distribution of the fission products cannot be covered completely by practical reasons. The full distribution must be estimated from the measurements at a few kinetic-energy values. This may introduce some systematic uncertainties.

Mass yields can be measured over the whole fission-product range. However, we mention that kinematic measurements of independent fission-product yields, fully determined in $A$ and $Z$, can only be performed for the lighter products. 
A combination of mass separation by the LOHENGRIN spectrometer and full nuclide identification by gamma spectroscopy, which has recently been introduced \cite{Martin14} but not yet used in the evaluations, is applicable also for the heavy fission products, but it depends again on the uncertainties of the gamma detection and the necessary gamma-spectroscopic information.

\subsection{Experiments with particle detectors in inverse kinematics} \label{S4.3}

\subsubsection{The method} \label{S4.3.1}
During the last years, an innovative experimental approach based on the use of inverse kinematics has been introduced \cite{Schmidt00,Caamano13,Boutoux13,Chatillon19}: The fissioning nucleus is prepared with high kinetic energies, and, thus, the fission products are emitted with velocities that are appreciably higher than those, which they get in the fission process in direct kinematics. 

\subsubsection{Strengths and weaknesses} \label{S4.3.2}
Excellent resolution in $A$ and $Z$ has been obtained, but, partly due to the insufficiently well defined initial excitation energy of the fissioning system, the results have not yet been exploited so much for extracting nuclear data for technical applications. Therefore, we mention this method only for completeness and for its growing importance in the future, but we will not consider this method here further.
We would like to mention that the GEF model could help to better interpret the measured data by providing an estimation of the energy dependence of the fission yields. 
An application of this method is used in section \ref{U238F} for the modeling of the mass yields of fast-neutron-induced fission of $^{238}$U.   

\section{Evaluations} \label{S5}
Evaluation assesses the measured data and their uncertainties, reconciles
discrepant experimental data and fills in missing data by exploiting systematic trends of the measured data in order to provide reliable nuclear data, primarily for applications in nuclear technology. Evaluation work is organized, and the resulting nuclear-data tables are disseminated by several nuclear-data centers under the auspices of the International Atomic Energy Agency. 

In the following, we will consider the evaluations ENDF/B-VII, JEFF-3.1.1 and JEFF-3.3. (Note that all fission-fragment yields used in this work have identical values in ENDF/B-VI.8, ENDF/B-VII.0, ENDF/B-VII.1, and ENDF/B-VIII.0 \cite{ENDF}.)   
The main sources of these evaluations are data from radiochemical measurements, supplemented by only a few data from LOHENGRIN experiments, in spite of their special advantages in accuracy. Theoretical fission models have been exploited only very little up to now.

\section{Comparative study} \label{S6}
Although the characteristics of the antineutrino emission in fission is specific to the fission-product nuclide, determined in $Z$ and $A$, it is meaningful, as a first step, to assure a good description of the mass yields, because these are usually measured with the highest accuracy. In the present section, we compare the fission-product mass distributions for thermal-neutron-induced fission of all systems, which are included in the ENDF/B-VII, the JEFF-3.1.1 or the JEFF-3.3 evaluation or for which experimental data from LOHENGRIN experiments are available, with the result of GEF-Y2019/V1.2. 
In addition, fast-neutron-induced fission of $^{238}$U is included.  
For a quick overview of the essential results, important conclusions and recommendations are given in italic. 
Throughout the paper we will use the following notations: n$_{\text{th}}$ means thermal
neutrons, n$_{\text{fast}}$ means fast neutrons, and n$_{\text{hi}}$ means neutrons of 14 MeV.

A systematic comparison of the independent yields of fully identified nuclides of the four systems $^{235}$U(n$_{\text{th}}$,f), $^{238}$U(n$_{\text{fast}}$,f), $^{239}$Pu(n$_{\text{th}}$,f), and $^{241}$Pu(n$_{\text{th}}$,f) from GEF and from JEFF-3.3 is presented in Appendix I.


\subsection{Overall impression} \label{S6.1}
In the present subsection \ref{S6.1}, the mass yields from the GEF code are compared with evaluated data or results from LOHENGRIN experiments, where at least satisfactory agreement has been obtained. At the same time, these are the systems that have been experimentally investigated the most intensively, and the data are expected to be the most reliable. Cases with larger deviations are discussed in section \ref{S6.2}. We concentrate here mostly on thermal-neutron-induced and fast-neutron-induced fission. A more general overview was given in Ref.~\cite{Schmidt16}, however with an older version of the GEF code.

\subsubsection{Illustrative cases}

\begin{figure}[h]
\centering
\includegraphics[width=0.36\textwidth]{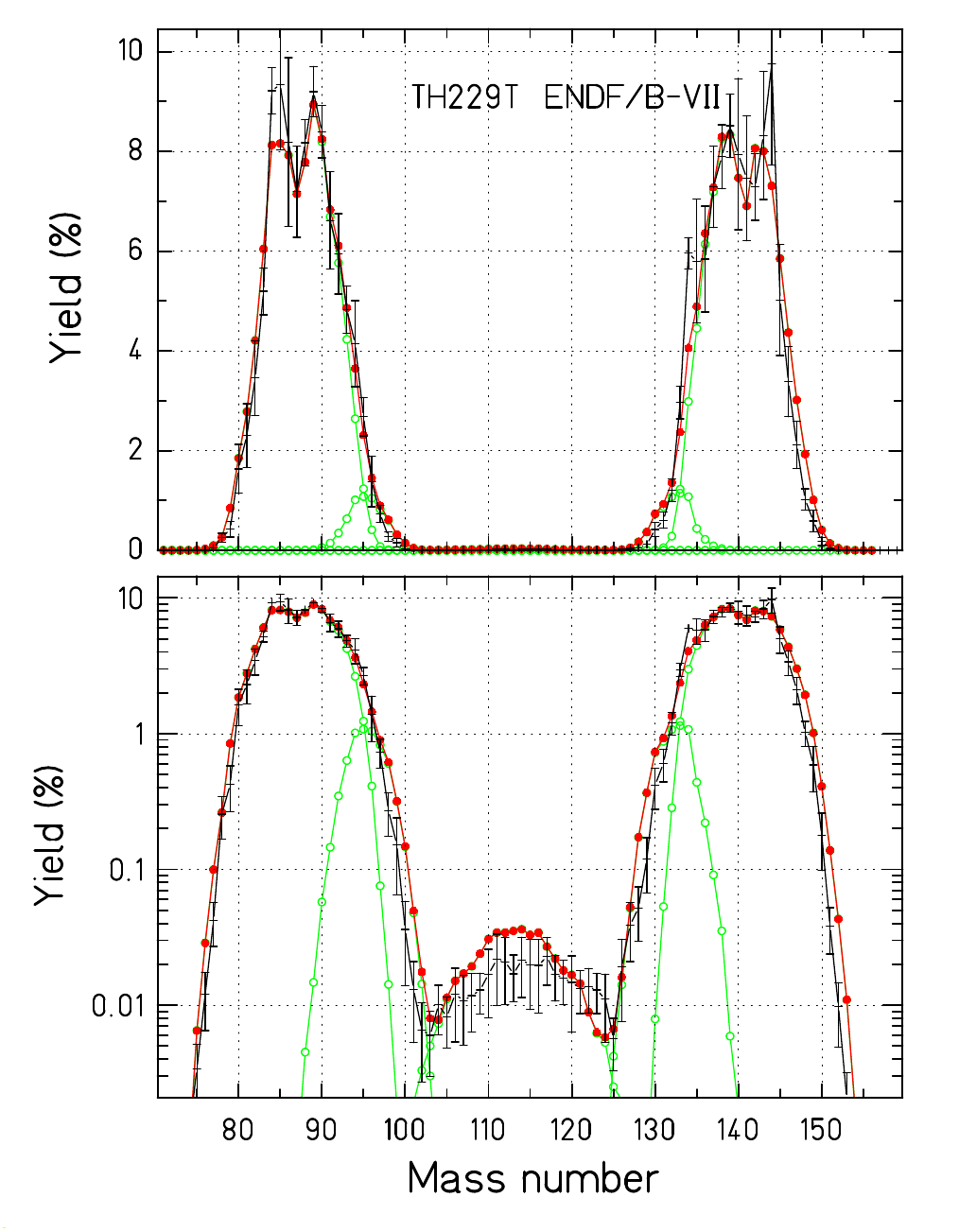}
\caption{Mass yields of $^{229}$Th(n$_{\text{th}}$,f), linear (upper frame) and logarithmic (lower frame) scale. GEF result (red points) in comparison with ENDF/B-VII (black symbols with error bars). Here and in the following figures, the green lines show the contributions of the different fission channels from GEF. } 
\label{TH229T-ENDF}       
\end{figure}

\begin{figure}[h]
\centering
\includegraphics[width=0.36\textwidth]{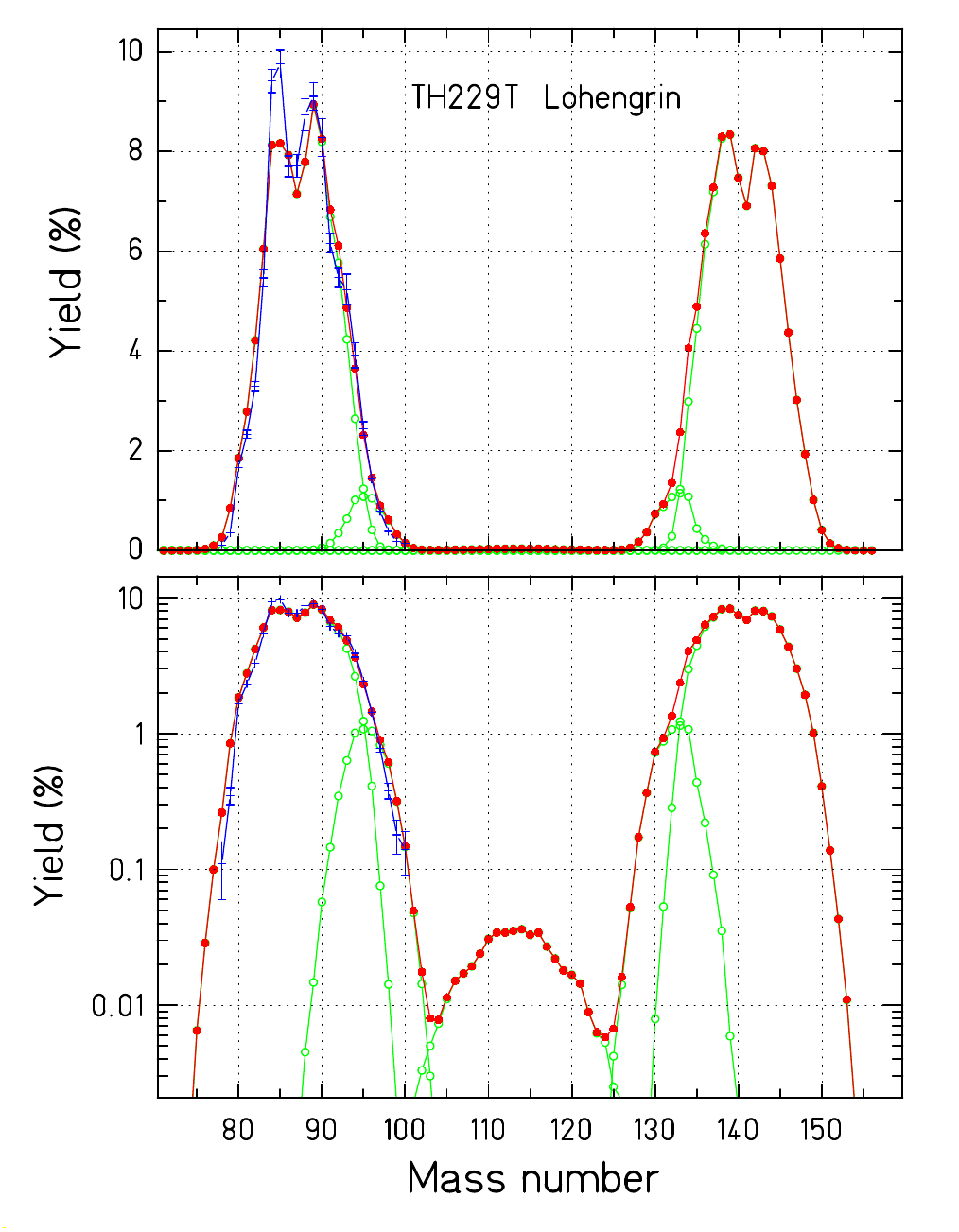}
\caption{Mass yields of $^{229}$Th(n$_{\text{th}}$,f), linear (upper frame) and logarithmic (lower frame) scale, GEF result (red points) in comparison with the data of a LOHENGRIN experiment (blue symbols). } 
\label{TH229T-LOHENGRIN}       
\end{figure}

\paragraph{Mass yields of $^{229}$Th(n$_{\text{th}}$,f):}
Figs. \ref{TH229T-ENDF} and \ref{TH229T-LOHENGRIN} show comparisons of the mass yields from GEF with the data from the ENDF/B-VII evaluation and from a LOHENGRIN experiment \cite{Bocquet90} for the system $^{229}$Th(n$_{\text{th}}$,f). There is fair agreement, except some underestimated intensities of the peaks near $A =$ 85 and $A =$ 144, which becomes significant in comparison with the LOHENGRIN data due to their appreciably higher accuracy. These deviations hint to a \emph{problem in the description of the S2 fission channel in GEF}. This problem might be cured by the introduction of a more complex shape of the S2 contribution to the mass yields. However, this is beyond the scope of the present status of GEF, because a higher degree of complexity and the corresponding introduction of additional model parameters might endanger the predictive power of the model. The symmetric yield is slightly overestimated.

\begin{figure}[h]
\centering
\includegraphics[width=0.36\textwidth]{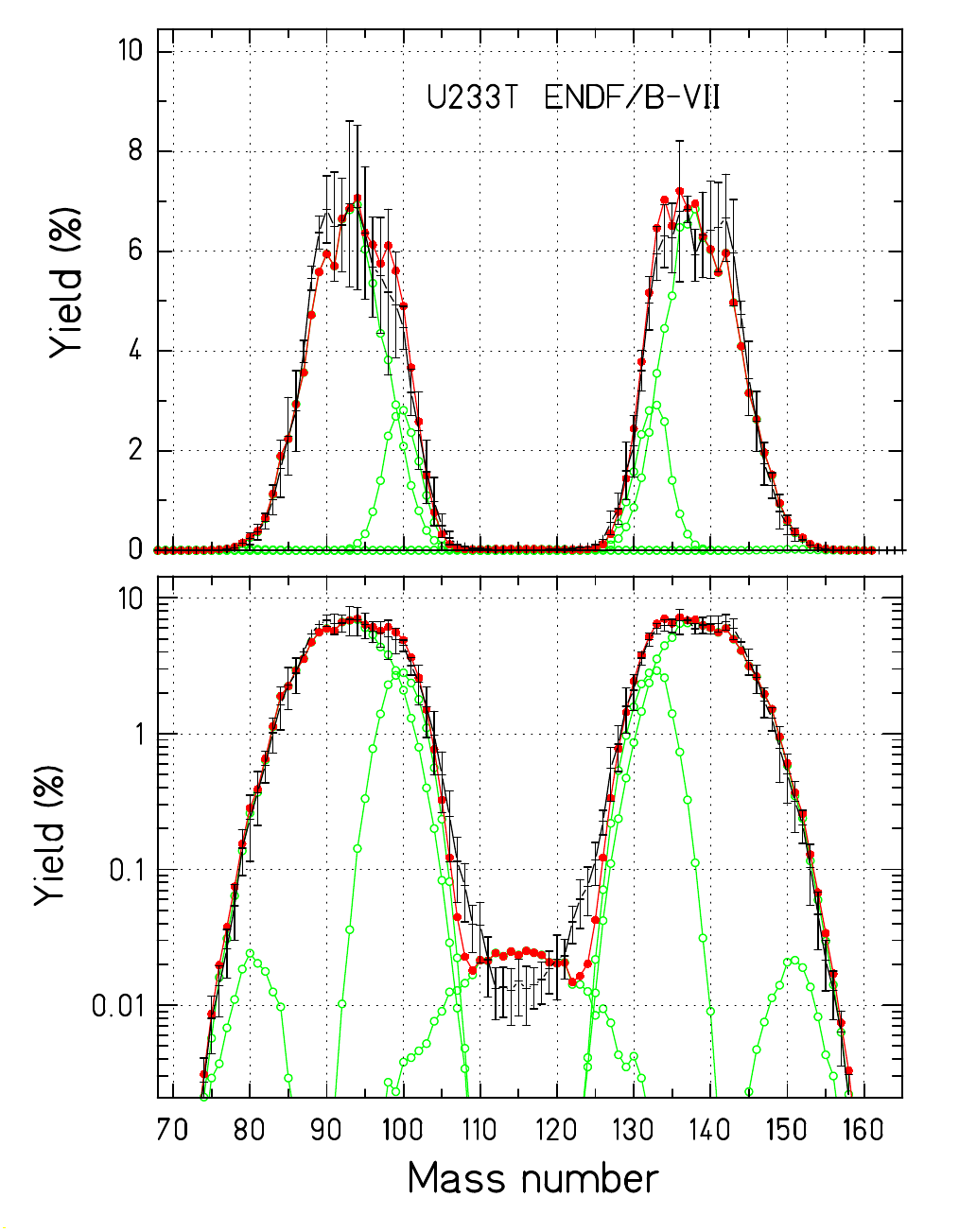}
\caption{Mass yields of $^{233}$U(n$_{\text{th}}$,f), linear (upper frame) and logarithmic (lower frame) scale. GEF result (red points) in comparison with ENDF/B-VII (black symbols).} 
\label{U233T-ENDF}       
\end{figure}
\begin{figure}[h]
\centering
\includegraphics[width=0.36\textwidth]{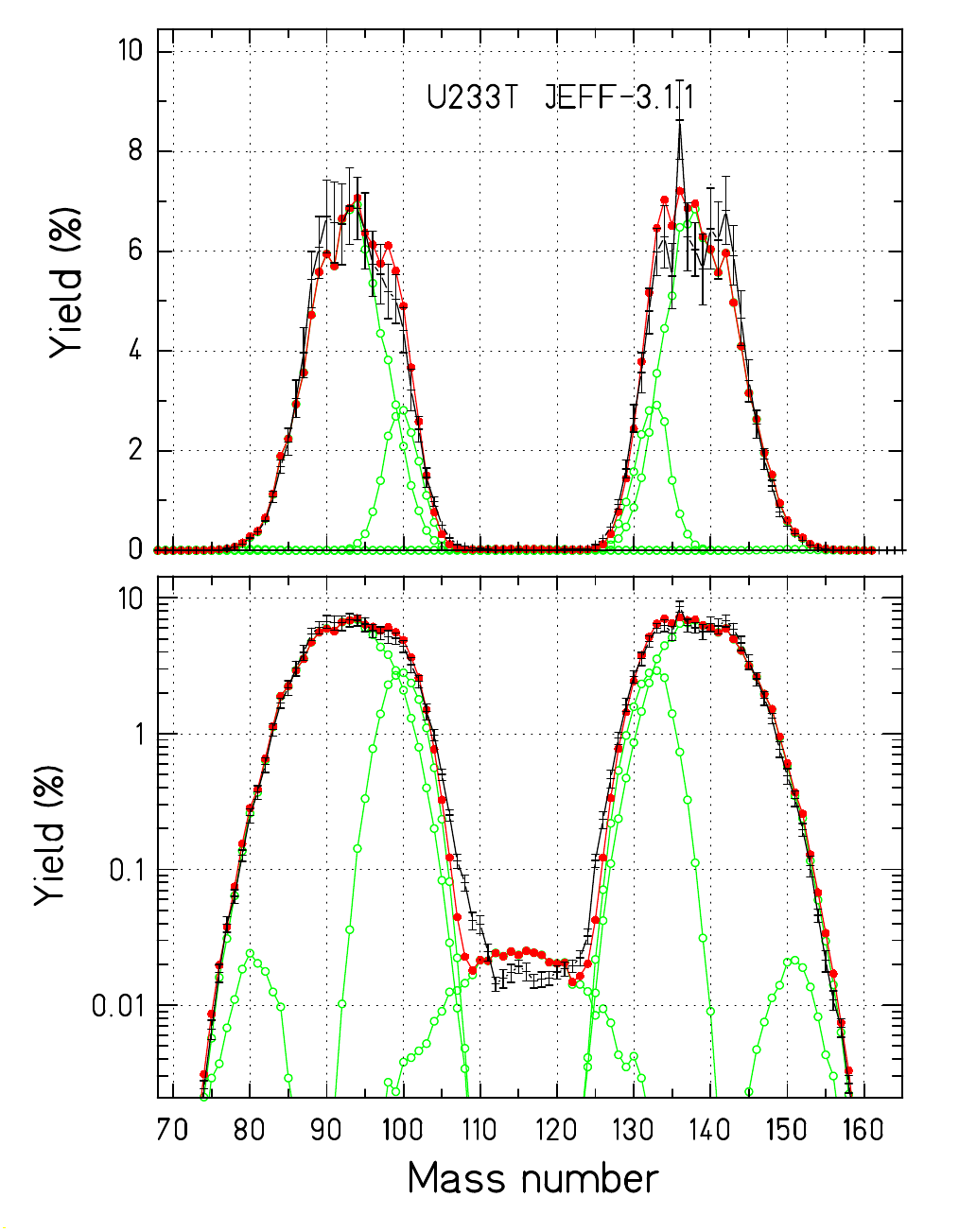}
\caption{Mass yields of $^{233}$U(n$_{\text{th}}$,f), linear (upper frame) and logarithmic (lower frame) scale. GEF result (red points) in comparison with JEFF-3.1.1 (black symbols).} 
\label{U233T-JEFF311}       
\end{figure}
\begin{figure}[h]
\centering
\includegraphics[width=0.36\textwidth]{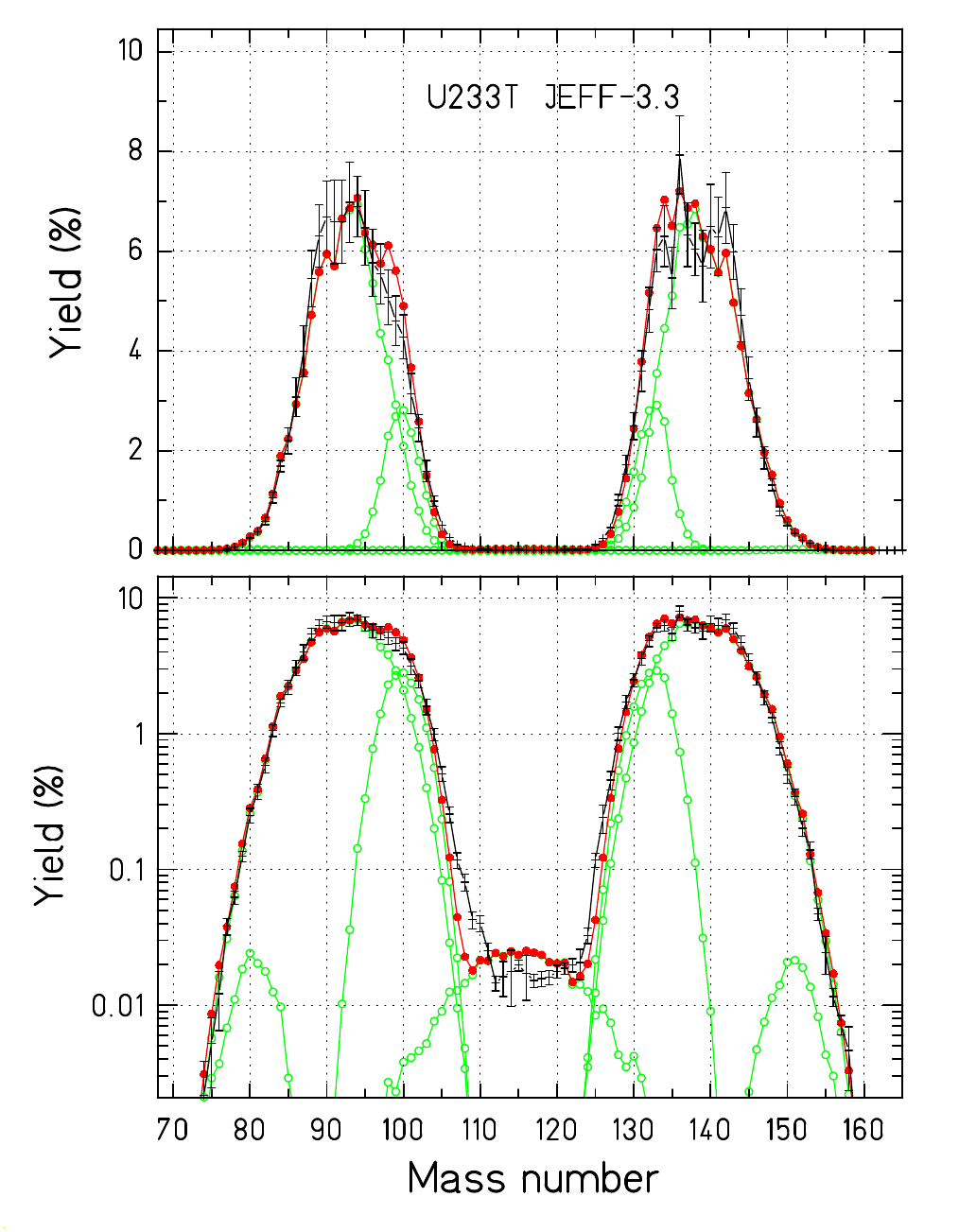}
\caption{Mass yields of $^{233}$U(n$_{\text{th}}$,f), linear (upper frame) and logarithmic (lower frame) scale. GEF result (red points) in comparison with JEFF-3.3 (black symbols).} 
\label{U233T-JEFF33}       
\end{figure}
\begin{figure}[h]
\centering
\includegraphics[width=0.36\textwidth]{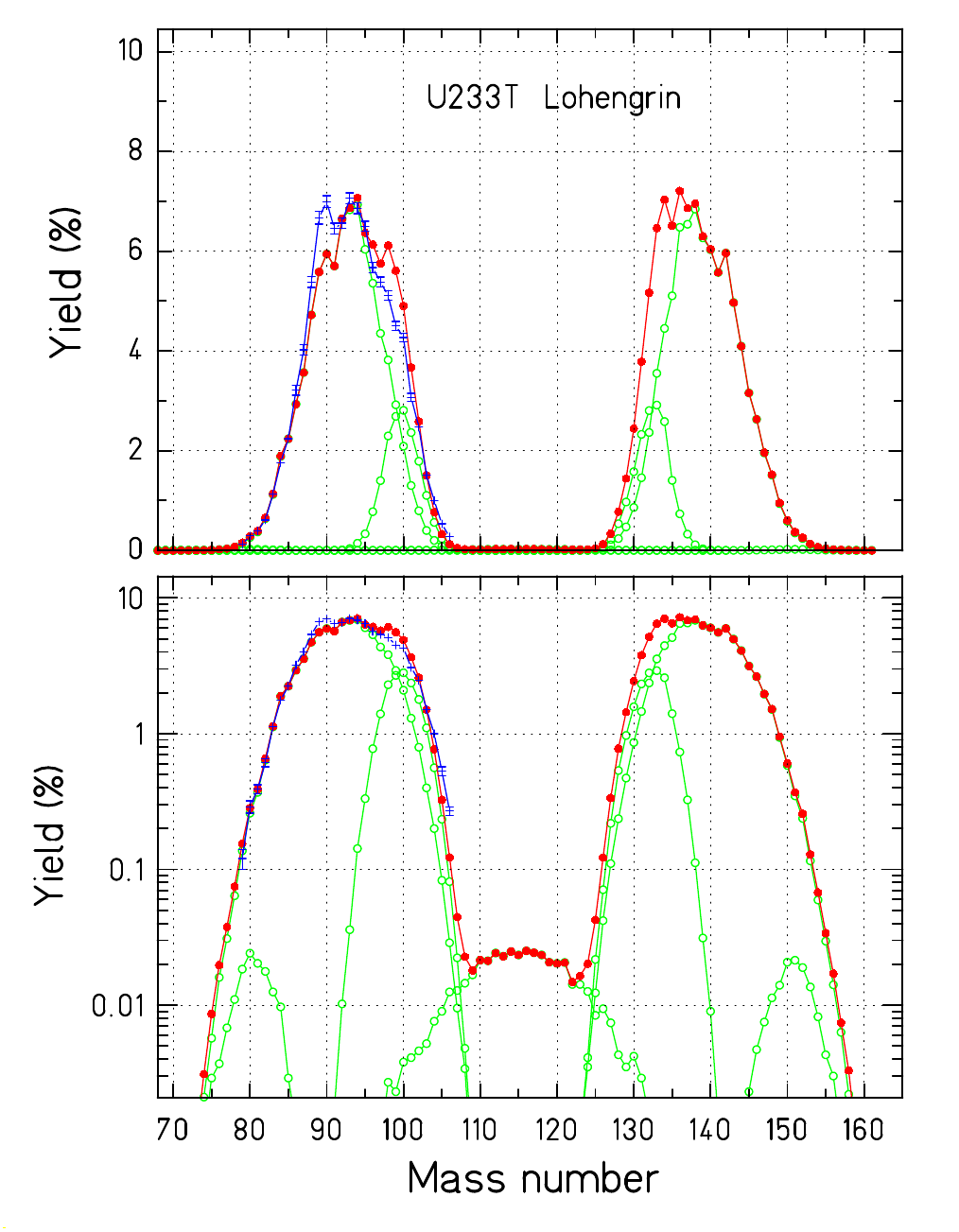}
\caption{Mass yields of $^{233}$U(n$_{\text{th}}$,f), linear (upper frame) and logarithmic (lower frame) scale. GEF result (red points) in comparison with LOHENGRIN data (blue symbols).} 
\label{U233T-LOHENGRIN}       
\end{figure}

\paragraph{Mass yields of $^{233}$U(n$_{\text{th}}$,f):}
Figs. \ref{U233T-ENDF}, \ref{U233T-JEFF311}, \ref{U233T-JEFF33}, and \ref{U233T-LOHENGRIN} show comparisons of the mass yields from GEF with the data from the ENDF/B-VII, JEFF-3.1.1 and JEFF-3.3 evaluations, as well as from a LOHENGRIN experiment \cite{Quade88} for the system $^{233}$U(n$_{\text{th}}$,f). There is fair agreement. However, the yields near A = 90 and A = 136 are somewhat underestimated, while the yields near A = 98 are somewhat overestimated. Again, this is most significant in comparison with the LOHENGRIN data. \emph{These discrepancies} appear with respect to the data from all sources. Thus, they \emph{must probably be attributed to deficiencies of GEF}, probably due to restrictions in the shape of the mass distribution of the asymmetric fission channel S2. This is in line with the observations for the system $^{229}$Th(n$_{\text{th}}$,f). 

\newpage

In addition, the yields in the inner wings of the asymmetric peaks are somewhat underestimated. 
In view of the good agreement between GEF and the evaluations in this mass region for the especially carefully studied system $^{235}$U(n$_{\text{th}}$,f), a common
deficiency of all 
these evaluations for $^{233}$U(n$_{\text{th}}$,f) due to some 
erroneous experimental data may be assumed.
A bump near A=124 in ENDF/B-VII does not appear in the JEFF evaluations anymore. In spite of differences between the evaluations, the yield at symmetry seems to be slightly overestimated. In addition, its shape is concave, while evaluations suggest a more flat, or even convex, pattern.

\begin{figure}[h]
\centering
\includegraphics[width=0.36\textwidth]{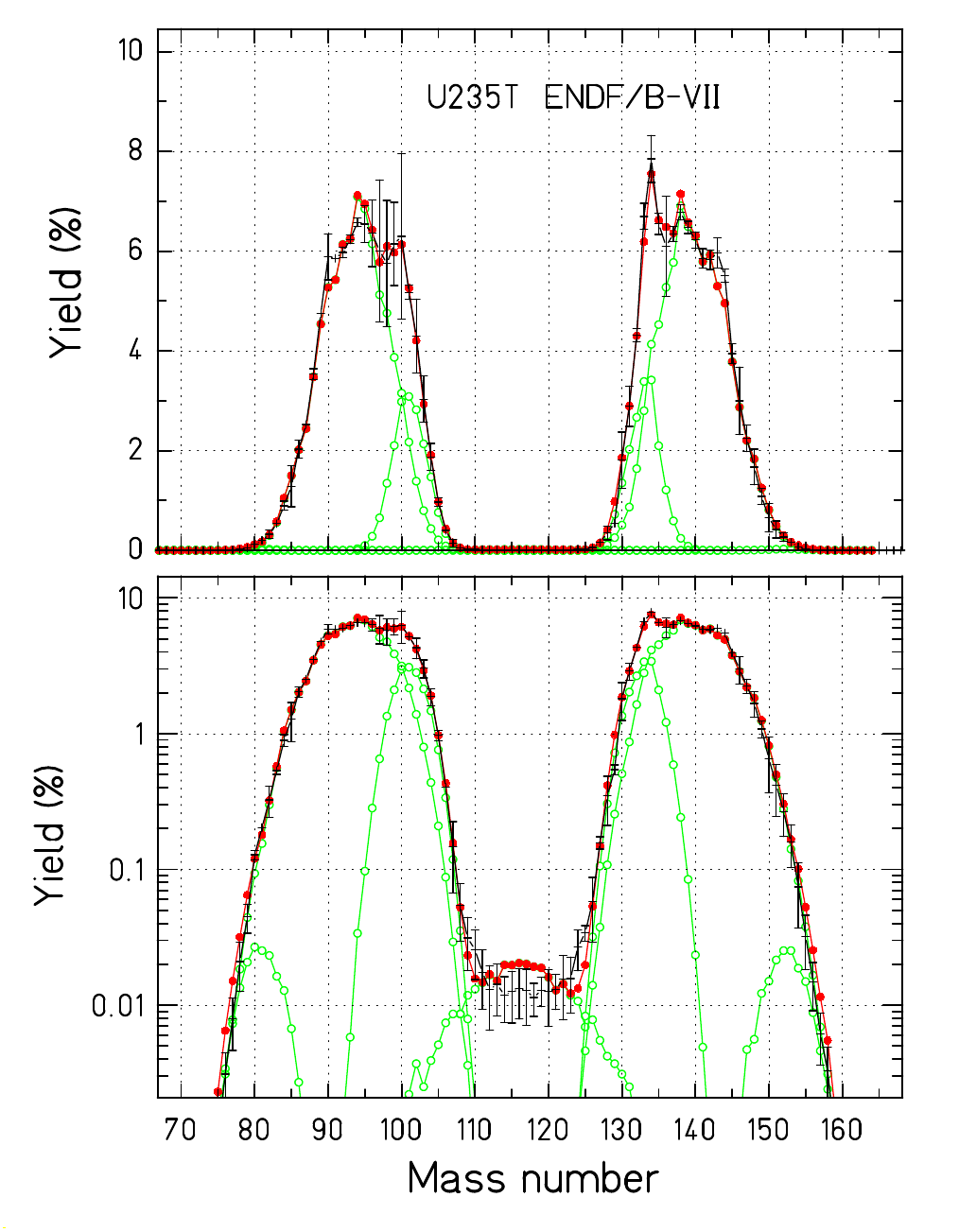}
\caption{Mass yields of $^{235}$U(n$_{\text{th}}$,f), linear (upper frame) and logarithmic (lower frame) scale. GEF result (red points) in comparison with ENDF/B-VII (black symbols).} 
\label{U235T-ENDF}       
\end{figure}



\begin{figure}[h]
\centering
\includegraphics[width=0.36\textwidth]{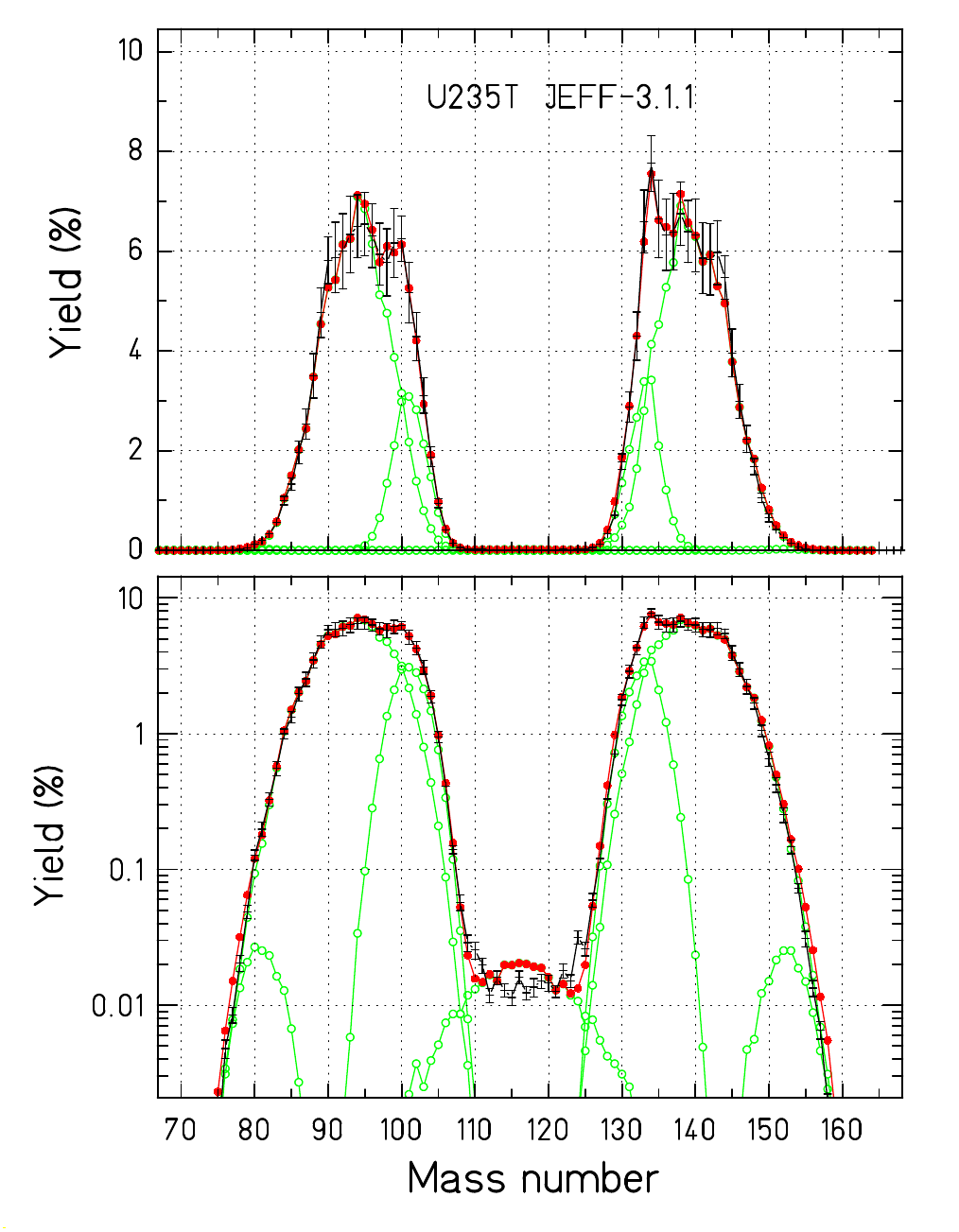}
\caption{Mass yields of $^{235}$U(n$_{\text{th}}$,f), linear (upper frame) and logarithmic (lower frame) scale. GEF result (red points) in comparison with JEFF-3.1.1 (black symbols).} 
\label{U235T-JEFF311}       
\end{figure}



\begin{figure}[h]
\centering
\includegraphics[width=0.36\textwidth]{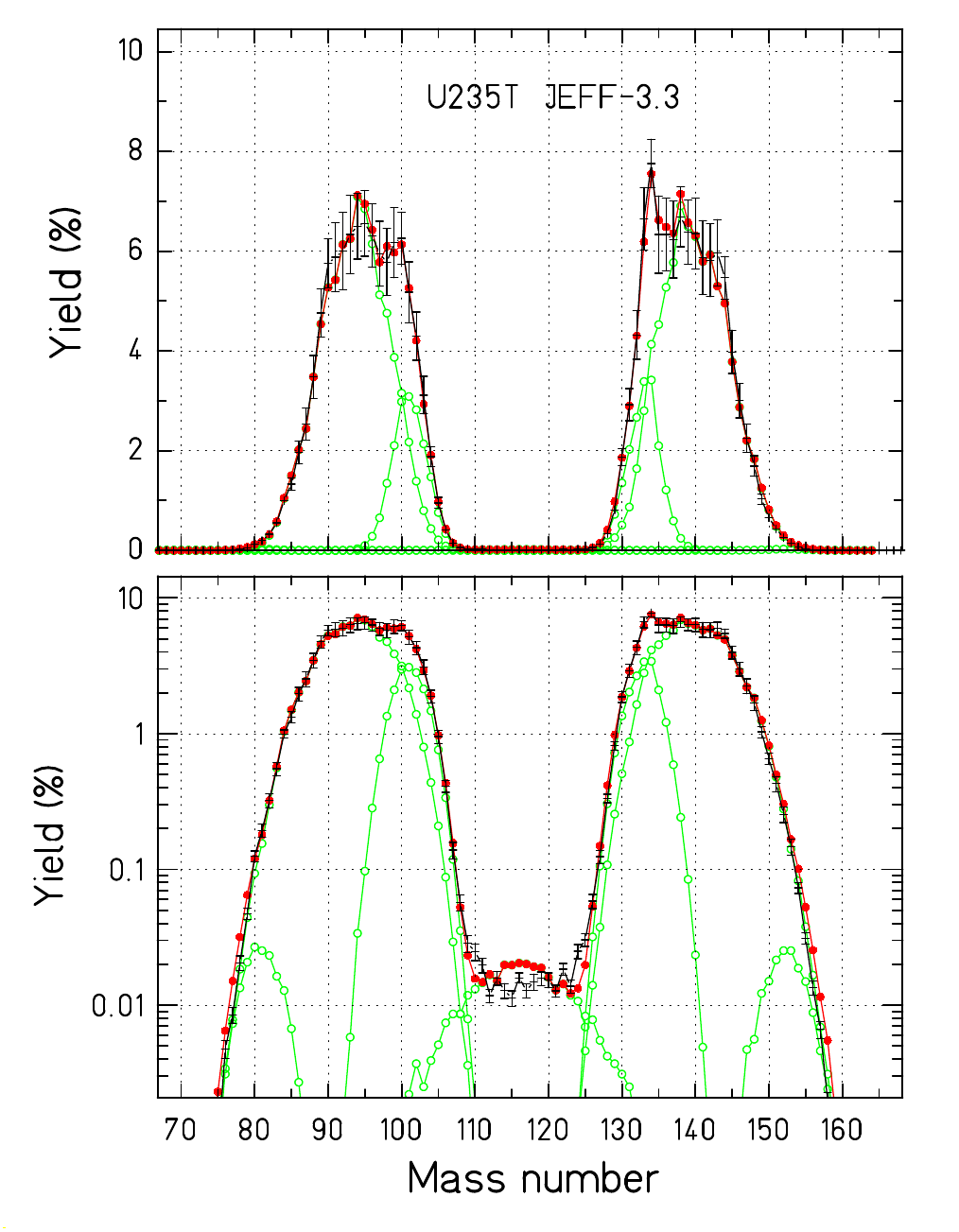}
\caption{Mass yields of $^{235}$U(n$_{\text{th}}$,f), linear (upper frame) and logarithmic (lower frame) scale. GEF result (red points) in comparison with JEFF-3.3 (black symbols).} 
\label{U235T-JEFF33}       
\end{figure}


\begin{figure}[h]
\centering
\includegraphics[width=0.36\textwidth]{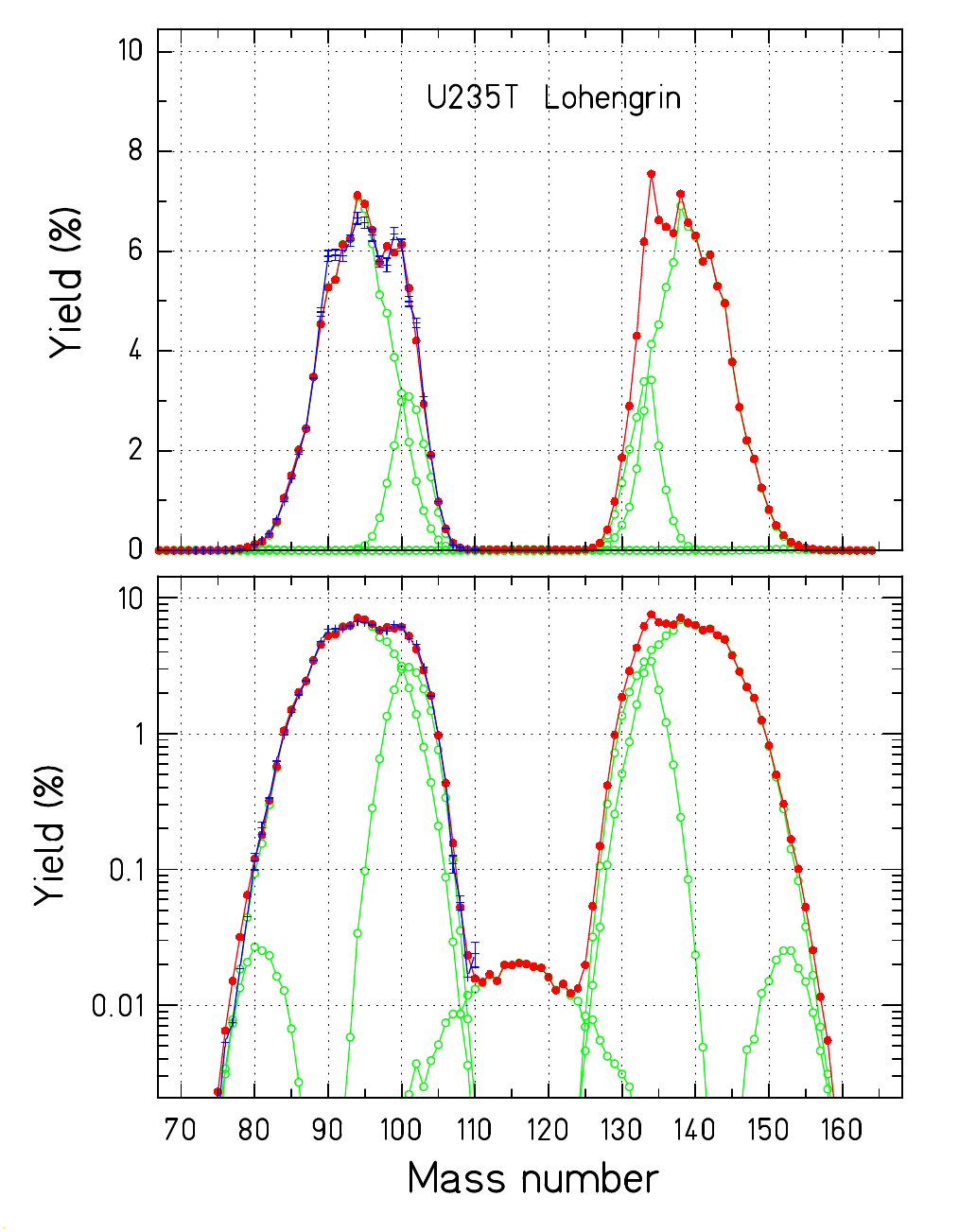}
\caption{Mass yields of $^{235}$U(n$_{\text{th}}$,f), linear (upper frame) and logarithmic (lower frame) scale. GEF result (red points) in comparison with LOHENGRIN data (blue symbols).} 
\label{U235T-LOHENGRIN}       
\end{figure}



\paragraph{Mass yields of $^{235}$U(n$_{\text{th}}$,f):}
Figs. \ref{U235T-ENDF}, \ref{U235T-JEFF311}, \ref{U235T-JEFF33}, and \ref{U235T-LOHENGRIN} show comparisons of the mass yields from GEF with the data from the ENDF/B-VII, JEFF-3.1.1 and JEFF-3.3 evaluations, as well as from LOHENGRIN experiments \cite{Lang80,Sida89} for the system $^{235}$U(n$_{\text{th}}$,f), which is the most intensively studied and best known system of all. 

The data of all evaluations are rather well reproduced. Deviations rarely exceed the uncertainties of the evaluations. The clearest picture is provided by the comparison with the LOHENGRIN data, which have by far the smallest uncertainties. 
Here some deviations appear in slightly underestimated yields around $A =$ 90 and slightly overestimated yields around $A =$ 94, which again hints to some \emph{shortcoming in the shape of the S2 fission channel in GEF}. The evaluations show similar deviations, but only the error bars of the ENDF evaluation are small enough in this mass region to make these significant. Moreover, some yields of the extremely asymmetric splits are overestimated, where the super-asymmetric fission channel dominates. We note that the yield at symmetry is very slightly overestimated.

\paragraph{Mass yields of $^{238}$Np(n$_{\text{th}}$,f):}

\begin{figure}[h]
\centering
\includegraphics[width=0.36\textwidth]{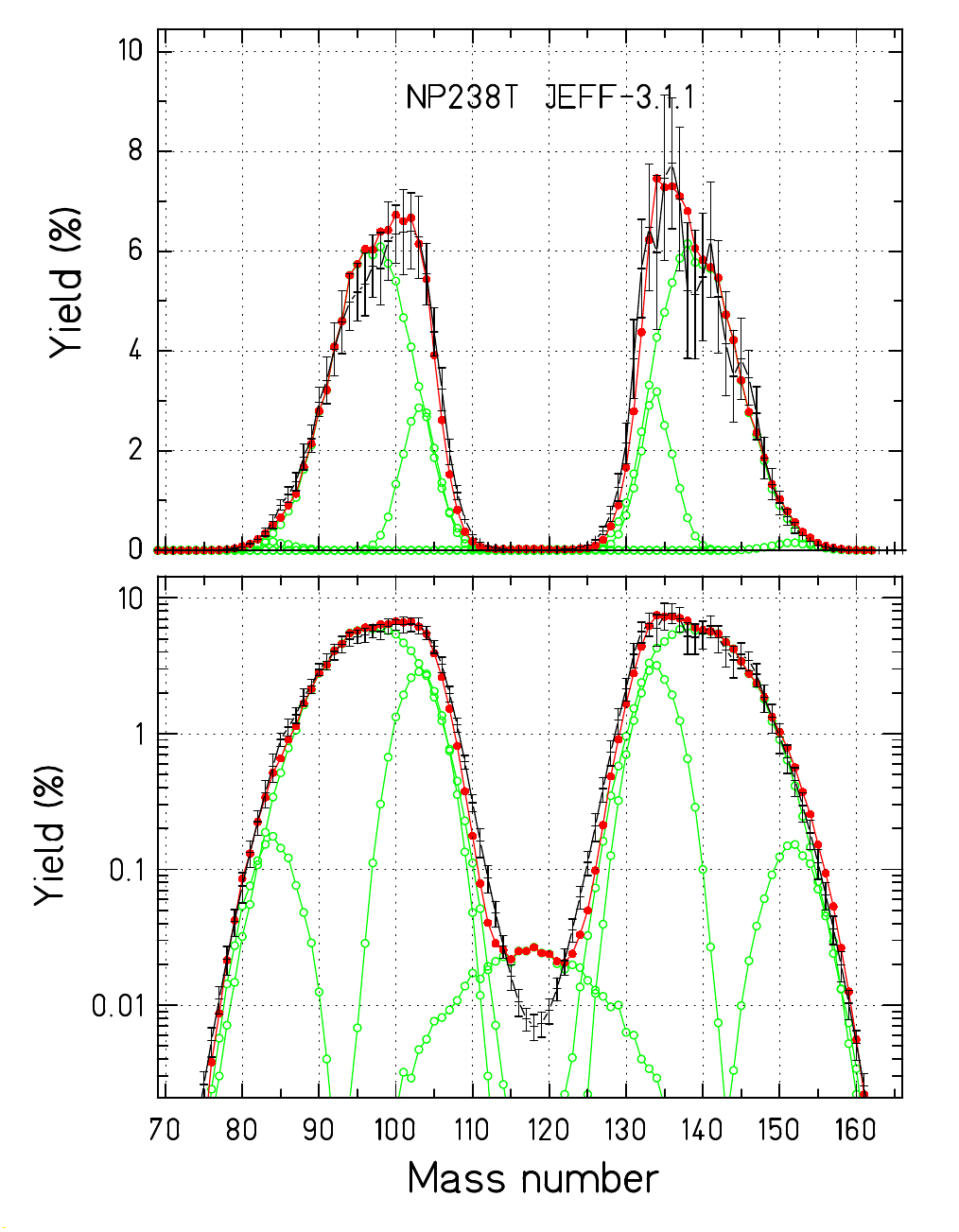}
\caption{Mass yields of $^{238}$Np(n$_{\text{th}}$,f), linear (upper frame) and logarithmic (lower frame) scale. GEF result (red points) in comparison with JEFF-3.1.1 (black symbols).} 
\label{NP238T-JEFF311}       
\end{figure}

\begin{figure}[h]
\centering
\includegraphics[width=0.36\textwidth]{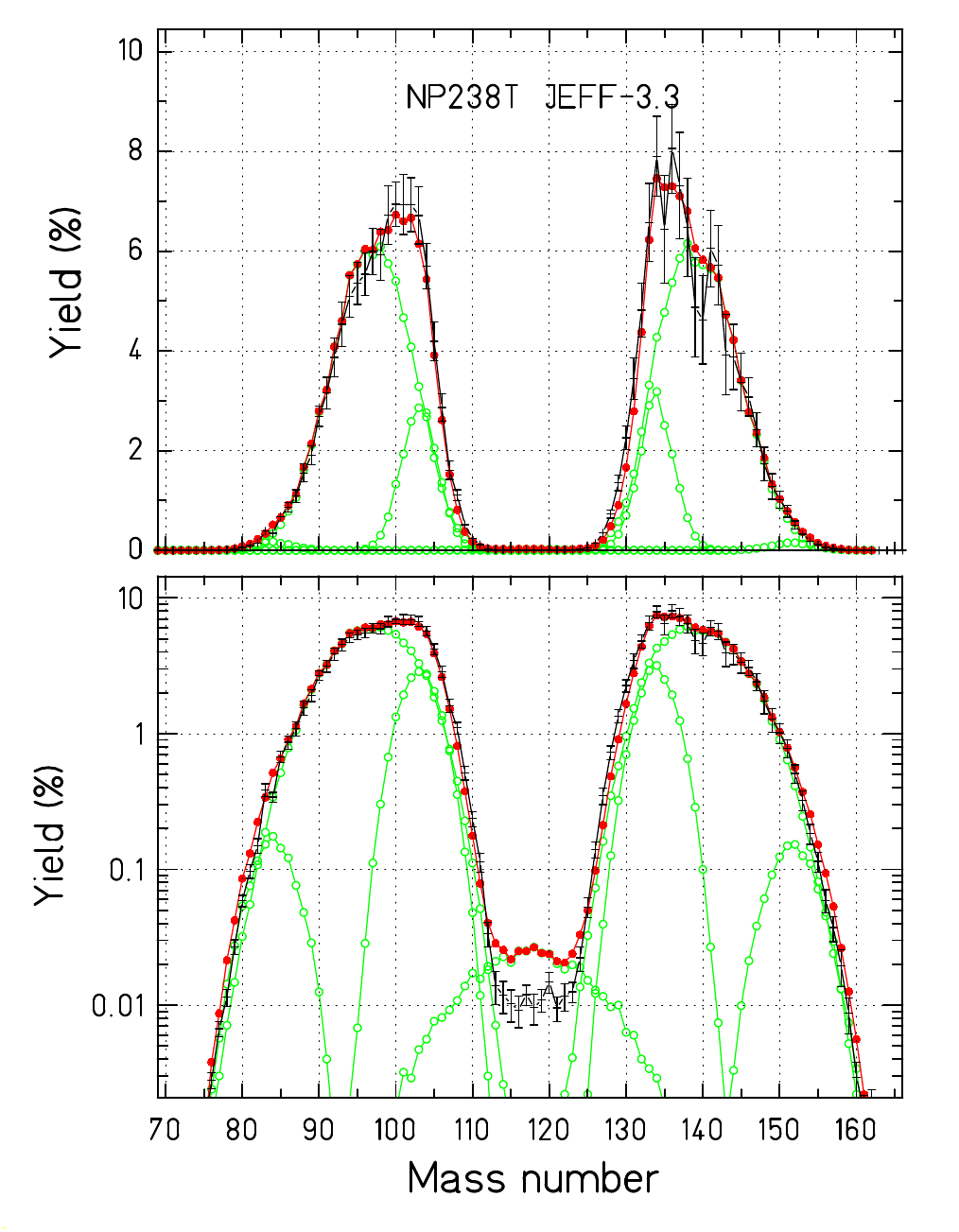}
\caption{Mass yields of $^{238}$Np(n$_{\text{th}}$,f), linear (upper frame) and logarithmic (lower frame) scale. GEF result (red points) in comparison with JEFF-3.3 (black symbols).} 
\label{NP238T-JEFF33}       
\end{figure}

\begin{figure}[h]
\centering
\includegraphics[width=0.36\textwidth]{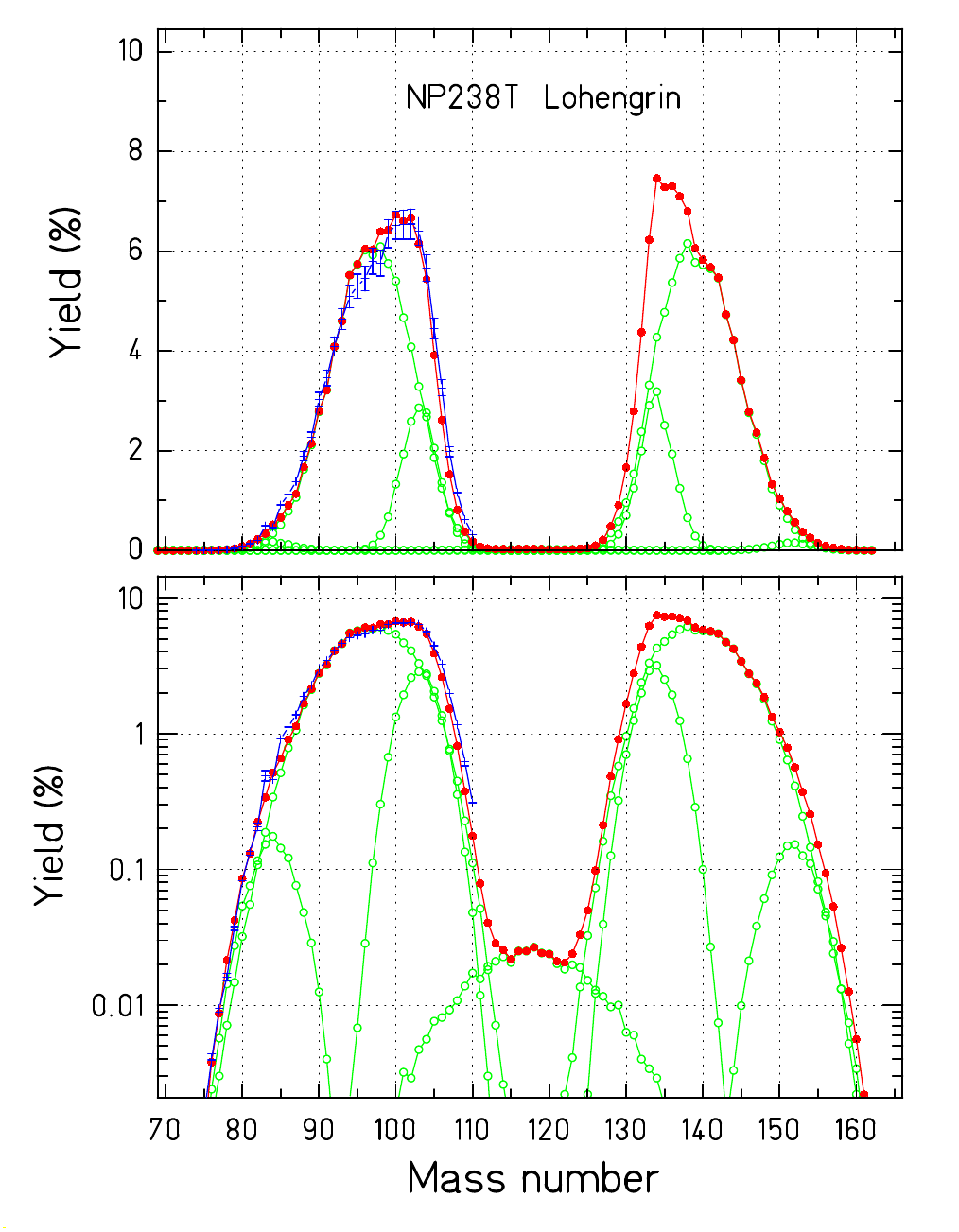}
\caption{Mass yields of $^{238}$Np(n$_{\text{th}}$,f), linear (upper frame) and logarithmic (lower frame) scale. GEF result (red points) in comparison with data from a LOHENGRIN experiment (blue symbols).} 
\label{NP238T-LOHENGRIN}       
\end{figure}

Figs. \ref{NP238T-JEFF311}, \ref{NP238T-JEFF33}, and \ref{NP238T-LOHENGRIN} show comparisons of the mass yields from GEF with the data from the JEFF-3.1.1 and JEFF-3.3 evaluations, as well as from LOHENGRIN experiments \cite{Martinez90,Tsekhanovich01} for the system $^{238}$Np(n$_{\text{th}}$,f) (by the $^{237}$Np(2n$_{\text{th}}$) reaction).
Again, the LOHENGRIN data have the smallest uncertainties. 
The data are quite well reproduced. Some deviations are found in the inner wings of the asymmetric peaks with JEFF-3.1.1, while there is good agreement with JEFF-3.3.
The shape of the distribution near symmetry of JEFF-3.1.1 shows a sharp minimum, while JEFF-3.3 and GEF show a plateau-like shape, however at different levels.  
The LOHENGRIN data do not reach above $A$=100. However, they show a slight underestimation of GEF near $A$=96.
The yield at symmetry is overestimated.

\paragraph{Mass yields of $^{238}$Pu(n$_{\text{th}}$,f):}

\begin{figure}[ht]
\centering
\includegraphics[width=0.36\textwidth]{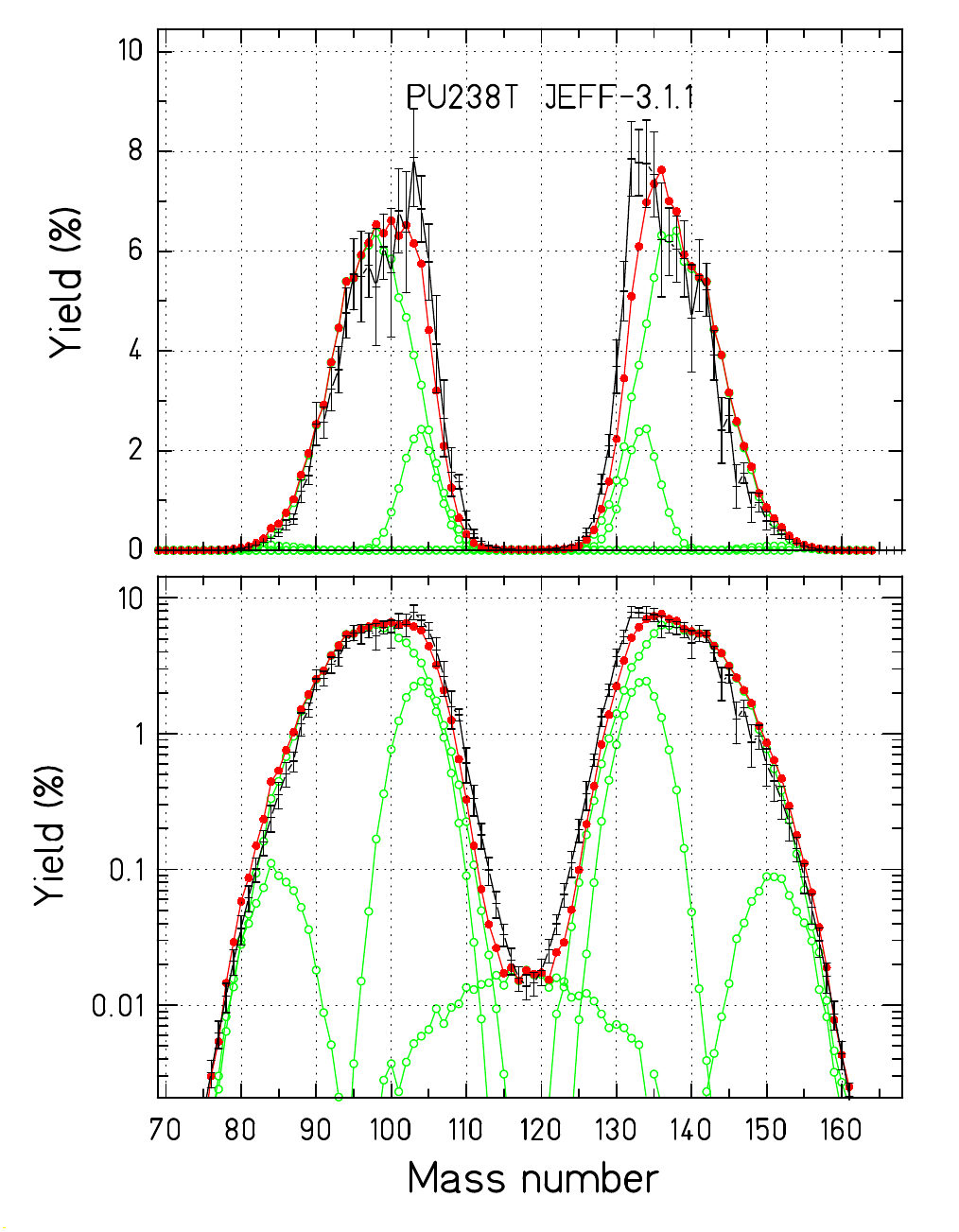}
\caption{Mass yields of $^{238}$Pu(n$_{\text{th}}$,f), linear (upper frame) and logarithmic (lower frame) scale. GEF result (red points) in comparison with JEFF-3.1.1 (black symbols).} 
\label{PU238T-JEFF311}       
\end{figure}

\begin{figure}[h]
\centering
\includegraphics[width=0.36\textwidth]{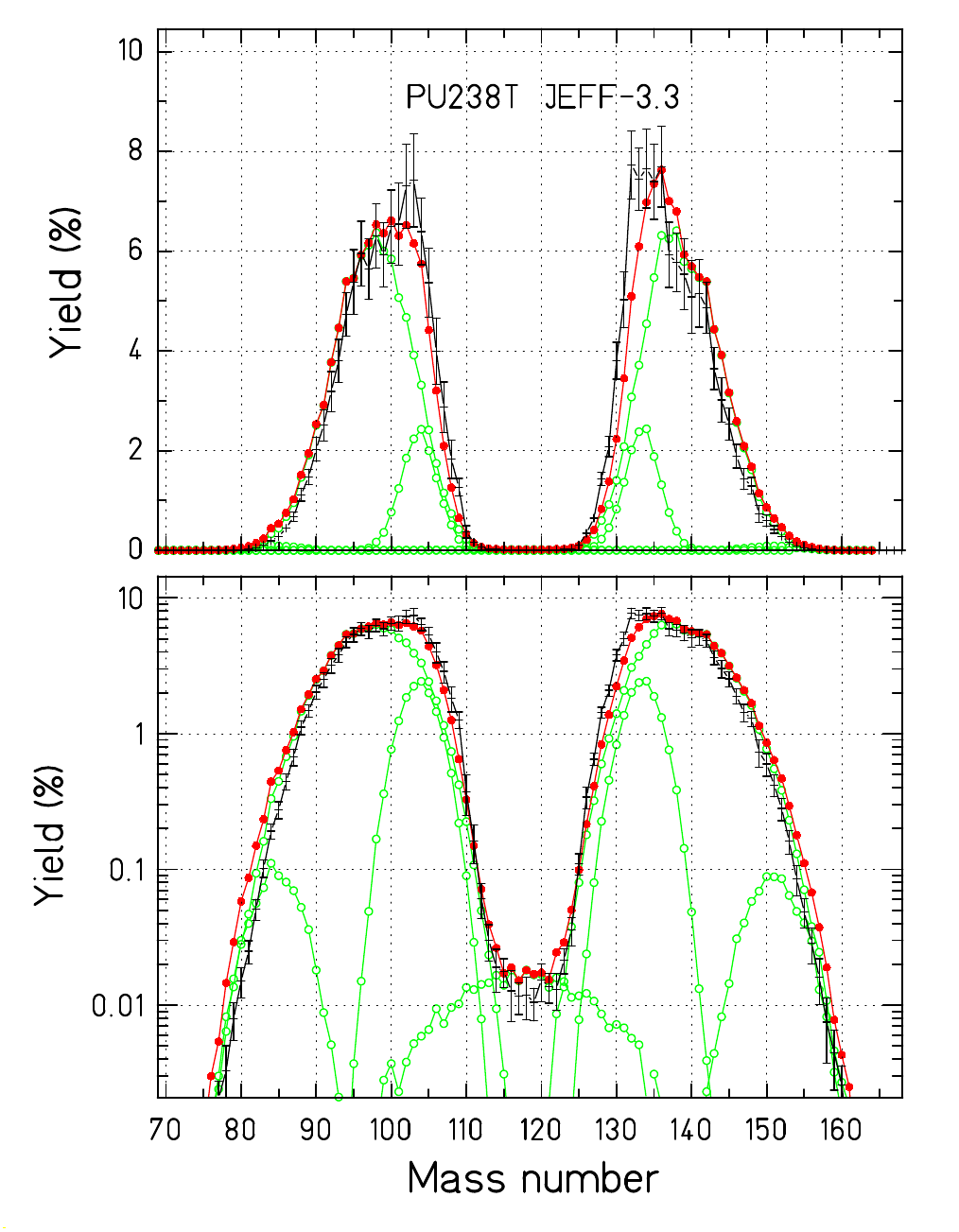}
\caption{Mass yields of $^{238}$Pu(n$_{\text{th}}$,f), linear (upper frame) and logarithmic (lower frame) scale. GEF result (red points) in comparison with JEFF-3.3 (black symbols).} 
\label{PU238T-JEFF33}       
\end{figure}

In Figs. \ref{PU238T-JEFF311} and \ref{PU238T-JEFF33}, the mass yields from GEF are compared with the data from the JEFF-3.1.1 and the JEFF-3.3 evaluations for the system $^{238}$Pu(n$_{\text{th}}$,f).

\clearpage

We note the reasonable description (height and shape) of the symmetric yield.
In the two peak regions, there are some deviations seen between both evaluations and GEF in linear scale: In GEF, the peaks are shifted to larger asymmetries. These deviations are astonishing, because the mass yields of the neighboring system $^{239}$Pu(n$_{\text{th}}$,f) are very well reproduced (see below). 
In the shoulders of the asymmetric peaks, there are systematic deviations seen in logarithmic scale: 
While GEF reproduces well the outer wings of JEFF-3.1.1 and the inner wings of JEFF-3.3, there are systematic shifts on the inner wings of JEFF-3.1.1 and the outer wings of JEFF-3.3.
A simultaneous reproduction of the mass yields of  $^{238}$Pu(n$_{\text{th}}$,f) and $^{239}$Pu(n$_{\text{th}}$,f)  from both JEFF-3.1.1 and JEFF-3.3 is in conflict with the regularities imposed by the physics of GEF. Considering in addition the rather limited data base for $^{238}$Pu(n$_{\text{th}}$,f) that was available for the evaluations, \emph{we tentatively recommend to use the GEF mass yields}. 

Finally, we would like to mention that $^{238}$Pu is a thermally not-fissile nucleus. Therefore, the fission-product yields measured in a pressurized-water reactor (PWR) originate to 42.1\% from neutrons with energies above 400 keV \cite{Kern12}. The expected enhancement of the evaluated yields at symmetry seems to be too weak to be seen.



\begin{figure}[t]
\centering
\includegraphics[width=0.36\textwidth]{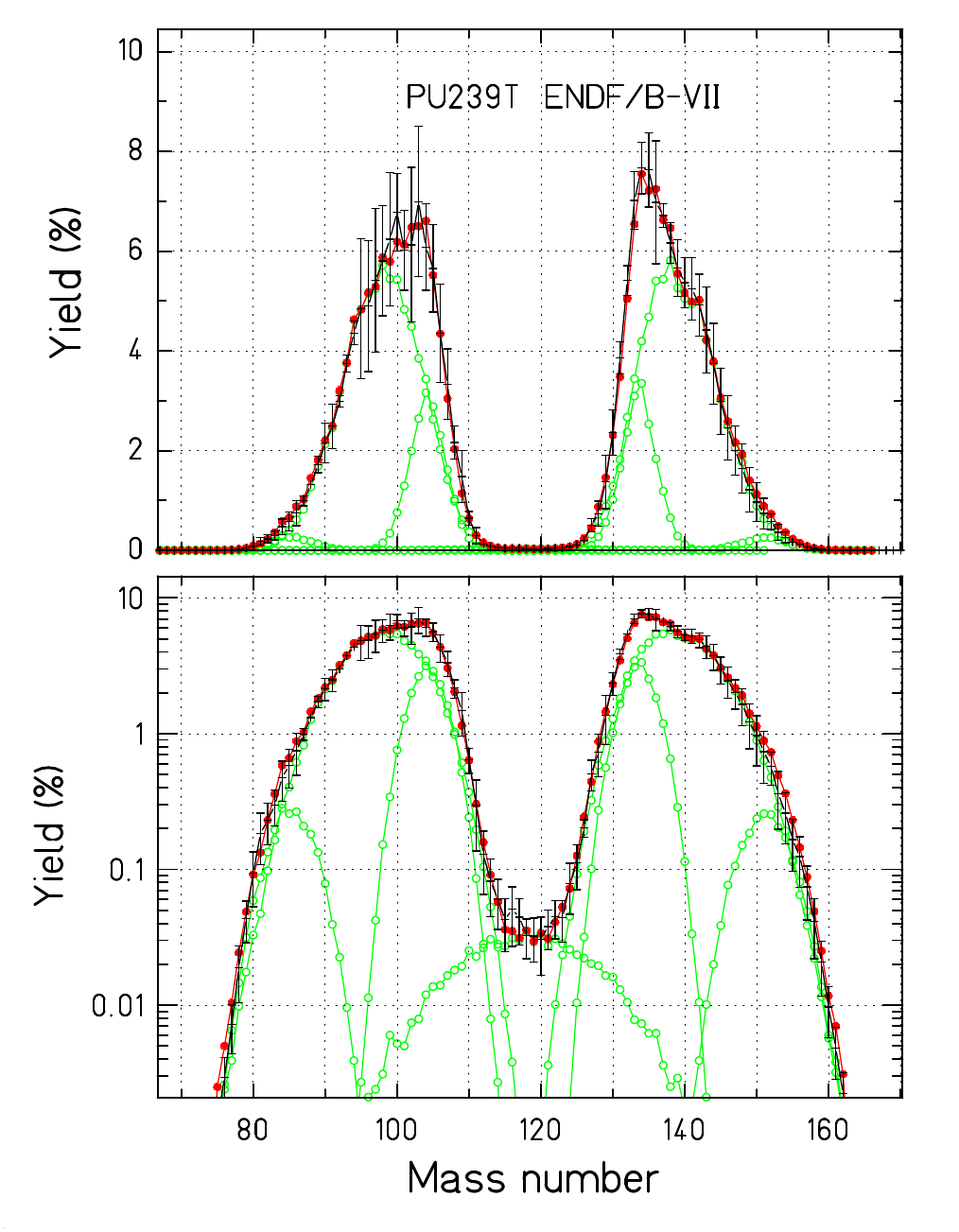}
\caption{Mass yields of $^{239}$Pu(n$_{\text{th}}$,f), linear (upper frame) and logarithmic (lower frame) scale, GEF result (red points) in comparison with ENDF/B-VII (black symbols).} 
\label{PU239T-ENDF}       
\end{figure}

\begin{figure}[h]
\centering
\includegraphics[width=0.36\textwidth]{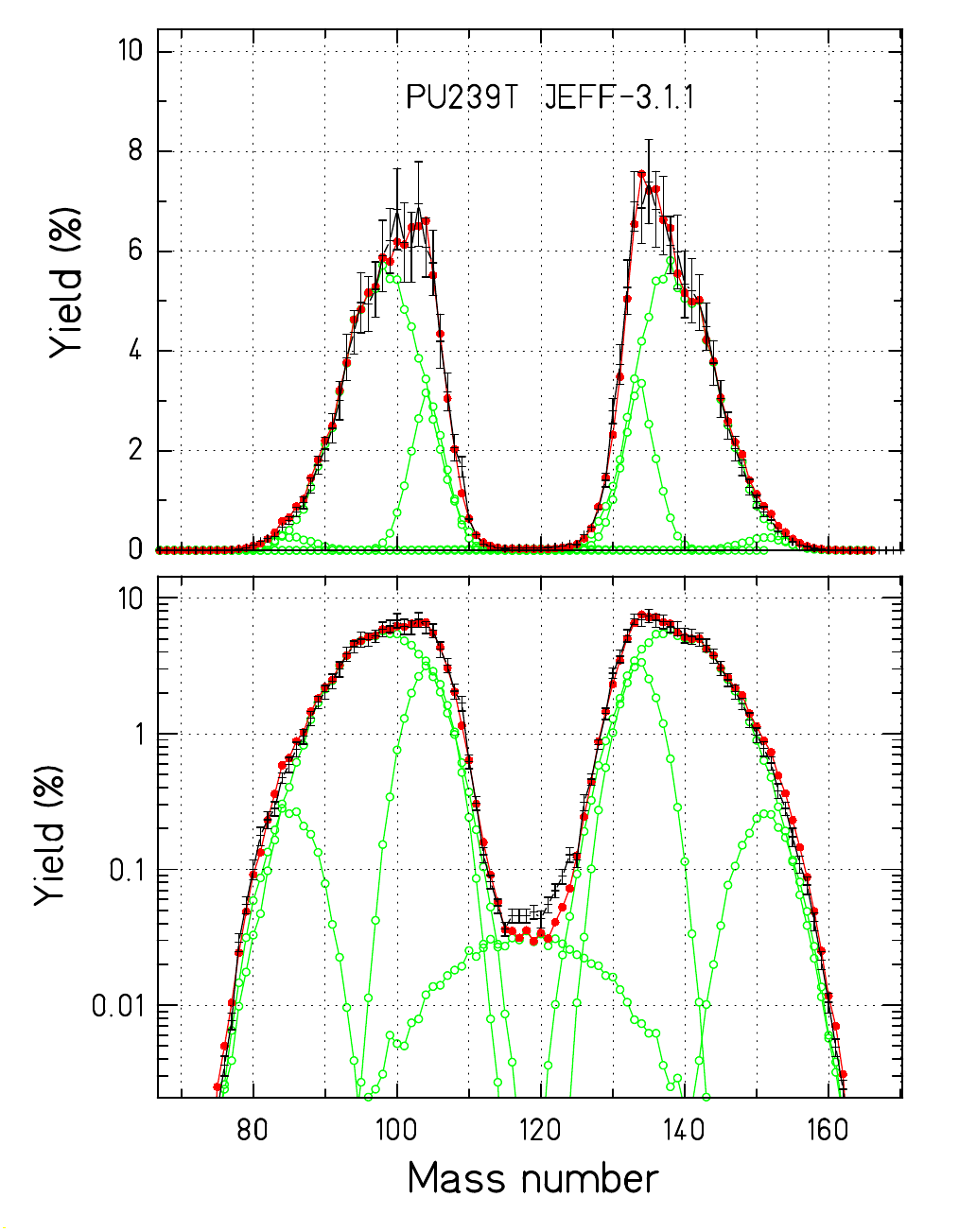}
\caption{Mass yields of $^{239}$Pu(n$_{\text{th}}$,f), linear (upper frame) and logarithmic (lower frame) scale. GEF result (red points) in comparison with JEFF-3.1.1 (black symbols).} 
\label{PU239T-JEFF311}       
\end{figure}

\begin{figure}[h]
\centering
\includegraphics[width=0.36\textwidth]{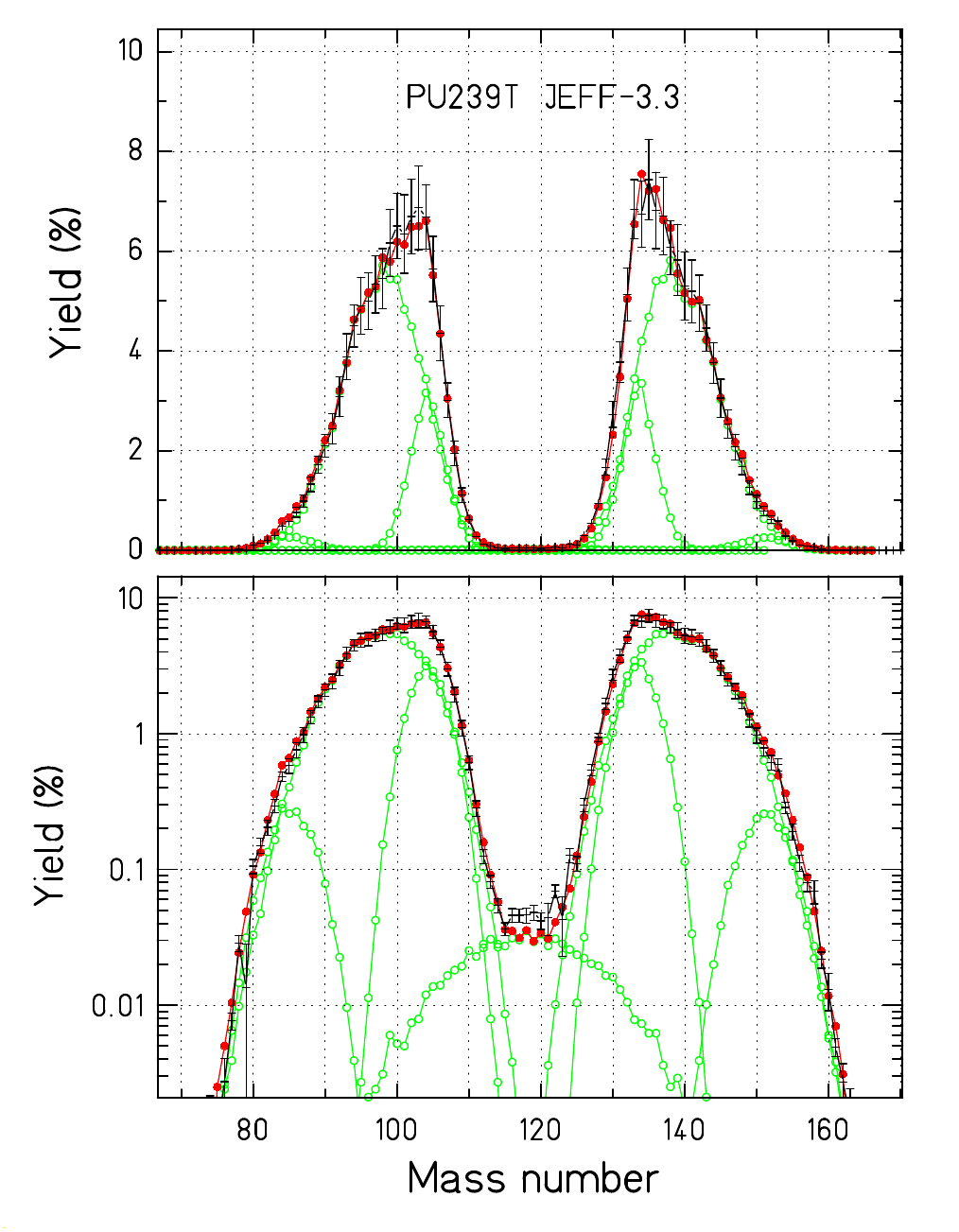}
\caption{Mass yields of $^{239}$Pu(n$_{\text{th}}$,f), linear (upper frame) and logarithmic (lower frame) scale. GEF result (red points) in comparison with JEFF-3.3 (black symbols).} 
\label{PU239T-JEFF33}       
\end{figure}

\begin{figure}[h]
\centering
\includegraphics[width=0.36\textwidth]{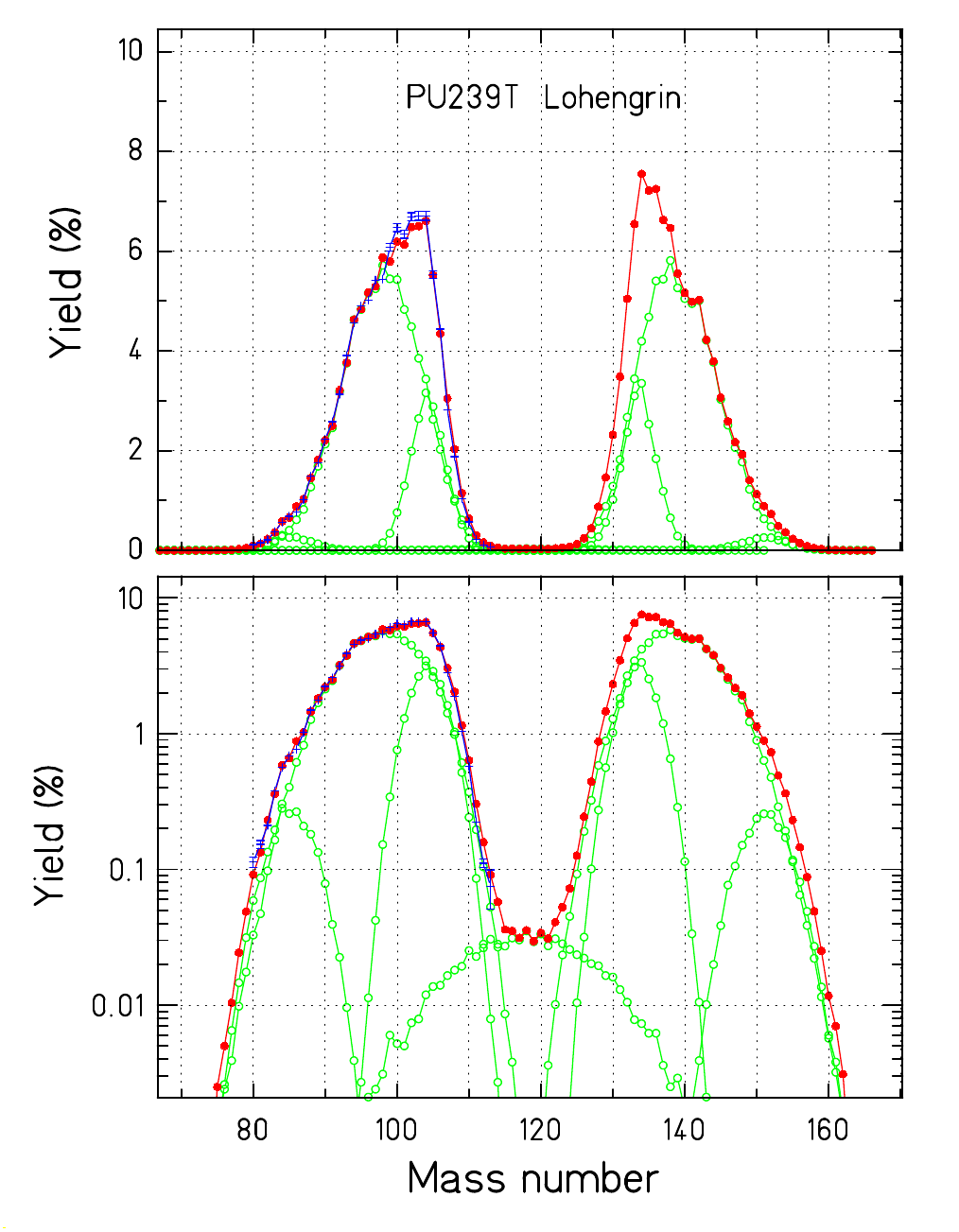}
\caption{Mass yields of $^{239}$Pu(n$_{\text{th}}$,f), linear (upper frame) and logarithmic (lower frame) scale. GEF result (red points) in comparison with data from a LOHENGRIN experiment (blue symbols).} 
\label{PU239T-LOHENGRIN}       
\end{figure}

\paragraph{Mass yields of $^{239}$Pu(n$_{\text{th}}$,f):}
Figs. \ref{PU239T-ENDF}, \ref{PU239T-JEFF311}, \ref{PU239T-JEFF33}, and \ref{PU239T-LOHENGRIN} show comparisons of the mass yields from GEF with the data from the ENDF/B-VII, JEFF-3.1.1 and JEFF-3.3 evaluations, as well as from a LOHENGRIN experiment \cite{Schmitt84} for the system $^{239}$Pu(n$_{\text{th}}$,f). The data of all evaluations are rather well reproduced. The smallest deviations are found with respect to the LOHENGRIN data, which have by far the smallest uncertainties. We would like to draw the attention to an interesting detail: In the LOHENGRIN data there appears a clear shoulder at $A =$ 84, dominated by $^{84}$Se, which is well reproduced by GEF. 
According to GEF, this shoulder marks the transition from the S2 to the super-asymmetric fission channel.
This shoulder does not appear in the evaluations. This shoulder is seen, less pronounced, also in the GEF results for $^{241}$Pu(n$_{\text{th}}$,f). We note the good description around symmetry, namely when compared to ENDF/B-VII.



\begin{figure}[h]
\centering
\includegraphics[width=0.36\textwidth]{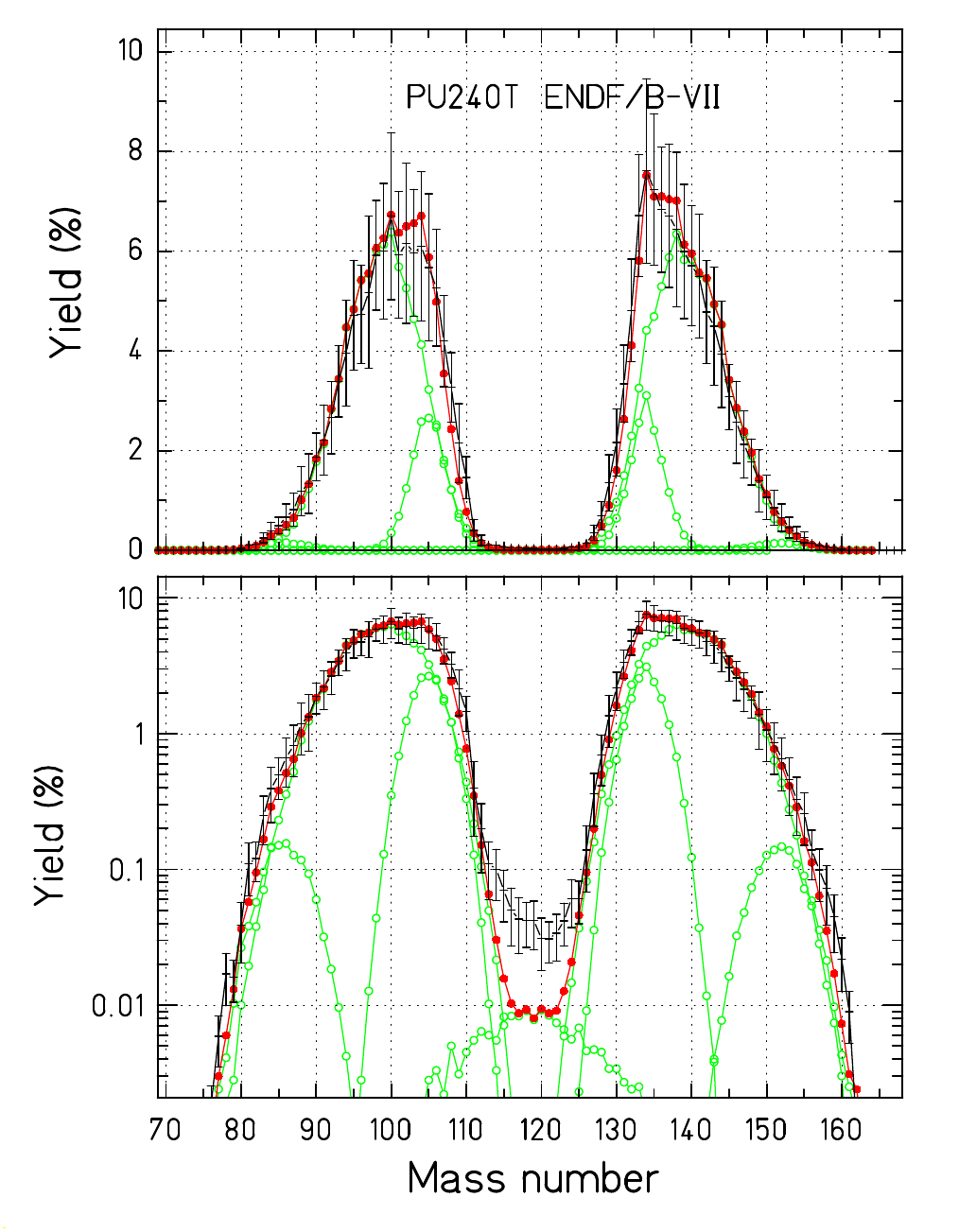}
\caption{Mass yields of $^{240}$Pu(n$_{\text{th}}$,f), linear scale (upper frame) and logarithmic (lower scale), GEF result (red points) in comparison with ENDF/B-VII (black symbols).} 
\label{PU240T-ENDF}       
\end{figure}

\paragraph{Mass yields of $^{240}$Pu(n$_{\text{th}}$,f):}
Fig. \ref{PU240T-ENDF} shows the comparison of the mass yields from GEF with the data from the ENDF/B-VII evaluation for the system $^{240}$Pu(n$_{\text{th}}$,f). The data are rather well reproduced, except near symmetry, where the yields from GEF are lower. 
This may be explained by the fact that $^{240}$Pu is a thermally not-fissile nucleus, like
$^{238}$Pu, which was discussed before. 
The evaluation shows an unexpected asymmetry in this region, which is not present
in the ENDF/B-VII and the JEFF-3.3 evaluations of the more intensively investigated
system $^{239}$Pu(n$_{\text{th}}$,f).
The GEF yields of the most asymmetric masses are slightly below the evaluation.

\begin{figure}[h]
\centering
\includegraphics[width=0.36\textwidth]{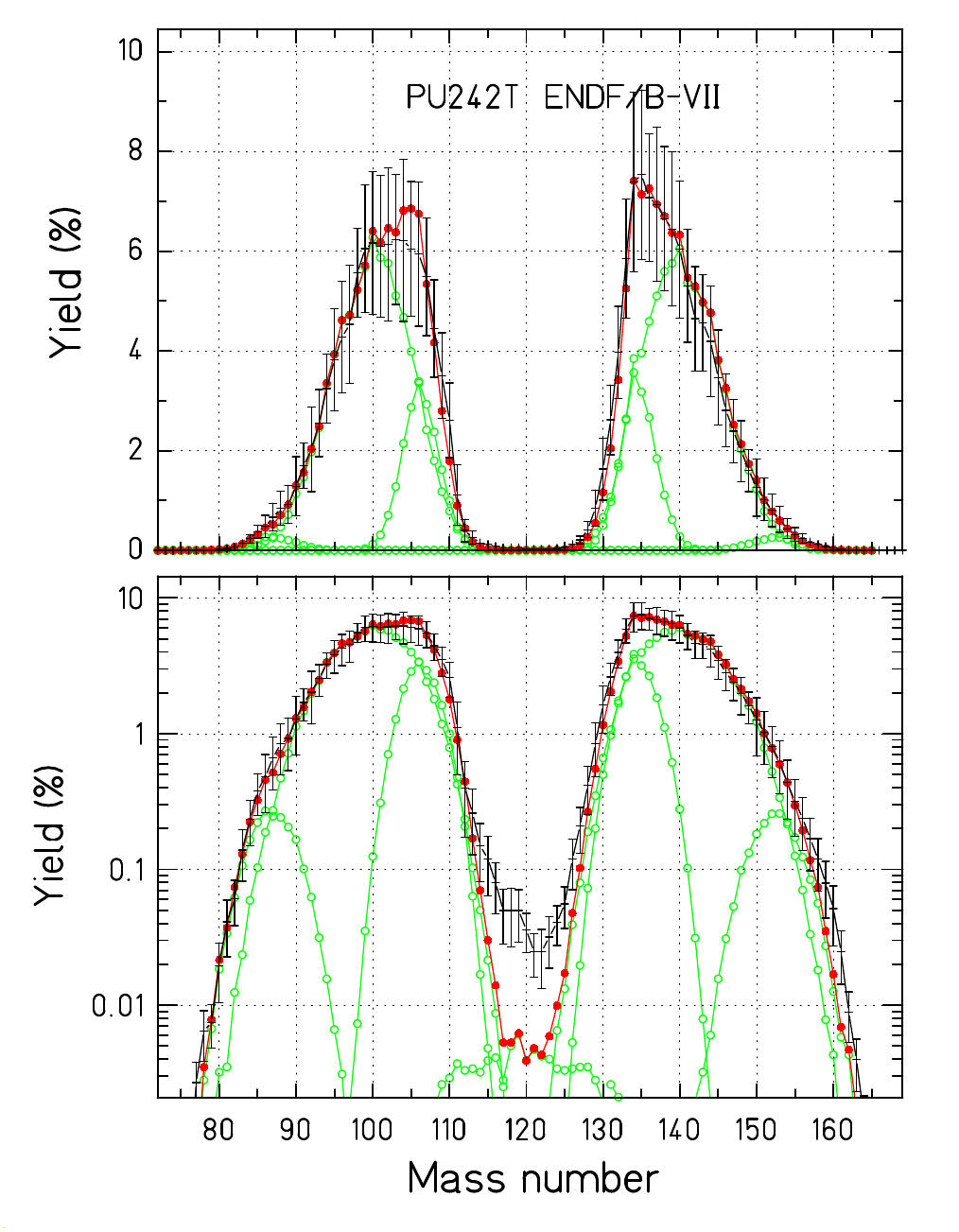}
\caption{Mass yields of $^{242}$Pu(n$_{\text{th}}$,f), linear (upper frame) and logarithmic (lower frame) scale. GEF result (red points) in comparison with ENDF/B-VII (black symbols).} 
\label{PU242T-ENDF}       
\end{figure}

\paragraph{Mass yields of $^{242}$Pu(n$_{\text{th}}$,f):}
Fig. \ref{PU242T-ENDF} shows the comparison of the mass yields from GEF with the data from the ENDF/B-VII for the system $^{242}$Pu(n$_{\text{th}}$,f). The data are rather well reproduced, except near symmetry and for $A$ around 160, where the yields from GEF are lower.
The underestimation of the mass yields at symmetry may again be explained by the fact the $^{242}$Pu is a thermally not-fissile nucleus, like $^{238}$Pu and $^{240}$Pu.
Again, there is an unexpected asymmetry around the symmetric valley in the evaluation.

\clearpage

\begin{figure}[h]
\centering
\includegraphics[width=0.36\textwidth]{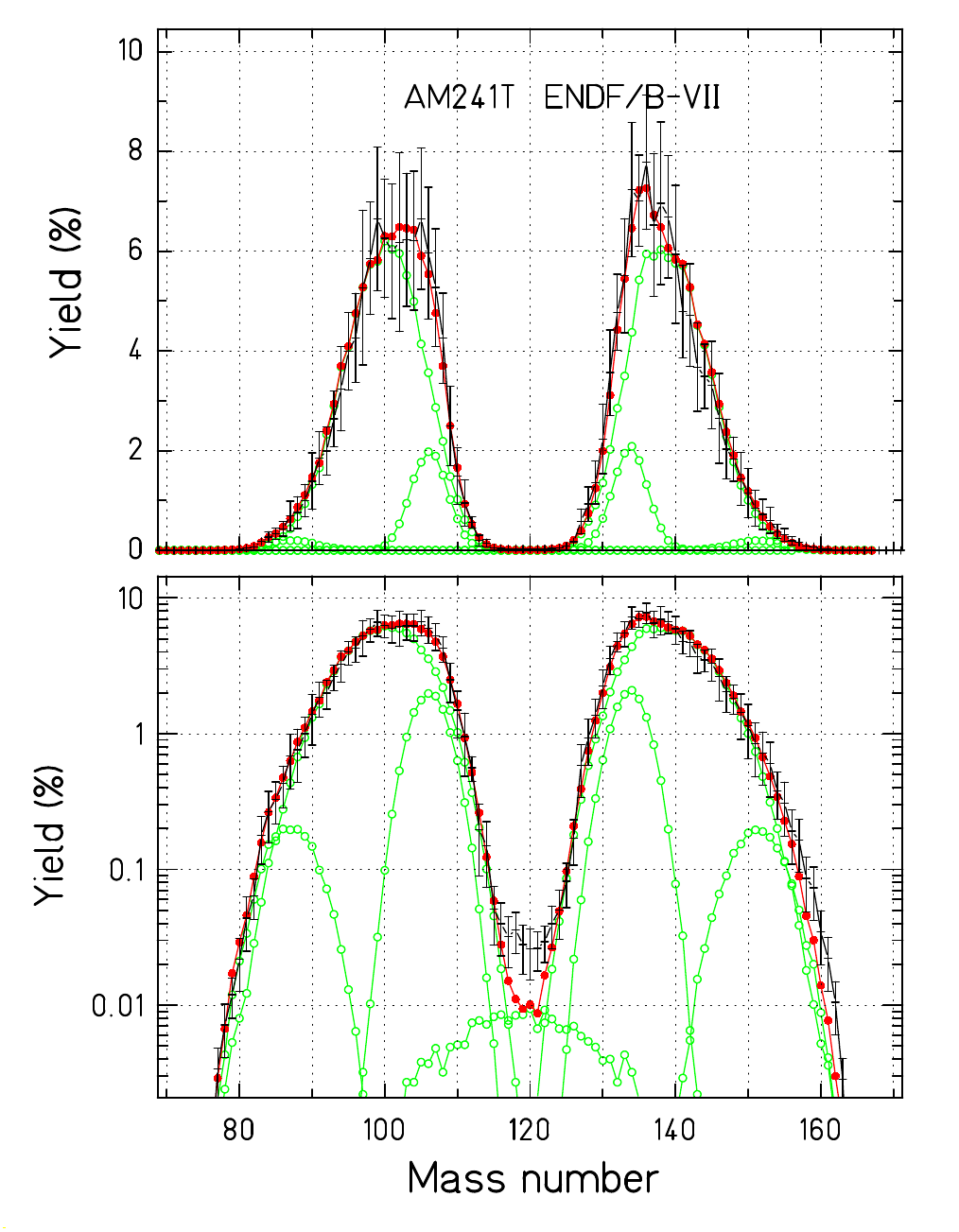}
\caption{Mass yields of $^{241}$Am(n$_{\text{th}}$,f), linear (upper frame) and logarithmic (lower frame) scale. GEF result (red points) in comparison with ENDF/B-VII (black symbols).} 
\label{AM241T-ENDF}       
\end{figure}

\begin{figure}[h]
\centering
\includegraphics[width=0.36\textwidth]{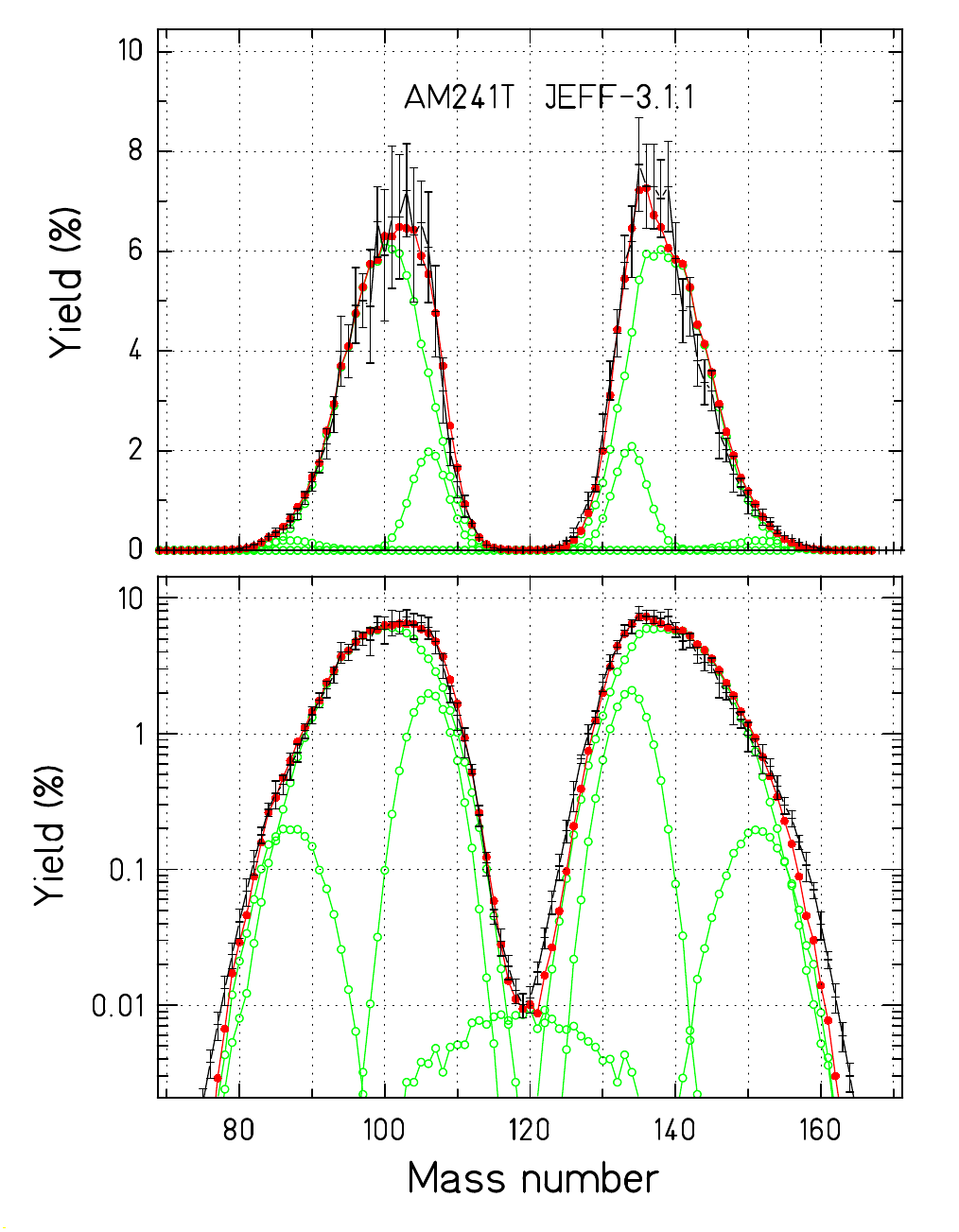}
\caption{Mass yields of $^{241}$Am(n$_{\text{th}}$,f), linear (upper frame) and logarithmic (lower frame) scale. GEF result (red points) in comparison with JEFF-3.1.1 (black symbols).} 
\label{AM241T-JEFF311}       
\end{figure}

\begin{figure}[h]
\centering
\includegraphics[width=0.36\textwidth]{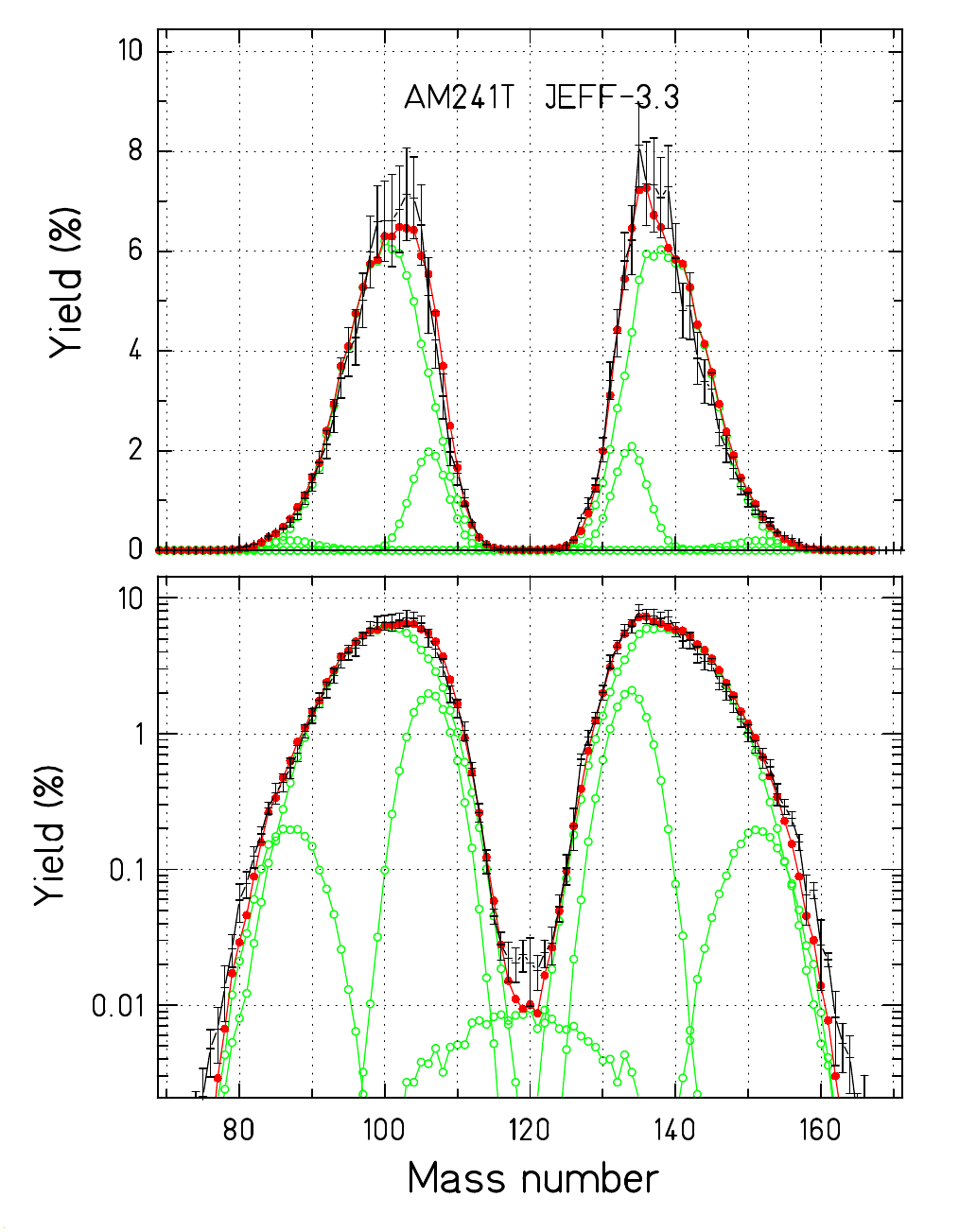}
\caption{Mass yields of $^{241}$Am(n$_{\text{th}}$,f), linear (upper frame) and logarithmic (lower frame) scale. GEF result (red points) in comparison with JEFF-3.3 (black symbols).} 
\label{AM241T-JEFF33}       
\end{figure}

\paragraph{Mass yields of $^{241}$Am(n$_{\text{th}}$,f):}
Figs. \ref{AM241T-ENDF}, \ref{AM241T-JEFF311}, and \ref{AM241T-JEFF33} show the comparison of the mass yields from GEF with the data from ENDF/B-VII, JEFF-3.1.1 and JEFF-3.3 for the system $^{241}$Am(n$_{\text{th}}$,f). The data are rather well reproduced.

Fission of $^{241}$Am in a PWR include a fraction of 65.4\% induced by neutrons with energies above 400 kV \cite{Kern12}, which could explain why the mass yields around symmetry from ENDF/B-VII and JEFF-3.3 are underestimated by GEF.  
Some more deviations appear in the upper wing of the distribution above $A=155$: 
The yields from GEF are systematically lower than the values from ENDF/B-VII and JEFF-3.1.1 and, to a lesser extent from JEFF-3.3, while there is rather good agreement in the lower wing of the distribution below $A=82$. 
\emph{It is not obvious to attribute the deviations in the upper wing to deficiencies of GEF}, because physics connects the yields in the two outer wings with the mass-dependent prompt-neutron multiplicities: A shift in the upper wing to higher masses with respect to GEF, while keeping the lower wing unchanged, as suggested by the evaluations for $^{241}$Am(n$_{\text{th}}$,f) demands a reduction of the prompt-neutron yields in the heavy-mass region with respect to the systematics of other systems, for example $^{239}$Pu(n$_{\text{th}}$,f), where the mass yields from GEF agree with the empirical data over the whole mass range.  





\begin{figure}[h]
\centering
\includegraphics[width=0.36\textwidth]{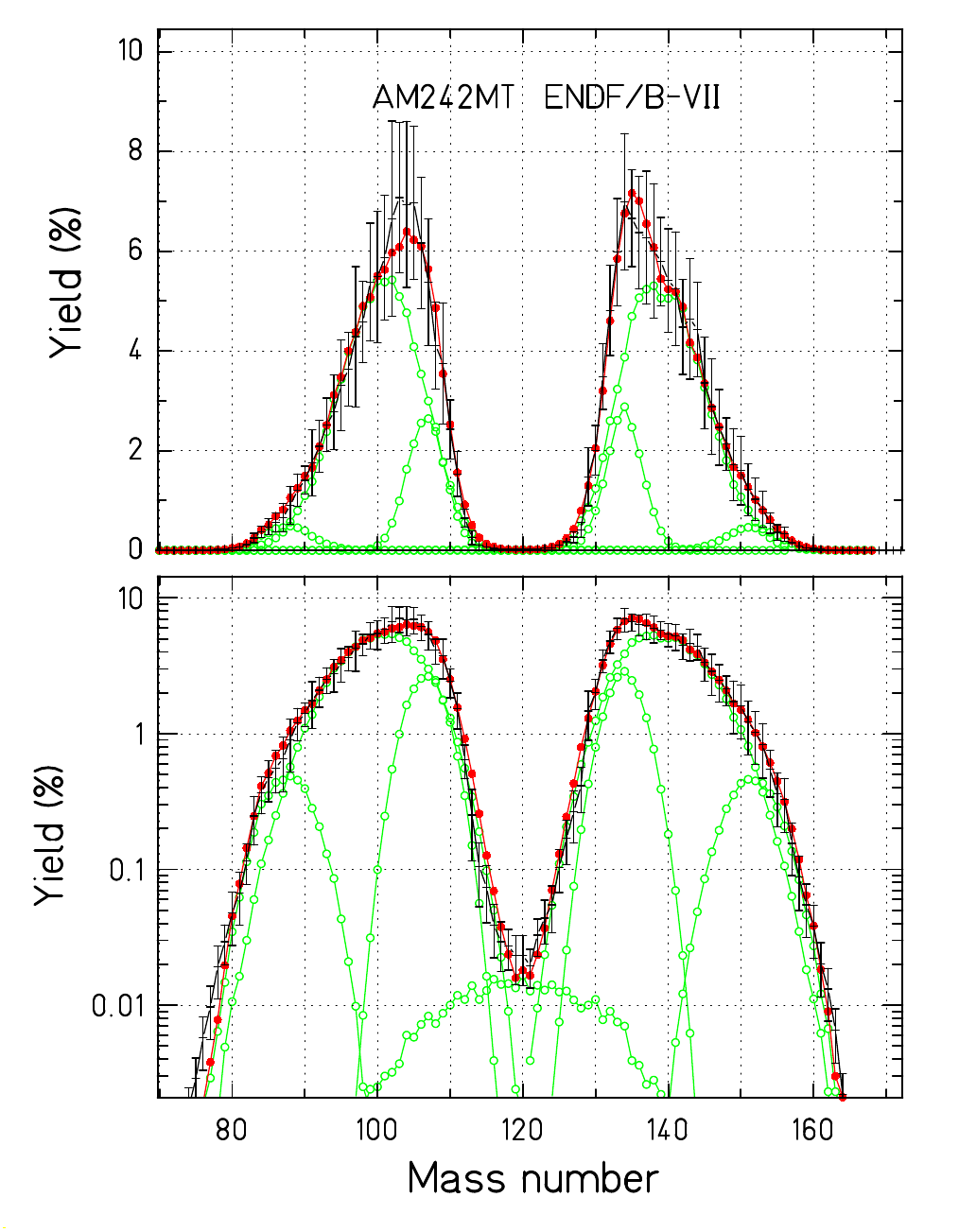}
\caption{Mass yields of $^{\text{242m}}$Am(n$_{\text{th}}$,f), linear (upper frame) and logarithmic (lower frame) scale, GEF result (red points) in comparison with ENDF/B-VII (black symbols).} 
\label{AM242T-ENDF}       
\end{figure}

\begin{figure}[h]
\centering
\includegraphics[width=0.36\textwidth]{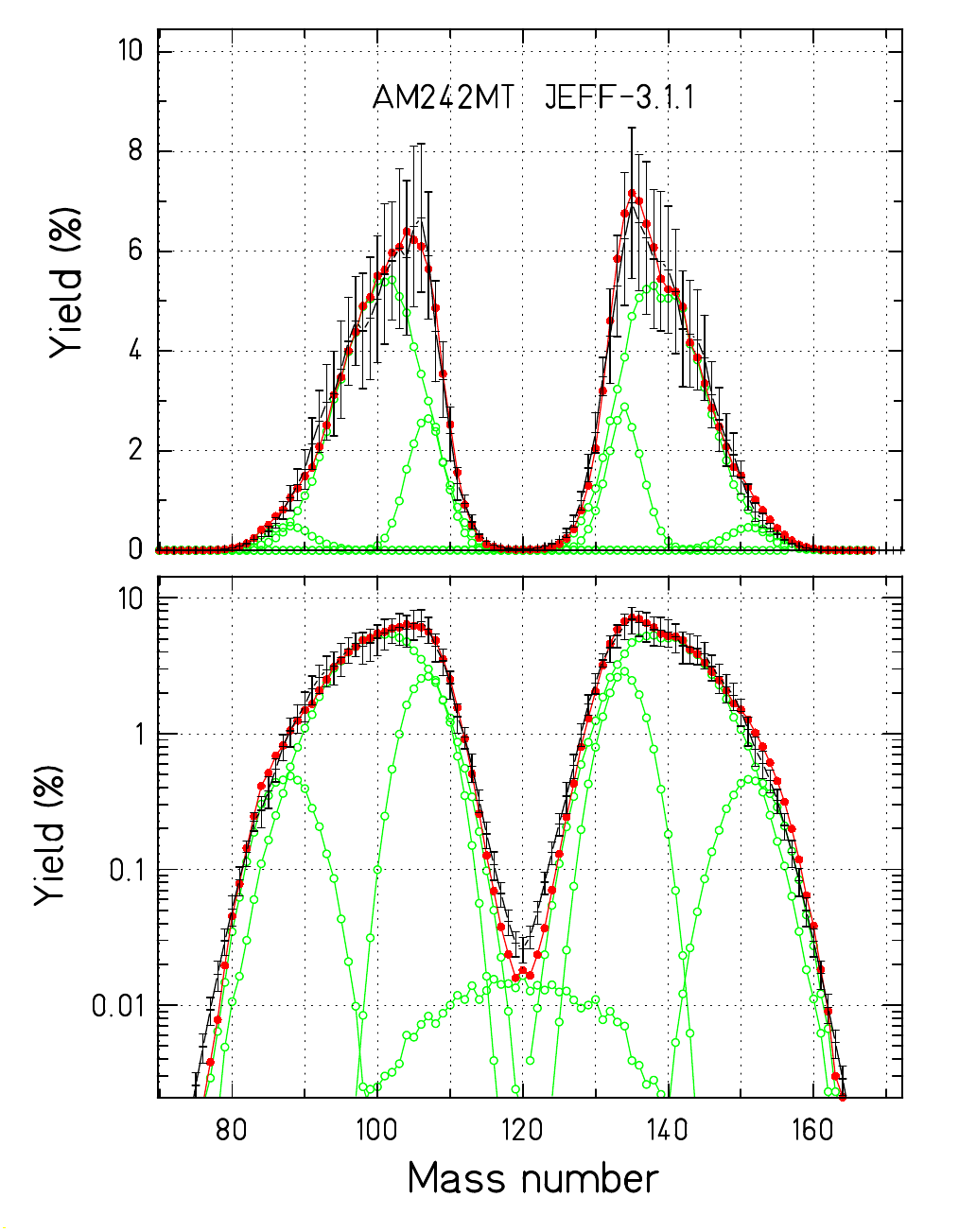}
\caption{Mass yields of $^{\text{242m}}$Am(n$_{\text{th}}$,f), linear (upper frame) and logarithmic (lower frame) scale. GEF result (red points) in comparison with JEFF-3.1.1 (black symbols).} 
\label{AM242T-JEFF311}       
\end{figure}

\paragraph{Mass yields of $^{242m}$Am(n$_{\text{th}}$,f):}
Figs. \ref{AM242T-ENDF} and \ref{AM242T-JEFF311} show the comparison of the mass yields from GEF with the data from ENDF/B-VII and JEFF-3.1.1 for the system $^{242m}$Am(n$_{\text{th}}$,f). The data of the evaluations are rather well reproduced by GEF with slightly underestimated yields at symmetry in the case of JEFF-3.1.1.

\begin{figure}[h]
\centering
\includegraphics[width=0.36\textwidth]{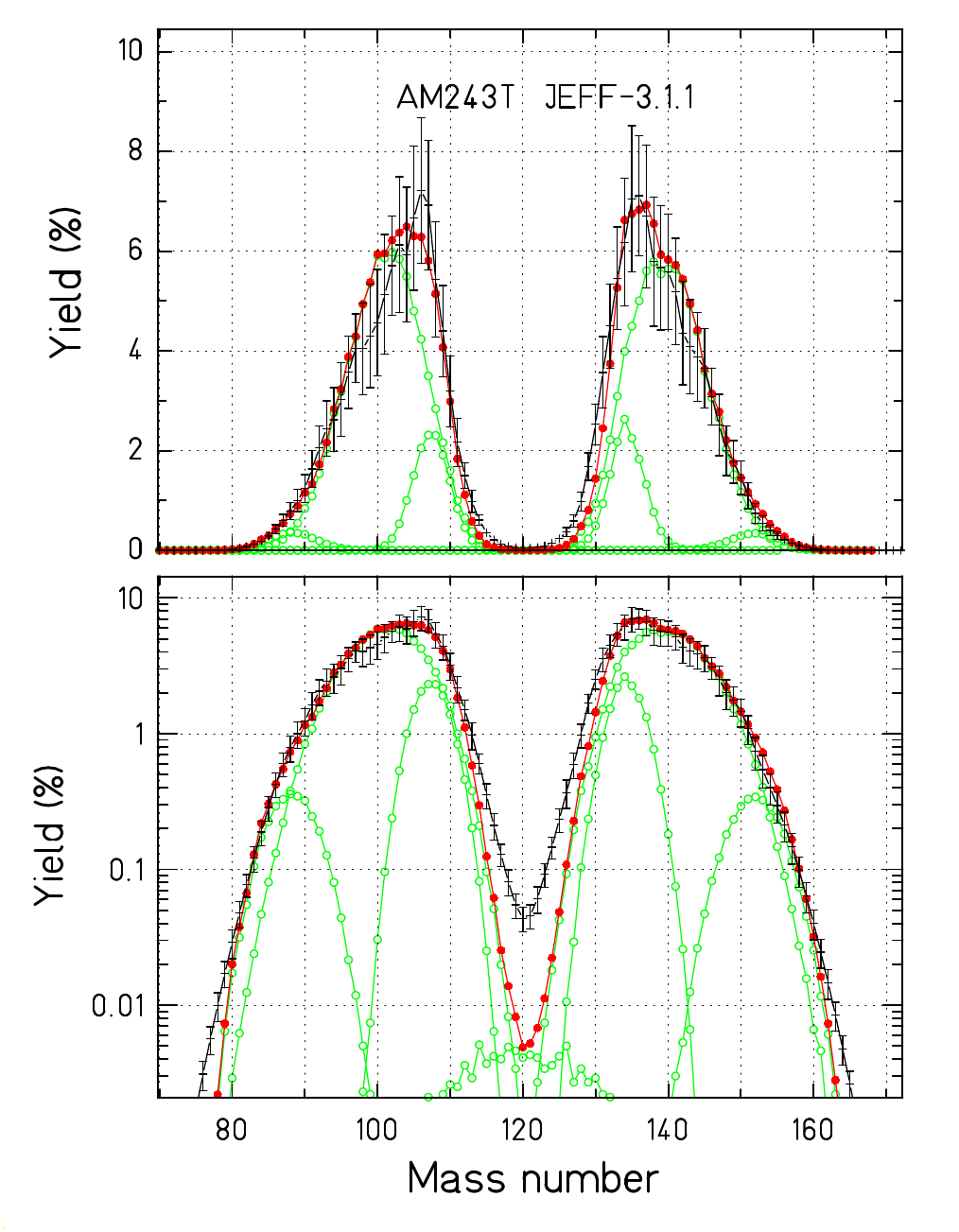}
\caption{Mass yields of $^{243}$Am(n$_{\text{th}}$,f), linear (upper frame) and logarithmic (lower frame) scale. GEF result (red points) in comparison with JEFF-3.1.1 (black symbols).} 
\label{AM243T-JEFF311}       
\end{figure}

\begin{figure}[h]
\centering
\includegraphics[width=0.36\textwidth]{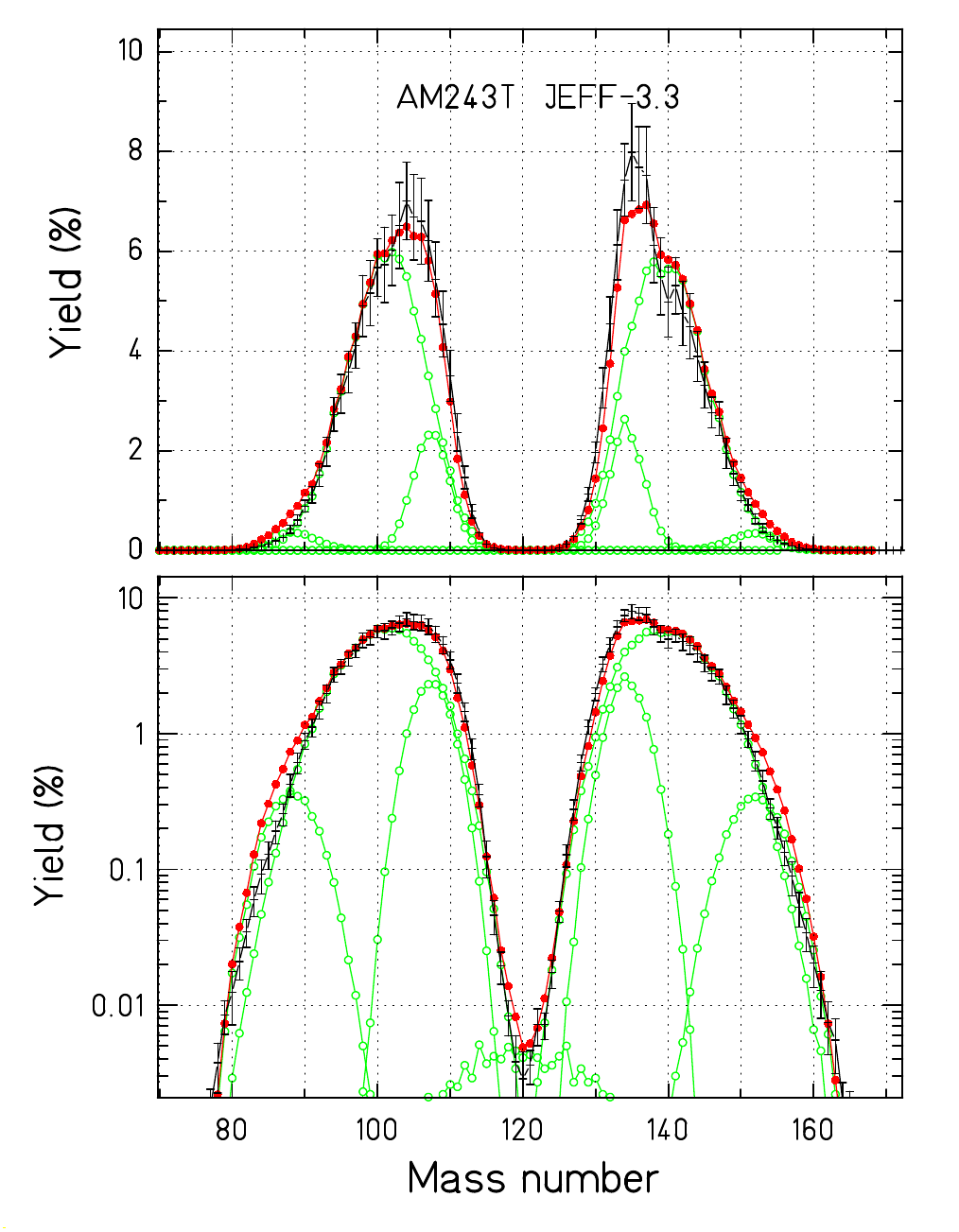}
\caption{Mass yields of $^{243}$Am(n$_{\text{th}}$,f), linear (upper frame) and logarithmic (lower frame) scale. GEF result (red points) in comparison with JEFF-3.3 (black symbols).} 
\label{AM243T-JEFF33}       
\end{figure}

\paragraph{Mass yields of $^{243}$Am(n$_{\text{th}}$,f):}
Figs. \ref{AM243T-JEFF311} and \ref{AM243T-JEFF33} show the comparison of the mass yields from GEF with the data from JEFF-3.1.1 and JEFF-3.3 for the system $^{243}$Am(n$_{\text{th}}$,f). There are discrepancies between the GEF results and JEFF-3.1.1 near symmetry and JEFF-3.3 in the outer wings, while GEF agrees well with JEFF-3.3 near symmetry and with JEFF-3.1.1 in the outer wings, which is a rather \emph{ambiguous result that calls for clarification}.

\begin{figure}[h]
\centering
\includegraphics[width=0.36\textwidth]{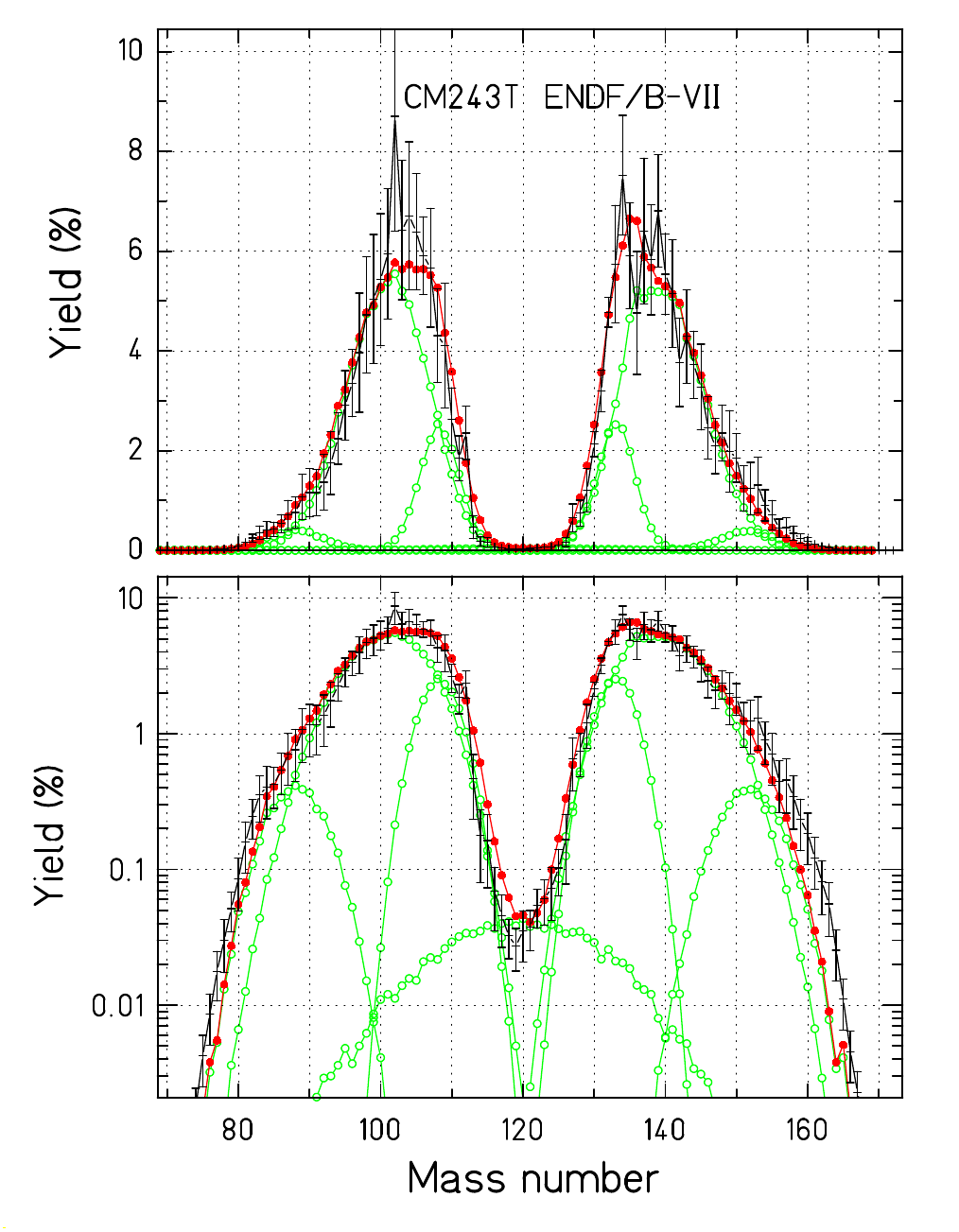}
\caption{Mass yields of $^{243}$Cm(n$_{\text{th}}$,f), linear (upper frame) and logarithmic (lower frame) scale. GEF result (red points) in comparison with ENDF/B-VII (black symbols).} 
\label{CM243T-ENDF}       
\end{figure}

\begin{figure}[h]
\centering
\includegraphics[width=0.36\textwidth]{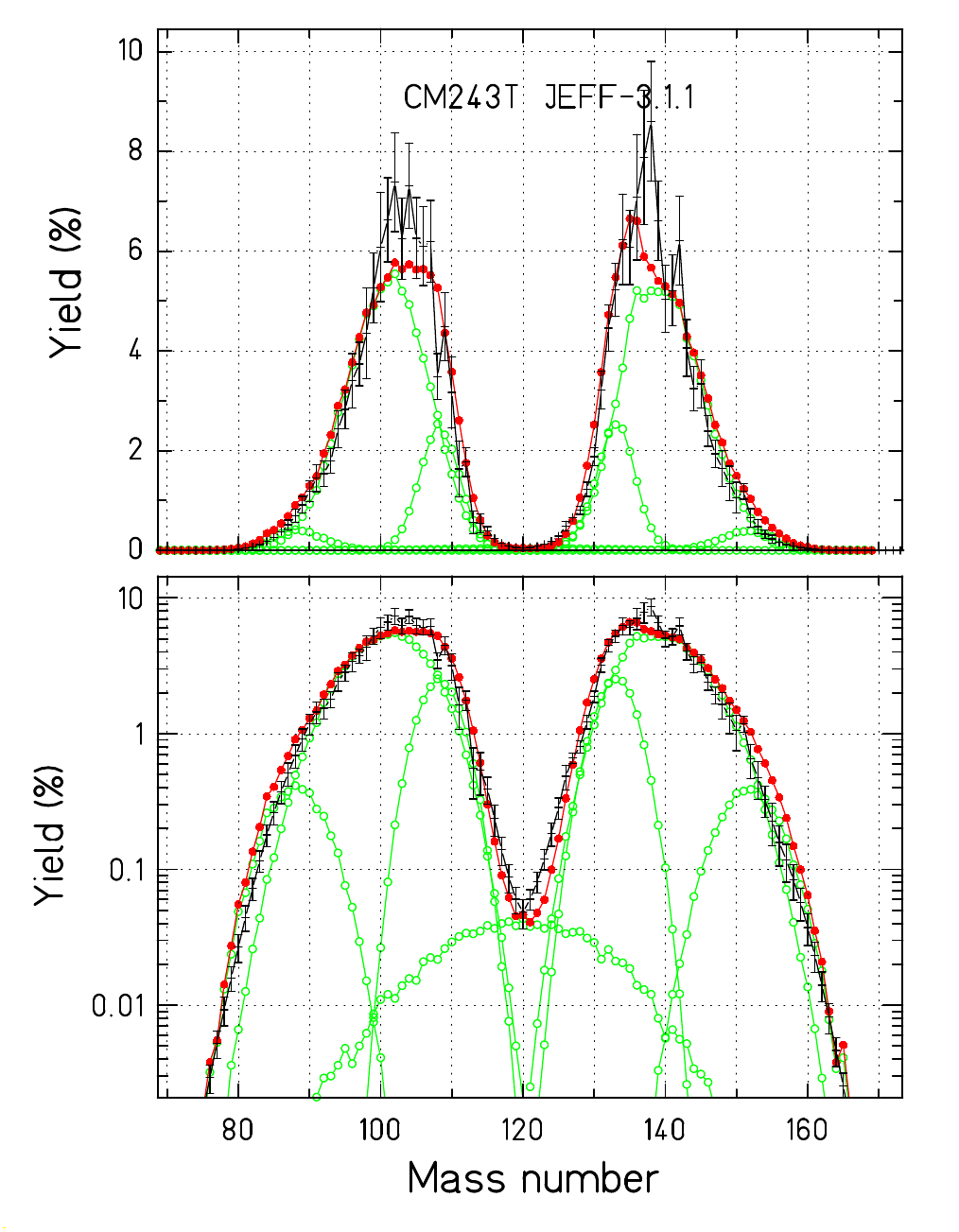}
\caption{Mass yields of $^{243}$Cm(n$_{\text{th}}$,f), linear (upper frame) and logarithmic (lower frame) scale. GEF result (red points) in comparison with JEFF-3.1.1 (black symbols).} 
\label{CM243T-JEFF311}       
\end{figure}

\begin{figure}[h]
\centering
\includegraphics[width=0.36\textwidth]{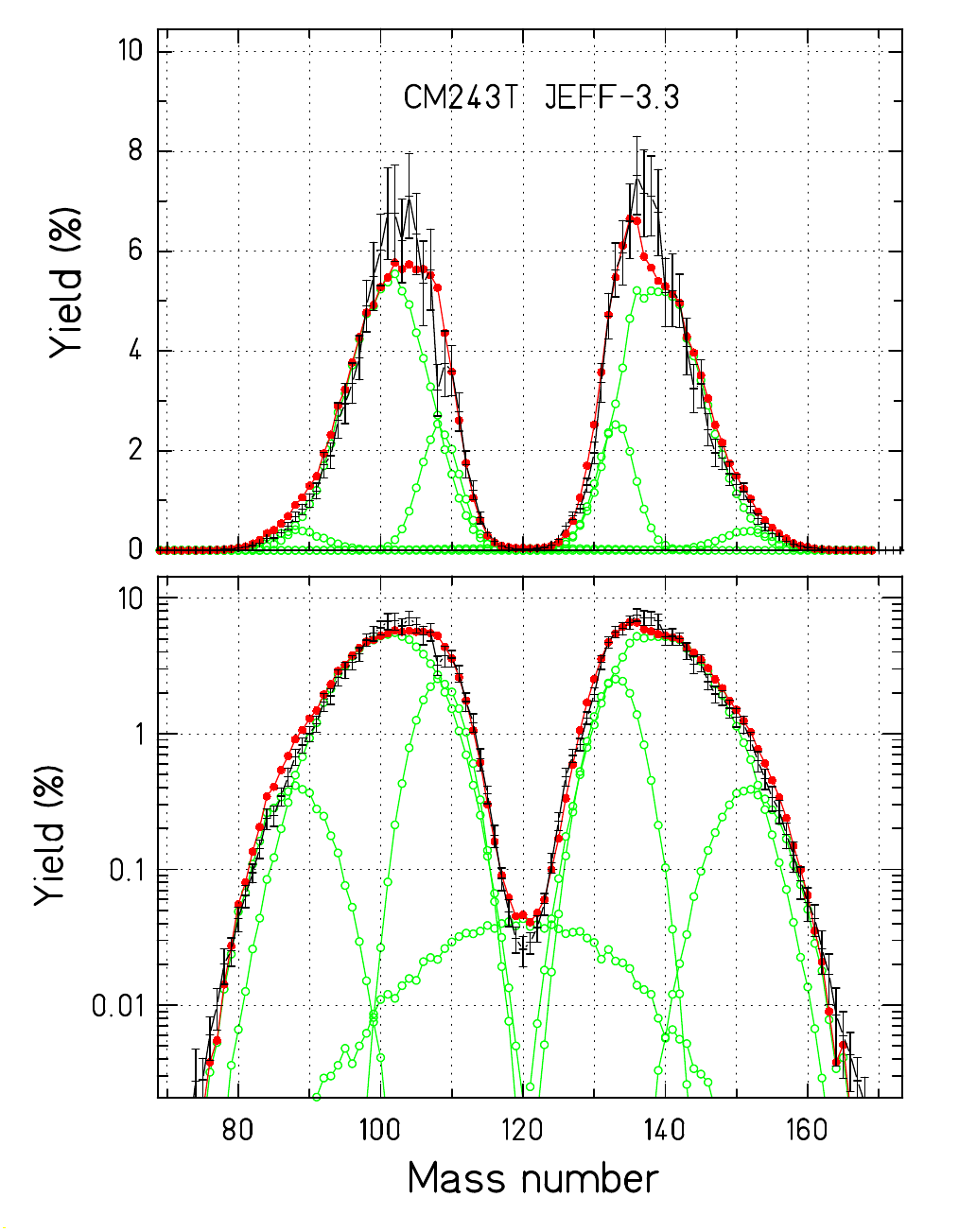}
\caption{Mass yields of $^{243}$Cm(n$_{\text{th}}$,f), linear (upper frame) and logarithmic (lower frame) scale. GEF result (red points) in comparison with JEFF-3.3 (black symbols).} 
\label{CM243T-JEFF33}       
\end{figure}

\paragraph{Mass yields of $^{243}$Cm(n$_{\text{th}}$,f):}
Figs. \ref{CM243T-ENDF}, \ref{CM243T-JEFF311} and \ref{CM243T-JEFF33} show the comparison of the mass yields from GEF with the data from ENDF/B-VII, JEFF-3.1.1 and JEFF-3.3 for the system $^{243}$Cm(n$_{\text{th}}$,f). The empirical distributions are fairly well reproduced. There are some deviations, in particular around the peaks, but it is difficult to deduce a systematic trend due to the large scattering of the evaluated yields between neighboring masses and between the different evaluations, and due to their large uncertainties.

\begin{figure}[h]
\centering
\includegraphics[width=0.36\textwidth]{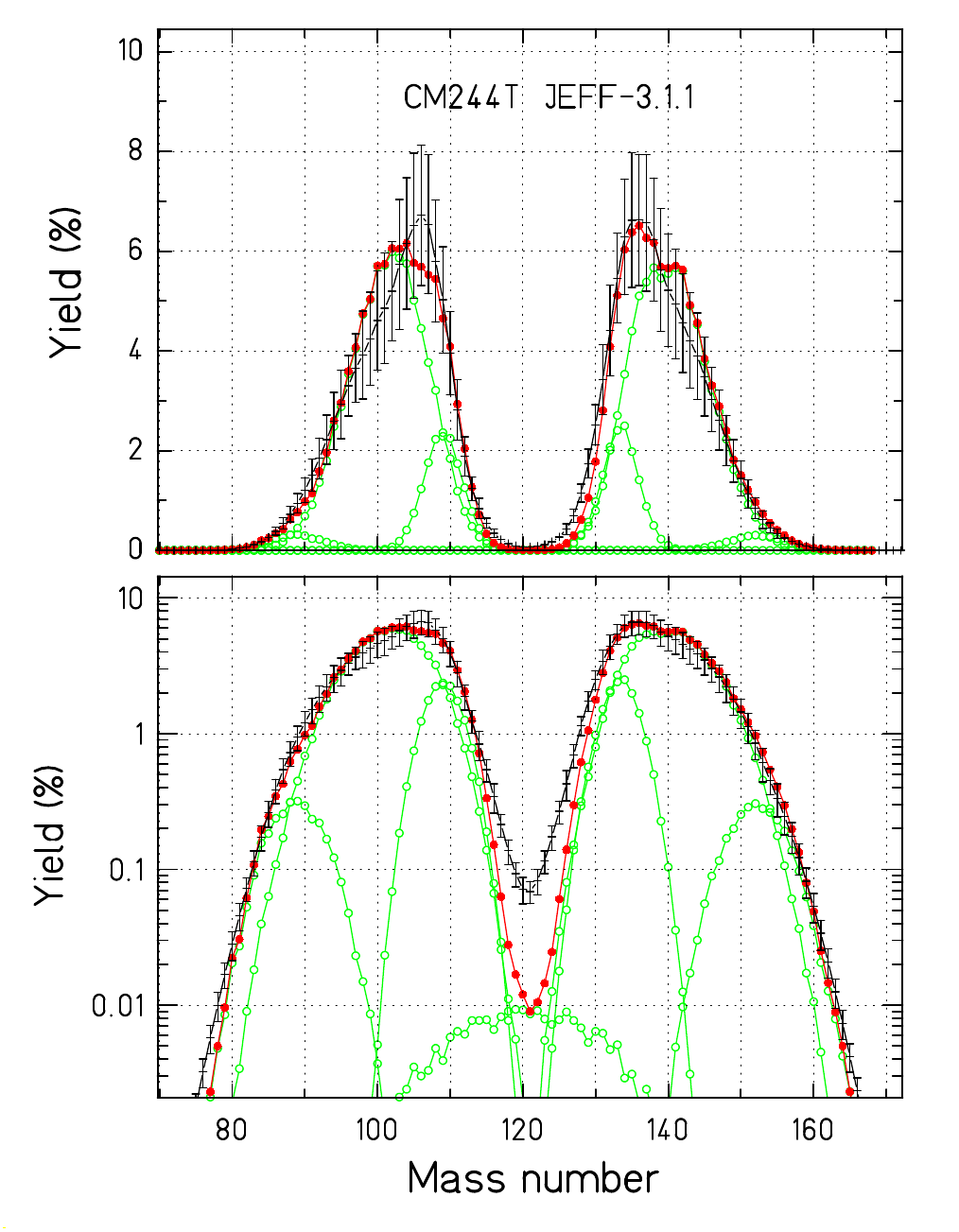}
\caption{Mass yields of $^{244}$Cm(n$_{\text{th}}$,f), linear (upper frame) and logarithmic (lower frame) scale. GEF result (red points) in comparison with JEFF-3.1.1 (black symbols).} 
\label{CM244T-JEFF311}       
\end{figure}

\begin{figure}[h]
\centering
\includegraphics[width=0.36\textwidth]{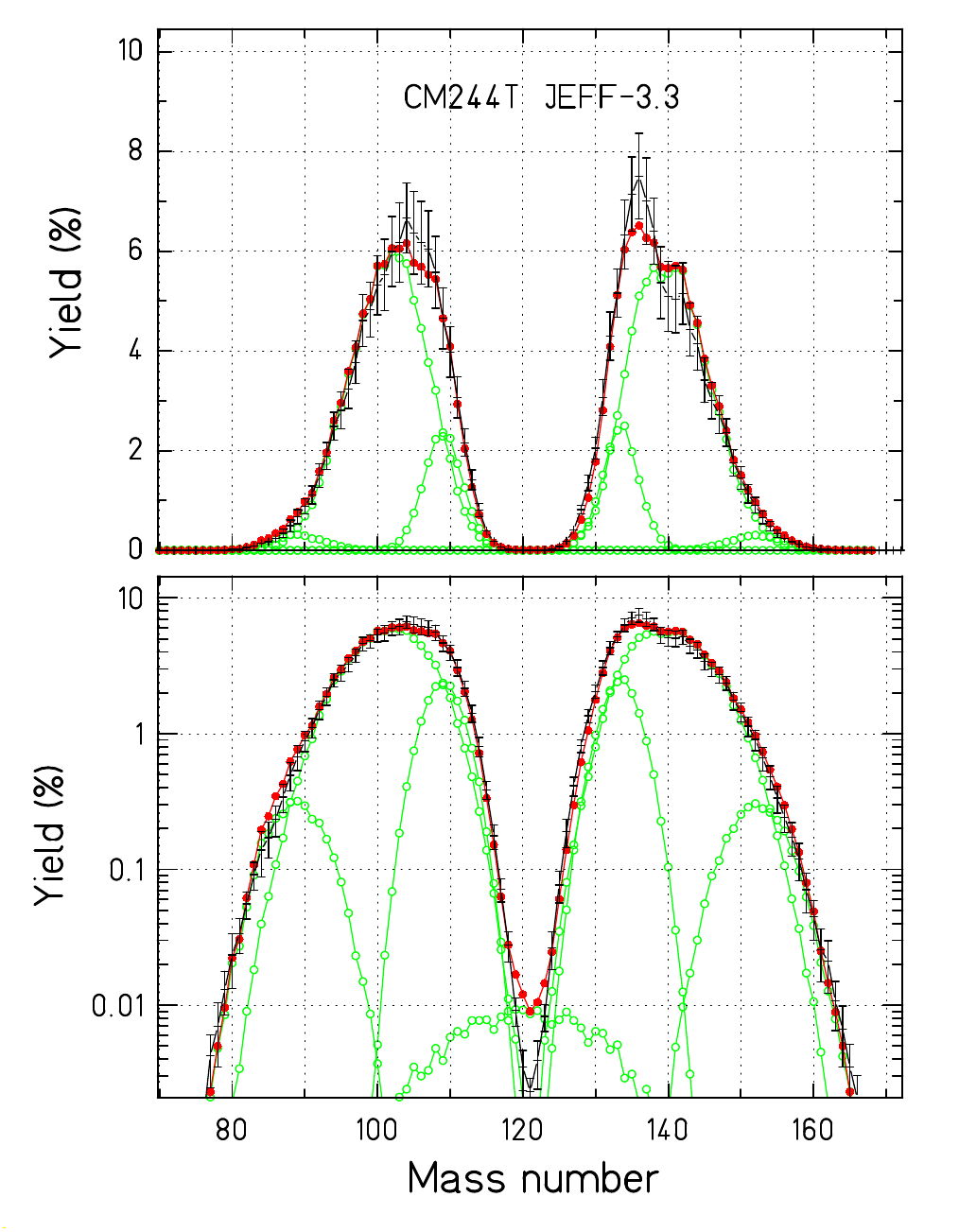}
\caption{Mass yields of $^{244}$Cm(n$_{\text{th}}$,f), linear (upper frame) and logarithmic (lower frame) scale. GEF result (red points) in comparison with JEFF-3.3 (black symbols).} 
\label{CM244T-JEFF33}       
\end{figure}

\paragraph{Mass yields of $^{244}$Cm(n$_{\text{th}}$,f):}
Figs. \ref{CM244T-JEFF311} and \ref{CM244T-JEFF33} show the comparison of the mass yields from GEF with the data from JEFF-3.1.1 and JEFF-3.3 for the system $^{244}$Cm(n$_{\text{th}}$,f). There are large deviations between the yields from JEFF-3.1.1 and the GEF results, namely at symmetry, but the discrepancies are appreciably reduced between the GEF yields and those of the more recent JEFF-3.3 evaluation.

\begin{figure}[h]
\centering
\includegraphics[width=0.36\textwidth]{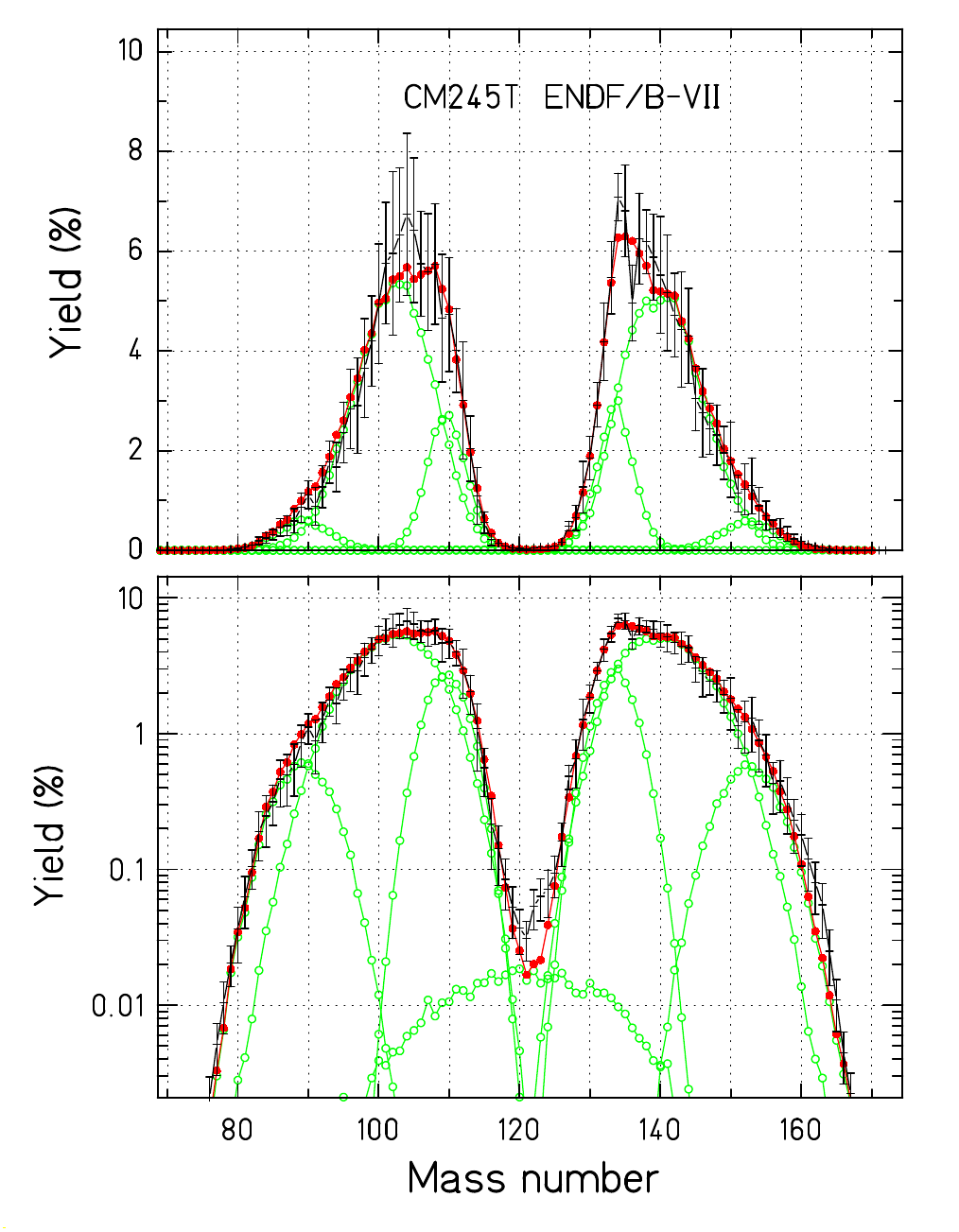}
\caption{Mass yields of $^{245}$Cm(n$_{\text{th}}$,f), linear (upper frame) and logarithmic (lower frame) scale. GEF result (red points) in comparison with ENDF/B-VII (black symbols).} 
\label{CM245T-ENDF}       
\end{figure}

\begin{figure}[h]
\centering
\includegraphics[width=0.36\textwidth]{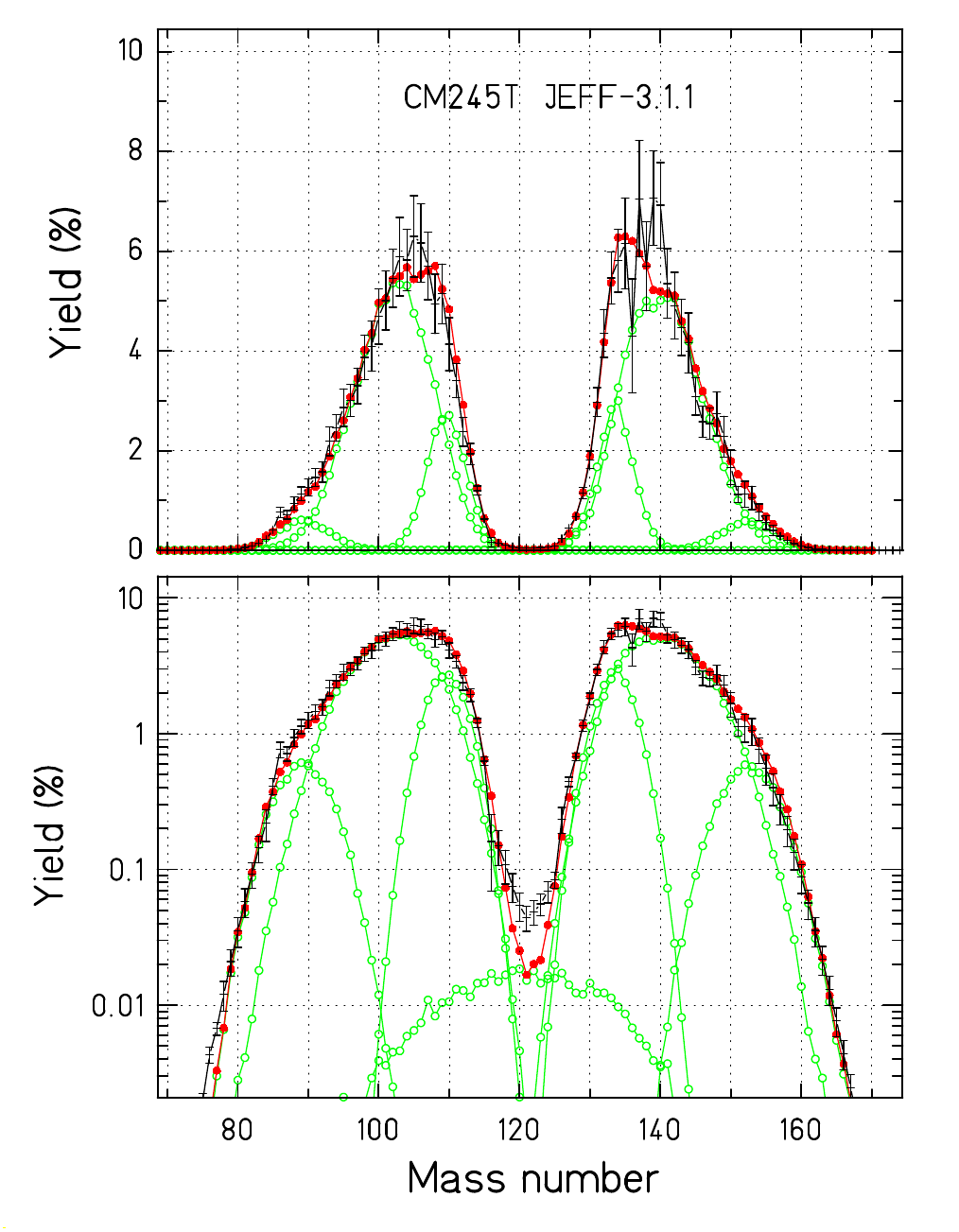}
\caption{Mass yields of $^{245}$Cm(n$_{\text{th}}$,f), linear (upper frame) and logarithmic (lower frame) scale. GEF result (red points) in comparison with JEFF-3.1.1 (black symbols). } 
\label{CM245T-JEFF311}       
\end{figure}

\begin{figure}[h]
\centering
\includegraphics[width=0.36\textwidth]{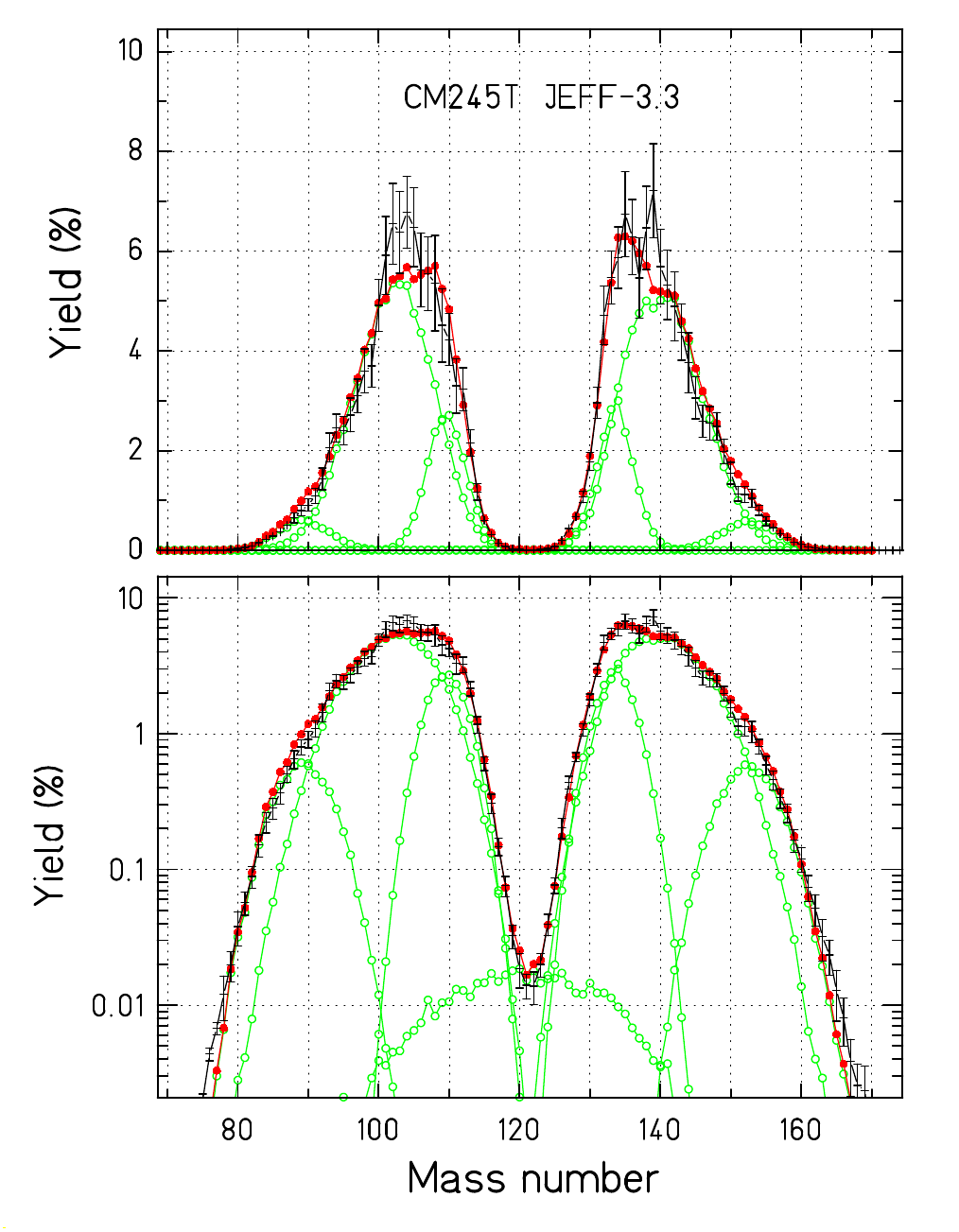}
\caption{Mass yields of $^{245}$Cm(n$_{\text{th}}$,f), linear (upper frame) and logarithmic (lower frame) scale. GEF result (red points) in comparison with JEFF-3.3 (black symbols). } 
\label{CM245T-JEFF33}       
\end{figure}

\paragraph{Mass yields of $^{245}$Cm(n$_{\text{th}}$,f):}
Figs. \ref{CM245T-ENDF}, \ref{CM245T-JEFF311} and \ref{CM245T-JEFF33} show the comparison of the mass yields from GEF with the data from ENDF/B-VII, JEFF-3.1.1 and JEFF-3.3 for the system $^{245}$Cm(n$_{\text{th}}$,f). The evaluated distributions are fairly well reproduced. There are some deviations between GEF and the one or the other evaluation, but they are not systematical, except that the two mass peaks from GEF are less sharp. Best agreement is found between GEF and the most recent JEFF-3.3 evaluation.

\paragraph{Mass yields of $^{249}$Cf(n$_{\text{th}}$,f):} 

\begin{figure}[h]
\centering
\includegraphics[width=0.36\textwidth]{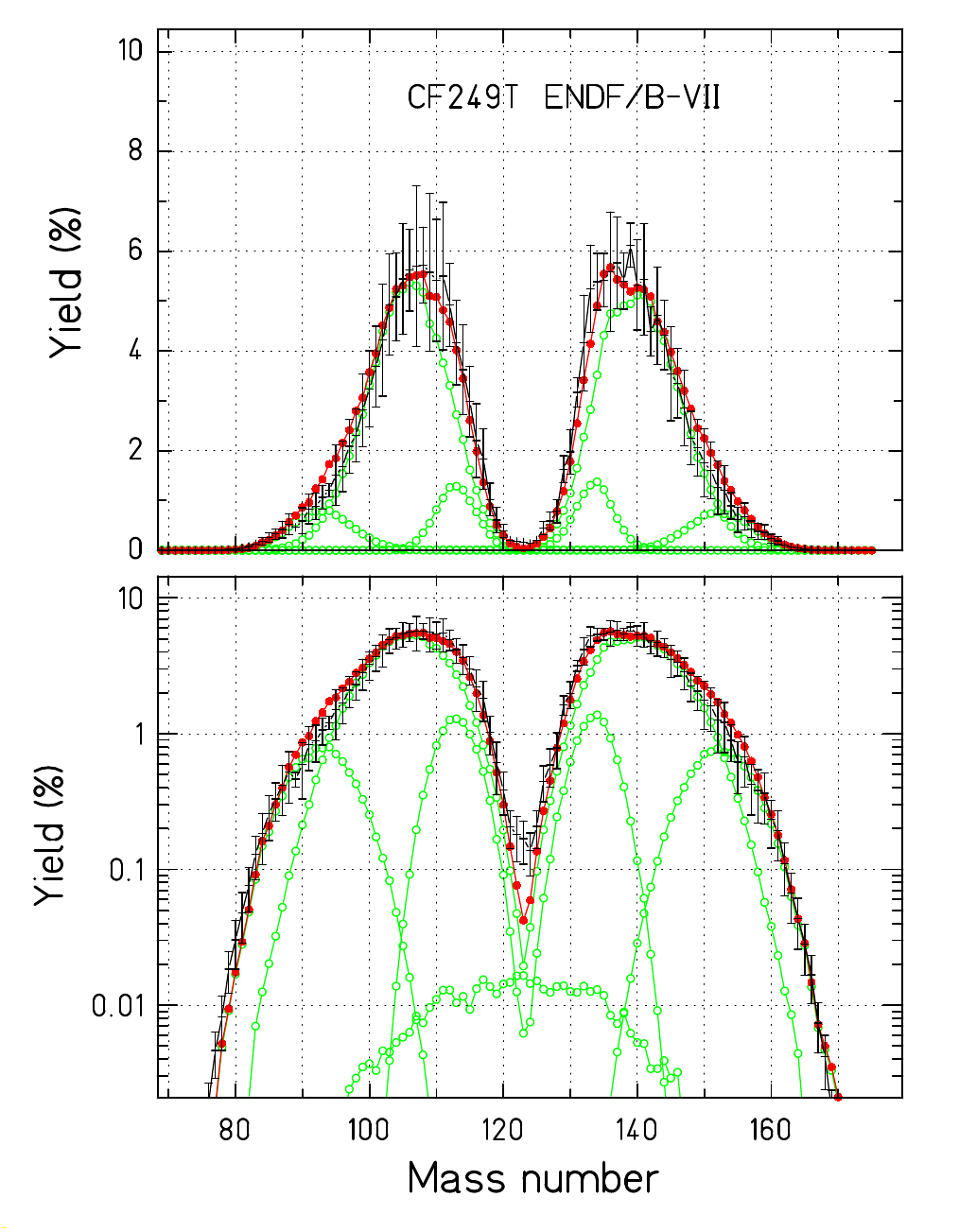}
\caption{Mass yields of $^{249}$Cf(n$_{\text{th}}$,f), linear (upper frame) and logarithmic (lower frame) scale.
GEF result (red points) in comparison with ENDF/B-VII (black symbols).} 
\label{CF249T-ENDF} 
\end{figure}

\begin{figure}[h]
\centering
\includegraphics[width=0.36\textwidth]{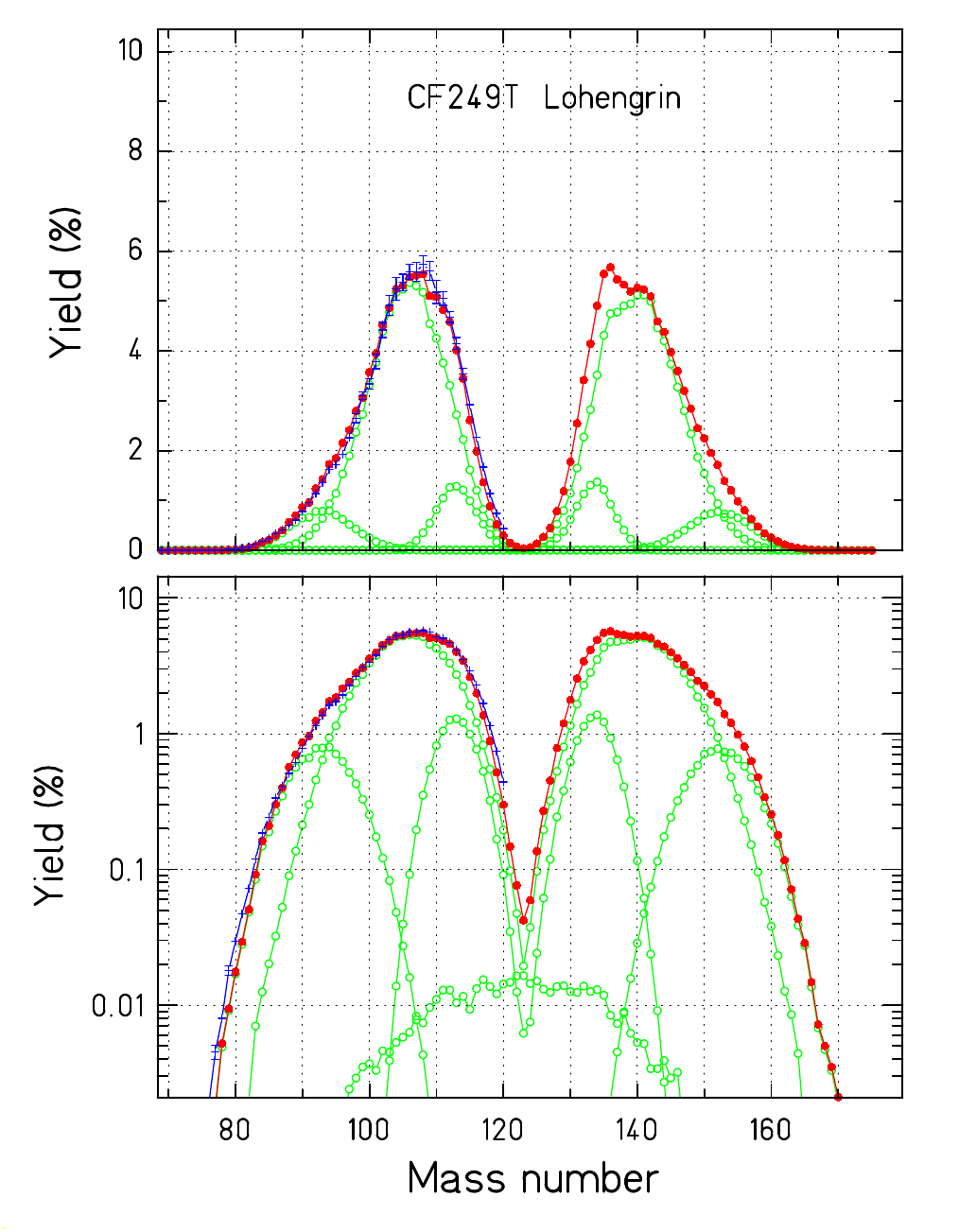}
\caption{Mass yields of $^{249}$Cf(n$_{\text{th}}$,f), linear (upper frame) and logarithmic (lower frame) scale, GEF result (red points) in comparison with 
the result from an experiment \cite{Djebara89,Hentzschel94} at LOHENGRIN (blue symbols). } 
\label{CF249T-LOHENGRIN}       
\end{figure}

The fission-product mass distributions of the system $^{249}$Cf(n$_{\text{th}}$,f) from both ENDF/B-VII and from a LOHENGRIN experiment are rather well reproduced by GEF, see
Figs. \ref{CF249T-ENDF} and  \ref{CF249T-LOHENGRIN}. One can observe a slight underestimation for the lightest masses in both figures. 
The LOHENGRIN experiment, which
covers only the light part, has provided data with very small uncertainties. Therefore, these data represent an especially stringent test case. There is a remarkably good agreement of the GEF result with the mass yields around the light peak, except the yield for A=109, which is somewhat underestimated in the calculation. 

\clearpage

\subsubsection{Treatment of energy distributions: the case of $^{238}$U(n$_{\text{fast}}$,f)} \label{U238F}
In some cases, the initial excitation energy of the fissioning nucleus extends over a range where the variation of the fission-product yields as a function of excitation energy cannot be neglected. 
A realistic description of the dependence of the fission process on the initial excitation energy is mandatory for obtaining reliable results.
In these cases, a series of theoretical calculations with a sequence of excitation energies must be performed, and the results must be added up with the appropriate weights. 

In the following, we present the fast-neutron-induced fission of $^{238}$U as an example for illustrating the procedural method. 
Note that also the other nuclei ($^{235}$U, $^{239,241}$Pu) that are considered in this work for the production of antineutrinos in a reactor are exposed to neutrons in a rather broad energy range. 
However, due to the large fission cross section of these fissile nuclei at low neutron energies, the fission yields are well represented by assuming thermal-neutron-induced fission. 

\begin{figure}[h]
\centering
\includegraphics[width=0.36\textwidth]{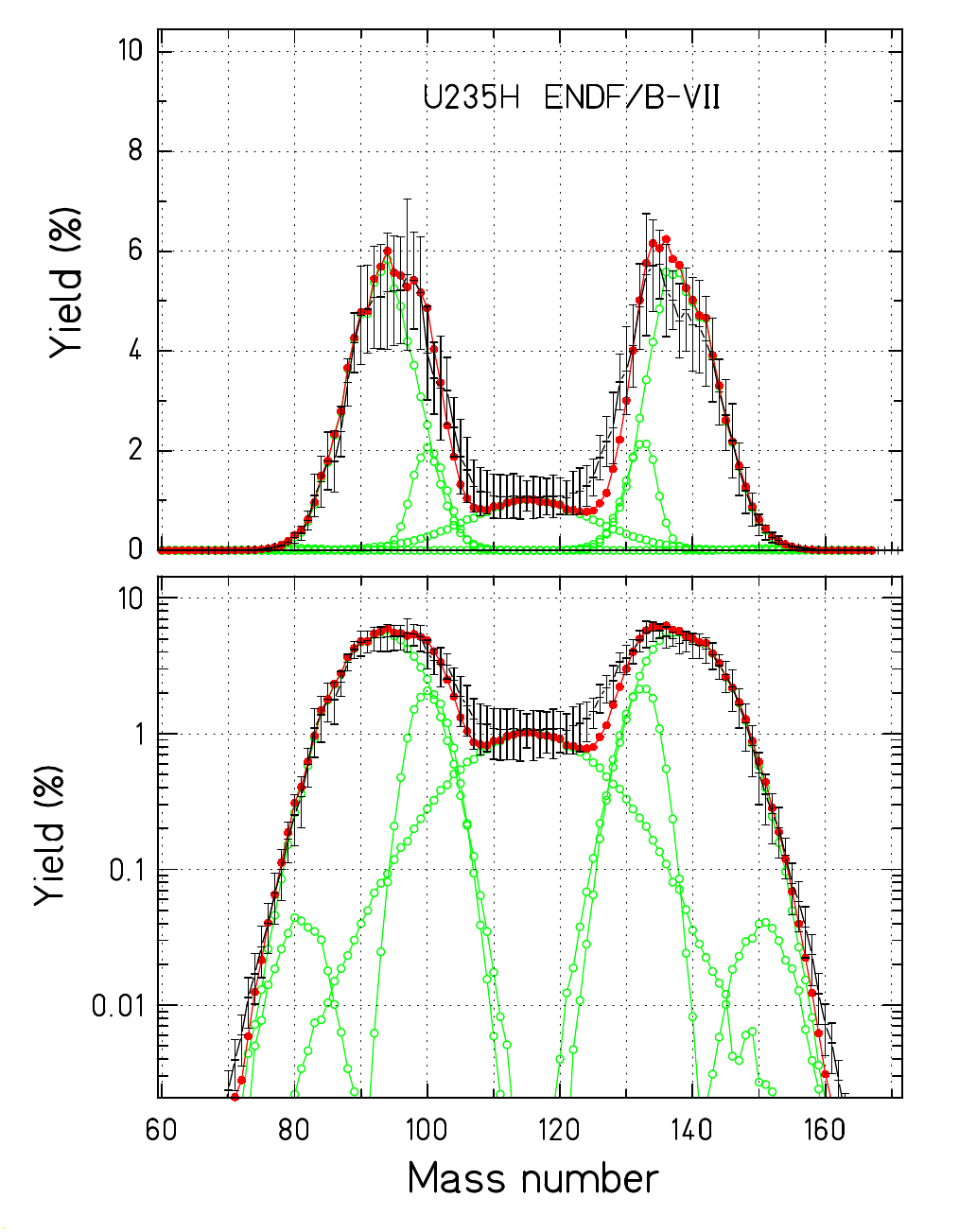}
\caption{Mass yields of $^{235}$U(n,f), $E_\text{n} = 14$ MeV, linear (upper frame) and logarithmic (lower frame) scale. GEF result (red points) in comparison with ENDF/B-VII (black symbols). } 
\label{U235H-ENDF}       
\end{figure}

\begin{figure}[h]
\centering
\includegraphics[width=0.36\textwidth]{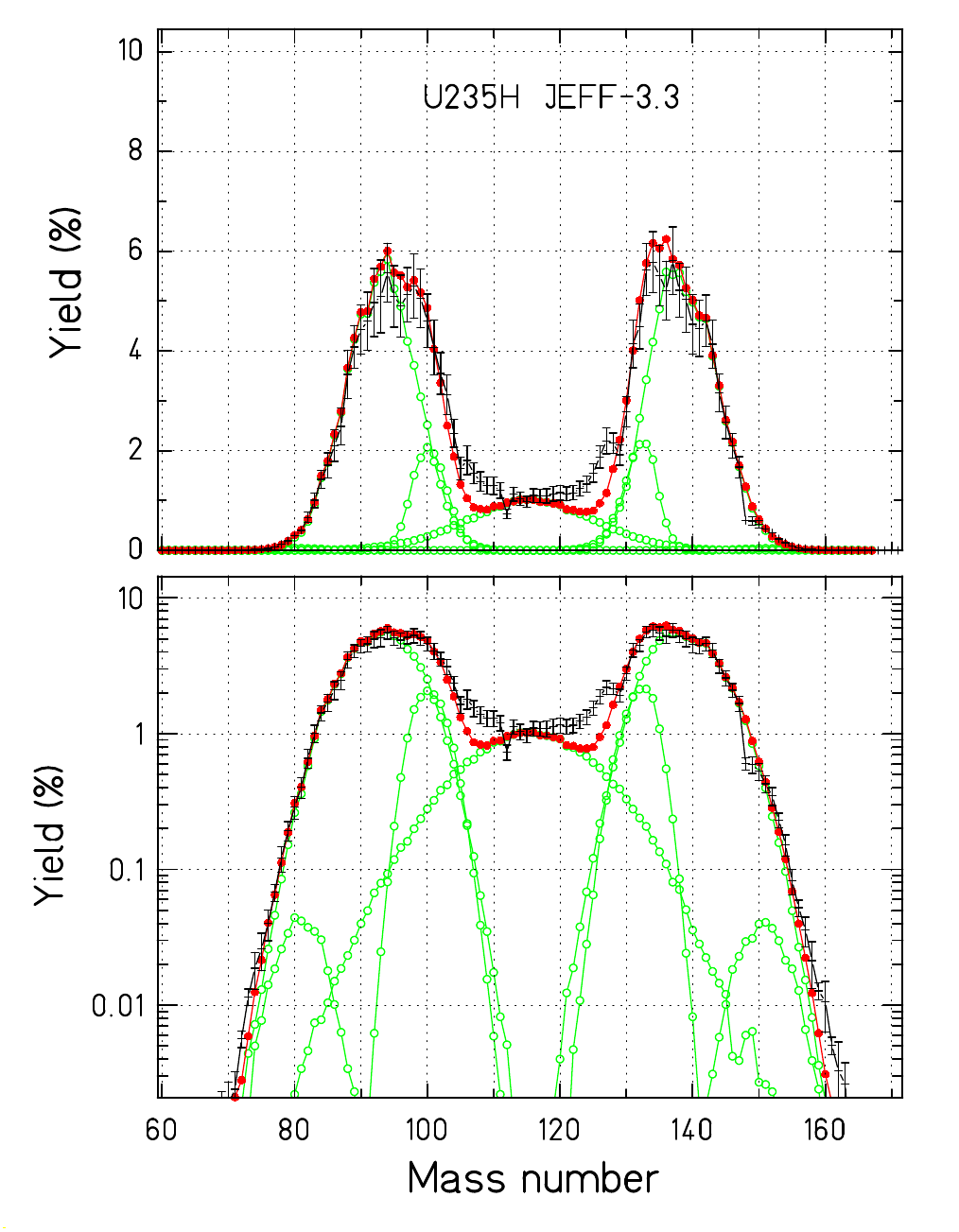}
\caption{Mass yields of $^{235}$U(n,f), $E_\text{n} = 14$ MeV, linear (upper frame) and logarithmic (lower frame) scale. GEF result (red points) in comparison with JEFF-3.3 (black symbols). } 
\label{U235H-JEFF33}       
\end{figure}

\begin{figure}[h]
\centering
\includegraphics[width=0.36\textwidth]{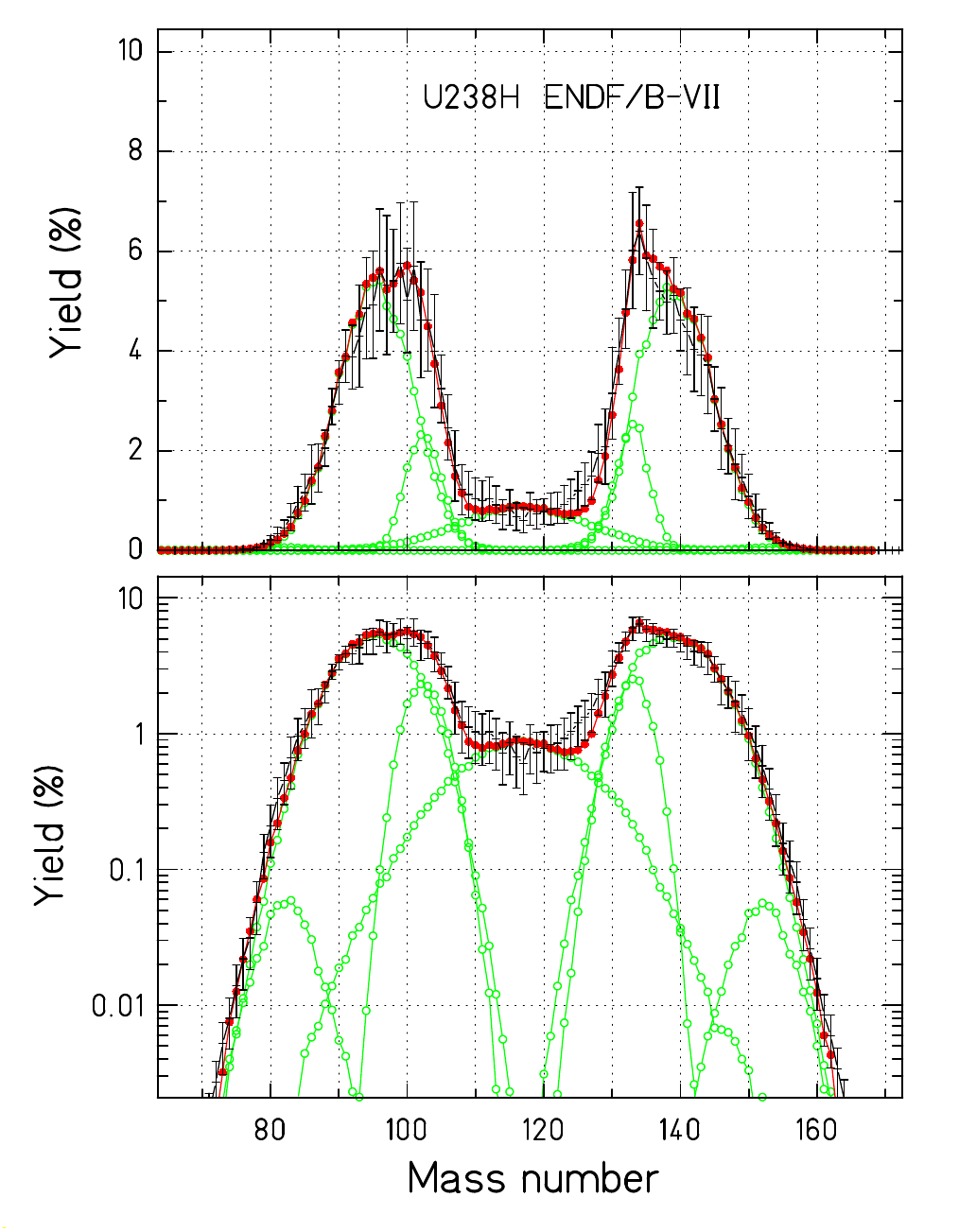}
\caption{Mass yields of $^{238}$U(n,f), $E_\text{n} = 14$ MeV, linear (upper frame) and logarithmic (lower frame) scale. GEF result (red points) in comparison with ENDF/B-VII (black symbols). } 
\label{U238H-ENDF}       
\end{figure}

\begin{figure}[h]
\centering
\includegraphics[width=0.36\textwidth]{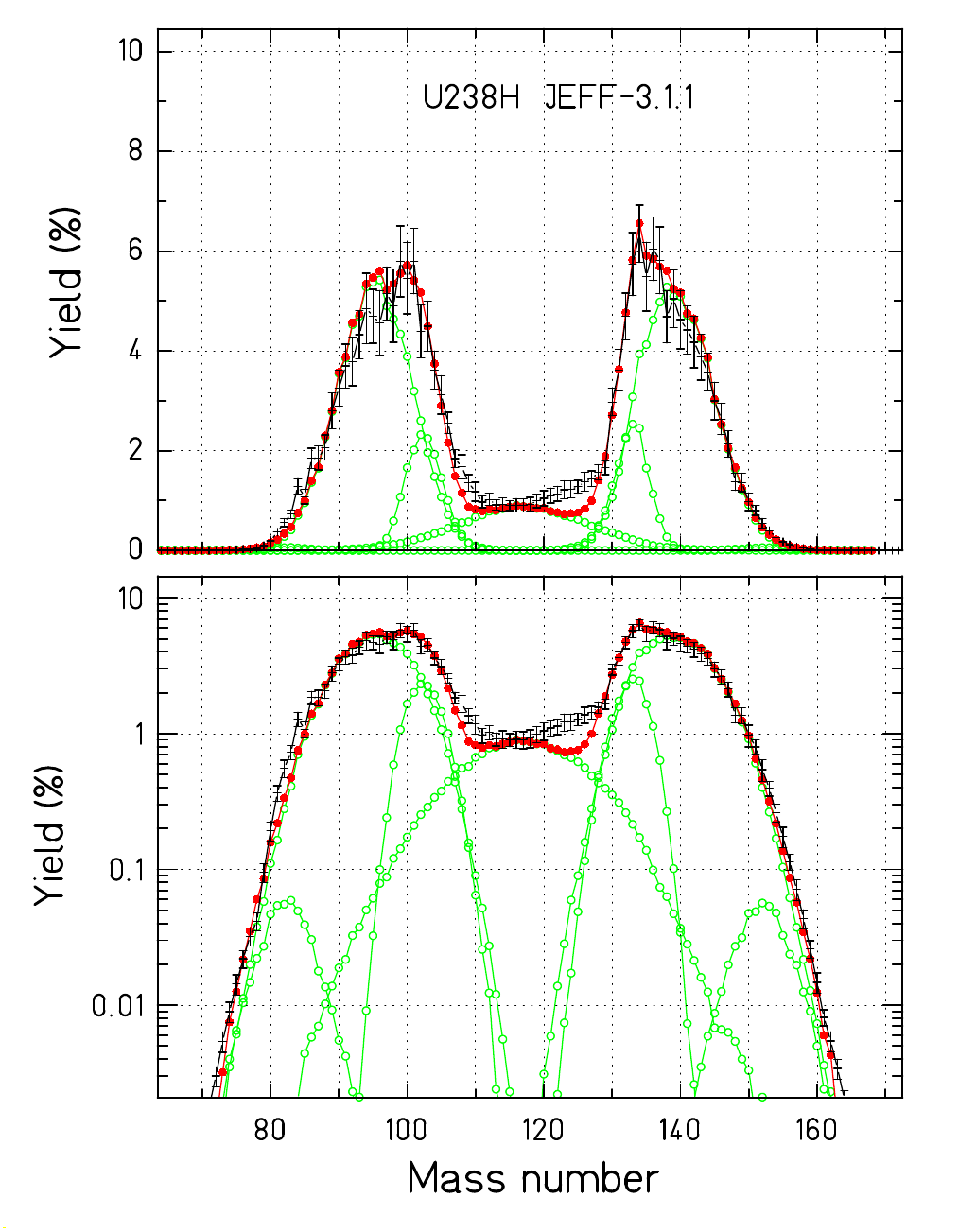}
\caption{Mass yields of $^{238}$U(n,f), $E_\text{n} = 14$ MeV, linear (upper frame) and logarithmic (lower frame) scale. GEF result (red points) in comparison with JEFF-3.1.1 (black symbols). } 
\label{U238H-JEFF311}       
\end{figure}

\begin{figure}[h]
\centering
\includegraphics[width=0.36\textwidth]{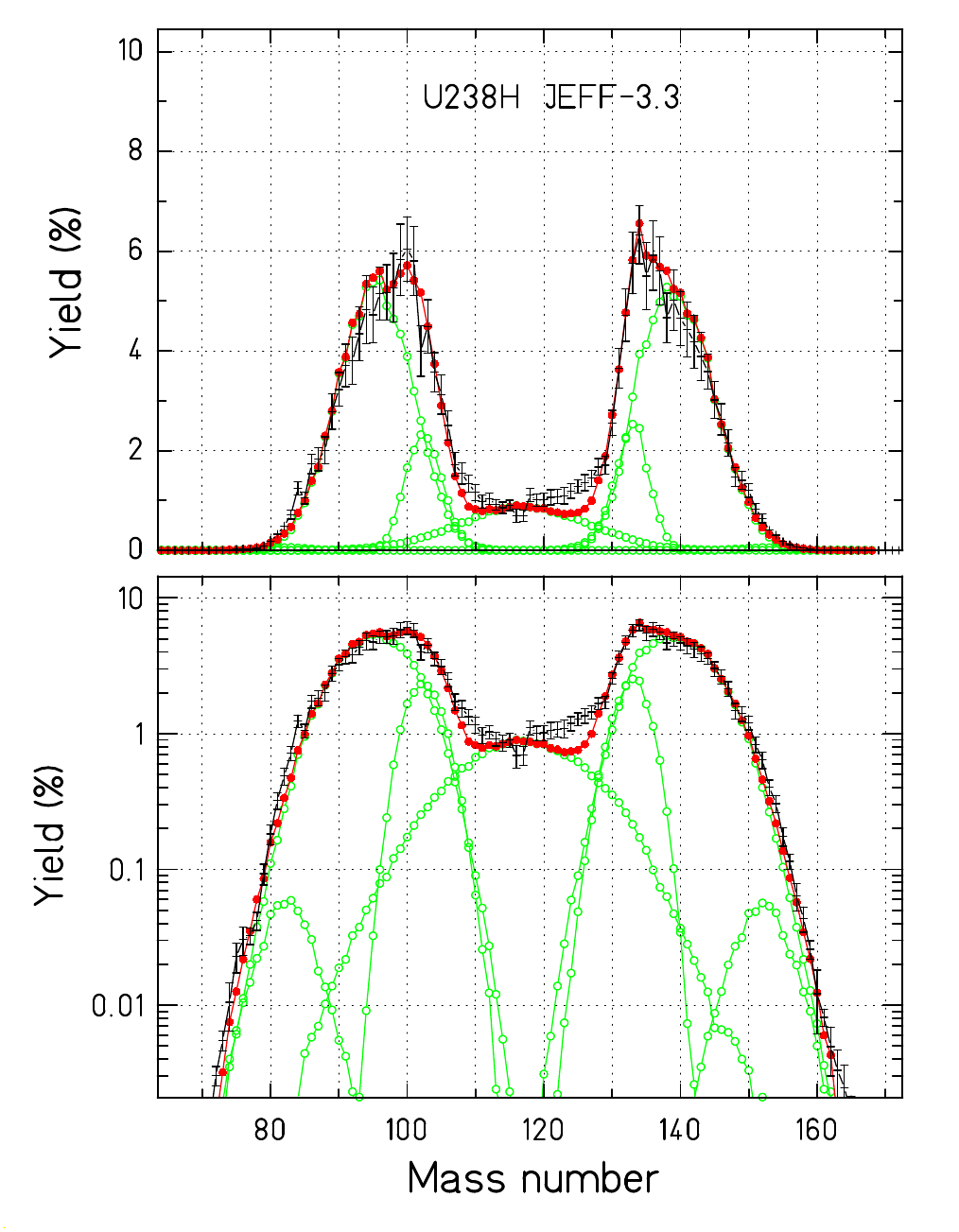}
\caption{Mass yields of $^{238}$U(n,f), $E_\text{n} = 14$ MeV, linear (upper frame) and logarithmic (lower frame) scale. GEF result (red points) in comparison with JEFF-3.3 (black symbols). } 
\label{U238H-JEFF33}       
\end{figure}

The following data with a sharp initial excitation energy were used to benchmark the excitation energy dependence of the fission yields in GEF: The mass yields of
$^{235}$U(n$_{\text{th}}$,f), shown above, and $^{235}$U(n,f) with $E_\text{n} = 14$ MeV, shown in Figs. \ref{U235H-ENDF} and \ref{U235H-JEFF33} from the ENDF/B-VII and the JEFF-3.3 evaluation, respectively, document well the variation of the fission yields from thermal energies to 14 MeV. 
In addition, the mass yields of $^{238}$U(n,f) with $E_\text{n} = 14$ MeV, shown in Figs. \ref{U238H-ENDF} to \ref{U238H-JEFF33} were used.

One can observe a rather good agreement between the evaluated mass yields and the GEF results at fixed $E_\text{n}$. The growth of the symmetric channel with increasing energy, as well as the shift towards symmetry and the broadening of the asymmetric modes are well reproduced by GEF.
The constraints of the theoretical framework do not allow to reproduce the data exactly, and some minor deviations can be observed. Moreover, also the evaluations do not agree with each other.
In $^{235}$U(n,f) with $E_\text{n} = 14$ MeV, GEF seems to slightly overestimate the yield of the symmetric mode, and its shape is not exactly reproduced. 
On the empirical side, in the mass yields of JEFF-3.3, there appear several apparently erratic deviations (at $A$ = 112, $A$ = 129, and $A$ = 148 for $^{235}$U(n,f) and around $A$ = 76, at $A$ = 85, $A$ = 102, $A$ = 116, and $A$ = 117 for $^{238}$U(n,f)) from the smooth behavior of the ENDF evaluation and of GEF, which are probably not realistic. 
For $^{238}$U(n,f) at the same neutron energy, GEF agrees with ENDF/B-VII, while the JEFF evaluations show an unexpected asymmetry near the symmetric valley. 
In summary, it often seems to be very difficult to decide whether one or the other evaluation or the GEF result provides the more reliable value for a specific mass yield in a specific case.


\begin{figure}[h]
\centering
\includegraphics[width=0.365\textwidth]{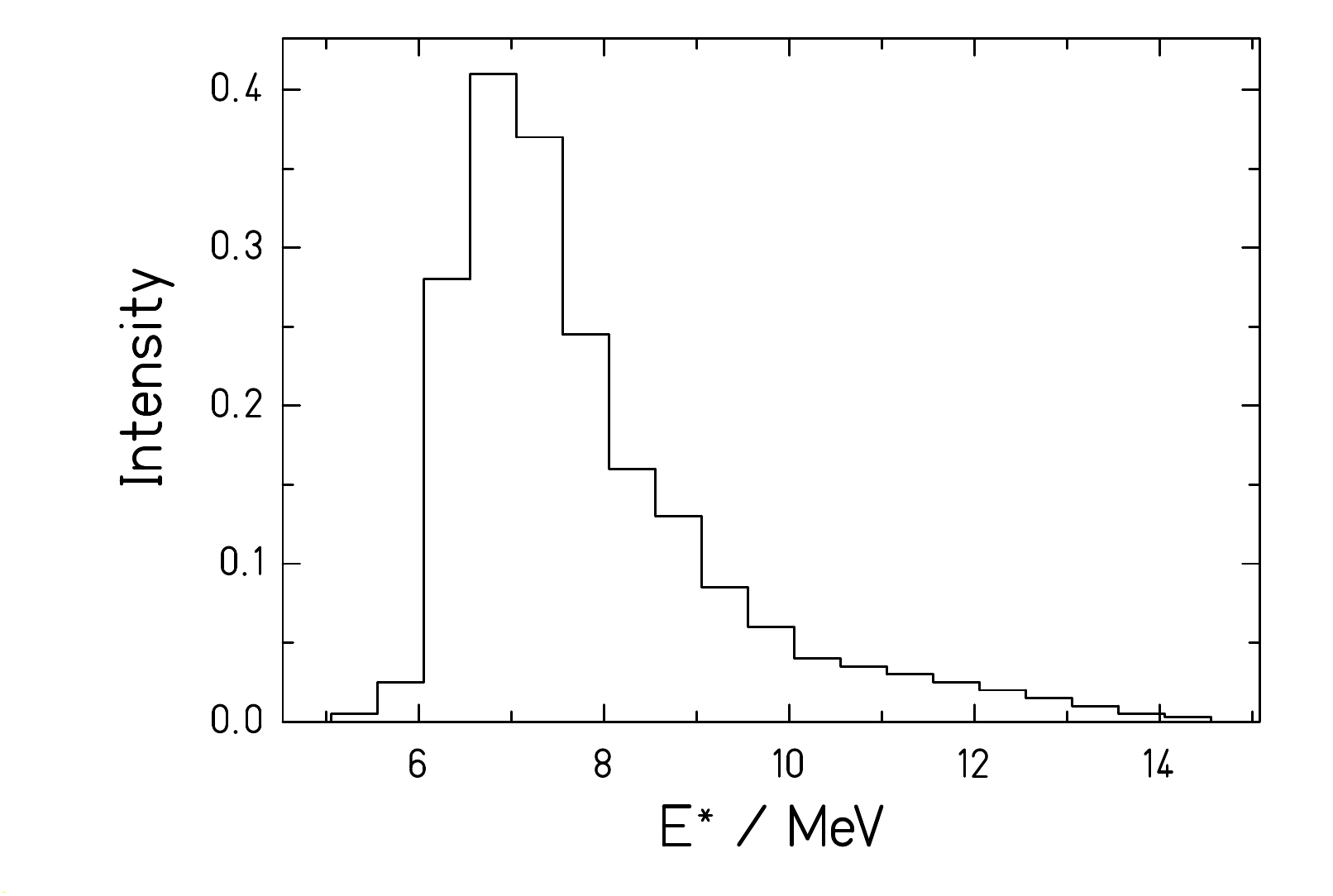}
\caption{Initial excitation energies of the fission events in fission of $^{238}$U in a PWR, rebinned from \cite{Kern12}. The GEF calculations were performed with a series of sharp energy values in the centres of the bins.} 
\label{EN-U238F}       
\end{figure}

The GEF calculation of the mass yields for the system $^{238}$U(n$_{\text{fast}}$,f) was performed with the distribution of initial neutron energies that lead to fission, taken from an estimation in Ref.~\cite{Kern12}. It is the spectrum of partly moderated fission neutrons in a PWR, multiplied with the corresponding fission cross section. The corresponding initial excitation energies are shown in figure \ref{EN-U238F}. 

\begin{figure}[h]
\centering
\includegraphics[width=0.36\textwidth]{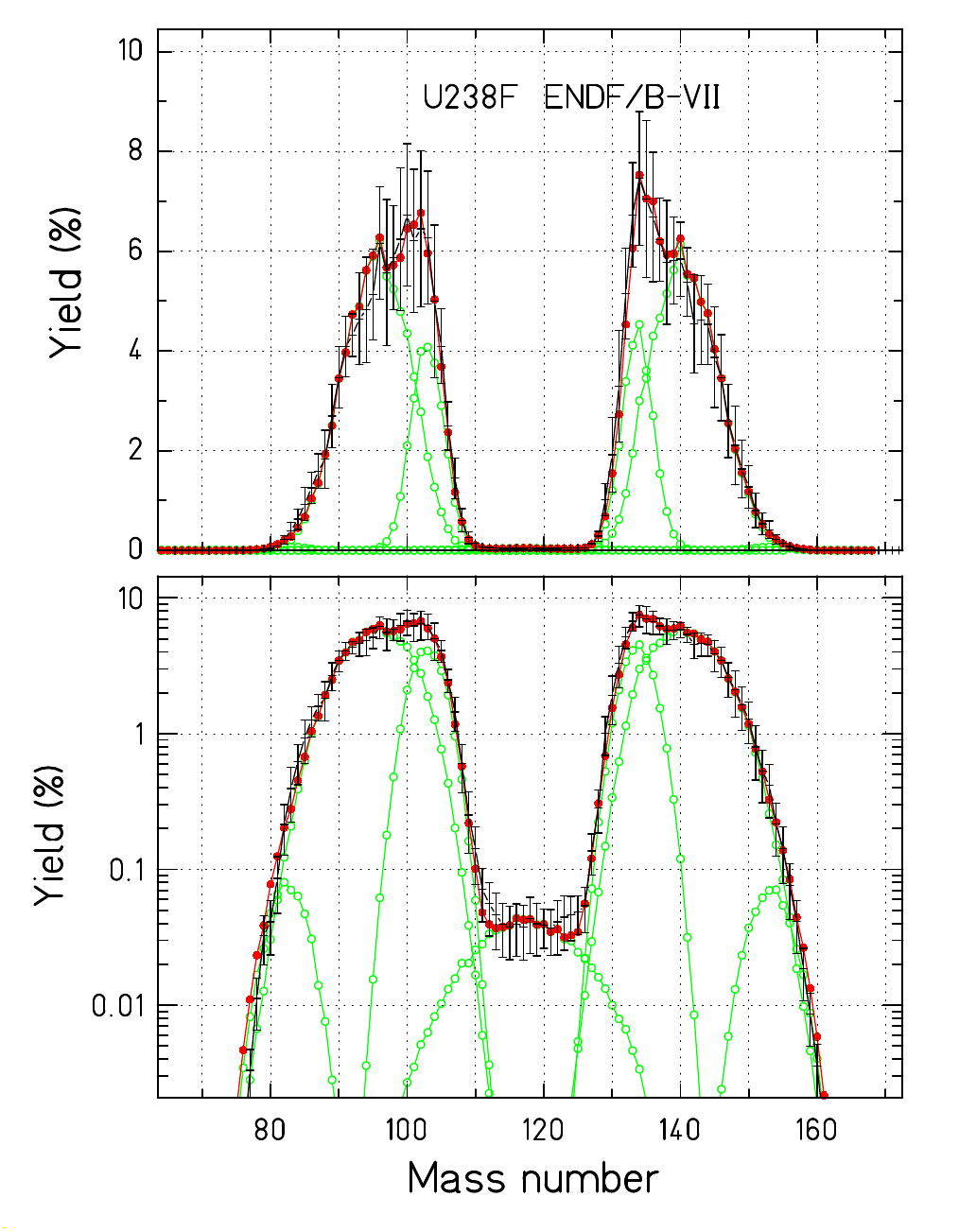}
\caption{Mass yields of $^{238}$U(n$_{\text{fast}}$,f), linear (upper frame) and logarithmic (lower frame) scale. GEF result (red points) in comparison with ENDF/B-VII (black symbols). } 
\label{U238F-ENDF}       
\end{figure}

\begin{figure}[h]
\centering
\includegraphics[width=0.36\textwidth]{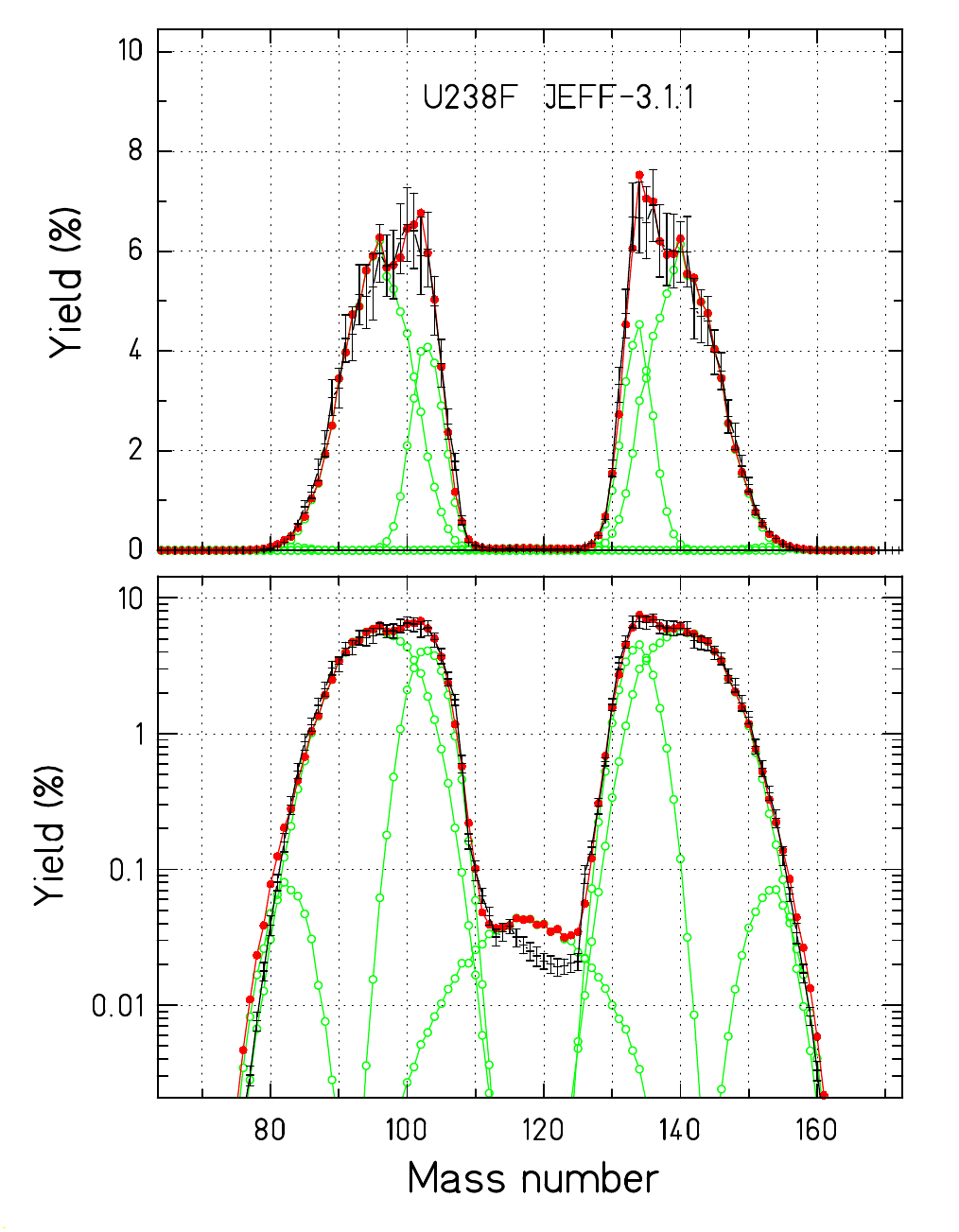}
\caption{Mass yields of $^{238}$U(n$_{\text{fast}}$,f), linear (upper frame) and logarithmic (lower frame) scale. GEF result (red points) in comparison with JEFF-3.1.1 (black symbols). } 
\label{U238F-JEFF311}       
\end{figure}

\begin{figure}[h]
\centering
\includegraphics[width=0.36\textwidth]{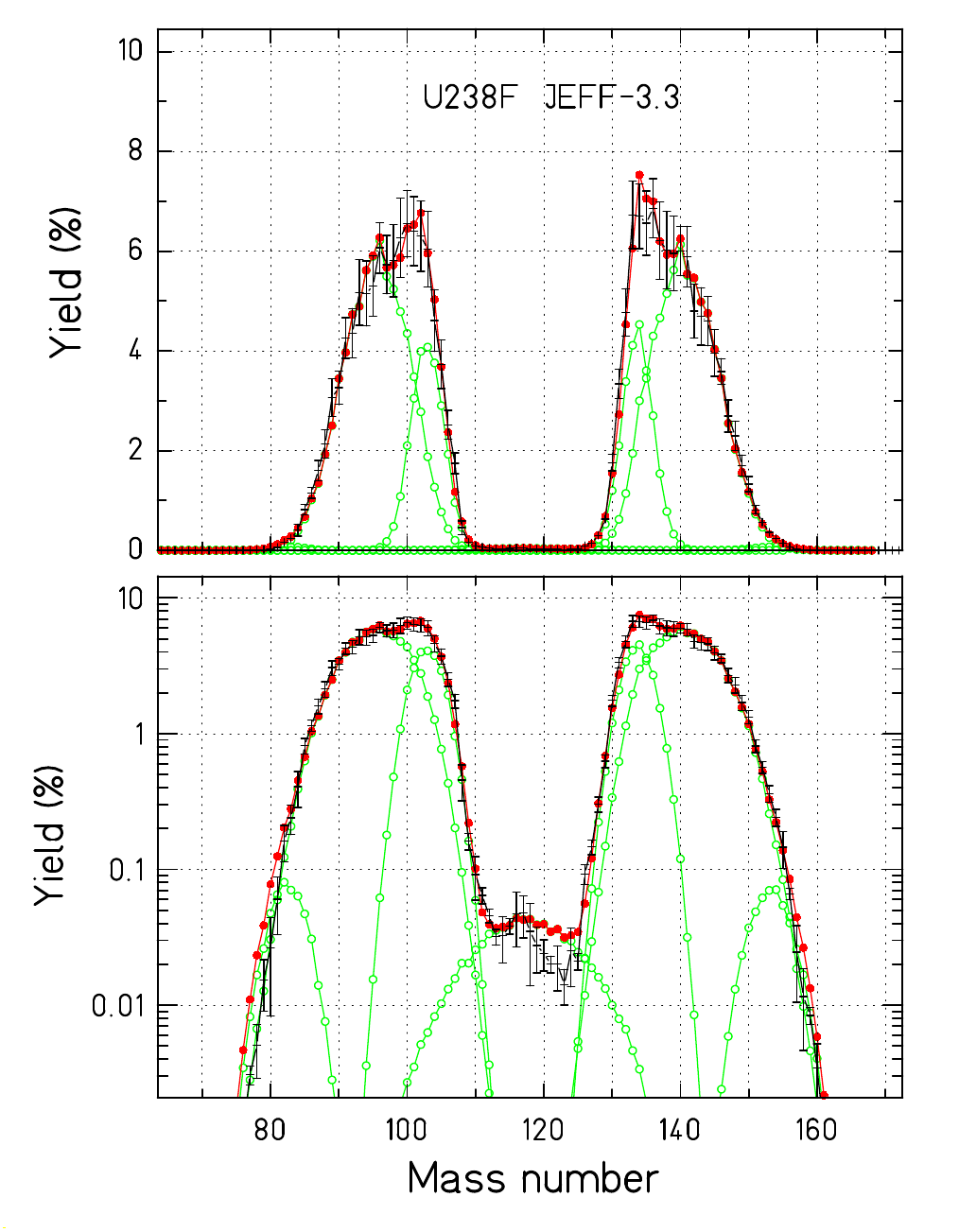}
\caption{Mass yields of $^{238}$U(n$_{\text{fast}}$,f), linear (upper frame) and logarithmic (lower frame) scale. GEF result (red points) in comparison with JEFF-3.3 (black symbols). } 
\label{U238F-JEFF33}       
\end{figure}

The result of the calculation for the mass yields of $^{238}$U(n$_{\text{fast}}$,f) is compared with different evaluations in figures \ref{U238F-ENDF},  \ref{U238F-JEFF311}, and \ref{U238F-JEFF33}. The yields of the different evaluations are rather well reproduced by the GEF calculation. (The discrepancies between the PROFIL experiment and GEF, reported in \cite{Noguere16}, do not appear anymore with the latest GEF version due to the new adjustment of the model parameters in GEF Y2019/V1.2.) There is some overestimation of the yields below 0.1\% in the low-mass tail of the distribution, where the super-asymmetric fission channel contributes appreciably, while the complementary high-mass tail is well reproduced. The low mass yields around symmetry from GEF agree well with the ENDF evaluation, while the JEFF evaluations show an unexpectedly strong slope.

We would like to stress that a calculation with a sharp "mean" or "representative" value of the incident neutron energy deviates appreciably from the "exact" result, obtained with the full energy distribution.


\subsection{Problems and proposed solutions} \label{S6.2}

In this section, we compare the fission yields from different evaluations and from some LOHENGRIN experiments with the GEF results in cases of severe discrepancies. The comparisons are shown for all evaluations among the three considered in this work, which are available for the respective system.

\subsubsection{$A$=129 yield of $^{235}$U(n$_{\text{th}}$,f)} \label{S6.2.1}

As a semi-empirical model, GEF relies on reliable and accurate data. 
The inclusion of erroneous data in the adjustment of GEF parameters leads to an aberrant behavior and to false predictions of the model. 
As an illustration for these difficulties, we have a closer view on the mass yield of $A = 129$ in the thermal-neutron-induced fission of $^{235}$U.

\begin{table}[h]
\begin{center}
\caption{Empirical values for the $A = 129$ yield of $^{235}$U(n$_{\text{th}}$,f).}
\label{A129yields}
~\\
\begin{tabular}{|c|c|c|} \hline \hline
Value         &  Uncertainty  & Reference           \\ \hline \hline
 0.610      &  4.9\%      &  \cite{Fowler74}       \\ \hline 
 0.804      &  5.0\%      &  \cite{Diiorio77}      \\ \hline
 0.817      &  5.8\%      &  \cite{Thierens76}     \\ \hline
 0.543      & 0.045 (8.3\%)        &   ENDF/B-VII.0           \\ \hline
 0.543      & 0.045 (8.3\%)       &   ENDF/B-VIII.0        \\ \hline
 0.706      & 0.037 (5.2\%)       &   JEFF-3.1.1           \\ \hline
 0.814      & 0.058 (7.1\%)       &   JEFF-3.3             \\ \hline
 0.978      & 0.18 (18\%)             &   GEF-2019/1.2                  \\ \hline
\end{tabular}
\small{~\\~\\Note: Selection of measured and evaluated mass yields for a case with large scattering.
The GEF estimation is listed in addition.}
\end{center}
\end{table}


Table \ref{A129yields} shows that the measured and the evaluated values scatter strongly: The highest value is larger by a factor of 1.5 than the smallest one, while the indicated uncertainties of the different values are in the order of 5\% to 10\%.
In such cases, the evaluator or the developer of a semi-empirical model must make a decision on how to treat these data. For example, the uncertainty could be increased, the data could be disregarded completely, or a personal choice on the basis of additional arguments could be performed. 
Therefore, in an evaluation as well as in a semi-empirical model, there is unevitably a portion of subjective influence and decision. 
In fact, GEF is less vulnerable than an evaluation, because the inherent regularities help to identify such problematic cases, like the one illustrated in Table \ref{A129yields}. 
In this specific case, a singular value deviates strongly from the GEF results, while the neighboring mass yields show good agreement. This behavior is in sharp conflict with the concept of fission channels, which extend over several masses and, thus, exludes sharp local fluctations of this kind.  
Following this reasoning, 
figures \ref{U235T-ENDF}, \ref{U235T-JEFF311} and \ref{U235T-JEFF33} suggest that \emph{the larger values given in Table \ref{A129yields} are the more reliable ones}.


In the determination of the parameters of GEF-Y2019/V1.2, the mass yield of $A=129$ in $^{235}$U(n$_{\text{th}}$,f) was disregarded.


\subsubsection{Mass yields of $^{227}$Th(n$_{\text{th}}$,f)} \label{S6.2.2}
\begin{figure}[h]
\centering
\includegraphics[width=0.36\textwidth]{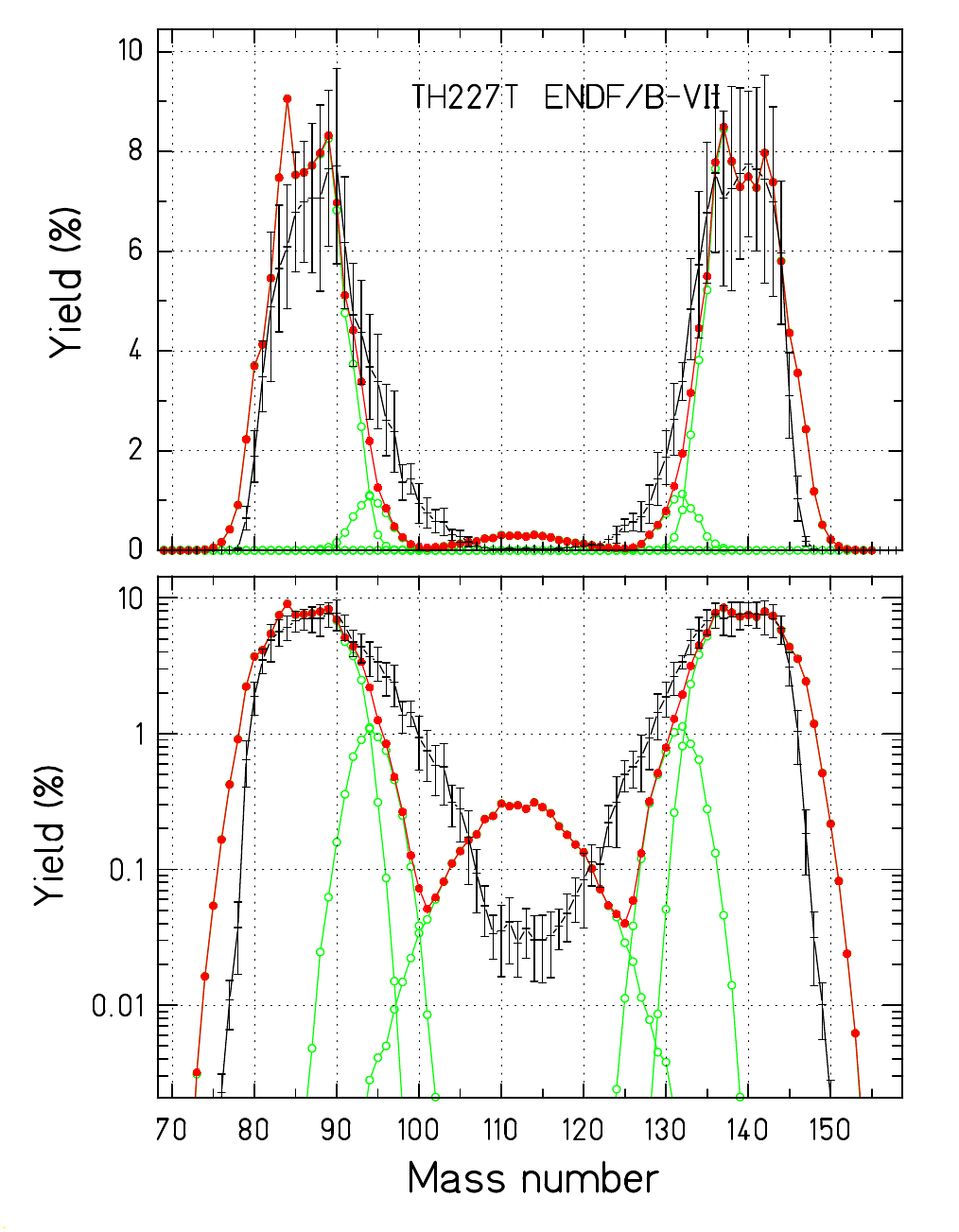}
\caption{Mass yields of $^{227}$Th(n$_{\text{th}}$,f), linear (upper frame) and logarithmic (lower frame) scale. GEF result (red points) in comparison with ENDF/B-VII (black symbols). } 
\label{TH227T-ENDF}       
\end{figure}

In figure \ref{TH227T-ENDF},
the mass yields of the system $^{227}$Th(n$_{\text{th}}$,f) in ENDF/B-VII deviate strongly from the GEF results almost over the whole distribution. In particular, in view of the relatively good reproduction of the mass yields of the close system $^{229}$Th(n$_{\text{th}}$,f), the shape proposed by ENDF/B-VII seems to be erroneous. \emph{We recommend to replace the mass yields, in particular between the asymmetric peaks, by the GEF results.} The relative yield of the symmetric fission channel, however, remains somewhat uncertain.

\subsubsection{Mass yields of $^{232}$U(n$_{\text{th}}$,f)} \label{S6.2.3}
\begin{figure}[h]
\centering
\includegraphics[width=0.36\textwidth]{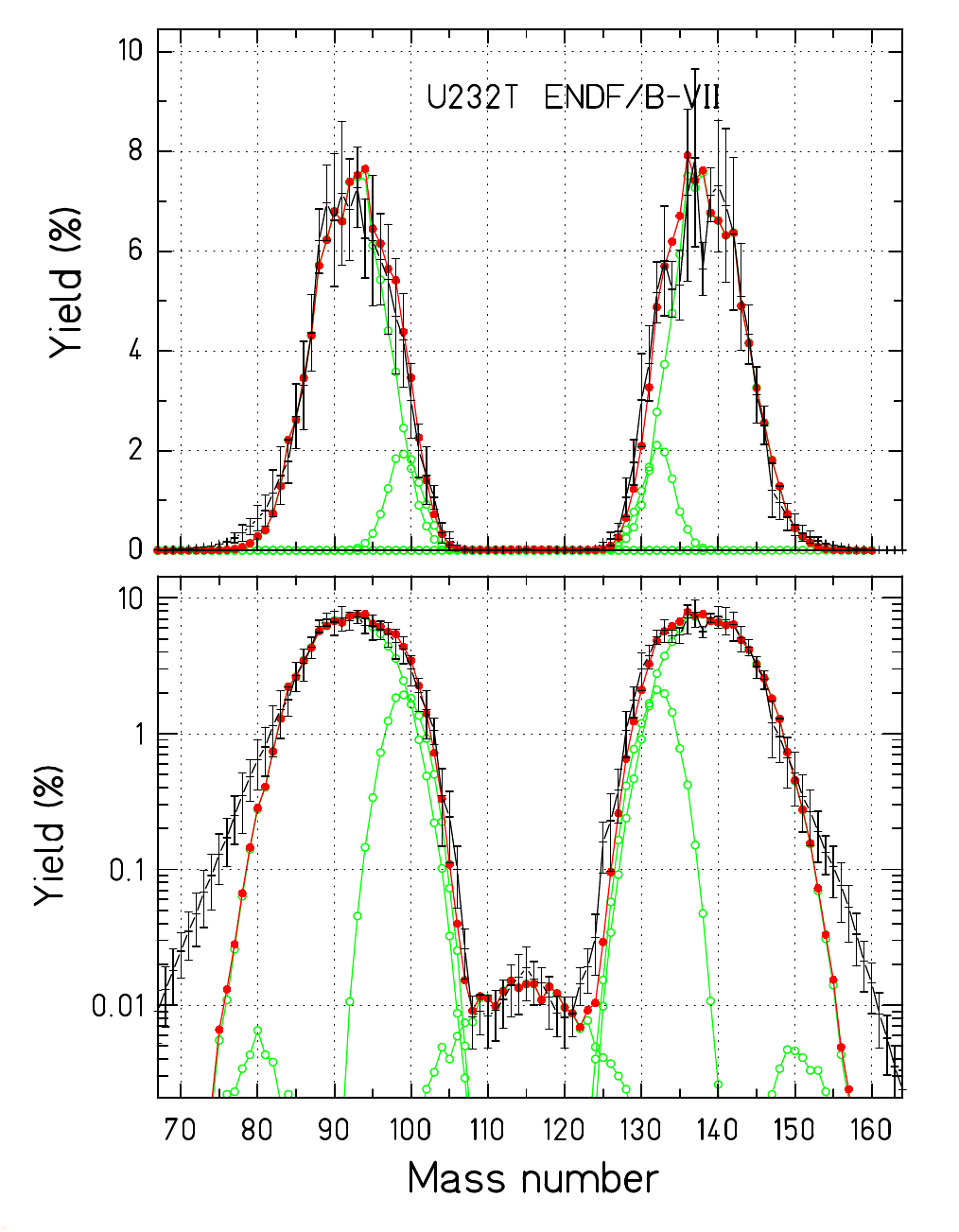}
\caption{Mass yields of $^{232}$U(n$_{\text{th}}$,f), linear (upper frame) and logarithmic (lower frame) scale, GEF result (red points) in comparison with ENDF/B-VII (black symbols). } 
\label{U232T-ENDF}       
\end{figure}

In figure \ref{U232T-ENDF},
the mass yields of the system $^{232}$U(n$_{\text{th}}$,f) in ENDF/B-VII deviate strongly from the GEF result in the wings at extreme mass asymmetry. \emph{We recommend to replace the mass yields for $A < 82$ and for $A > 150$ by the GEF results.}

\subsubsection{Mass yields of $^{236}$U(n$_{\text{th}}$,f)} \label{S6.2.4}

\begin{figure}[h]
\centering
\includegraphics[width=0.36\textwidth]{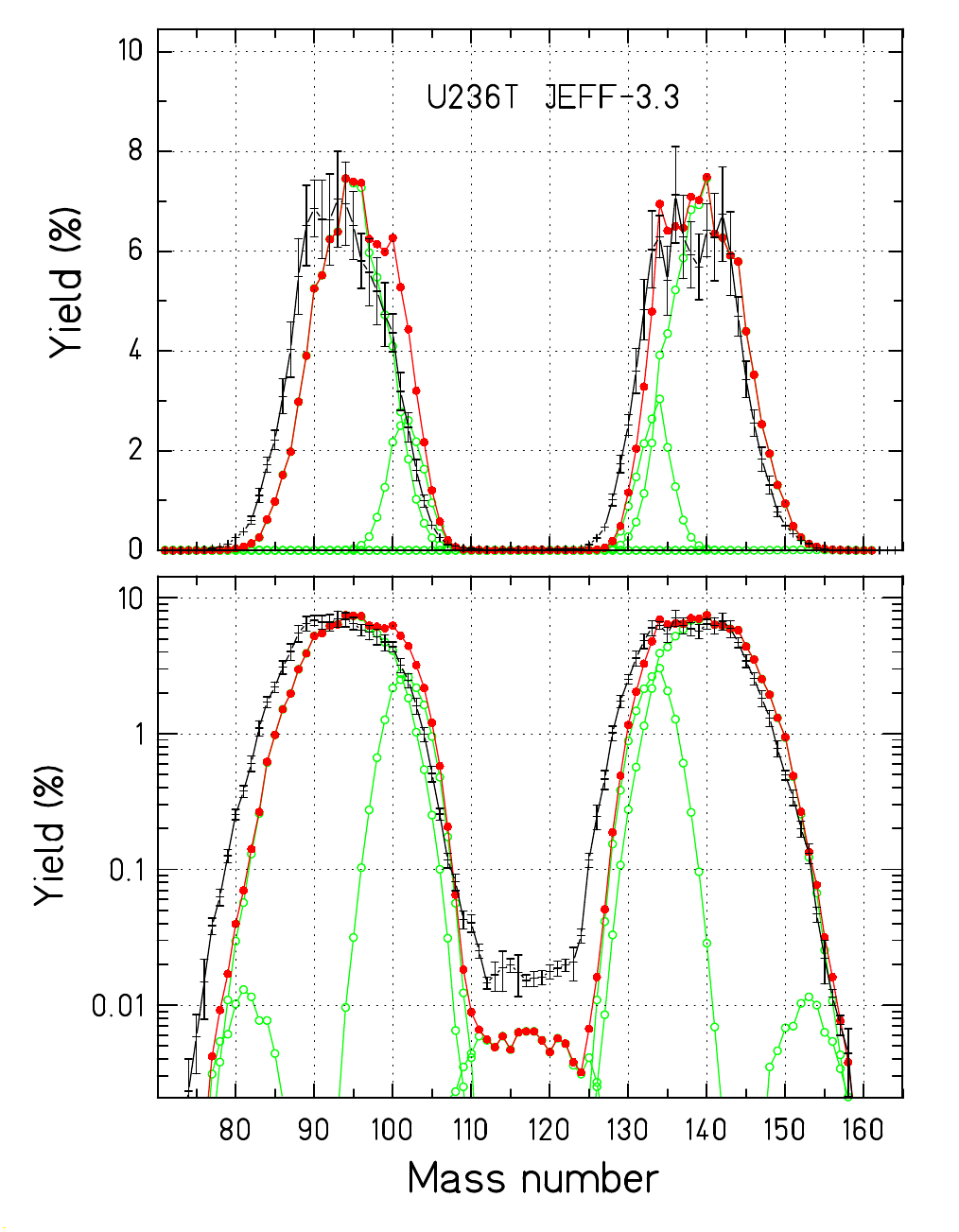}
\caption{Mass yields of $^{236}$U(n$_{\text{th}}$,f), linear (upper frame) and logarithmic (lower frame) scale, GEF result (red points) in comparison with JEFF-3.3 (black symbols). } 
\label{U236T-JEFF33}       
\end{figure}

In figure \ref{U236T-JEFF33}, the mass yields of JEFF-3.3 are compared with the GEF results. 
Apart from the discrepancy of the mass yields near symmetry, which may be explained by the fact that $^{236}$U is again thermally not fissile, 
there is a clear shift in almost all the wings of the mass-yield distribution of the system $^{236}$U(n$_{\text{th}}$,f). \emph{We recommend to replace the discrepant values by the GEF results.}

\subsubsection{Mass yields of $^{237}$Np(n$_{\text{th}}$,f)} \label{S6.2.5}

\begin{figure}[h]
\centering
\includegraphics[width=0.36\textwidth]{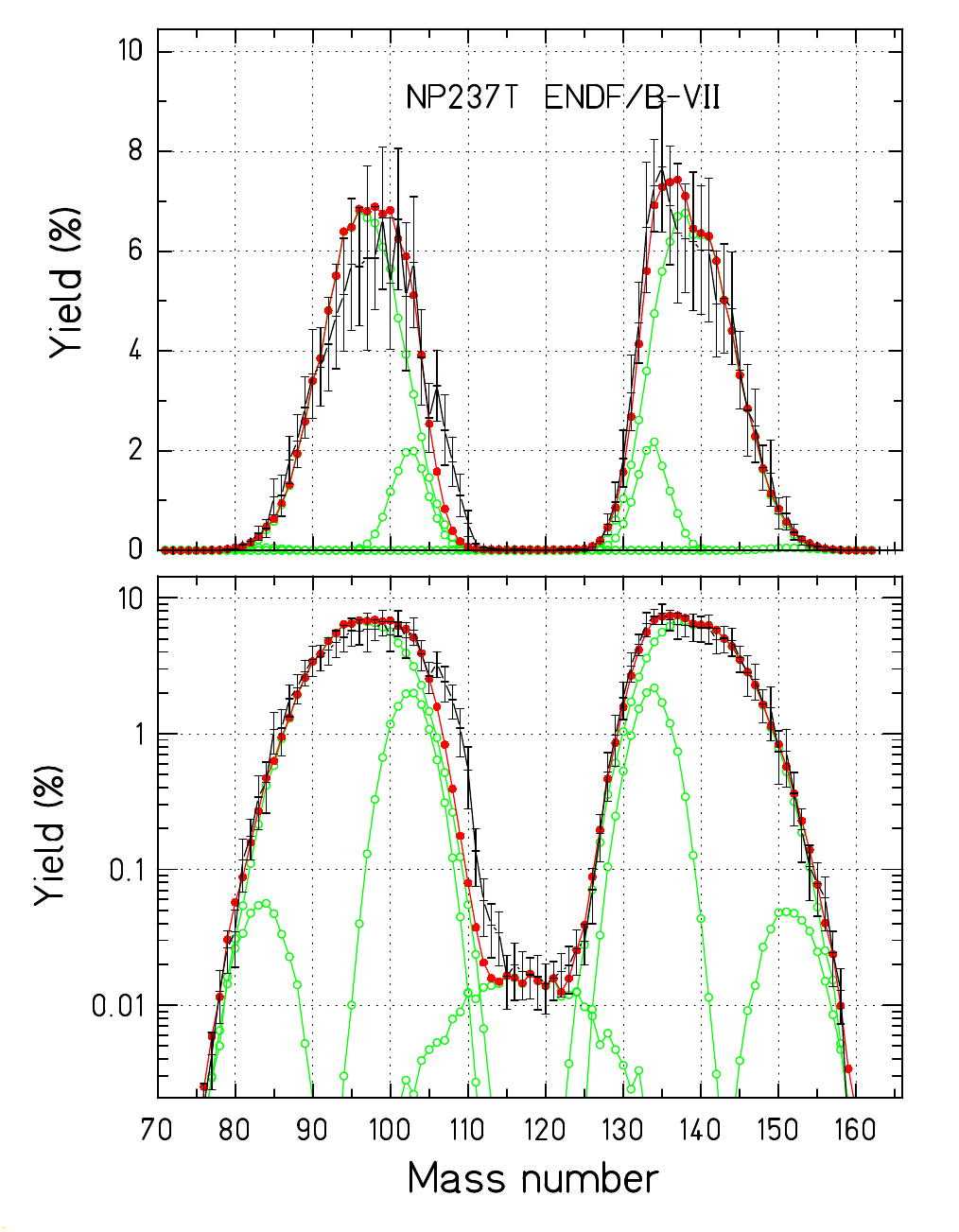}
\caption{Mass yields of $^{237}$Np(n$_{\text{th}}$,f), linear (upper frame) and logarithmic (lower frame) scale. GEF result (red points) in comparison with ENDF/B-VII (black symbols). } 
\label{NP237T-ENDF}       
\end{figure}

\begin{figure}[h]
\centering
\includegraphics[width=0.36\textwidth]{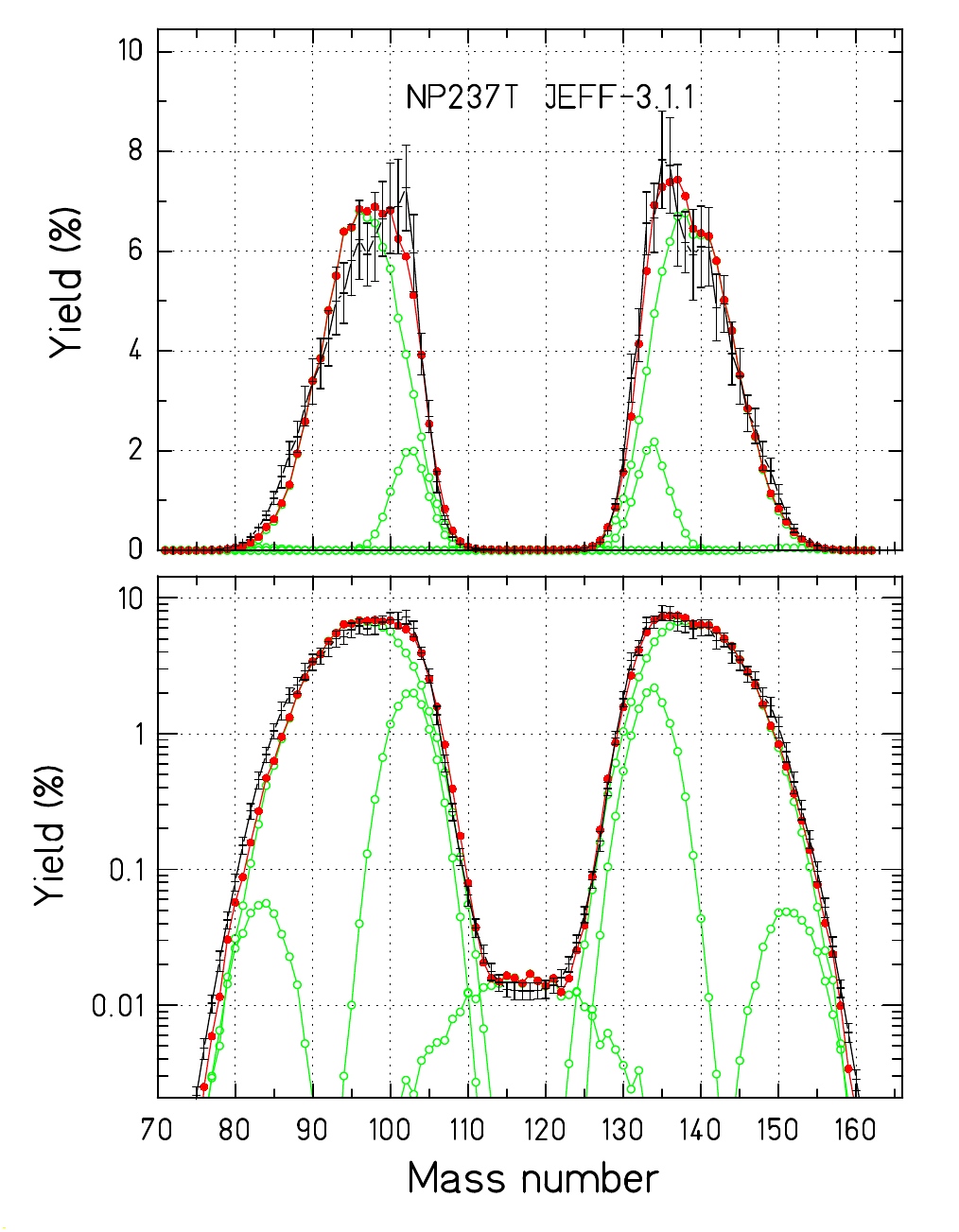}
\caption{Mass yields of $^{237}$Np(n$_{\text{th}}$,f), linear (upper frame) and logarithmic (lower frame) scale. GEF result (red points) in comparison with JEFF-3.1.1 (black symbols). } 
\label{NP237T-JEFF311}       
\end{figure}

\begin{figure}[h]
\centering
\includegraphics[width=0.36\textwidth]{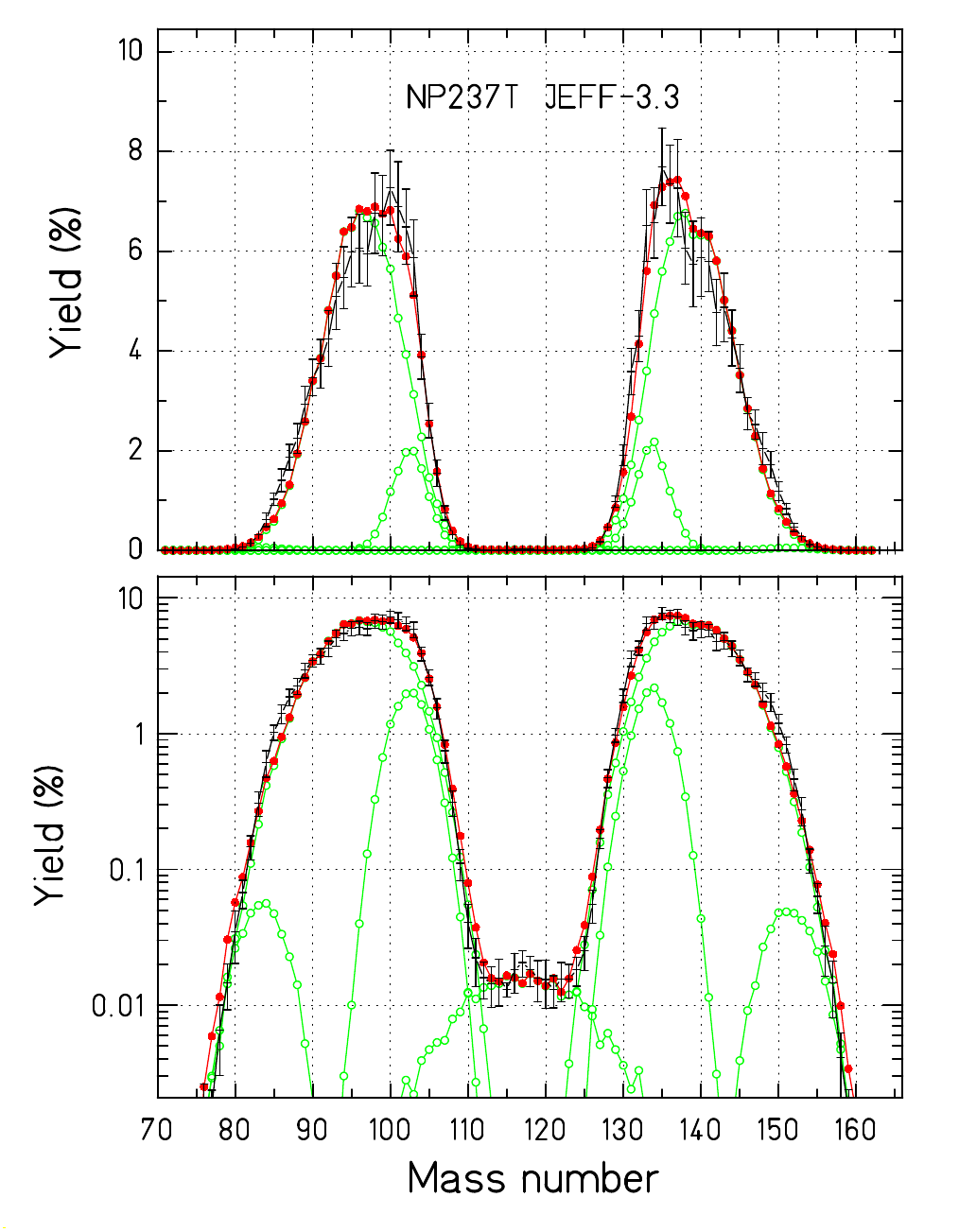}
\caption{Mass yields of $^{237}$Np(n$_{\text{th}}$,f), linear (upper frame) and logarithmic (lower frame) scale, GEF result (red points) in comparison with JEFF-3.3 (black symbols). } 
\label{NP237T-JEFF33}       
\end{figure}

In figure \ref{NP237T-ENDF}, there is a clear shift in the right wing of the light peak between GEF and ENDF/B-VII in the mass-yield distribution of the system $^{237}$Np(n$_{\text{th}}$,f) and some discrepancy in the whole light peak. This problem has already been mentioned in Ref.~\cite{Schmidt16}. It has been attributed to a target contamination, probably of $^{239}$Pu. Figures \ref{NP237T-JEFF311} and \ref{NP237T-JEFF33} show that \emph{this problem does not appear in the JEFF evaluations}, probably by the use of some more recent data.

\subsubsection{Mass yields of $^{241}$Pu(n$_{\text{th}}$,f)} \label{S6.2.6}

\begin{figure}[h]
\centering
\includegraphics[width=0.36\textwidth]{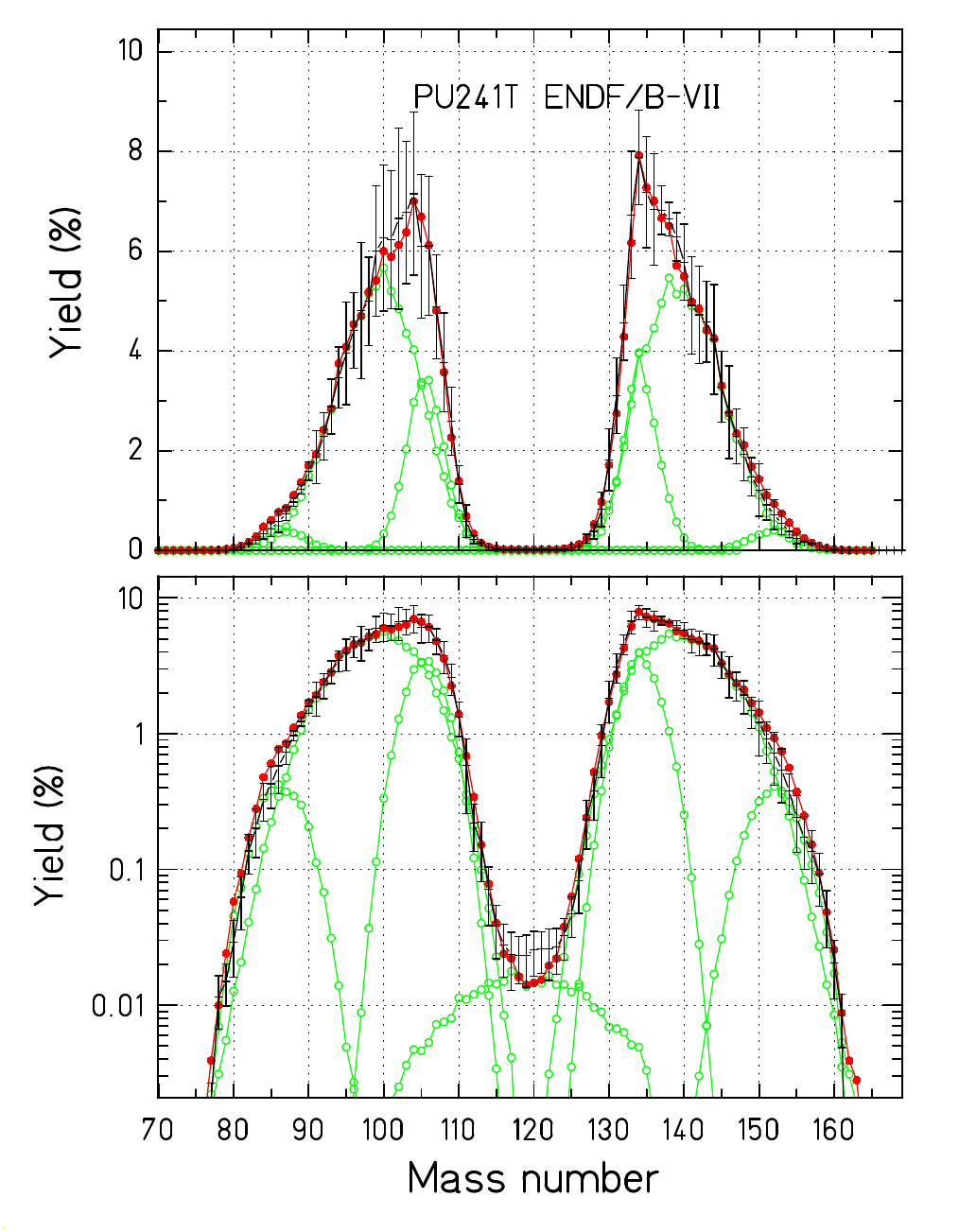}
\caption{Mass yields of $^{241}$Pu(n$_{\text{th}}$,f), linear (upper frame) and logarithmic (lower frame) scale. GEF result (red points) in comparison with ENDF/B-VII (black symbols). } 
\label{PU241T-ENDF}       
\end{figure}

\begin{figure}[h]
\centering
\includegraphics[width=0.36\textwidth]{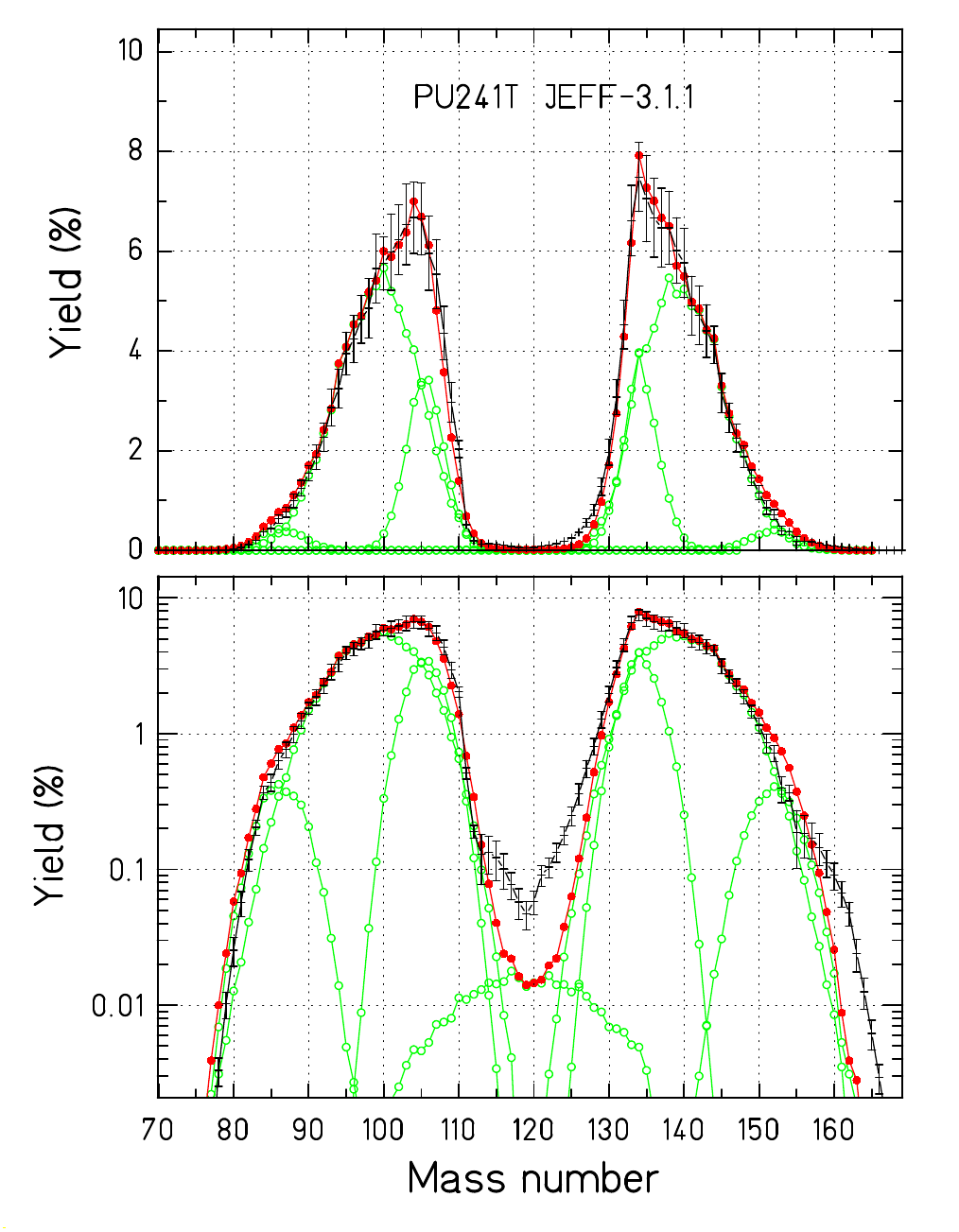}
\caption{Mass yields of $^{241}$Pu(n$_{\text{th}}$,f), linear (upper frame) and logarithmic (lower frame) scale, GEF result (red points) in comparison with JEFF-3.1.1 (black symbols). } 
\label{PU241T-JEFF311}       
\end{figure}

\begin{figure}[h]
\centering
\includegraphics[width=0.36\textwidth]{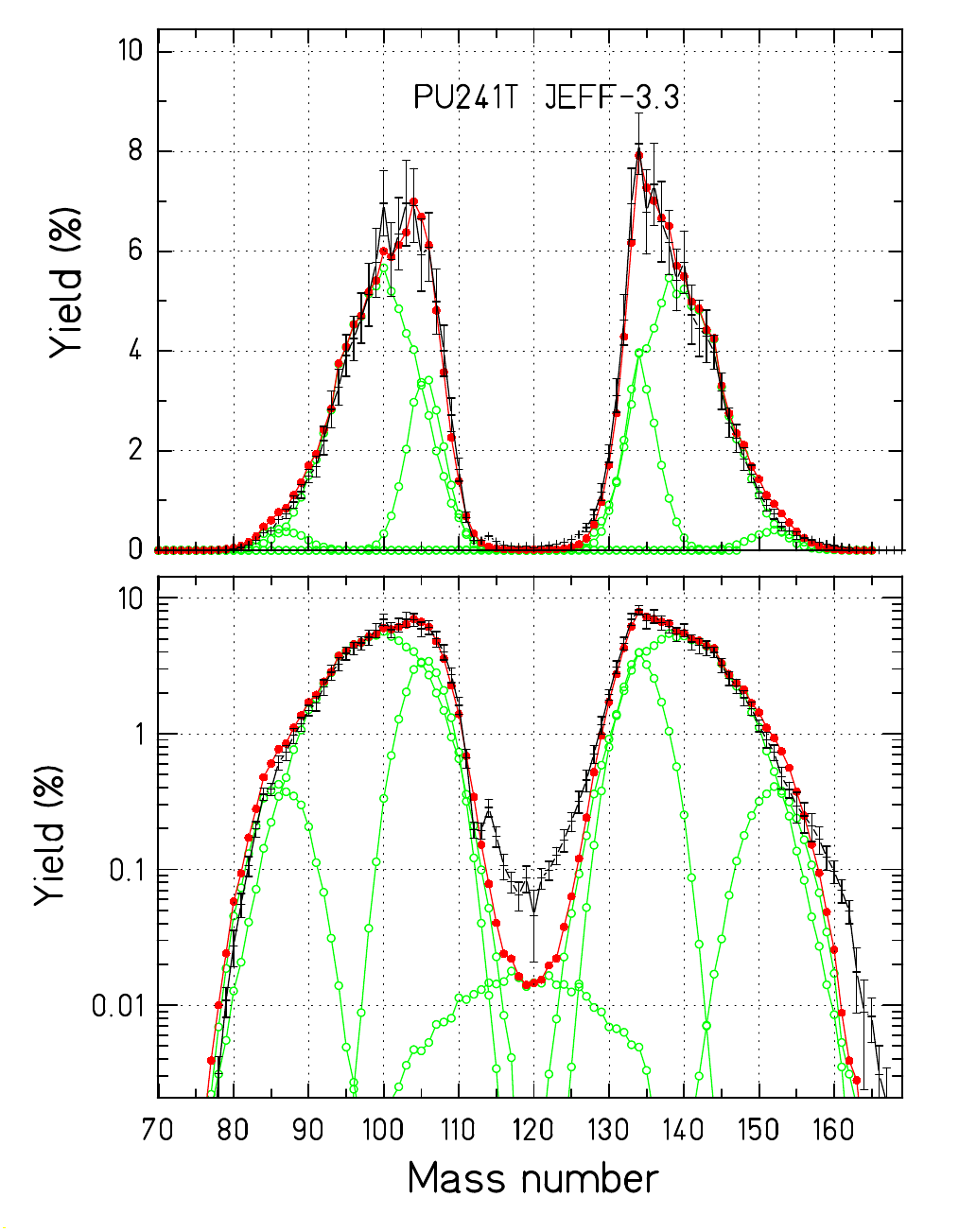}
\caption{Mass yields of $^{241}$Pu(n$_{\text{th}}$,f), linear (upper frame) and logarithmic (lower frame) scale. GEF result (red points) in comparison with JEFF-3.3 (black symbols). } 
\label{PU241T-JEFF33}       
\end{figure}

In figures \ref{PU241T-ENDF}, \ref{PU241T-JEFF311}, and \ref{PU241T-JEFF33} there is good agreement between the GEF results and ENDF/B-VII for mass yields of the system $^{241}$Pu(n$_{\text{th}}$,f). However, the evaluations JEFF-3.1.1 and JEFF-3.3 show strong discrepancies near symmetry and in the upper wing. \emph{We recommend to use the ENDF/B-VII compilation or the GEF results.}

\subsubsection{Mass yields of $^{251}$Cf(n$_{\text{th}}$,f)} \label{S6.2.7} 

\begin{figure}[h]
\centering
\includegraphics[width=0.36\textwidth]{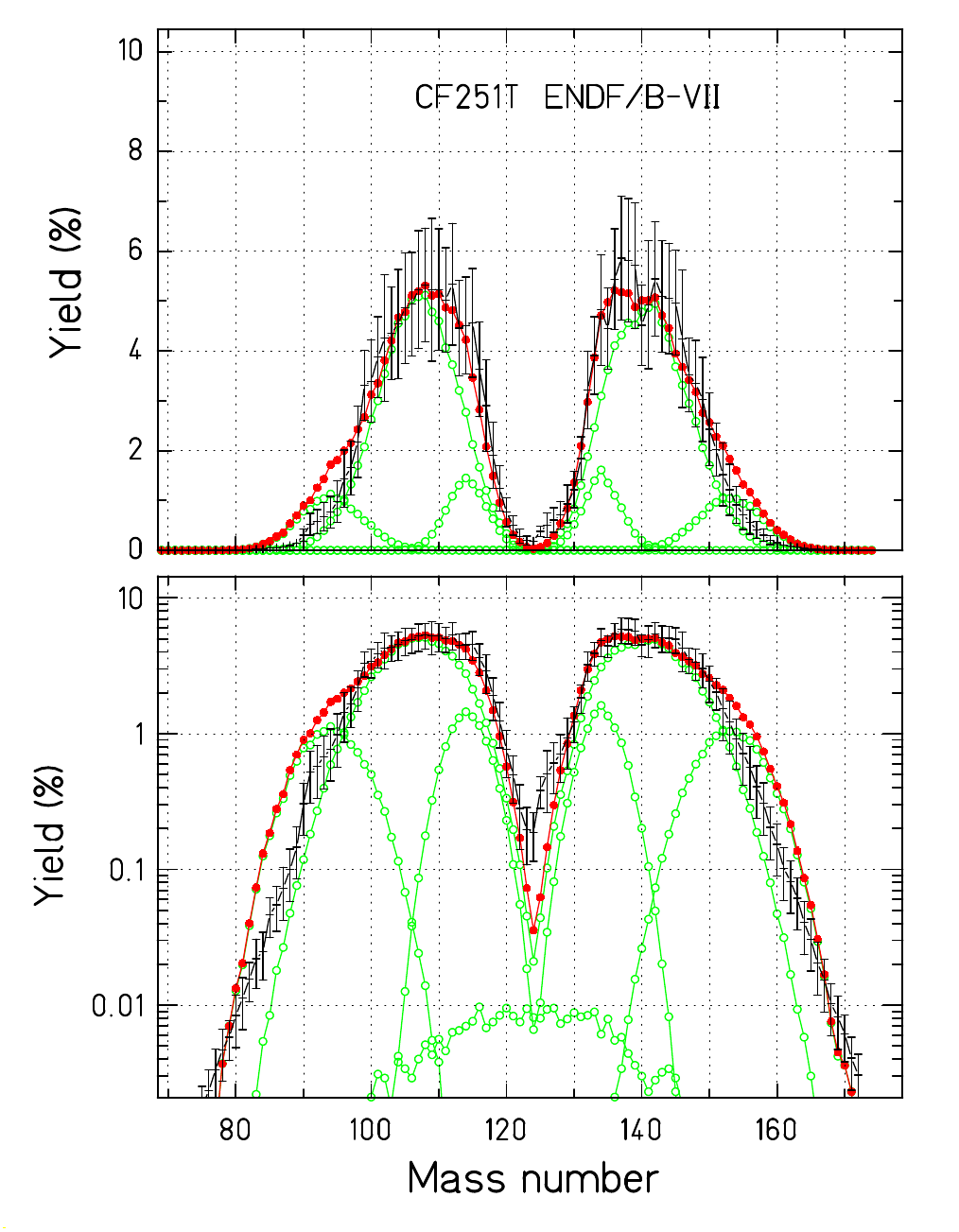}
\caption{Mass yields of $^{251}$Cf(n$_{\text{th}}$,f), linear (upper frame) and logarithmic (lower frame) scale, GEF result (red points) in comparison with ENDF/B-VII (black symbols). } 
\label{CF251T-ENDF}       
\end{figure}

\begin{figure}[h]
\centering
\includegraphics[width=0.36\textwidth]{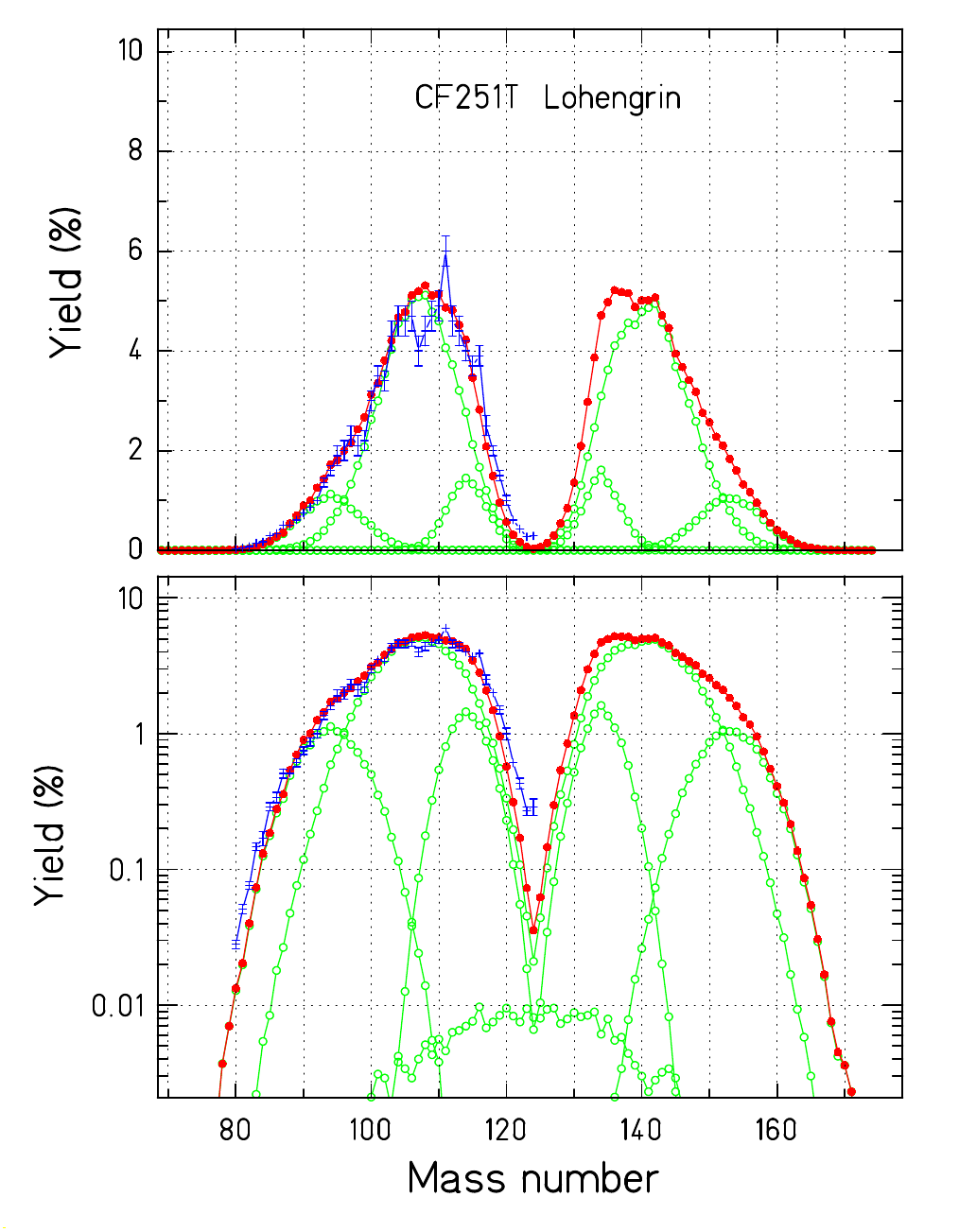}
\caption{Mass yields of $^{251}$Cf(n$_{\text{th}}$,f), linear (upper frame) and logarithmic (lower frame) scale. GEF result (red points) in comparison with a LOHENGRIN \cite{Birgersson07} experiment (blue symbols). } 
\label{CF251T-LOHENGRIN}       
\end{figure}

In figure \ref{CF251T-ENDF}, there are important discrepancies between the mass yields of ENDF/B-VII and the GEF results for the system $^{251}$Cf(n$_{\text{th}}$,f), while in figure \ref{CF251T-LOHENGRIN} the data of the LOHENGRIN experiment \cite{Birgersson07} agree on a coarse scale quite well with the GEF results. On a finer scale, however, the LOHENGRIN data show erratic fluctuations, which are much larger than the given uncertainties. Such fluctuations are not found in the fission yields of any other system. Therefore, we attribute the fluctuations to difficulties in the experiment or in the data analysis. \emph{We recommend to use the GEF results.}

\subsubsection{Mass yields of $^{254}$Es(n$_{\text{th}}$,f)} \label{6.2.9}
\begin{figure}[h]
\centering
\includegraphics[width=0.36\textwidth]{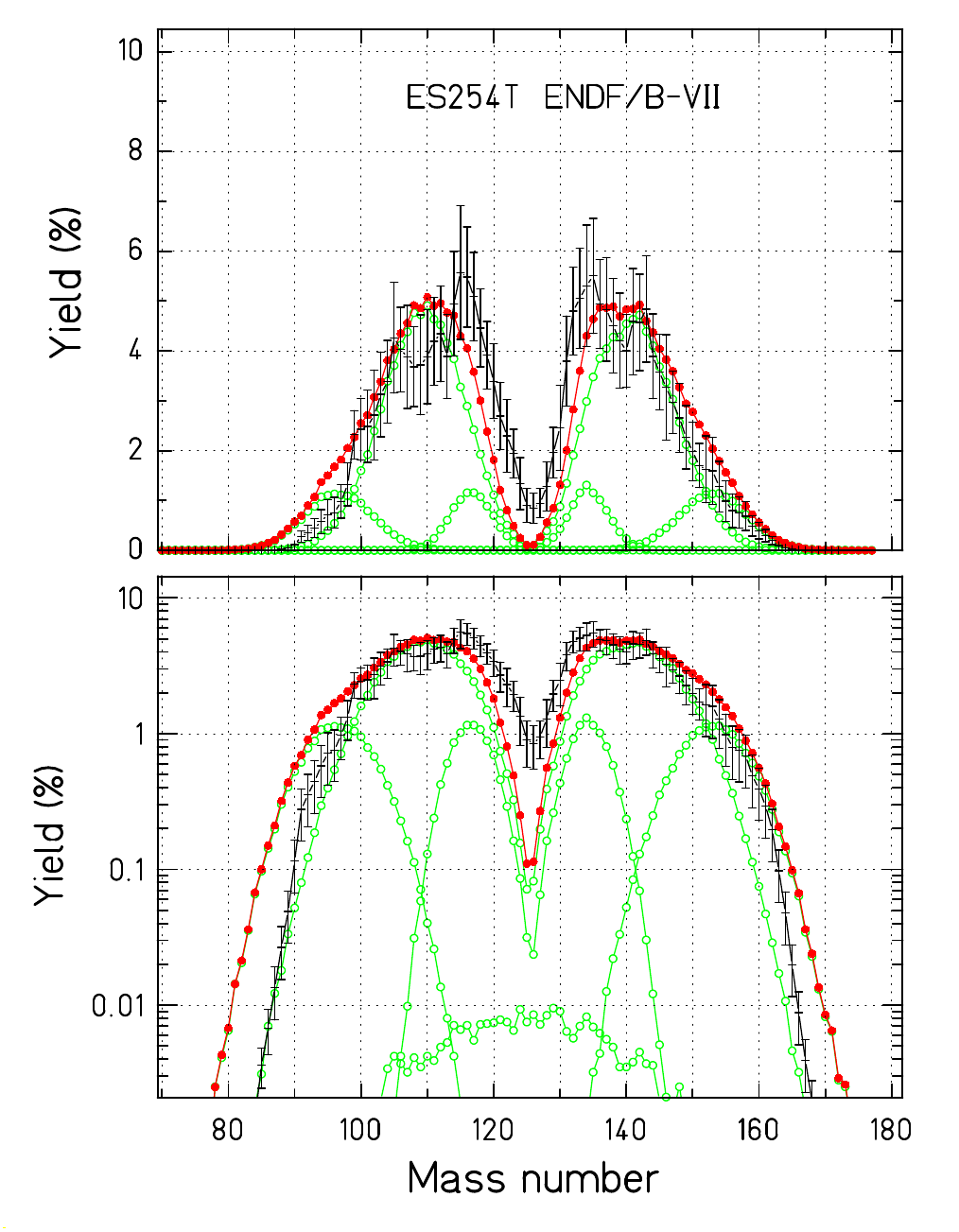}
\caption{Mass yields of $^{254}$Es(n$_{\text{th}}$,f), linear (upper frame) and logarithmic (lower frame) scale, GEF result (red points) in comparison with ENDF/B-VII (black symbols). } 
\label{ES254T-ENDF}       
\end{figure}

In figure \ref{ES254T-ENDF}, there are strong discrepancies in the whole mass distribution between GEF and ENDF/B-VII for $^{254}$Es(n$_{\text{th}}$,f). \emph{It is rather speculative to argue, which set of mass yields is more reliable. 
It is, however, rather difficult to reconcile the fission yields from ENDF/B-VII with the inherent regularities of the GEF model.}

\begin{figure}[h]
\centering
\includegraphics[width=0.36\textwidth]{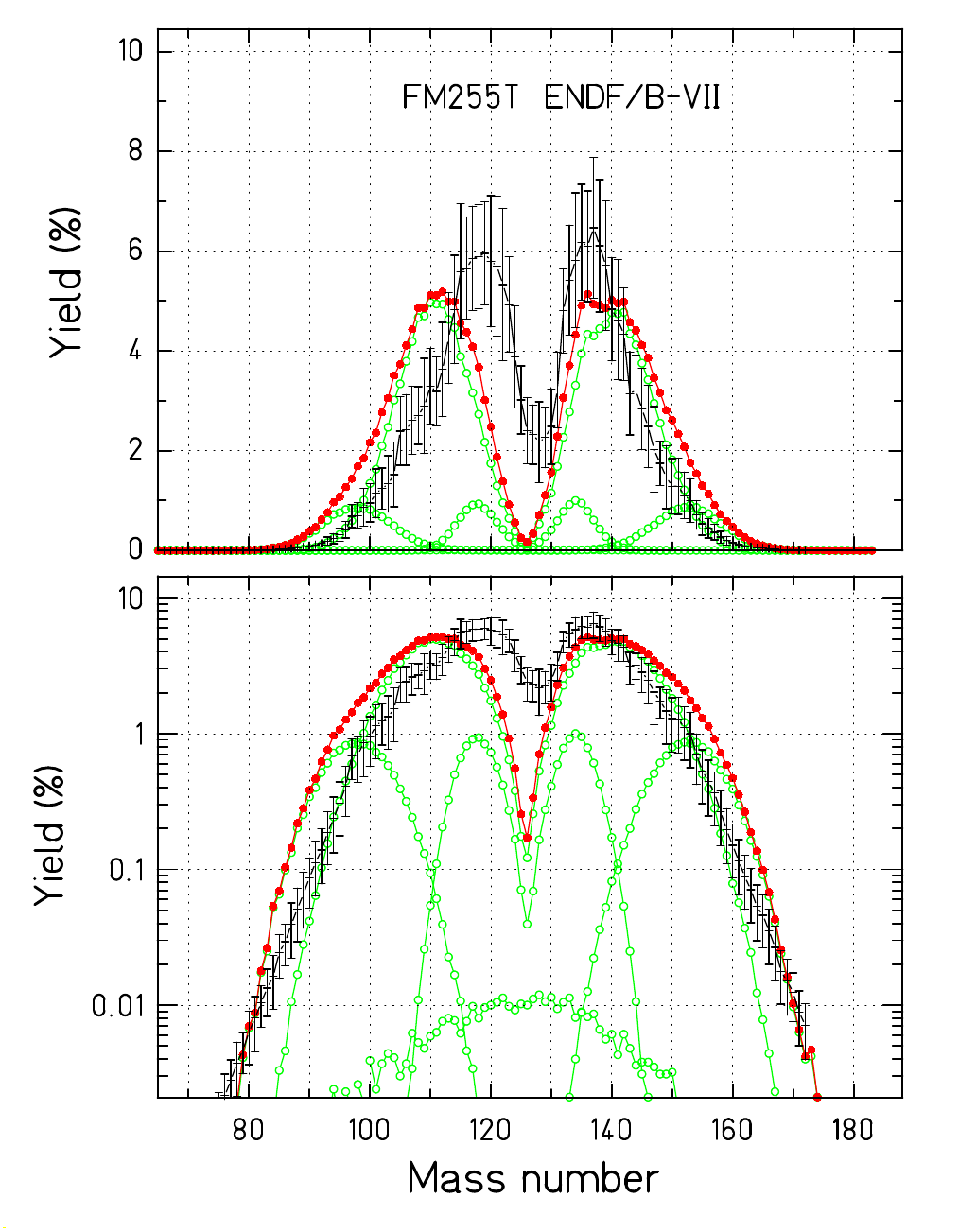}
\caption{Mass yields of $^{255}$Fm(n$_{\text{th}}$,f), linear (upper frame) and logarithmic (lower frame) scale. GEF result (red points) in comparison with ENDF/B-VII (black symbols). } 
\label{FM255T-ENDF}       
\end{figure}


\clearpage

\subsubsection{Mass yields of $^{255}$Fm(n$_{\text{th}}$,f)} \label{S6.2.8}
In figure \ref{FM255T-ENDF} there are strong discrepancies in the whole mass distribution between GEF and ENDF/B-VII for $^{255}$Fm(n$_{\text{th}}$,f). In particular, the mean value of the distribution is shifted by about 3 units. This entails a drastic difference in the mean number of prompt neutrons, where the deduced ENDF value deviates strongly from the systematics, see Ref.~\cite{Schmidt16}. \emph{We recommend to replace the whole distribution by the GEF results.}

\subsection{Quantitative analysis} \label{S6-3}

The preceding sections revealed numerous discrepancies between the mass yields from different evaluations, selected kinematical experiments and the results of the GEF code. 
By analyzing the graphical presentations of the full mass distributions, conjectures on the origin of many observed deviations were given.
As a complementary quantitative but lumped information, Table \ref{CHI_square} presents the RMS values of the deviations per degree of freedom (reduced chi-square values) between fission-product mass yields from GEF and values from different evaluations and from LOHENGRIN experiments.   


The values extend over a large range. 
The highest ones are found for cases, where indications for severe shortcomings of the evaluated or measured data were found, for example for thermal-neutron-induced fission of $^{227}$Th, $^{236}$U, $^{254}$Es, $^{255}$Fm and, to a lesser degree, $^{241}$Pu, $^{249}$Cf and $^{251}$Cf. 
The observed problems of GEF in reproducing the yields in the light peak for fission of thorium and uranium isotopes, which are given with rather small uncertainties, are reflected by relatively large chi-square values of up to about 5.

However, a few peculiarities must be taken into account when interpreting the numbers. For example, it seems that the uncertainties of the ENDF/B-VII evaluation are estimated rather conservatively, leading to rather low chi-square values. On the other hand, the uncertainties attributed to the yields measured at LOHENGRIN are very small, which leads to relatively high chi-square values. Maybe these uncertainties do not fully consider the systematic uncertainties of this method. In addition, the use of symmetric uncertainties on a linear scale in our calculation, also for very small yields, is not realistic. Asymmetric uncertainties or symmetric uncertainties on a logarithmic scale would have been more adapted.

\begin{table}[h]
\begin{center}
\caption{RMS values of the deviations between fission-product mass yields from GEF
and values from the evaluations ENDF/B-VII, JEFF-3.1.1 and JEFF-3.3 as well as from LOHENGRIN experiments.}
\label{CHI_square}
~\\
\begin{tabular}{|c|c|c|c|c|} \hline \hline
  System                 & ENDF      & JEFF-3.1.1 & JEFF-3.3  & Lohengrin  \\ \hline \hline
   $^{227}$Th(n$_{\text{th}}$,f)  & 420/140   &   ---      &  ---      & ---        \\ \hline
   $^{229}$Th(n$_{\text{th}}$,f)  & 4.9/3.6   &   ---      &  ---      & 17/4.2     \\ \hline
   $^{232}$U(n$_{\text{th}}$,f)   & 1.2/31    &   ---      &  ---      & ---        \\ \hline
   $^{233}$U(n$_{\text{th}}$,f)   & 1.3/4.0   & 5.1/3.7    & 4.2/3.7   & 26/4.5     \\ \hline
   $^{235}$U(n$_{\text{th}}$,f)   & 3.8/1.1   & 5.4/1.0    & 4.4/0.96  & 4.8/0.93   \\ \hline
   $^{235}$U(n$_{\text{hi}}$,f)   & 0.51/8.3  & 3.7/10.8   & 3.7/9.3   & ---        \\ \hline
   $^{236}$U(n$_{\text{th}}$,f)   & 24/50     & ---        & ---       & ---        \\ \hline
   $^{238}$U(n$_{\text{fast}}$,f)   & 0.34/1.6  & 5.1/1.6    & 2.3/1.5   & ---        \\ \hline
   $^{238}$U(n$_{\text{hi}}$,f)   & 0.27/4.4  & 2.2/7.6    & 2.3/6.8   & ---        \\ \hline
   $^{237}$Np(n$_{\text{th}}$,f)   & 0.69/21.7 & 2.5/4.0   & 1.4/3.5   & ---        \\ \hline
   $^{238}$Np(n$_{\text{th}}$,f)   & ---       & 2.2/4.6   & 5.7/2.5   & 8.8/3.2    \\ \hline
   $^{238}$Pu(n$_{\text{th}}$,f)   & ---       & 3.7/9.5   & 13.1/7.1  & ---        \\ \hline
   $^{239}$Pu(n$_{\text{th}}$,f)   & 0.37/1.4  & 2.0/2.2   & 1.7/1.5   & 9.2/0.74   \\ \hline
   $^{240}$Pu(n$_{\text{th}}$,f)   & 0.59/4.9  & ---        & ---       & ---        \\ \hline
   $^{241}$Pu(n$_{\text{th}}$,f)   & 0.53/1.1  & 6.3/15.7   & 6.1/17.2  & ---        \\ \hline
   $^{242}$Pu(n$_{\text{th}}$,f)   & 0.49/4.06 & ---        & ---       & ---        \\ \hline
   $^{241}$Am(n$_{\text{th}}$,f)   & 3.7/3.0   & 2.2/4.4    & 1.6/3.6   & ---        \\ \hline
   $^{242}$Am(n$_{\text{th}}$,f)   & 0.5/1.7   & 1.2/2.4    & ---       & ---        \\ \hline
   $^{243}$Am(n$_{\text{th}}$,f)   & ---       & 2.6/14.7   & 5.2/2.3   & ---        \\ \hline
   $^{243}$Cm(n$_{\text{th}}$,f)   & 1.4/14    & 4.1/13     & 2.4/7.3   & ---        \\ \hline
   $^{244}$Cm(n$_{\text{th}}$,f)   & ---       & 2.7/15.4   & 0.6/2.0   & ---        \\ \hline
   $^{245}$Cm(n$_{\text{th}}$,f)   & 0.51/4.1  & 1.5/5.1    & 1.1/3.5   & ---        \\ \hline
   $^{249}$Cf(n$_{\text{th}}$,f)   & 1.0/5.6   & ---        & ---       & 27/1.8     \\ \hline
   $^{251}$Cf(n$_{\text{th}}$,f)   & 10/19     & ---        & ---       & 16/21      \\ \hline
   $^{254}$Es(n$_{\text{th}}$,f)   & 33/74     & ---        & ---       & ---        \\ \hline
   $^{255}$Fm(n$_{\text{th}}$,f)   & 13/42     & ---        & ---       & ---        \\ \hline
\end{tabular}
\small{~\\~\\Note: Two numbers are given. The first one uses the uncertainties of the evaluation, respectively experiment. The second one uses the uncertainties from GEF. Masses with calculated or evaluated, respectively experimental, yields below 0.01 percent are not considered. $n_{\text{th}}$} means thermal neutrons, $n_{\text{fast}}$ means fast neutrons, and $n_{\text{hi}}$ means neutrons of 14 MeV.
\end{center}
\end{table}

\subsection{Summary}  \label{S6.4}
The comparative study of the preceding sections gives the following result for thermal-neutron-induced fission:
In fifteen cases, good or at least satisfactory agreement is obtained between the mass yields from GEF and the empirical data. 
In eight cases, severe discrepancies appeared, most of them hinting to erroneous evaluations, according to our analysis.
The yields at symmetry in the low-energy fission of the actinides show deviations in several systems. They pose specific difficulties to both the evaluations and the GEF code due to large experimental uncertainties in the measurement of low yields and due to the influence of weak shells on the depth of the symmetric fission valley, respectively.
Additional problems, especially at mass symmetry, arize from the broad energy distribution of "thermal" reactor neutrons, in particular for thermally not-fissile nuclei. 
Most of the LOHENGRIN experiments seem to be much more accurate than the evaluations. The agreement of the mass yields with the GEF results tends to confirm the small indicated uncertainties of these experiments, except in the case of $^{251}$Cf(n$_{\text{th}}$,f). 
The LOHENGRIN data form a backbone for determining the parameters of GEF. 
However, this is not a direct adjustment. On the contrary, the compatibility of the LOHENGRIN results with the regularities and constraints of the theoretical framework of basic concepts and laws of general validity in the GEF model tends to corroborate both the LOHENGRIN data and the GEF model.
Thus, the evaluations can be improved by including the LOHENGRIN data to a greater extent. 
The remaining deviations between empirical mass yields and GEF results reveal some deficiencies of both the evaluations and GEF, depending on the case.
Local deviations for individual systems hint more to a problem in the evaluations, while general deviations for several neighboring systems hint more to a problem in GEF.
In many cases where satisfactory agreement with the GEF result is found, but the uncertainties of the evaluations are very large, the GEF results may be included in the evaluation process and help to improve the accuracy of the evaluated mass yields. 
Thus, the present comparative study can be exploited to improve the evaluations leading to enhance the quality of nuclear data. It also provides information on a few remaining deficiencies of the GEF code, which call for further refinements. This is a very important issue for the estimation of the characteristics of the antineutrino production, where the requirements on accuracy are extremely high.

We have learned that the adjustment of GEF parameters to empirical data is a rather difficult task. 
Indeed, performing a least-squares fit to all data does not lead to a satisfactory result, because many evaluated values are erroneous. In some cases, this is evident, but in the majority of cases a careful analysis and a systematic comparison between data from different sources and evaluations and with GEF is needed to sort out the more reliable and the less trustworthy values. 


Eventually, we came up with the following rules:

1. Radiochemical data have very different quality. As demonstrated in section \ref{S6}, by far the most reliable ones are the FY for $^{235}$U(n$_{\text{th}}$,f), followed by $^{239}$Pu(n$_{\text{th}}$,f) and very few other systems. The data of all other systems are less trustworthy due to large uncertainties of the measured yields or lacking data for a number of masses. The quality is not always reflected by the error bars. 

2. Mass yields from LOHENGRIN experiments are much more accurate than those from radiochemical measurements (with one exception, see paragraph \ref{S6.2.7}).

3. Indirect information on FY (antineutrino spectrum, decay heat etc.) are extremely sensitive probes for the overall quality of the FYs for specific systems.

4. It is important to primarily adjust the parameters of GEF to the most trustworthy data. The regularities of GEF help to recognize faulty data of other systems.

The results of applying them have been shown above.


\section{Predictions of antineutrino energy spectra based on the GEF fission yields}  \label{S7}

\subsection{Beta-decay emitters} \label{S3.3}

Using the summation method and GEF fission yields, the nuclei contributing mainly to the antineutrino energy spectra of the most dominant isotopes in a reactor core can be extracted in bins of antineutrino energies as it was done in the past from~\cite{Report-NDS-676} with the summation method of~\cite{Zak} obtained with the JEFF3.1.1 fission yields.

\begin{table}
\caption{List of the 20 nuclides contributing most importantly to the 
the $^{235}$U antineutrino spectrum in the 4 to 5 MeV bin ordered by importance of contribution obtained with the fission yields of JEFF-3.1.1. 
In the second and third columns are indicated the absolute relative discrepancy of the GEF fission yields to the JEFF-3.1.1 and JEFF-3.3 evaluated cumulative yields.
}
~\\
\label{tab:tableMainContribGEF}       
\begin{tabular}{l c c}
\hline\noalign{\smallskip}
Nuclide & \parbox[c]{3cm}{\centering Rel. Dif. GEF vs JEFF-3.1.1}
        & \parbox[c]{3cm}{\centering Rel. Dif. GEF vs JEFF-3.3} \\
\noalign{\smallskip}\hline\noalign{\smallskip}
39-Y-95  	&  5.4\% &  6.5\% \\
39-Y-94  	&   9.4\%&  9.5\% \\
38-Sr-93  	&   0.6\%&  0.4\% \\
55-Cs-139  	&   2.8\%&  3.9\% \\
 55-Cs-140 	&   2.8\%&  0.23\% \\
57-La-142 	&   2.1\%&  2.4\% \\
41-Nb-98 	&   5.7\%&  5.8\% \\
37-Rb-91  	&   9.4\%&  5.7\% \\
41-Nb-100  	&   1.2\%&  2.8\% \\
 57-La-144 	&   9.0\%&  9.0\% \\
38-Sr-95 	&   7.6\%&  6.7\% \\
54-Xe-139	&   3.6\%&  5.1\% \\
41-Nb-101	&   0.2\%&  3.3\% \\
36-Kr-90 & 12.3\%& 8.9\% \\
 55-Cs-141  & 2.0\%& 3.1\% \\
37-Rb-92	&  0.02\% &  10.5\% \\
39-Y-96 & 25.9\% & 27.5\% \\
37-Rb-89	& 4.2\%  &  4.8\% \\
36-Kr-89 	&  3.9\% &  4.0\% \\
37-Rb-90 & 6.0\% & 2.6\% \\
\end{tabular}
\end{table}

%

Table~\ref{tab:tableMainContribGEF} shows the relative discrepancies between the GEF, JEFF-3.1.1 and JEFF-3.3 cumulative yields for the top twenty of the largest contributions to the $^{235}$U antineutrino energy spectrum obtained with the summation method~\cite{Estienne}. The agreement reached after the complex tuning of the GEF model on the available datasets for a wide set of fissioning systems is quite satisfactory, and mainly constrained by the small uncertainties of the LOHENGRIN data. As shown in section~\ref{S6}, the LOHENGRIN fission yields are in good agreement with the JEFF evaluated fission yields in the case of $^{235}$U, but their uncertainties are smaller, which lets us think that the uncertainties of the JEFF yields could be reduced.


\begin{figure}[h]
\centering
\includegraphics[width=0.5\textwidth]
{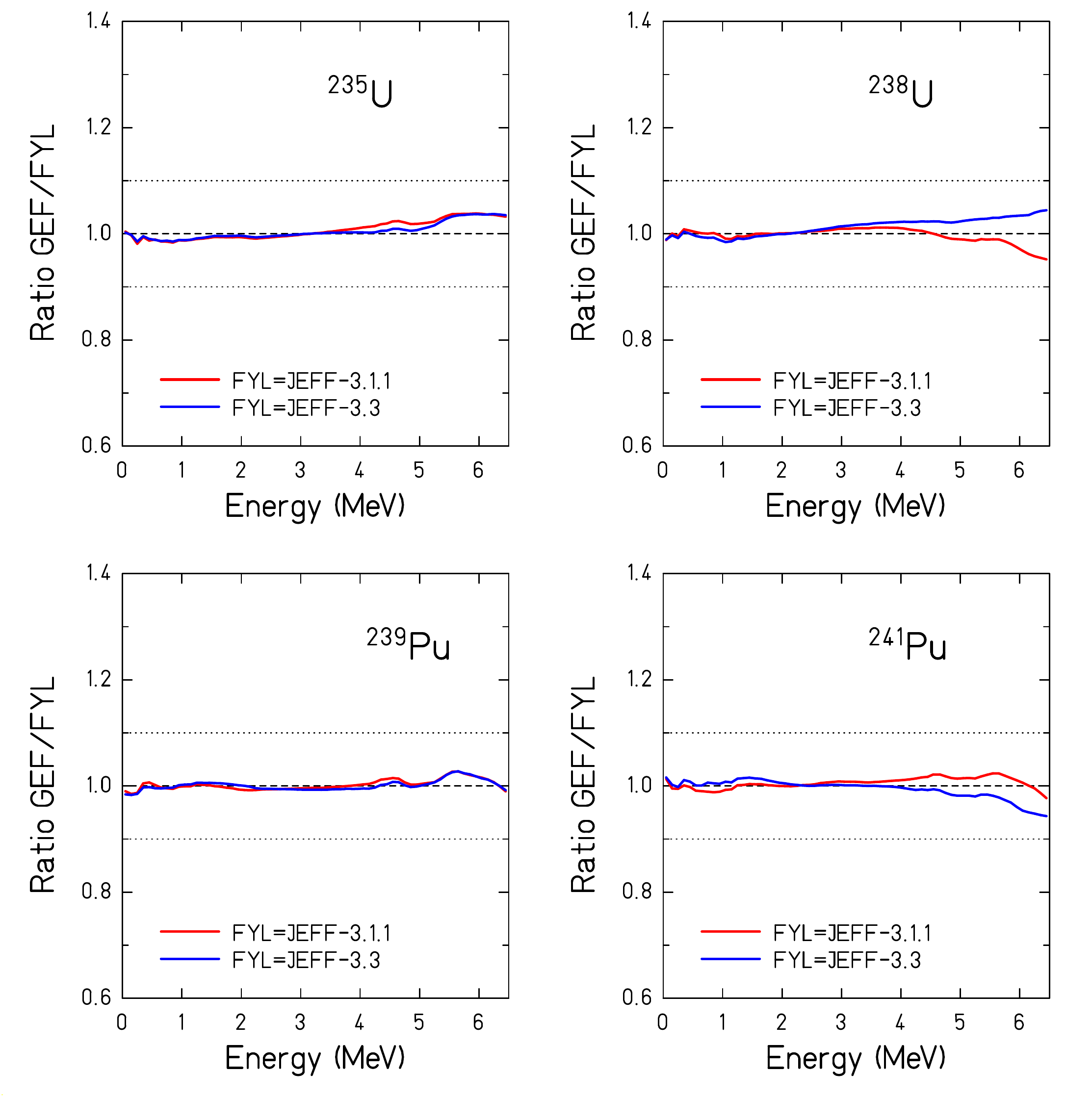}
\caption{Ratio of the antineutrino spectra calculated with yields from GEF
and from the fission-yield libraries (FYL) JEFF-3.1.1, respectively JEFF-3.3, after tuning. } 
\label{ratio_tuned}       
\end{figure}
Figure~\ref{ratio_tuned} shows the relative ratio of the antineutrino energy spectra of $^{235}$U, $^{239}$Pu, $^{241}$Pu and $^{238}$U obtained with the cumulative yields computed using the GEF code in its latest version to those obtained either with the cumulative fission yields of JEFF-3.1.1(red line) or JEFF-3.3 (blue line). 
An agreement at the 2\% level is observed with JEFF-3.1.1 up to 4~MeV in the four cases. The agreement is also good with JEFF-3.3 though it deviates by 3\% above 3~MeV in the case of $^{238}$U. Above 4~MeV, larger deviations can be observed reaching 4\% around  5.5~MeV in the $^{235}$U and 3\% in the $^{239}$Pu ratios.
In the cases of $^{241}$Pu and $^{238}$U, the discrepancies between the two sets of JEFF fission yields are noticeable above 3~MeV, with the largest deviation reached in the case of the JEFF-3.3 yields of the $^{241}$Pu. 
Special care was taken for using a realistic effective energy distribution of the impinging neutrons for the calculation of the mass yields for the fission of $^{238}$U induced by fast neutrons.

Overall, the level of agreement now reached between the spectra obtained with the GEF predictions and that obtained with the evaluated fission yields has been greatly improved by the  
adjustment of GEF to empirical data performing a survey on the FYs of all the systems.


\begin{figure}[h]
\centering
\includegraphics[width=0.36\textwidth]
{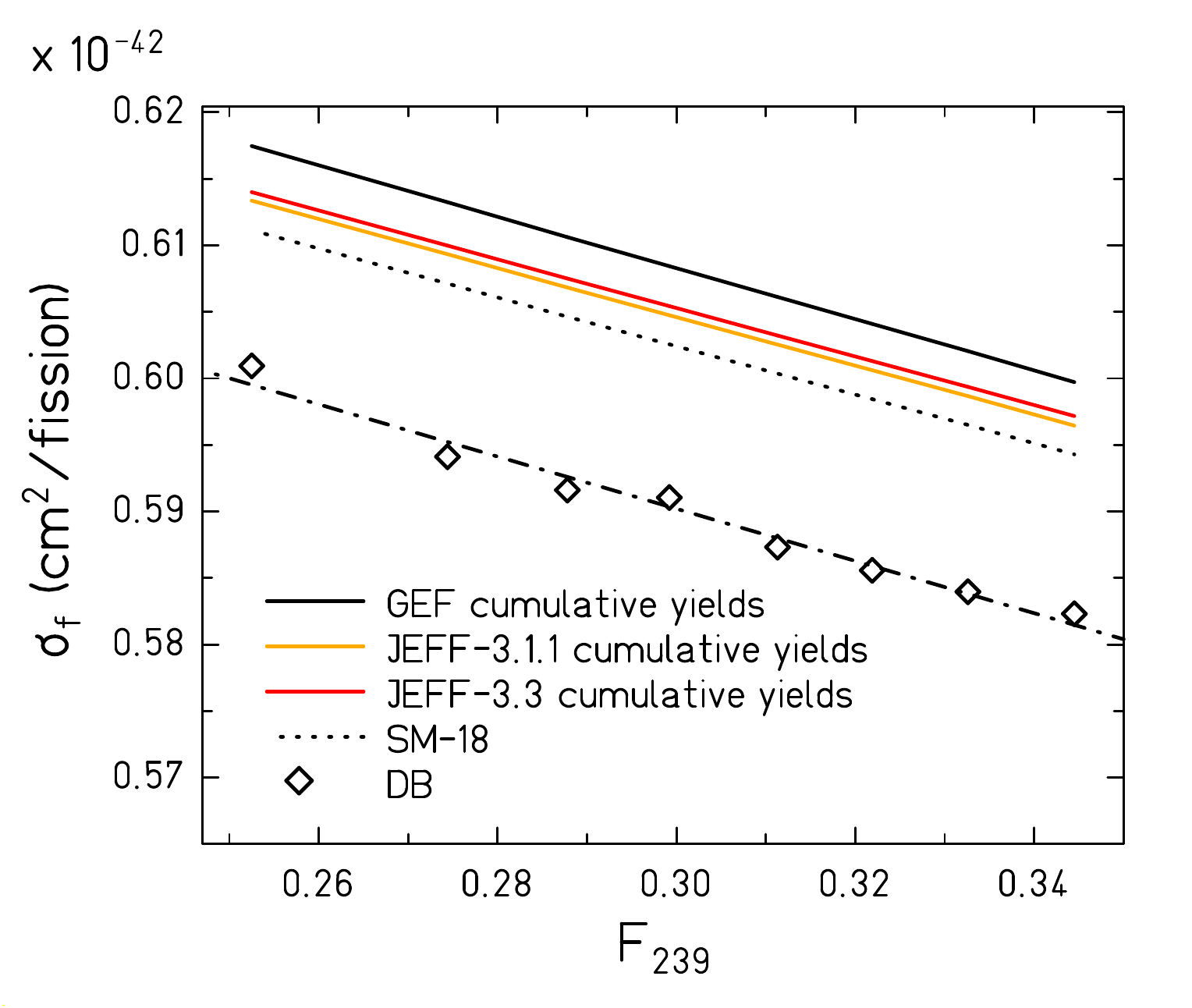}
\caption{Comparison of the Inverse Beta Decay yields as a function of the fission fraction of $^{239}$Pu obtained by the Daya Bay experiment~\cite{DayaBay17} (see text) with summation-model predictions in which the decay data are those of~\cite{Estienne} and the fission yields are the cumulative ones from the new version of the GEF code presented here, from the JEFF-3.1.1 database and from the JEFF-3.3 database. } 
\label{IBDYield_DB}       
\end{figure}

In order to compare the GEF prediction of antineutrino spectra with reactor neutrino experiments, it is necessary to fold the spectra with the detection reaction, the Inverse Beta Decay (IBD) process, historically used in the first experimental evidence of the neutrino by Reines and Cowan~\cite{ReinesCowan}. The IBD process is the result of the interaction of an $\bar{\nu_e}$ with a proton, producing a positron and a neutron (reaction threshold 1.8 MeV).
In Figure~\ref{IBDYield_DB}, the inverse beta decay yield as a function of the amount of fission events in the reactor coming from $^{239}$Pu (F$_{239}$, called ``fission fraction") published by the Daya Bay collaboration~\cite{DayaBay17} and defined following the equation 
\begin{equation}
\sigma_{\text{f}} (F_{239}) = \bar{\sigma}_{\text{f}} + \frac{d\sigma_{\text{f}}}{dF_{239}} (F_{239} - \bar{F}_{239}) \\
\end{equation}
 is displayed with open diamonds (the data points include error bars). In this formula, the average IBD yield $\bar{\sigma}_{\text{f}}$ is obtained by folding the IBD cross-section with the total antineutrino energy spectrum computed by weighting the $^{235}$U, $^{238}$U, $^{239}$Pu and $^{241}$Pu spectra by their average fission fractions provided in~\cite{DayaBay17}. 
$\bar{F}_{239}$ is the average $^{239}$Pu fission fraction, and $\frac{d\sigma_{\text{f}}}{dF_{239}}$ is the change of the IBD yield per unit $^{239}$Pu fission fraction.

In~\cite{Estienne}, the latest predictions performed with the summation model (SM) updated with the TAGS measurements performed during the last decade predict an IBD yield located only 1.9\% above the IBD yield measured by Daya Bay, and a slope of its evolution with $F_{239}$ in very good agreement with the experimental one. In this SM, the individual fission yields from the JEFF-3.1.1 database are evolved at 450 days of irradiation with the MURE code (dashed line called SM-18 in Fig.~\ref{IBDYield_DB}), taking into account the evolution due to the various half-lives of the fission products and neutron capture.
In order to see how the predictions of the SM using the same beta decay data but cumulative yields from the evaluated JEFF database in its 3.1.1 and 3.3 versions and the cumulative yields from the new version of the GEF code presented in this article would compare with the Daya Bay results, we have computed the associated IBD yields. They are displayed in Fig.~\ref{IBDYield_DB} with respectively orange, red and black lines. 
The discrepancy between the SM-18 and the yellow line (cumulative yields from JEFF-3.1.1) arises only from the use of cumulative yields instead of evolved individual fission yields. Indeed in the cumulative yields, a few very long-lived nuclei are assumed to have reached equilibrium whereas they should not have after 450 days. This gives rise to about a 0.4\% over-estimate of the IBD yields, due to the use of cumulative yields. In addition, one can see that using the JEFF-3.3 cumulative yields, the IBD yields obtained are slightly above the ones obtained with the JEFF-3.1.1 version of the fission yields database.

\subsection{Antineutrino energy spectra}  \label{S8}

\begin{figure}[h]
\centering
\includegraphics[width=0.36\textwidth]{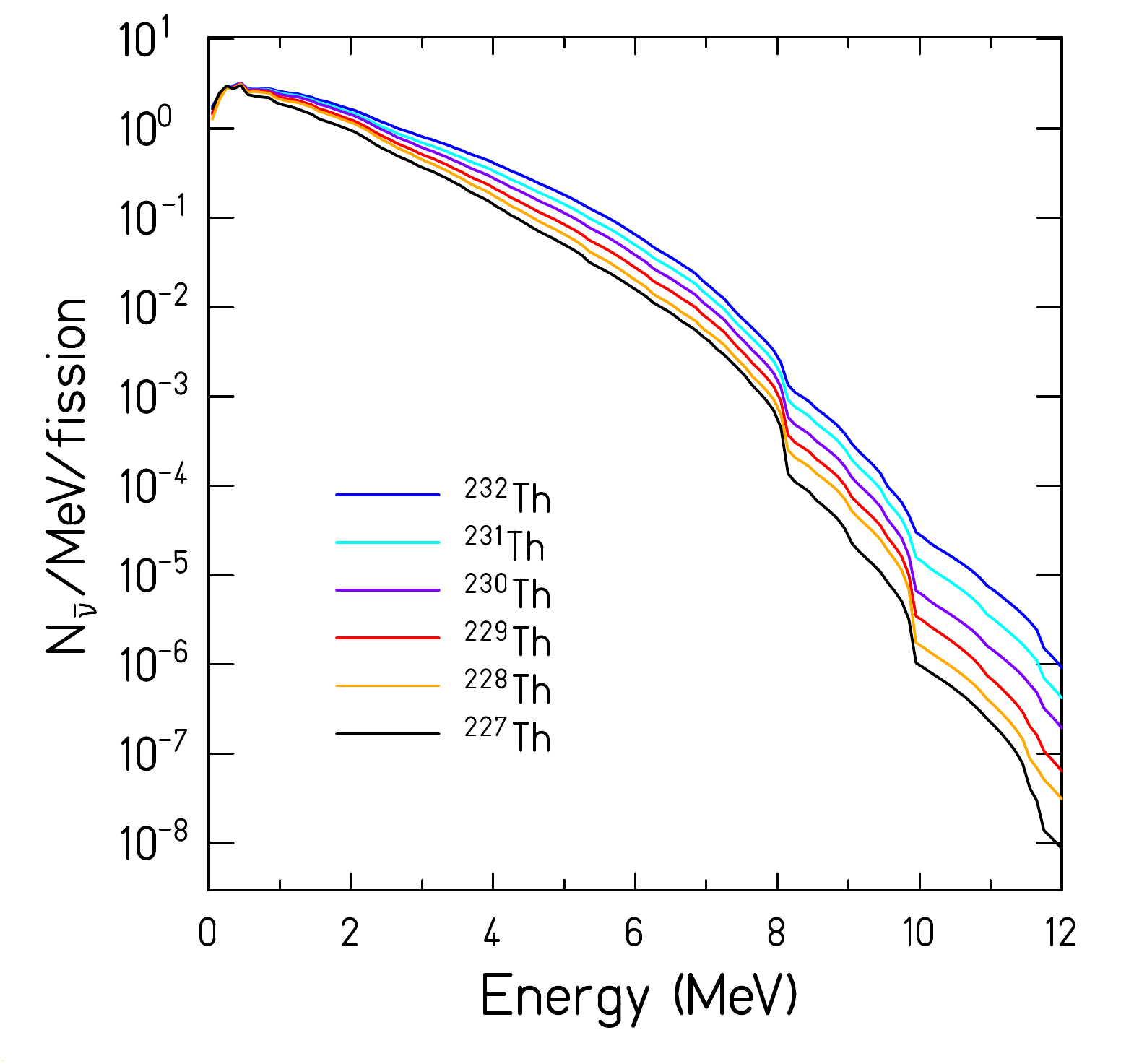}
\caption{Calculated antineutrino spectra from GEF combined with the selection of decay data of~\cite{Estienne} for 6 isotopes of Th ($Z$ = 90) in the case of thermal fission.}
\label{spectresThorium}       
\end{figure}

We have mentioned in the previous sections that there are indeed discrepancies between the two versions of the JEFF fission yields database, which are not always well understood. This reflects into the antineutrino spectra (see for instance the differences observed for $^{238}$U and $^{241}$Pu). It is not obvious to us that the fission yields of JEFF-3.3 should be used systematically instead of the ones of JEFF-3.1.1. For instance in the case of $^{241}$Pu for which inconsistencies in the evaluation of the fission yields have been evidenced in section~\ref{S6.2.6}, the antineutrino energy spectrum computed with the JEFF-3.3 fission yields departs from the one computed with GEF above 4 MeV (see Fig.~\ref{ratio_tuned}).
Overall, the SM model using the cumulative yields from GEF lies about 1\% above the SM-18 prediction (but one would have to correct the result for the impact of the very long-lived nuclei, i.e. a -0.4\% effect on the IBD yield) and about 0.7\% above the JEFF-3.1.1 cumulative yields. This result shows that GEF is an excellent model for the prediction of fission yields for antineutrino fundamental and applied physics, accurate enough to be competitive with evaluated databases even in the case of the most well known thermal fissioning systems and to be compared with neutrino experiments.

\begin{figure}[h]
\centering
\includegraphics[width=0.36\textwidth]{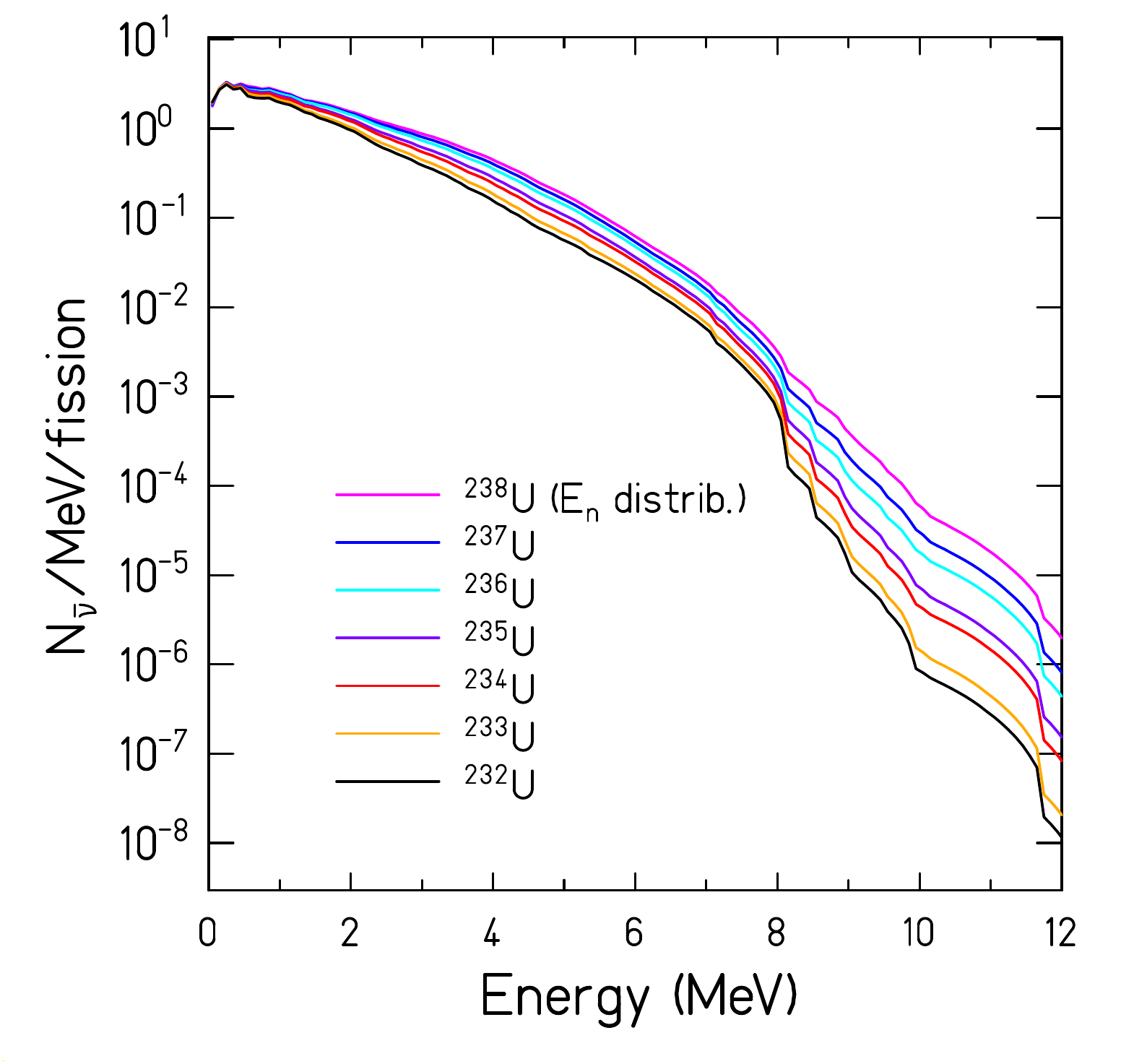}
\caption{Calculated antineutrino spectra from GEF combined with the selection of decay data of~\cite{Estienne} for seven isotopes of U ($Z$ = 92) in the case of thermal fission except for $^{238}$U. The prediction for the fast-neutron-induced fission of $^{238}$U is given for an input neutron energy distribution which follows the prescription of~\cite{Kern12}. 
}
\label{spectresUranium}       
\end{figure}

\begin{figure}[h]
\centering
\includegraphics[width=0.36\textwidth]{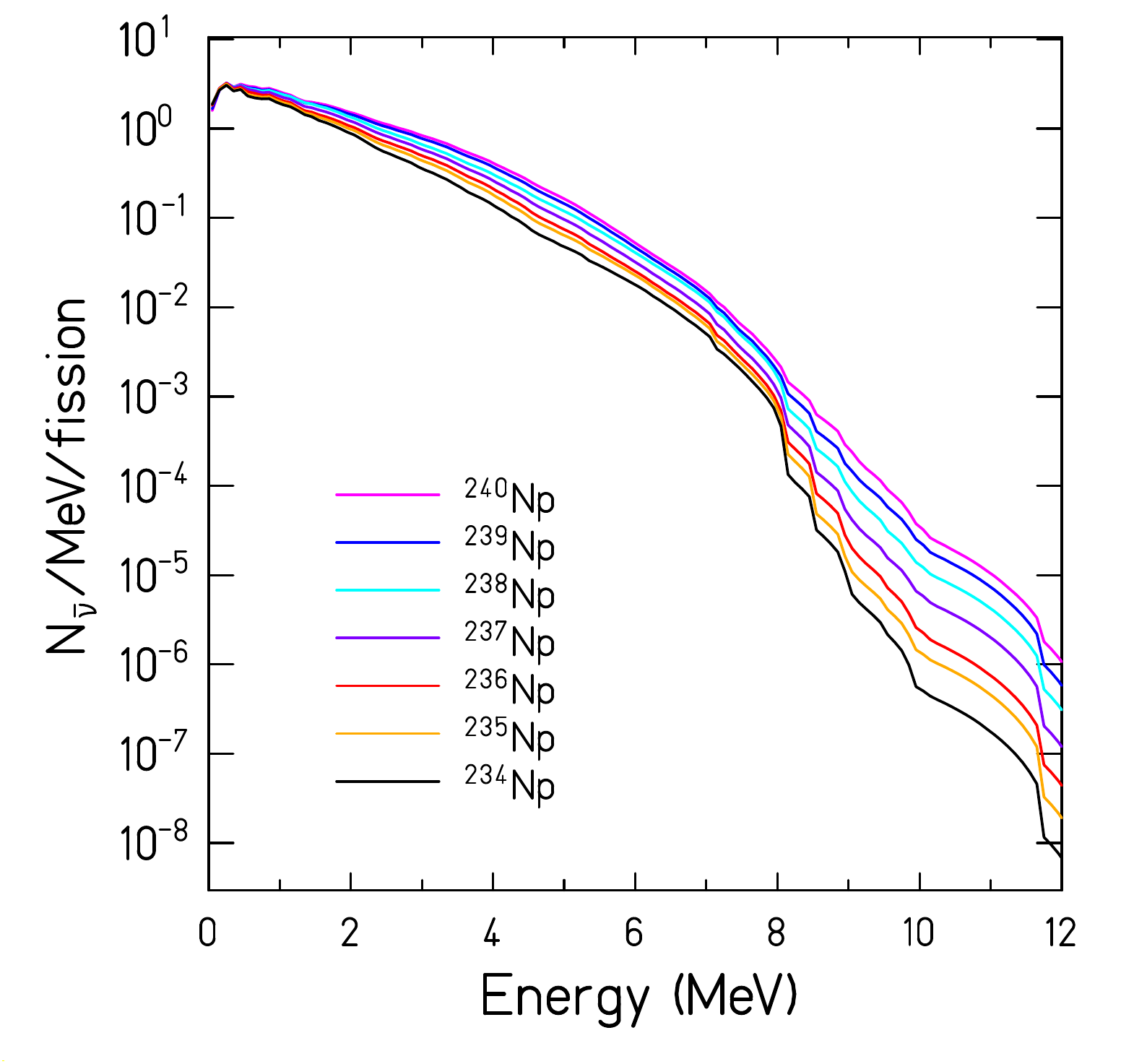}
\caption{Calculated antineutrino spectra from GEF combined with the selection of decay data of~\cite{Estienne} for seven isotopes of Np ($Z$ = 93) in the case of thermal fission.}
\label{spectresNeptunium}       
\end{figure}

\begin{figure}[h]
\centering
\includegraphics[width=0.36\textwidth]{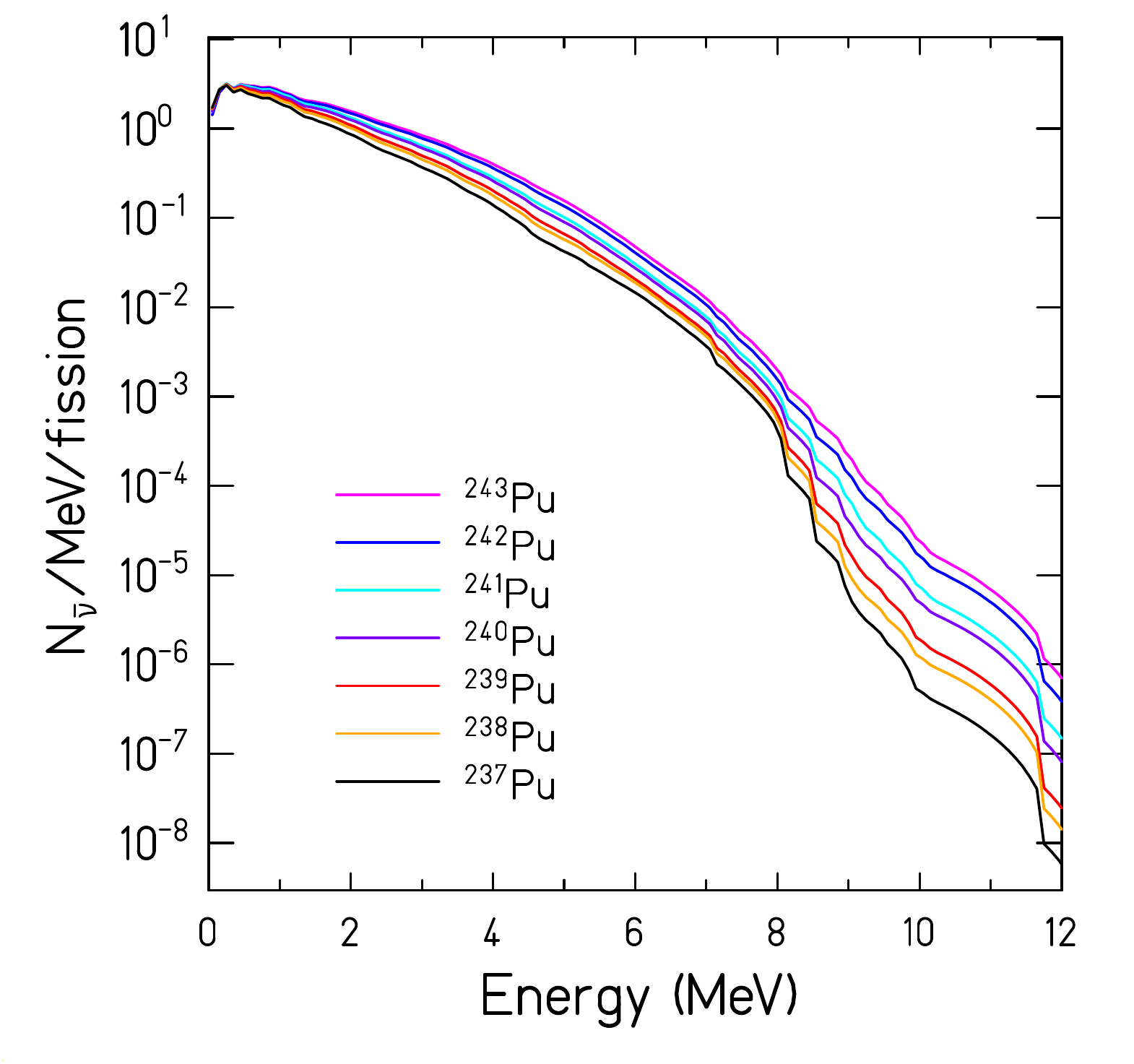}
\caption{Calculated antineutrino spectra from GEF combined with the selection of decay data of~\cite{Estienne} for seven isotopes of $Z$ = 94 in the case of thermal fissions.}
\label{spectresPlutonium}       
\end{figure}

\begin{figure}[h]
\centering
\includegraphics[width=0.36\textwidth]{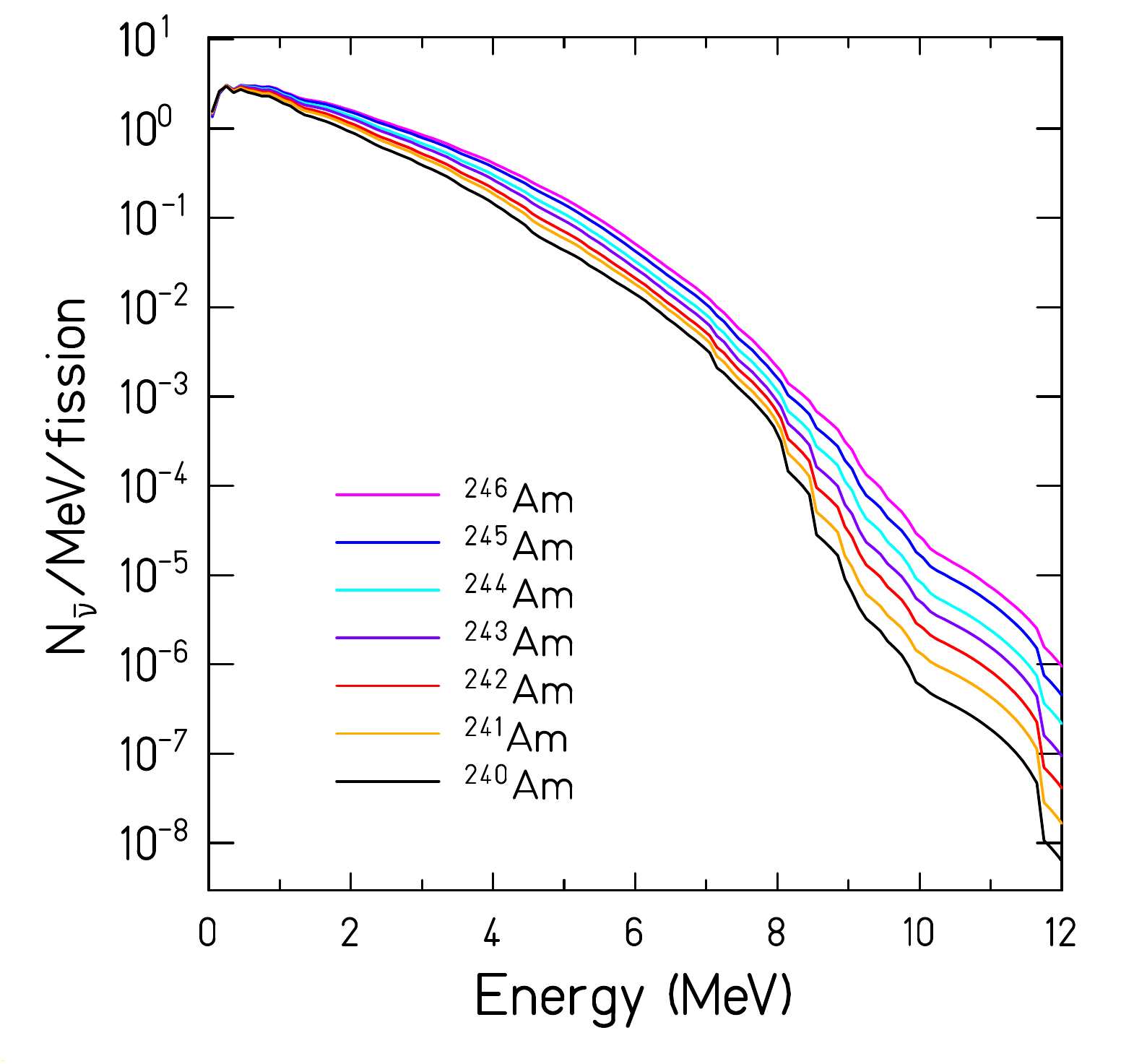}
\caption{Calculated antineutrino spectra from GEF combined with the selection of decay data of~\cite{Estienne} for seven isotopes of $Z$ = 95 in the case of thermal fissions.}
\label{spectresAmericium}       
\end{figure}

\begin{figure}[h]
\centering
\includegraphics[width=0.36\textwidth]{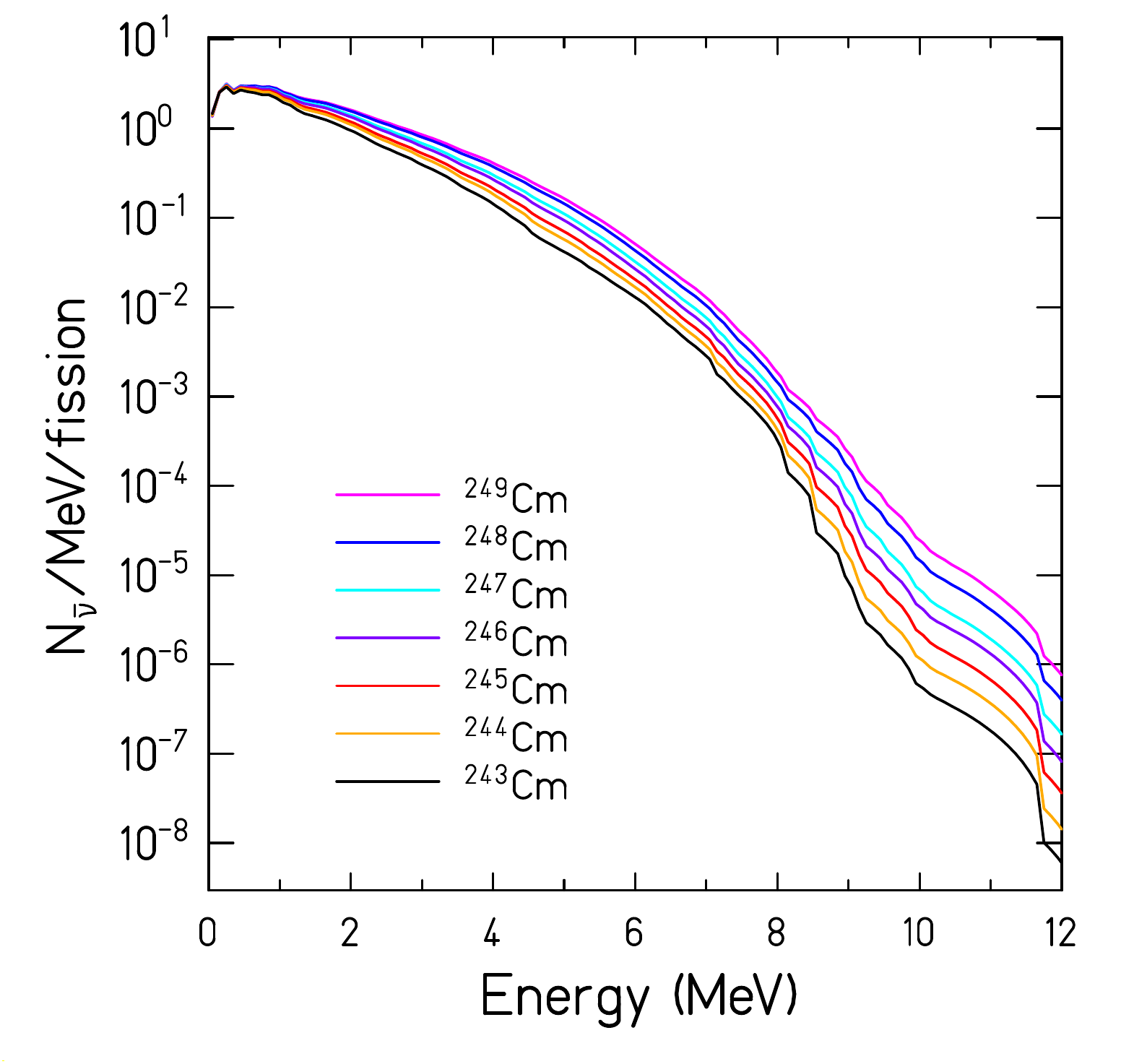}
\caption{Calculated antineutrino spectra from GEF combined with the selection of decay data of~\cite{Estienne} for seven isotopes of $Z$ = 96 in the case of thermal fissions.}
\label{spectresCurium}       
\end{figure}

\begin{figure}[h]
\centering
\includegraphics[width=0.36\textwidth]{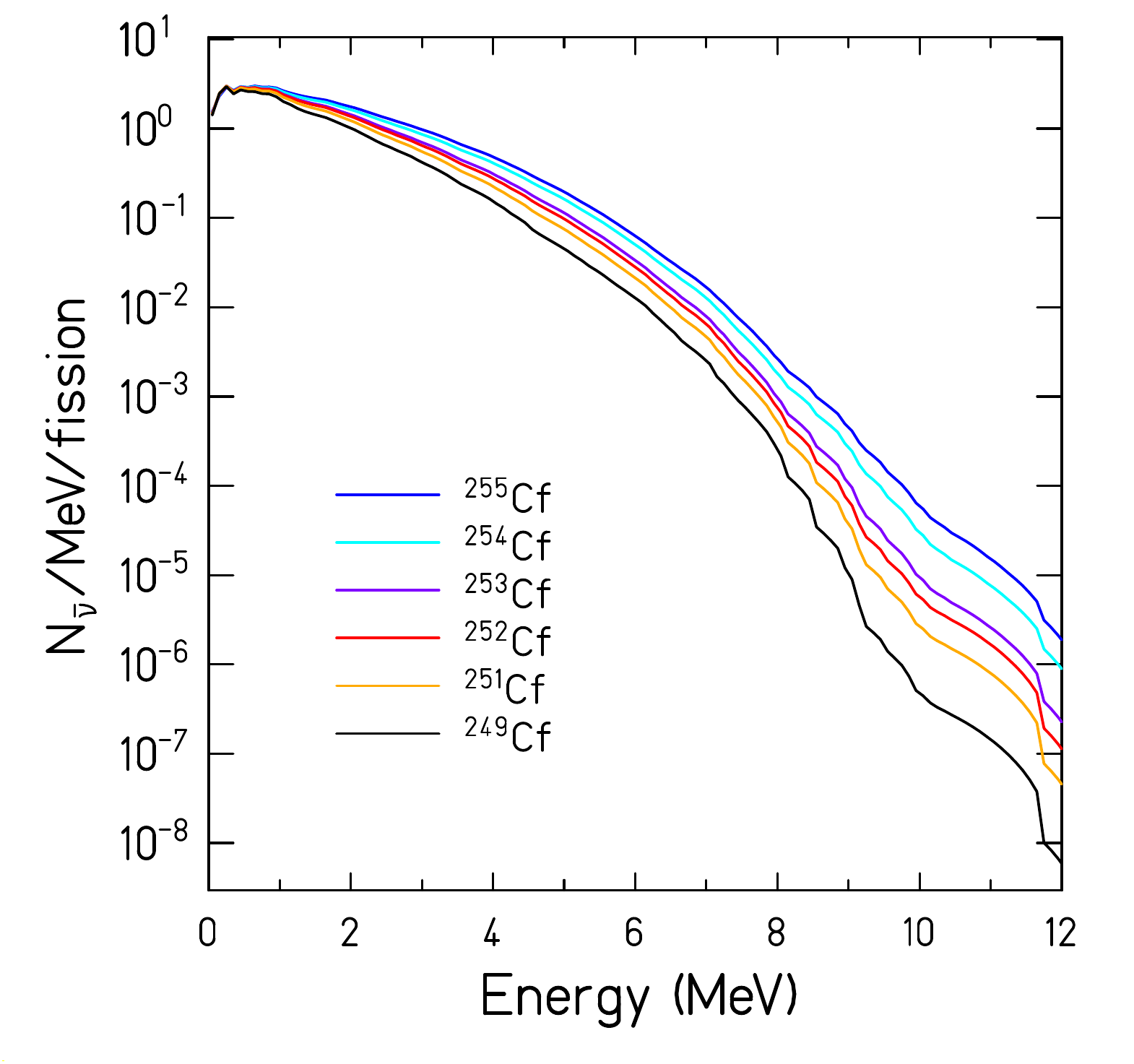}
\caption{Calculated antineutrino spectra from GEF combined with the selection of decay data of~\cite{Estienne} for six isotopes of $Z$ = 98 in the case of thermal fissions.}
\label{spectresCalifornium}       
\end{figure}

The antineutrino energy spectra for the systems listed in the first part of this article have been computed using the GEF cumulative fission yields combined with the set of beta decay data described above~\cite{Estienne}. They are shown in Figs.~\ref{spectresThorium},~\ref{spectresUranium},~\ref{spectresNeptunium},~\ref{spectresPlutonium},~\ref{spectresAmericium},~\ref{spectresCurium},~\ref{spectresCalifornium} for the systems $Z$ = 90, $Z$ = 92, $Z$ = 93, $Z$ = 94, $Z$ = 95, $Z$ = 96 and $Z$ = 98, respectively, in case of thermal fission. 

For the specific case of $^{238}$U, which undergoes fast-neutron-induced fission, we have considered two cases: fission with 2 MeV incident neutrons and fission with an energy distribution of the incident neutrons, which follows the prescription of~\cite{Kern12} (cf. Fig. \ref{EN-U238F}). 
The calculations in this section for all other target nuclei are performed assuming fission induced by thermal neutrons, also for those target nuclei, which are thermally not-fissile. 
A detailed view on the additional effect of energy-dependent fission cross sections, which depends on the reactor type, is beyond the scope of this work. 
(From section \ref{S6} it may be assumed that it is small in most cases, except for $^{238}$U.)
This way, a clear view on the influence of global and structural features of the fission process itself and of the radioactive-decay properties on the production of antineutrinos and other particles is obtained.

The corresponding tables containing the datapoints of the spectra are provided as supplemental material to this paper. These spectra can be used to compute the antineutrino energy spectra of future reactors loaded with various types of fuels in the frame of non-proliferation scenario studies. A big advantage of this set of spectra is the consistency brought by the use of the same model to compute the fission yields of all fissioning systems. This consistency could not be attained with the current evaluated fission yield datasets for the variety of fissioning systems that are needed for non-proliferation studies because of the lack of underlying experimental data and because of the remaining problems in the data listed in the paragraphs above. In the frame of the comparisons of diversion scenarios versus legitimate use of a given type of nuclear reactor~\cite{CormonND}, the fission yield covariance matrices provided by GEF make it possible to computate the uncertainties for the corresponding antineutrino emissions.

With the sets of fissioning systems provided in this article, studies of the antineutrino emission of reactors loaded with thorium/uranium fuel such as CANDU reactors, or studies of Generation-IV breeders including blankets loaded with minor actinides or of ADS loaded with assemblies containing nuclear wastes for transmutation purposes are possible.
The spectra provided in this article are computed for fission induced by thermal neutrons. The study of the fast-neutron-induced fission yields of $^{238}$U presented in the section~\ref{U238F} has shown that a more accurate determination of the fission yields could be obtained by taking into account their dependence on the neutron energy spectrum in the reactor in the fission process, a dependenceF that has an influence on the antineutrino emission as we could observe in the case of the fast-neutron-induced fission of $^{238}$U. This is the reason why we would recommend to readers interested in fast-neutron-induced fission to use the neutron energy distribution of the studied reactor design in the calculation of the corresponding fission yields with GEF. 


We have performed a quick systematic study of the antineutrino spectra presented in the Figs.~\ref{spectresThorium},~\ref{spectresUranium},~\ref{spectresNeptunium},~\ref{spectresPlutonium},~\ref{spectresAmericium},~\ref{spectresCurium} and~\ref{spectresCalifornium}. In the following, the specific case of $^{238}$U is considered to be beyond the scope of the present work, which focuses on thermal fission. It will be addressed in a future publication. 

\begin{figure}[h]
\centering
\includegraphics[width=0.36\textwidth]{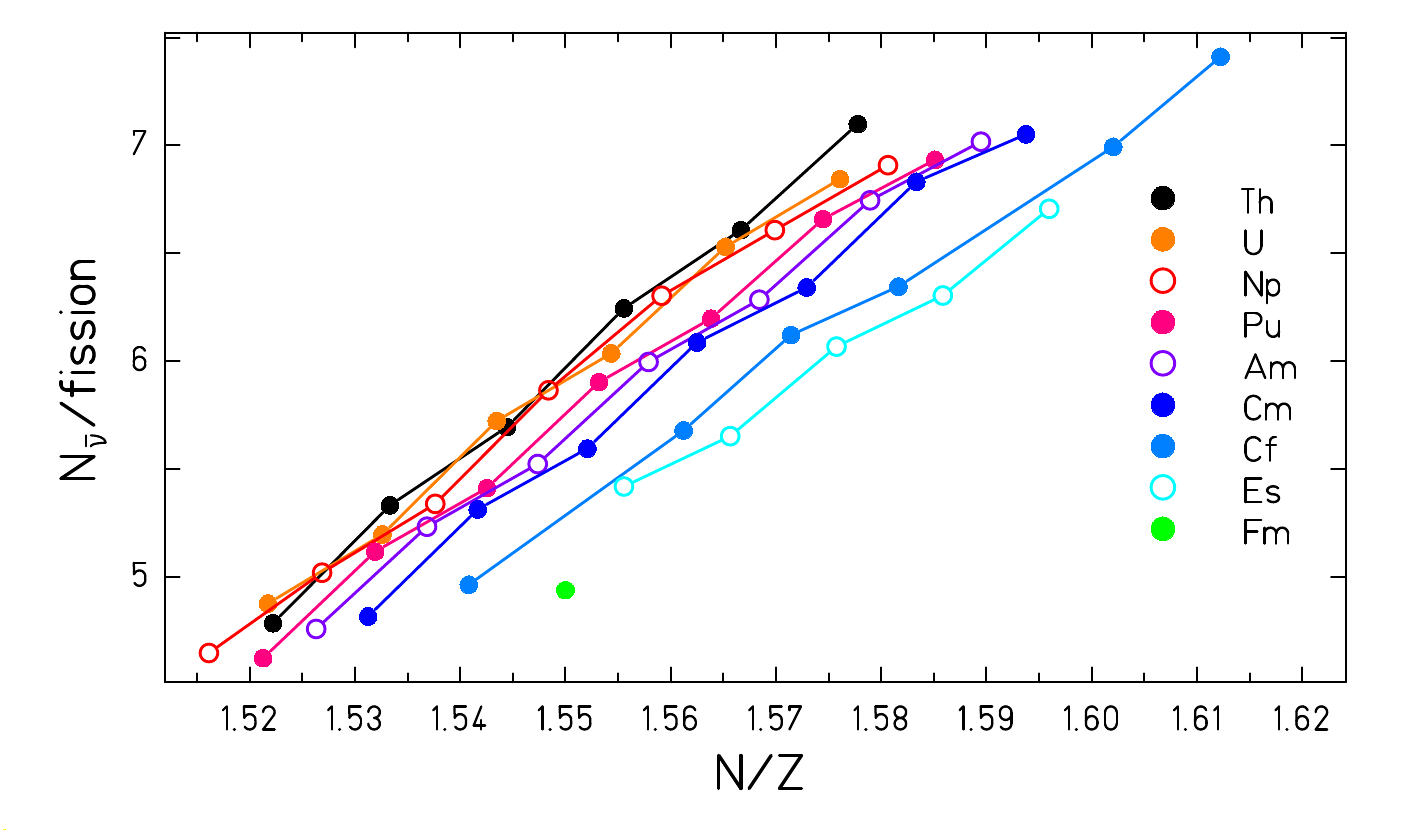}
\includegraphics[width=0.36\textwidth]{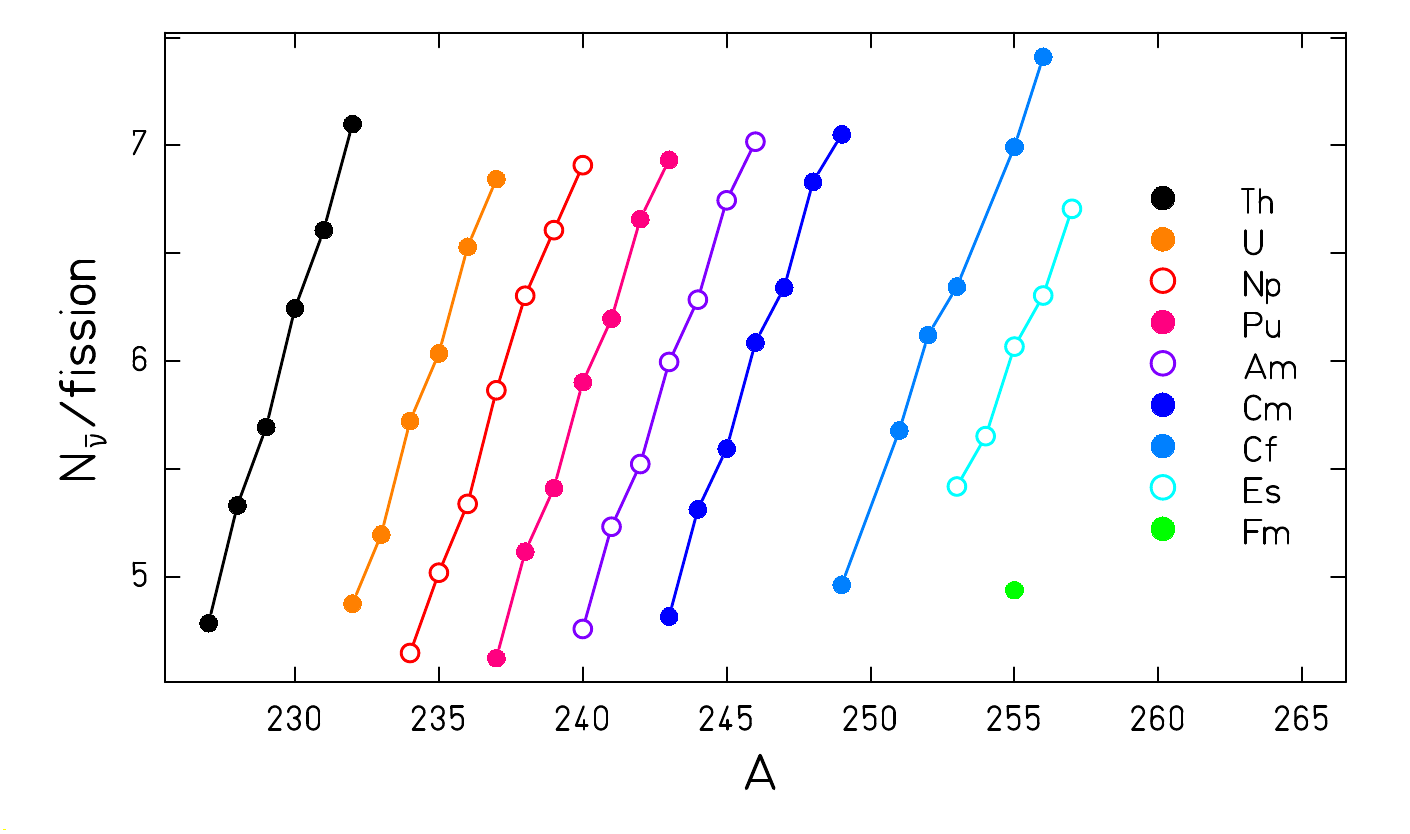}

\caption{Top: emitted antineutrino flux corresponding to the spectra of Fig.~\ref{spectresThorium},~\ref{spectresUranium},~\ref{spectresNeptunium},~\ref{spectresPlutonium},~\ref{spectresAmericium},~\ref{spectresCurium},~\ref{spectresCalifornium} obtained with the fission yields from GEF combined with the selection of decay data of~\cite{Estienne} for different systems as a function of the $N$ over $Z$ ratio of the fissioning system.
Bottom: same but plotted as a function of $A$.}
\label{dependencyNoverZ}       
\end{figure}

The antineutrino emitted fluxes per fission corresponding to isotopic chains of the fissioning systems are presented in Fig.~\ref{dependencyNoverZ} as a function of the $A$ over $Z$ ratio of the fissioning nucleus. They all show a generic linear trend on which an odd-even effect is superposed. A linear trend has already been shown in Fig. 4 of~\cite{Sonzogni} where the detected antineutrino flux was plotted as a function of (3$Z$-$A$) for a set of fissioning systems. In this figure, the detected antineutrino fluxes nearly align, but some deviations could be observed. The odd-even effect observed in Fig.~\ref{dependencyNoverZ} is one of the reasons of the deviations observed in~\cite{Sonzogni}. In Fig.~\ref{dependencyNoverZ}, two linear fits could be performed for the sets of odd or even isotopes of each element. In addition, the emissions of some isotopic chains are very close to each other as a function of $N$ over $Z$. This is the case for thorium, uranium and neptunium, and then plutonium, americium and curium. 
The bottom plot of Fig.~\ref{dependencyNoverZ} shows the same antineutrino fluxes but plotted as a function of $A$, which makes the figure easier to read. The linear trend for each isotopic chain appears more clearly as the scale makes the odd-even effect appearing smaller. The lines are quasi-parallel showing that the increase of the neutron number of the fissioning nucleus directly impacts the neutron number of the fission products. The number of emitted antineutrinos per fission follows directly the difference in $N$ over $Z$ between the post-neutron fission products and the stability valley. This implies a strong correlation with the $N$ over $Z$ ratio of the fissioning system, a correlation that we find to be linear.   
From this generic study, it is easy to extrapolate the values of antineutrino fluxes to fissioning systems for which neither data nor calculation exist.

\begin{figure}[h]
\centering
\includegraphics[width=0.36\textwidth]{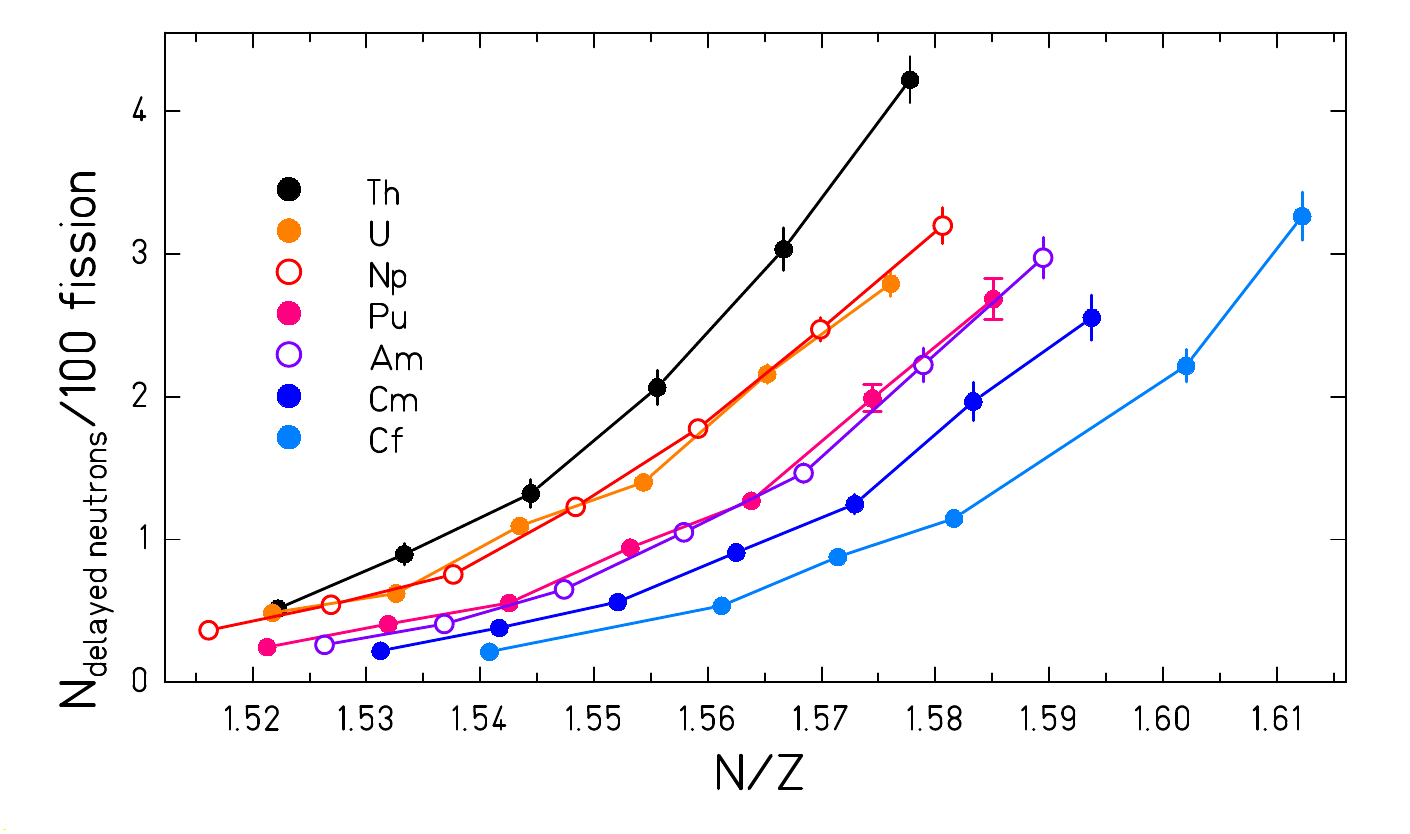}
\caption{Delayed-neutron fractions computed with GEF for different systems as a function of the $N$ over $Z$ ratio of the fissioning system.}
\label{dependencydnNoverZ}       
\end{figure}

\begin{figure}[h]
\centering
\includegraphics[width=0.36\textwidth]{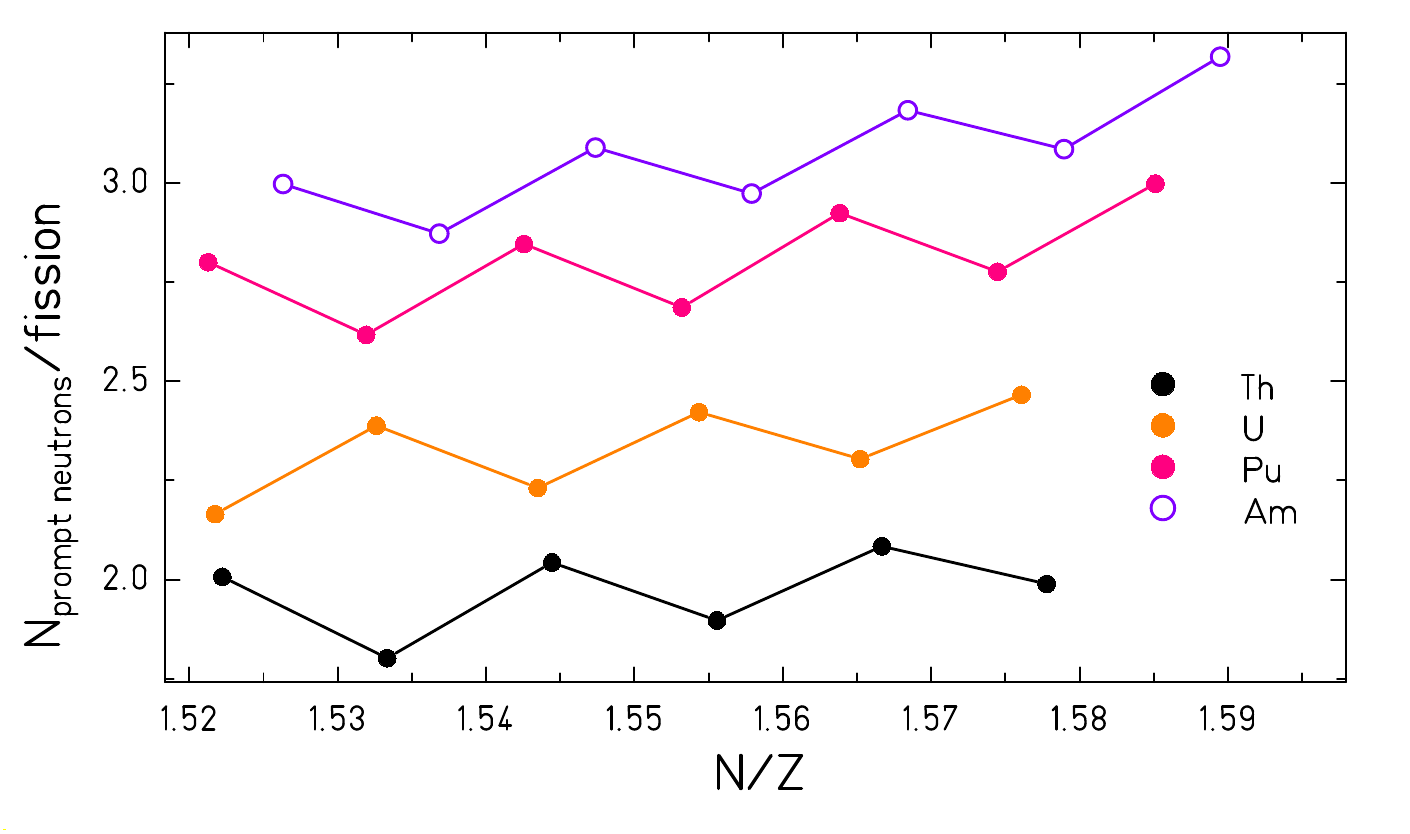}
\caption{Prompt neutron multiplicities computed with GEF for different systems as a function of the $N$ over $Z$ ratio of the fissioning system.}
\label{dependencypnNoverZ}       
\end{figure}

For comparison, the delayed-neutron fractions and the number of prompt neutrons per fission associated with different fissioning systems were computed with GEF as well.
In Fig.~\ref{dependencydnNoverZ}, the delayed-neutron fractions per fission show a behavior similar to that of antineutrinos, except that their increase with $N$ over $Z$ is not linear. This could be explained by the fact that only the neutron-richest nuclei are $\beta$-n precursors. Moreover, the odd-even effect along isotopic chains is less pronounced.

\begin{figure*}[h]
\centering
\includegraphics[width=1.0\textwidth]{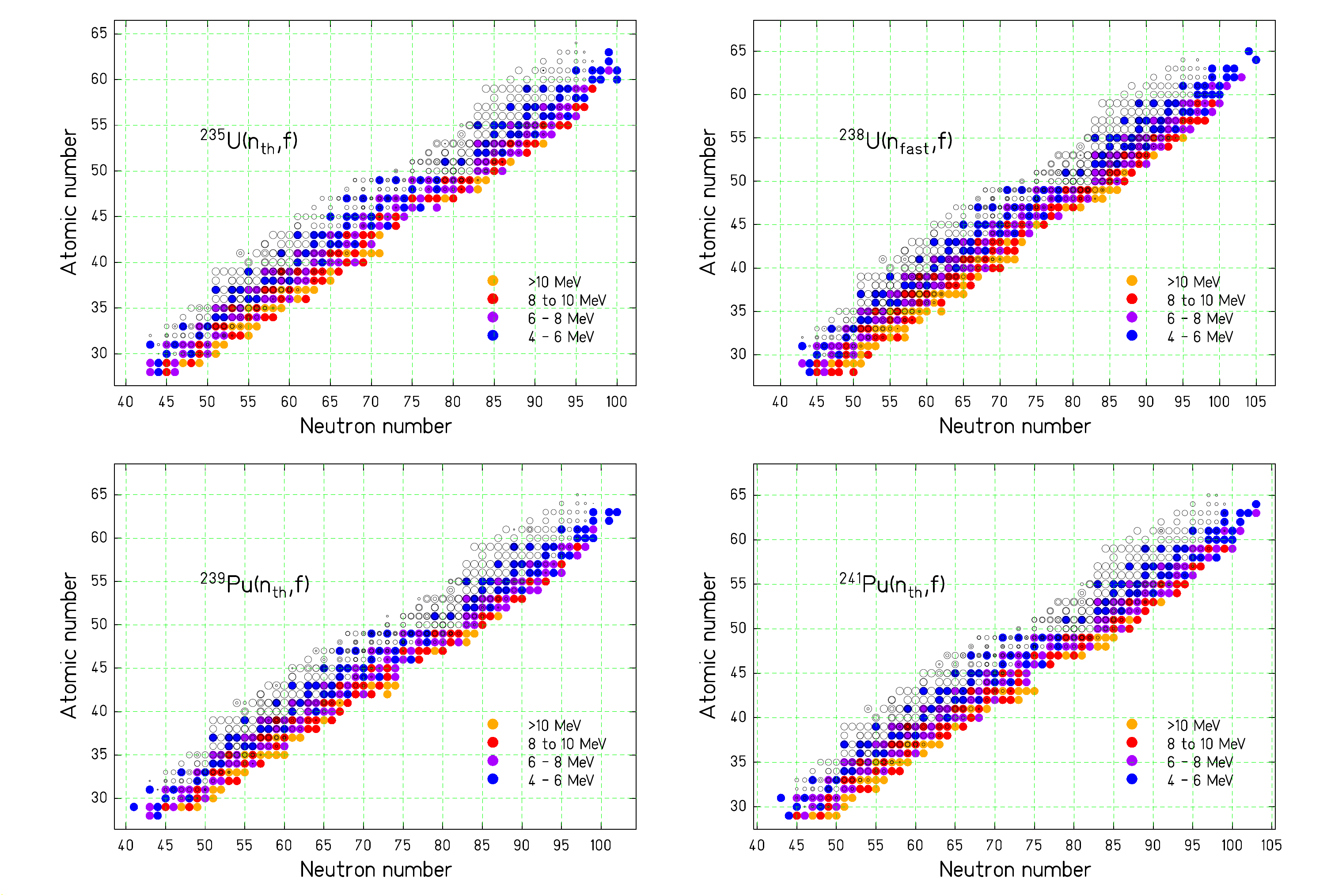}
\caption{Individual beta decay contributions of fission products from the reactions $^{235}$U(n$_{\text{th}}$,f), $^{238}$U(n$_{\text{fast}}$,f), $^{239}$Pu(n$_{\text{th}}$,f) and $^{241}$Pu(n$_{\text{th}}$,f) calculated with the GEF code. The size of the black circles, which shrink to points towards increasing neutron excess, is proportional to the logarithm of the magnitude over four orders of magnitude, and the color indicates the Q-value range.} 
\label{Q-U235T}       
\end{figure*}

\clearpage

The delayed-neutron fractions of uranium and neptunium isotopes, as well as those of plutonium and americium are almost identical. 
This is explained by the fact that the odd-even effect in fission-product $Z$ is much larger for even-$Z$ fissioning systems. Thus, the relative production of odd-$Z$ fission products is enhanced for odd-$Z$ fissioning systems. 

Due to the enhanced beta Q value of odd-$Z$ fission products, the delayed-neutron fractions for odd-$Z$ fissioning systems are so much enhanced that the average decrease of the delayed-neutron fraction for fixed $N$ over $Z$ by the increase of $Z$ by one unit is just compensated. 
This way the values of the even-$Z$ uranium and the neighboring odd-$Z$ neptunium as well as the values of the even-$Z$ plutonium and the neighboring odd-$Z$ americium become very close.  
In more detail, also the enhanced fission Q values for odd-$Z$ fissioning systems that lead to slightly less neutron-rich fission products and to a reduction of the delayed-neutron fractions (see next paragraph) must be considered.

The number of prompt neutrons per fission for four isotopic chains are plotted in Fig.~\ref{dependencypnNoverZ}. Their behavior deviates from that of antineutrinos and delayed neutrons. The prompt-neutron multiplicity is most sensitive to the $Z$ of the fissioning system and shows only a moderate increase along the isotopic chains. 
This corresponds to the linear increase of the total fission-fragment excitation energy with $Z^2/A^{1/3}$, postulated by Asghar and Hasse \cite{Asghar84} on the basis of macroscopic nuclear properties.
The odd-even staggering of the prompt-neutron multiplicity as a function of neutron number is clearly evidenced. 
The largest values of prompt neutrons per fission are obtained for odd-$N$ fission targets. 

The staggering can be explained as follows:
With respect to the smooth variation, even-$N$ target isotopes are more bound than odd-$N$ isotopes by the pairing energy $\Delta_N$, and, in addition, their neutron separation energy is higher by $2 \Delta_N$. 
This leads to a systematically higher fission Q value for odd-$N$ target nuclei by $\Delta_N$ in thermal-neutron-induced fission and a corresponding larger number of prompt neutrons. 
It is assumed that most of this staggering appears in the total excitation energy of the fission fragments, although some odd-even staggering might also be present in the total kinetic energy by the breaking of nucleon pairs during the fission process according to an idea of Ref. \cite{Lang80}. This effect is included in GEF. 
However, measured prompt-neutron multiplities may show weaker staggering, because many even-$N$ nuclei are thermally not-fissile.
In these cases, the observed fission events may be induced by neutrons of higher energies, depending on the experimental conditions, like discussed in section \ref{U238F} in the case of $^{238}$U.

Since delayed neutrons and antineutrinos originate both from the radioactive decay of the fission products while the prompt neutrons arise from the de-excitation of the fission fragments, a similar trend is awaited in the behavior of the two first observables. The main difference between the two is that the delayed-neutron precursors are less numerous than the antineutrino emitters, making the antineutrino emission more sensitive to the fission yield distribution in its entirety. 

The comparative view on the characteristics of antineutrinos, prompt and delayed neutrons demonstrates the complexity of global trends and structural properties of these three observables. In particular, it calls into question any linear interpolation or extrapolation of trends deduced from scarce or incomplete experimental data.

Altogether, the results in this section demonstrate that the antineutrino observable is directly linked to the fission process, and an improved experimental knowledge of the antineutrino emission could help understanding the fission process itself.



\subsection{Sensitivity to the fission product distributions from different systems}

In addition to calculating the so-far presented yields of the secondary fission products resulting from the de-excitation of the primary fragments produced at scission, the GEF model can compute their radioactive decay whenever it applies. Hence, the code can provide a complete overview of the contributions of the various fission products to the Q value distribution of beta decays. 
Fig. \ref{Q-U235T}
shows the calculated intensities and Q values of the beta decays for the four systems $^{235}$U(n$_{\text{th}}$,f), $^{238}$U(n$_{\text{fast}}$,f), $^{239}$Pu(n$_{\text{th}}$,f) and $^{241}$Pu(n$_{\text{th}}$,f) on the chart of the nuclides.

The highest decay energies are generally found in the light fission-product group with an odd-even staggering that enhances the decay energies of the odd elements. 
A detailed analysis of the results of the calculation shows that 
high decay energies (above 9 MeV) and presumably also the high-energy part of the antineutrino spectrum are dominated by contributions of the odd-$Z$ elements from $Z$ = 33 to $Z$ = 37. 

\begin{figure}[h]
\centering
\includegraphics[width=0.5\textwidth]{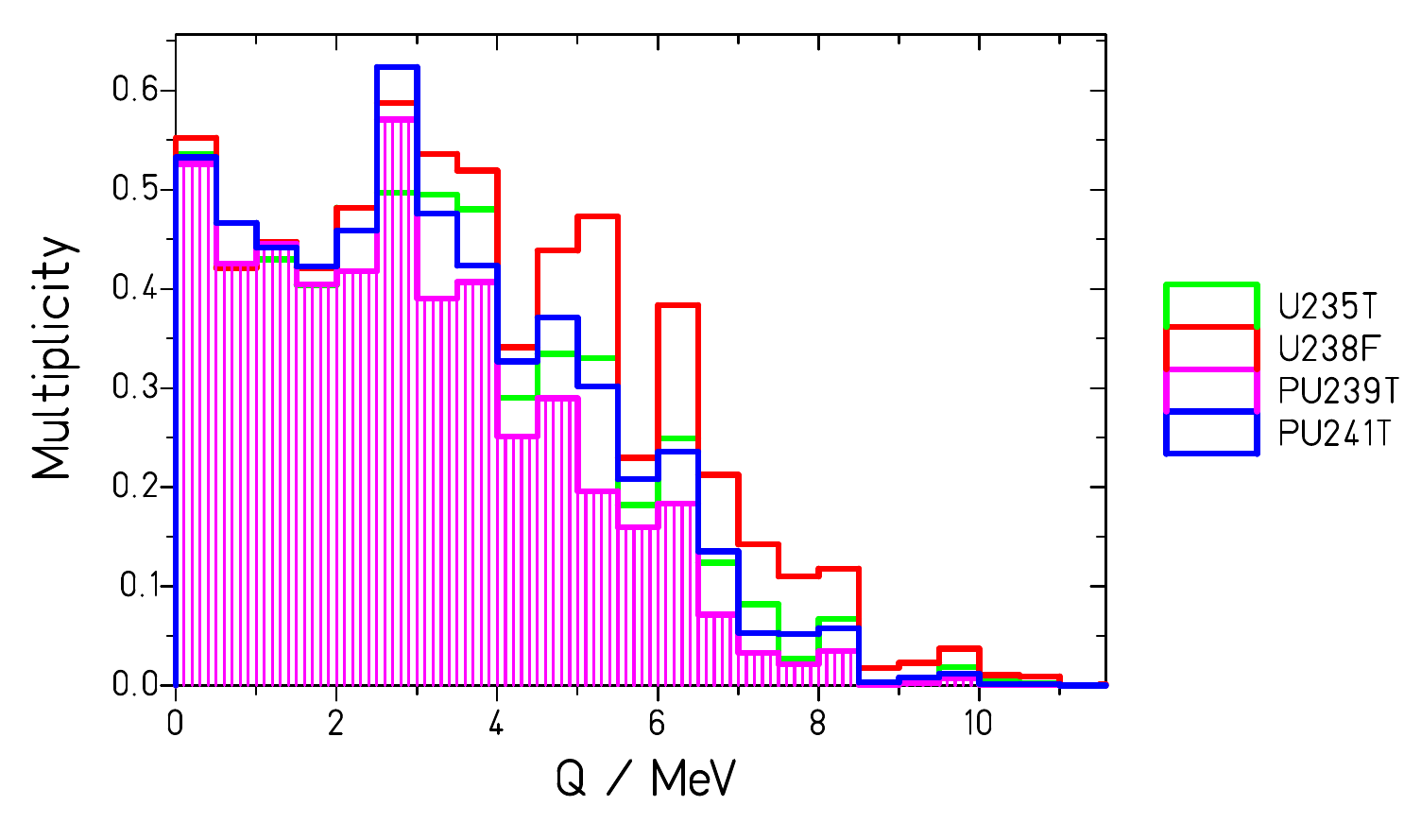}
\caption{Antineutrino multiplicity as a function of the Q value of the consecutive beta decays of the fission products for the systems $^{235}$U(n$_{\text{th}}$,f), $^{238}$U(n$_{\text{fast}}$,f), $^{239}$Pu(n$_{\text{th}}$,f), and $^{241}$Pu(n$_{\text{th}}$,f), calculated with the GEF code. For clarity, the spectrum is shown with a coarse binning of 500 keV.} 
\label{QBETA}       
\end{figure}

As expected, one observes a shift to more exotic nuclides with a tendency to longer beta-decay chains with increasing neutron excess of the fissioning system.
This goes in line with an enhancement of higher decay energies.
$^{238}$U(n$_{\text{fast}}$,f) provides the highest contributions to the high-energy part of the spectrum, because it is the most neutron-rich system.
A detailed comparison of this piece of information with that on the accuracy of the fission-product yields, discussed in the preceding sections, provides a good basis for revealing the contributions of individual fission-yield uncertainties to the uncertainties of calculated antineutrino spectra.



The possible application of antineutrino spectroscopy for reactor monitoring depends essentially on the sensitivity of the antineutrino energy spectrum to the fissioning system.
A first glance on this sensitivity can be obtained by accumulating the Q values of the consecutive beta decays of the fission products with their respective appearances. 
This signature has the advantage of not being influenced by the branchings of the beta decay to excited levels. 
This could introduce a bias, because the experimental knowledge on these branchings is systematically less detailed for the more neutron-rich nuclei.

Fig. \ref{QBETA} shows the accumulated distribution of Q-values for 
$^{235}$U(n$_{\text{th}}$,f), $^{238}$U(n$_{\text{fast}}$,f), $^{239}$Pu(n$_{\text{th}}$,f), and $^{241}$Pu(n$_{\text{th}}$,f). 
Obviously, there is a systematic and rather important increase of the antineutrino multiplicity, in particular at higher energies, with increasing $A/Z$ of the fissioning system as was observed in the section~\ref{S8}. 
Because the relative enhancement is energy dependent, the shape of the antineutrino energy spectrum is sensitive to the relative contributions of the different fissioning systems. 
Combining this information with the expected uncertainty of the measured antineutrino energy spectrum will provide a good estimation of the sensitivity of antineutrino spectroscopy for reactor monitoring in specific cases.
A more detailed quantitative study is beyond the scope of the present work, which is limited to prove that GEF provides the necessary tools for such a work.


\section{Conclusion}

The calculation of the antineutrino production in fission product decay is presently one of the most demanding applications of nuclear data due to the required high accuracy. This is true for both
antineutrino physics and spectroscopy for reactor monitoring.
In this work, it was shown that the presently reached quality of related nuclear data, in particular of the fission yields, can appreciably be improved by exploiting and combining different approaches: traditional radiochemical experiments, kinematic experiments and suitable theoretical models. 
For the first time, a careful analysis and a systematic comparison between data from different sources and evaluations and with GEF have been performed to sort out the more reliable and the less trustworthy values, thus assisting the evaluation process.

Examples were shown of how erroneous data in different evaluations, up to very recent ones, can be detected and rather credible estimations of un-measured values can be performed.
In a number of cases, our recommendations were given to replace apparently erroneous data by more realistic estimations.
%

As a result of this work, the level of agreement attained on the antineutrino energy spectra computed with the new GEF fission yields in comparison with the JEFF evaluated fission yields has been remarkably improved in the case of the four main fissioning systems in actual reactors. 
Predictions performed with the summation method using the GEF cumulative fission yields show that the new version of GEF (GEF-Y2019/V1.2) has reached a level of predictiveness of the Inverse Beta Decay yields at the percent level with respect to the one of the JEFF-3.1.1 evaluated fission yields and only 2.5\% above the Daya Bay IBD yields once corrected from the contribution of the very long-lived nuclei. 
This excellent agreement with the results obtained using the JEFF-3.1.1 and JEFF-3.3 fission yields as well as the indications for realistic GEF-based uncertainty estimates for the most important fissioning systems open the new possibility to propagate the latter from GEF to the antineutrino spectra.

A systematics of calculated intensities and beta Q values of all fission products for the four most important fissioning systems, contributing to the antineutrino production in a fission reactor, reveals some prevailing characteristics of the underlying fission and radioactive-decay processes. 
These are crucial for estimating the sensitivity of a possible application of antineutrino production to reactor monitoring.  
Predictions of antineutrino energy spectra based on the GEF fission yields combined with the most recent decay data sets from~\cite{Estienne} are provided for a list of fissioning systems which could be used in the frame of such sensitivity studies.

By extending the GEF calculations, presented in this work, with explicit calculations of the beta-decay energies, including error propagation and correlations, one obtains a powerful tool for identifying the specific problems and limitations of the summation method that determine the quality that can presently be reached. 
This can also be used for establishing a list of most urgent improvements of the quality of underlying nuclear data.

\section*{Acknowledgement}
This work was supported by the University of Nantes, the CNRS/in2p3, and the NACRE project of the NEEDS challenge by financing several research visits in 2018 and 2019. 
K.-H. S. thanks the SUBATECH laboratory for warm hospitality. 
 
\newpage 
 
\section{Appendix} 







\section*{Observations}

Four systems were selected for a detailed comparison of the independent yields from the JEFF-3.3 evaluation and the GEF results. 
These systems contribute most strongly to the antineutrino production in presently operating fission reactors.

\subsection*{$^{235}$U(n$_{\text{th}}$,f)}
\begin{figure}[h]
\centering
\includegraphics[width=1.0\textwidth]{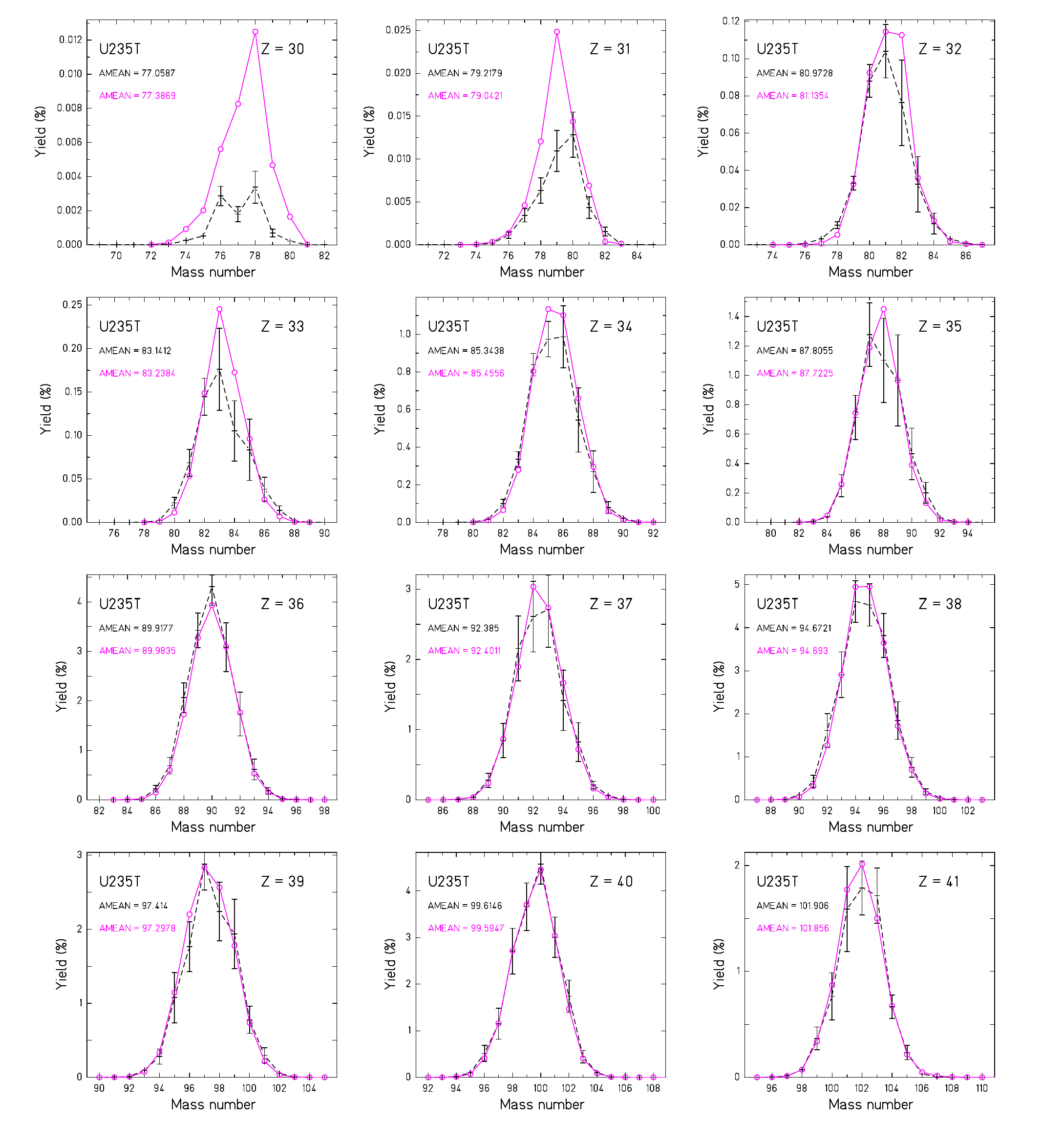}
\caption{Isotopic distributions of $^{235}$U(n$_{\text{th}}$,f) fission-product yields for different elements, comparison of JEFF-3.3 (black symbols and error bars) and GEF (magenta symbols), linear scale.
The mean mass values are given in the appropriate color.} 
\label{SUB-ZA_1}       
\end{figure}

\begin{figure}[h]
\centering
\includegraphics[width=1.0\textwidth]{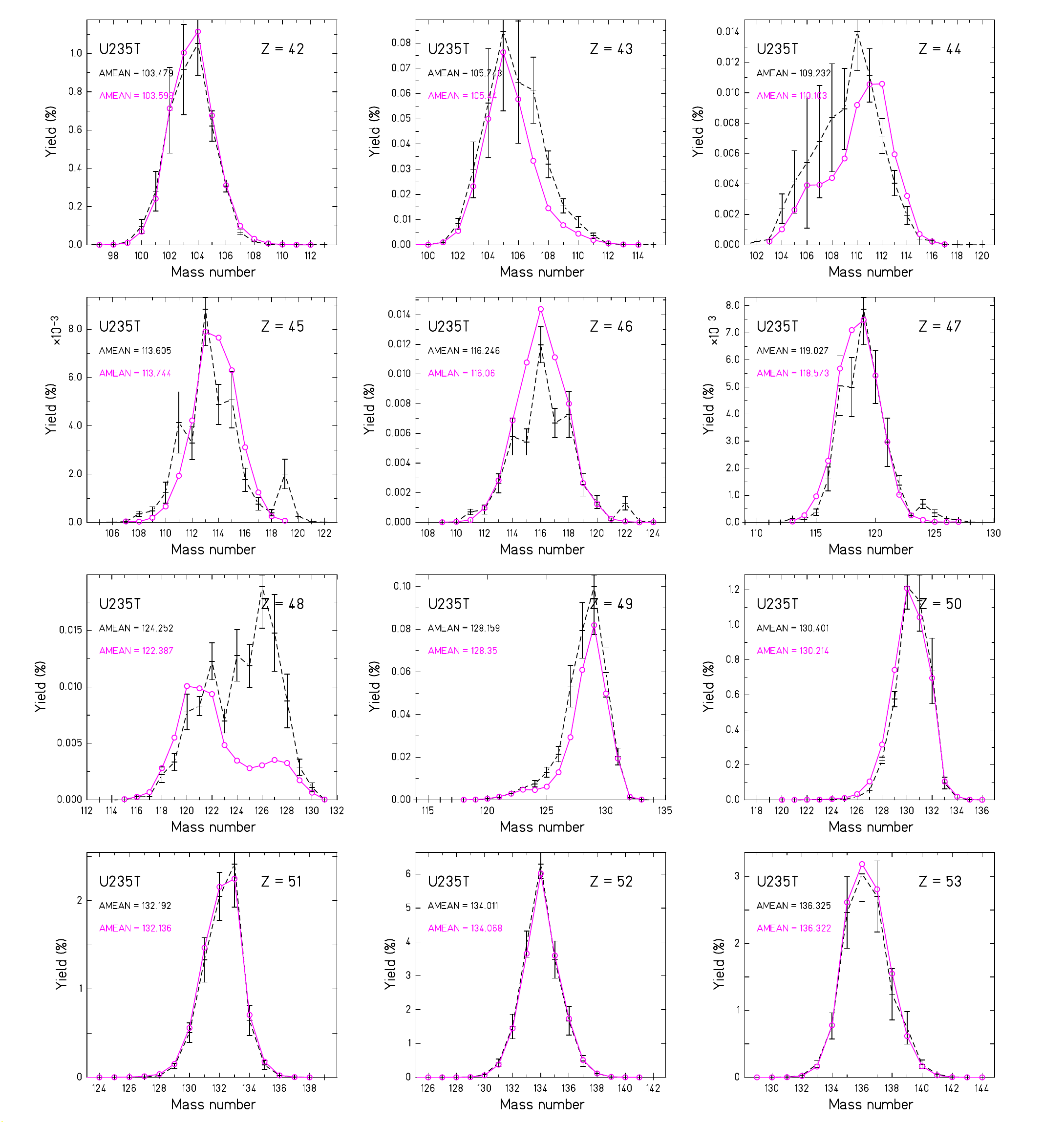}
\caption{Isotopic distributions of $^{235}$U(n$_{\text{th}}$,f) fission-product yields, comparison of JEFF-3.3 and GEF, linear scale.} 
\label{SUB-ZA_2}       
\end{figure}

\begin{figure}[h]
\centering
\includegraphics[width=1.0\textwidth]{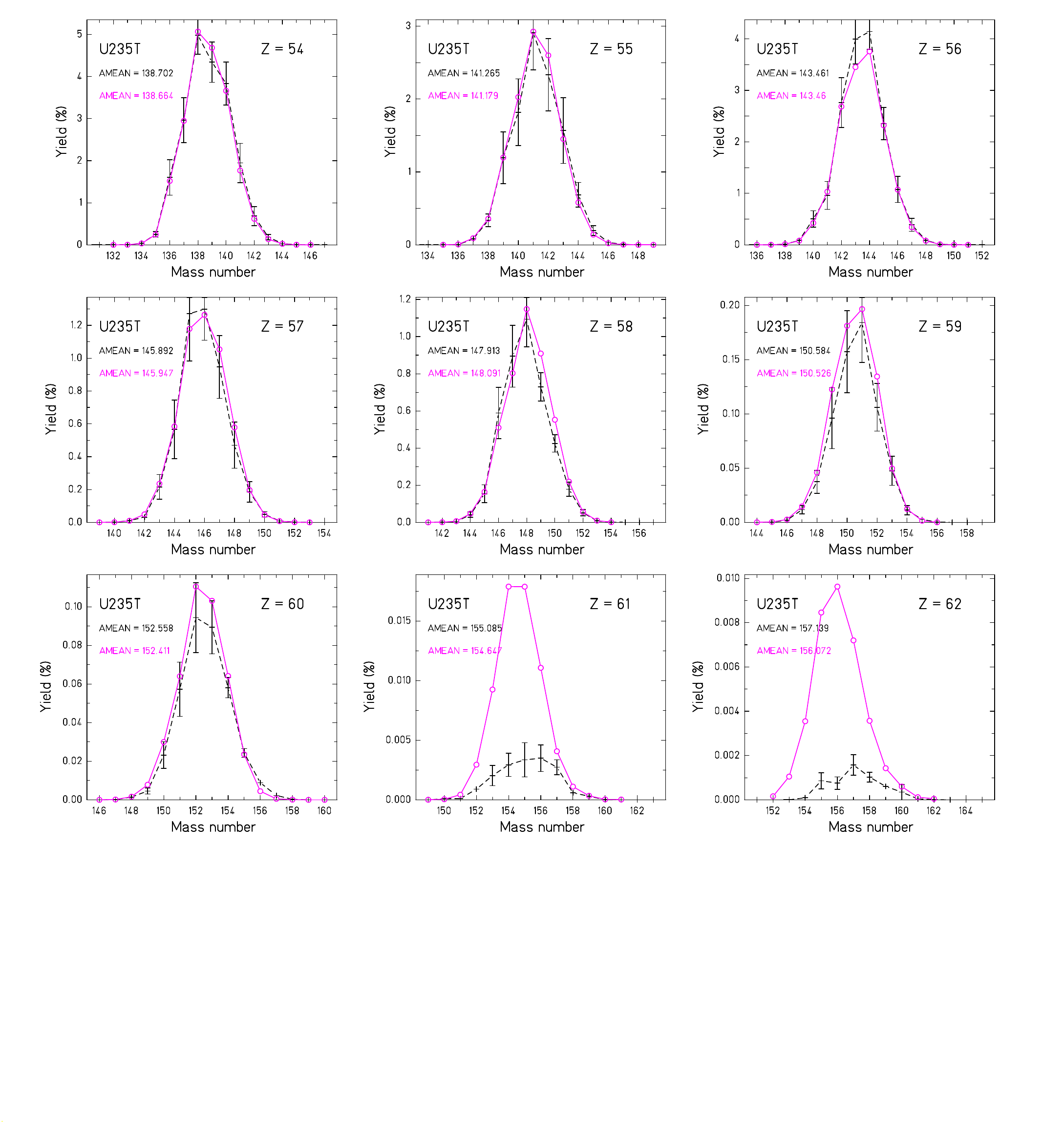}
\caption{Isotopic distributions of $^{235}$U(n$_{\text{th}}$,f) fission-product yields, comparison of JEFF-3.3 and GEF, linear scale.} 
\label{SUB-ZA_3}       
\end{figure}

\begin{figure}[h]
\centering
\includegraphics[width=1.0\textwidth]{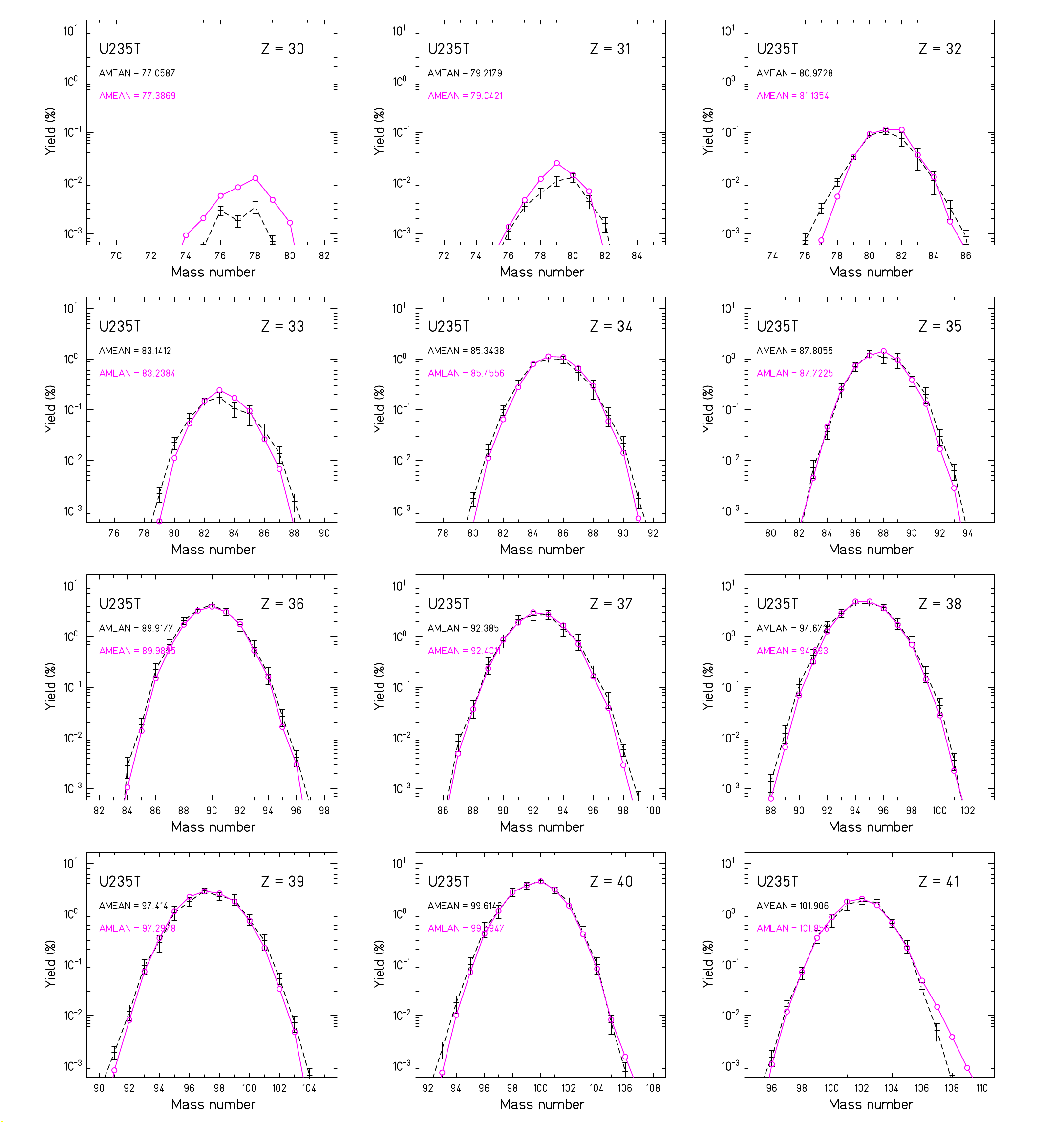}
\caption{Isotopic distributions of $^{235}$U(n$_{\text{th}}$,f) fission-product yields, comparison of JEFF-3.3 and GEF, logarithmic scale.} 
\label{SUB-ZA_7}       
\end{figure}

\begin{figure}[h]
\centering
\includegraphics[width=1.0\textwidth]{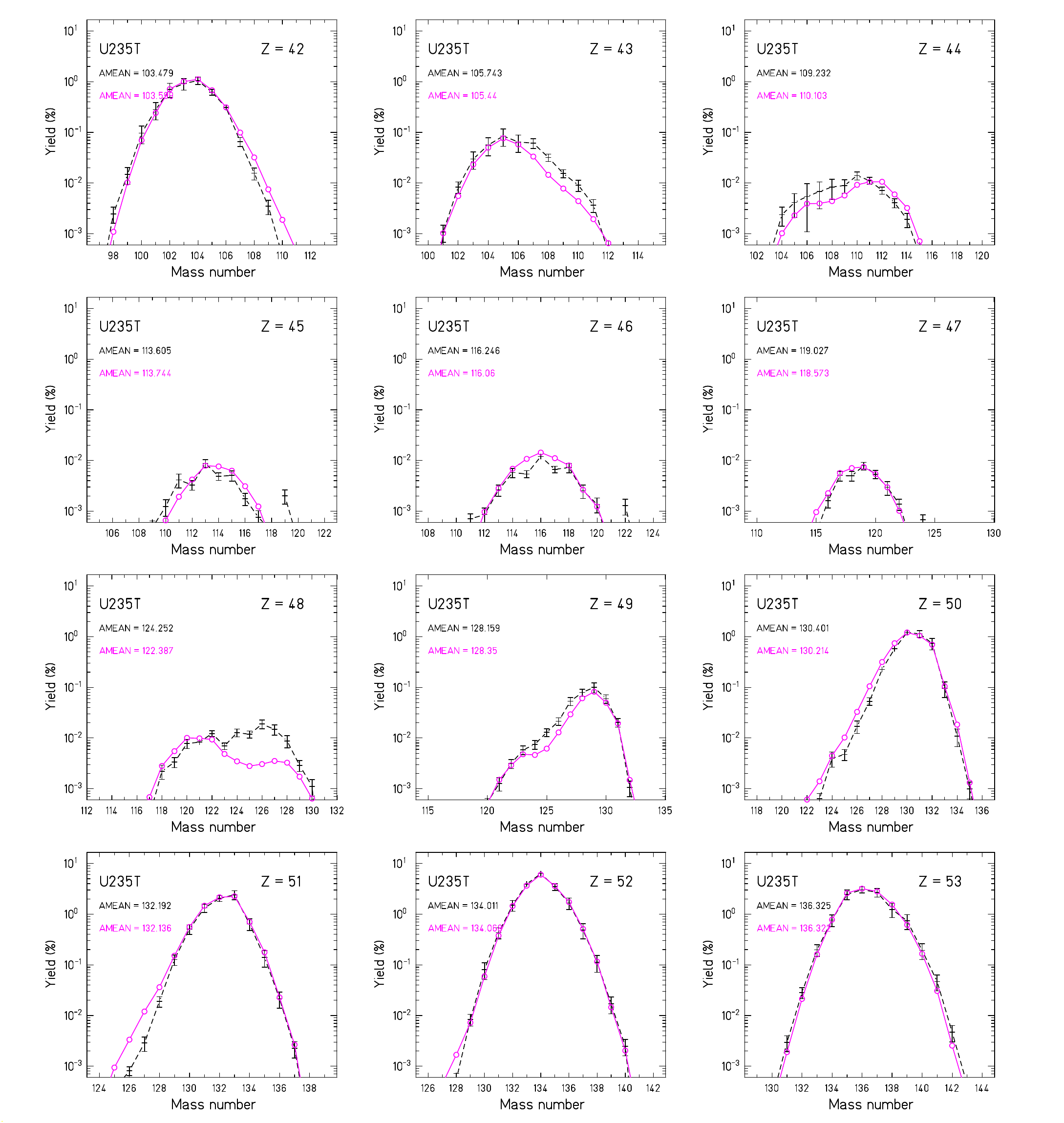}
\caption{Isotopic distributions of $^{235}$U(n$_{\text{th}}$,f) fission-product yields, comparison of JEFF-3.3 and GEF, logarithmic scale.} 
\label{SUB-ZA_8}       
\end{figure}

\begin{figure}[h]
\centering
\includegraphics[width=1.0\textwidth]{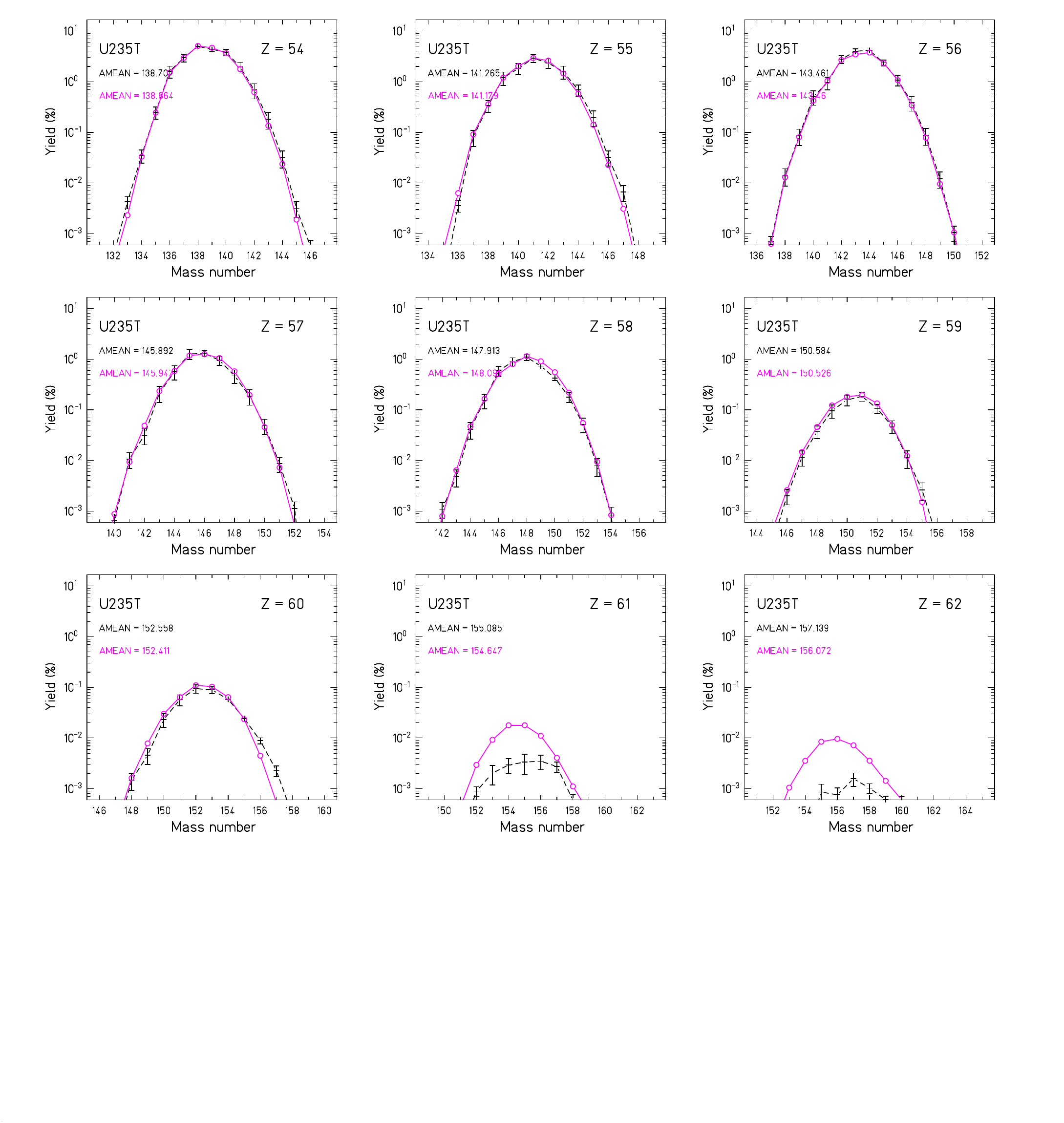}
\caption{Isotopic distributions of $^{235}$U(n$_{\text{th}}$,f) fission-product yields, comparison of JEFF-3.3 and GEF, logarithmic scale.} 
\label{SUB-ZA_9}       
\end{figure}

$^{235}$U(n$_{\text{th}}$,f) is the most intensively studied reaction. 
Thus, the evaluated yields for this case are expected to be the most reliable. Figs. \ref{SUB-ZA_1} to \ref{SUB-ZA_9} show almost perfect agreement between JEFF-3.3 and GEF for the elements with peak yields above 1\%.
There are some issues in the most asymmetric wings, where the super-asymmetric (S3) mode contributes. 
Severe discrepancies appear for $Z < 32$ and $Z > 60$. In both cases, the yields are overestimated; in the second case, the isotopic distributions are shifted towards lighter isotopes in addition. 
The distributions near symmetry are rather well reproduced. 
However, the shape of the distribution of the $Z = 48$ isotopic yields from GEF does not agree with the one from the evaluation: the height of the right peak is strongly underestimated. An underestimation of the left wing of the mass distributions for $Z = 44$ is also seen, although by a much smaller amount. However in both the evaluation and the calculation the conservation laws are fulfilled, by imposing it in the first case and by the consistency of the model in the second case.
The differences in shape of the different distributions (for the light and the heavy fission product from GEF and from JEFF-3.1.1) can be explained by the influence of the prompt-neutron-multiplicity distribution as a function of the fission-product $A$ and $Z$ in these four cases.

It is known that the symmetric mode is characterized by a small charge polarization and a low TKE, corresponding to a large prompt-neutron multiplicity, while the asymmetric modes (S1 and S2) in this $Z$ range are characterized by a large charge polarization, favoring the production of neutron-rich heavy fragments at scission, and a high TKE, corresponding to a small prompt-neutron multiplicity. With this information, one can attribute the left peak in the isotopic distribution of $Z$ = 48 to the symmetric mode and the right peak to the asymmetric component, consisting of the S1 and S2 fission channels.
Thus the contribution of the symmetric mode to the $Z$ = 48 yield is correctly calculated by GEF, while the contribution of the asymmetric component is underestimated.  
In view of the good reproduction of the distributions of $Z = 50$ and higher, which fixes the shape of the heavy part of the asymmetric component, the shape of the distribution of $Z = 48$ indicates the presence of a further-reaching tail of the asymmetric component towards symmetry.
This problem is already visible in the distributions from $Z = 45$ to $Z = 47$. However, the almost constant intensity of the right side-peak in these distributions from JEFF-3.3 is very difficult to reconcile with the inherent regularities of the GEF model.
The solution of this problem is not obvious.
Our previous study \cite{Schmitt18} on fission-product yields from fission at higher excitation energies revealed the very same problem in the isotopic distribution of $Z = 49$ for the electromagnetic-induced fission of $^{238}$U.  

In summary, the isotopic distributions with peak yields above 0.1\% are fairly or well reproduced, except the problem near symmetry. 
There is a need for re-considering the S3 fission channel and the competition between symmetric and asymmetric fission channels for $Z = 48$. 
Attempts for solving these problems have not yet been successful because GEF is not a direct fit to the fission yields. 
The inherent regularities of the GEF model and the reproduction of other types of data, for example the mass-dependent prompt-neutron multiplicities, see ref. \cite{Terrell62}, impose additional constraints. 
Finally, one must always be aware that some evaluated yields might be erroneous, in particular in the low-yield regions.

\subsection*{$^{238}$U(n$_{\text{fast}}$,f)}
\begin{figure}[h]
\centering
\includegraphics[width=1.0\textwidth]{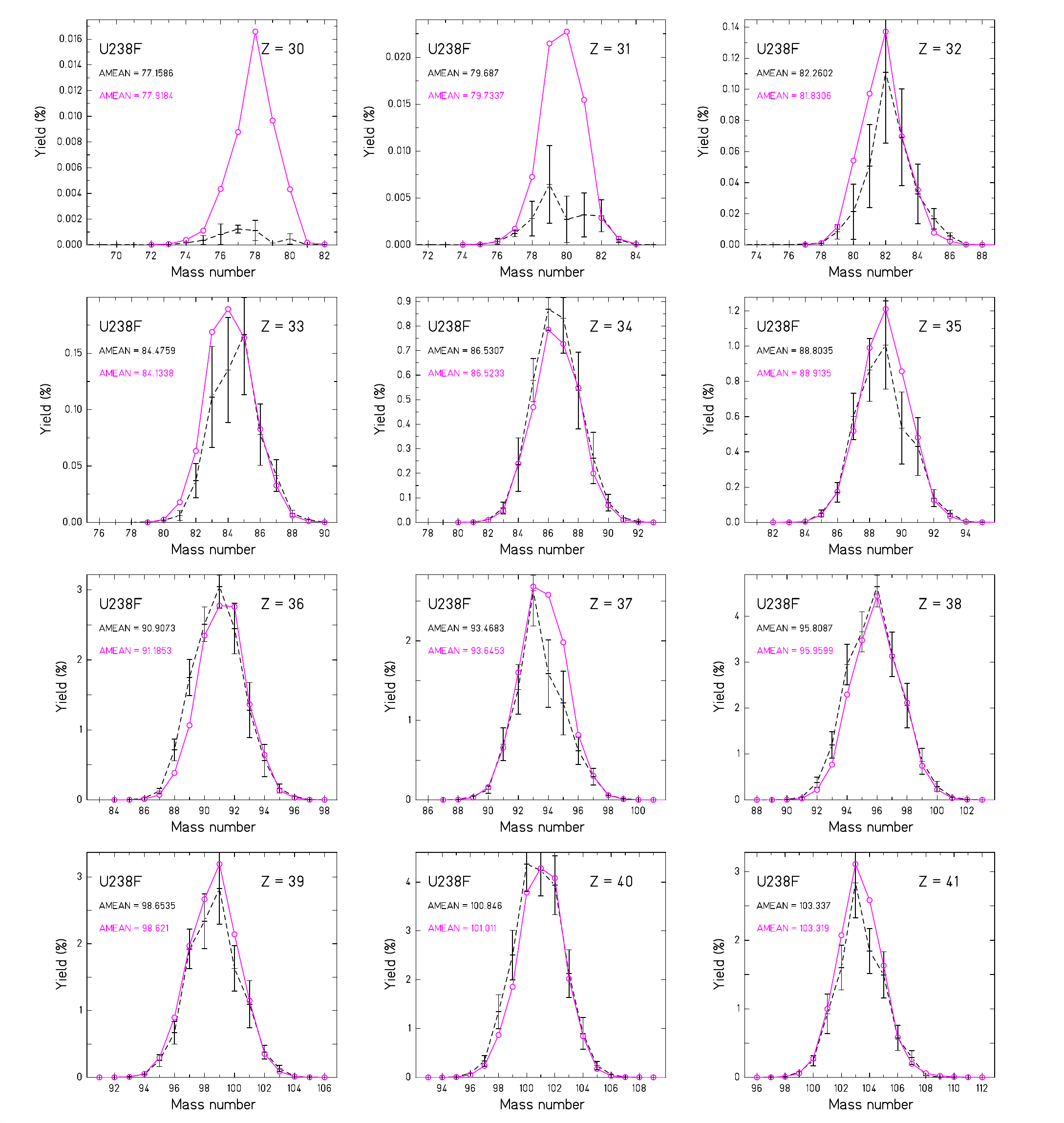}
\caption{Isotopic distributions of $^{238}$U(n$_{\text{fast}}$,f) fission-product yields, comparison of JEFF-3.3 and GEF, linear scale.} 
\label{SUB-ZA_10}       
\end{figure}

\begin{figure}[h]
\centering
\includegraphics[width=1.0\textwidth]{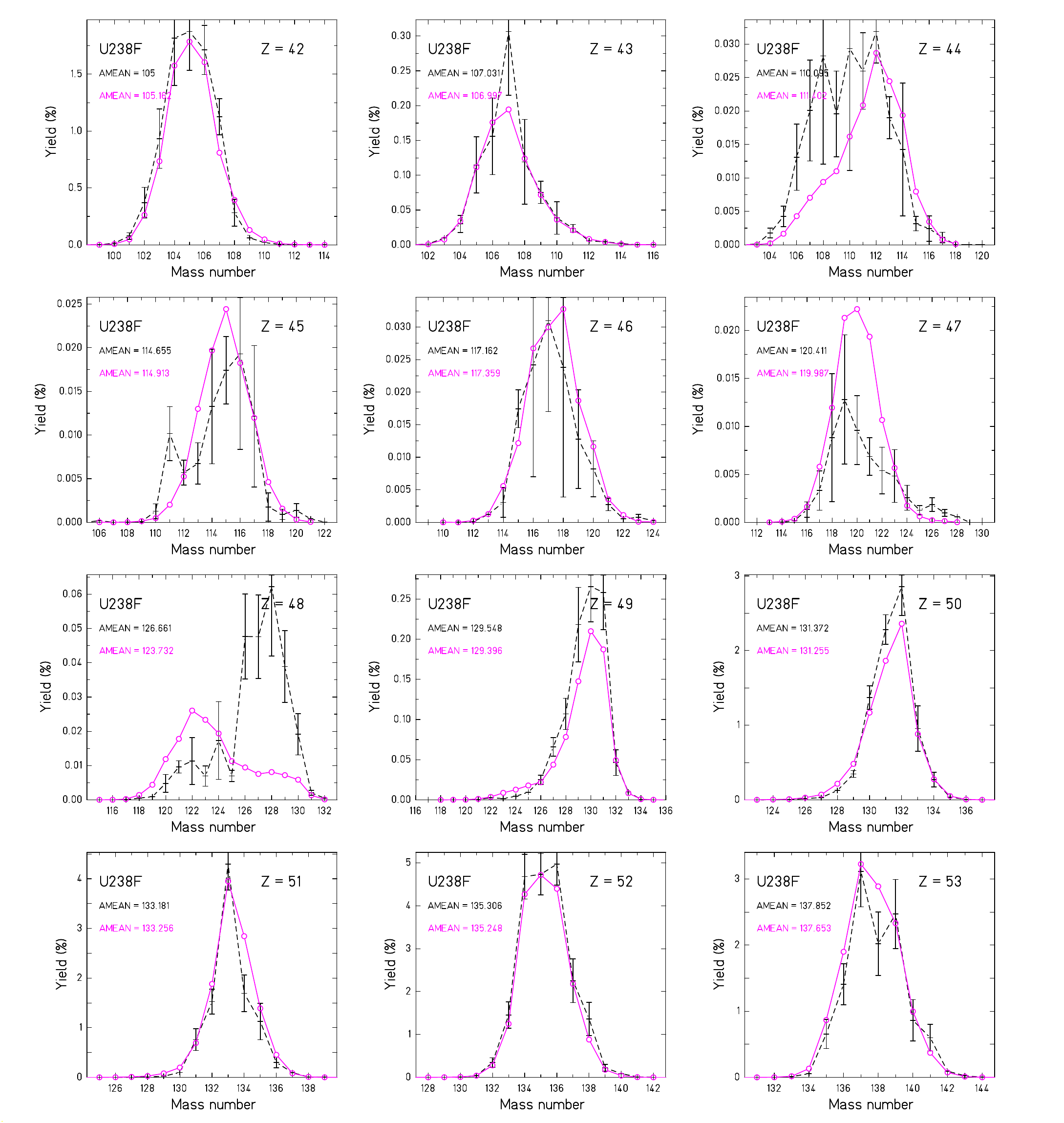}
\caption{Isotopic distributions of $^{238}$U(n$_{\text{fast}}$,f) fission-product yields, comparison of JEFF-3.3 and GEF, linear scale.} 
\label{SUB-ZA_11}       
\end{figure}

\begin{figure}[h]
\centering
\includegraphics[width=1.0\textwidth]{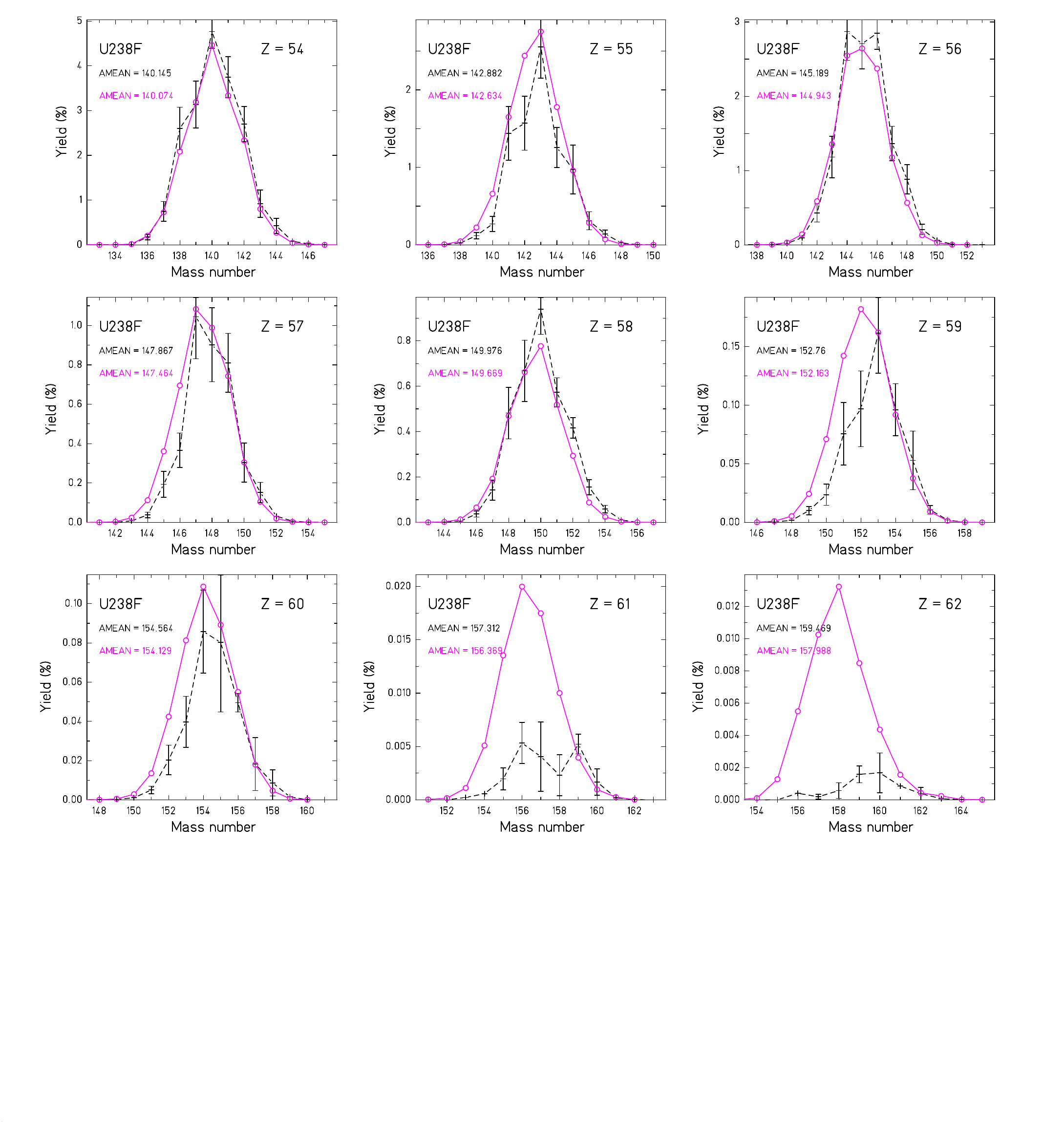}
\caption{Isotopic distributions of $^{238}$U(n$_{\text{fast}}$,f) fission-product yields, comparison of JEFF-3.3 and GEF, linear scale.} 
\label{SUB-ZA_12}       
\end{figure}

\begin{figure}[h]
\centering
\includegraphics[width=1.0\textwidth]{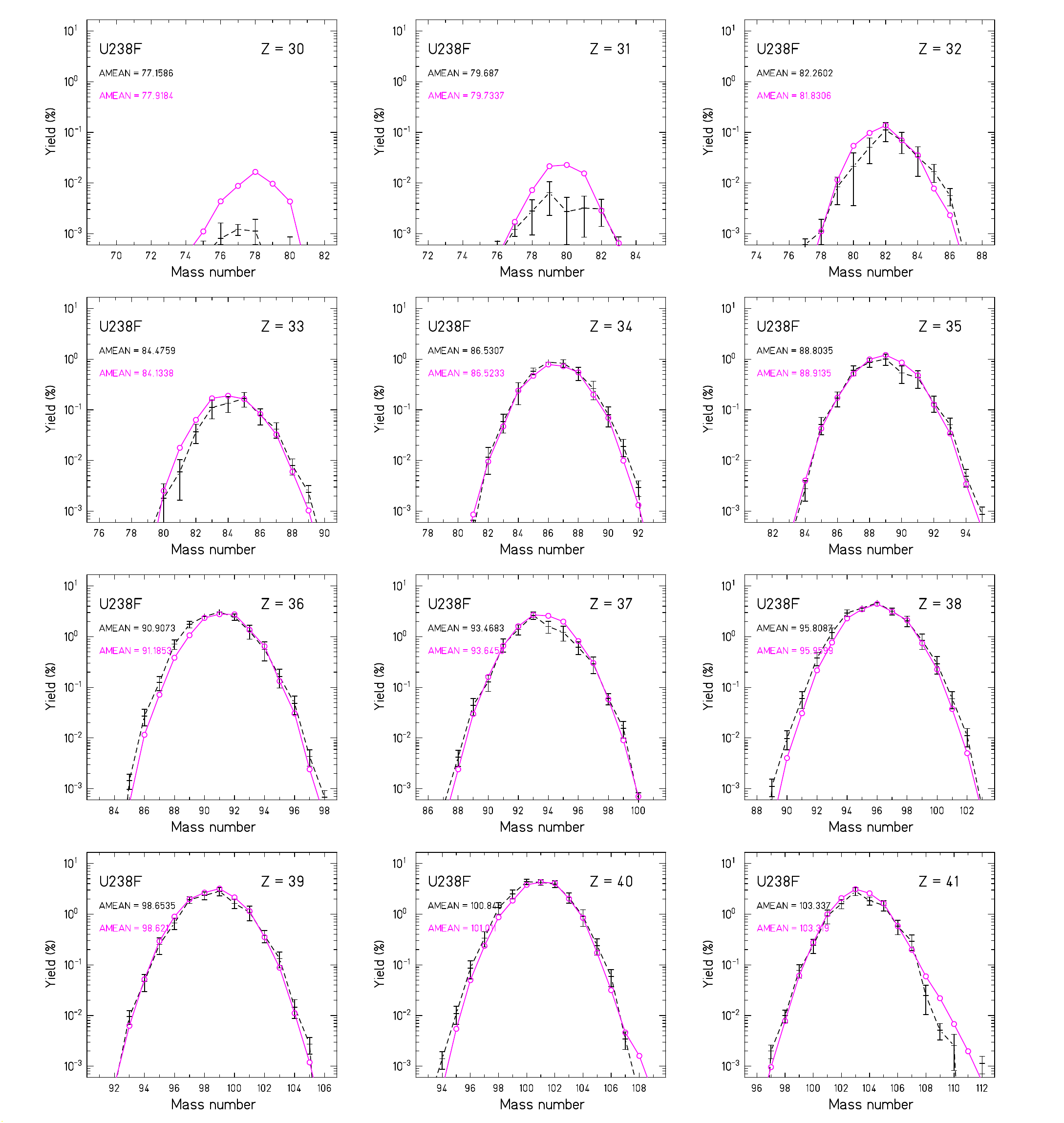}
\caption{Isotopic distributions of $^{238}$U(n$_{\text{fast}}$,f) fission-product yields, comparison of JEFF-3.3 and GEF, logarithmic scale.} 
\label{SUB-ZA_16}       
\end{figure}

\begin{figure}[h]
\centering
\includegraphics[width=1.0\textwidth]{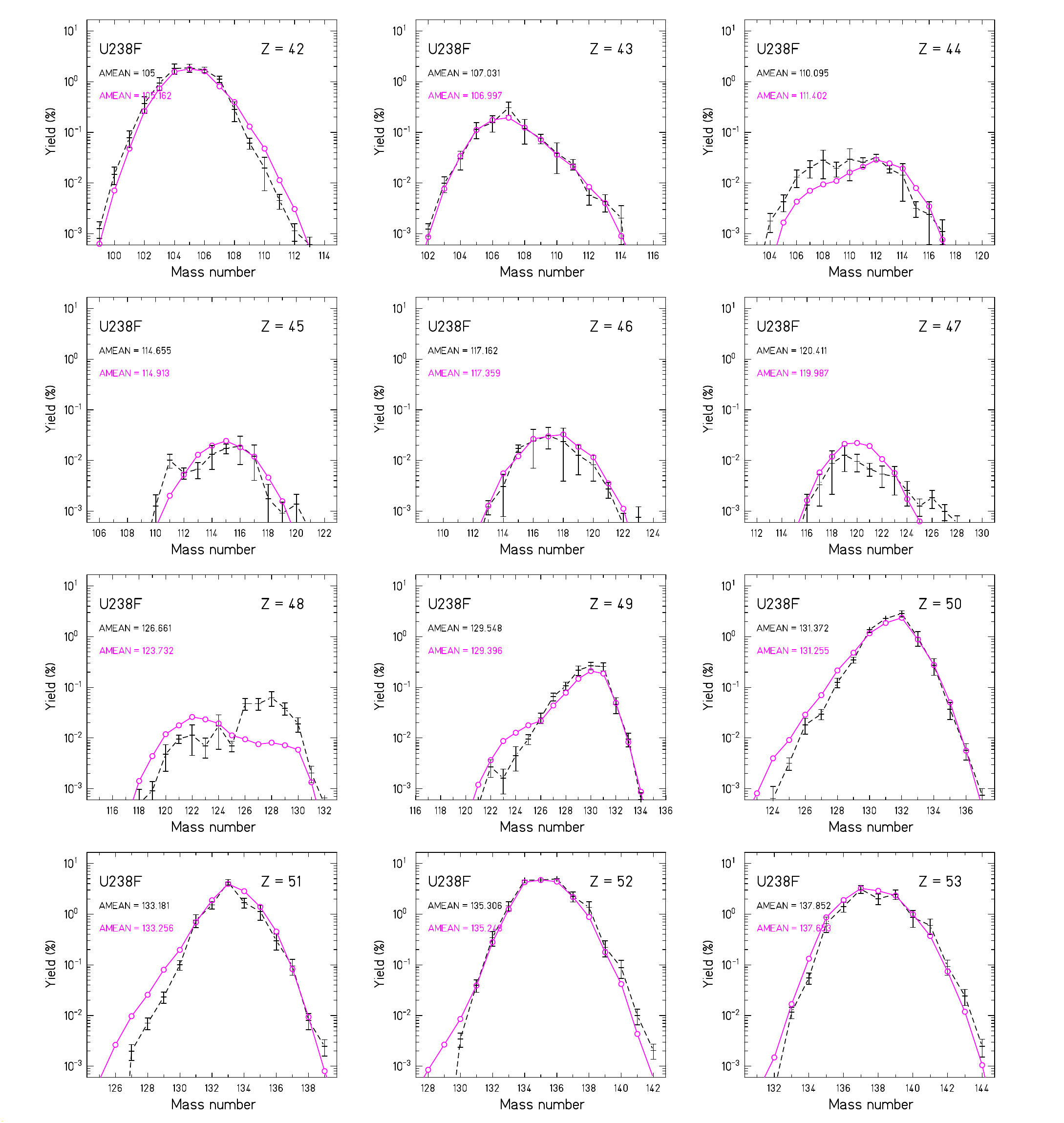}
\caption{Isotopic distributions of $^{238}$U(n$_{\text{fast}}$,f) fission-product yields, comparison of JEFF-3.3 and GEF, logarithmic scale.} 
\label{SUB-ZA_17}       
\end{figure}

\begin{figure}[h]
\centering
\includegraphics[width=1.0\textwidth]{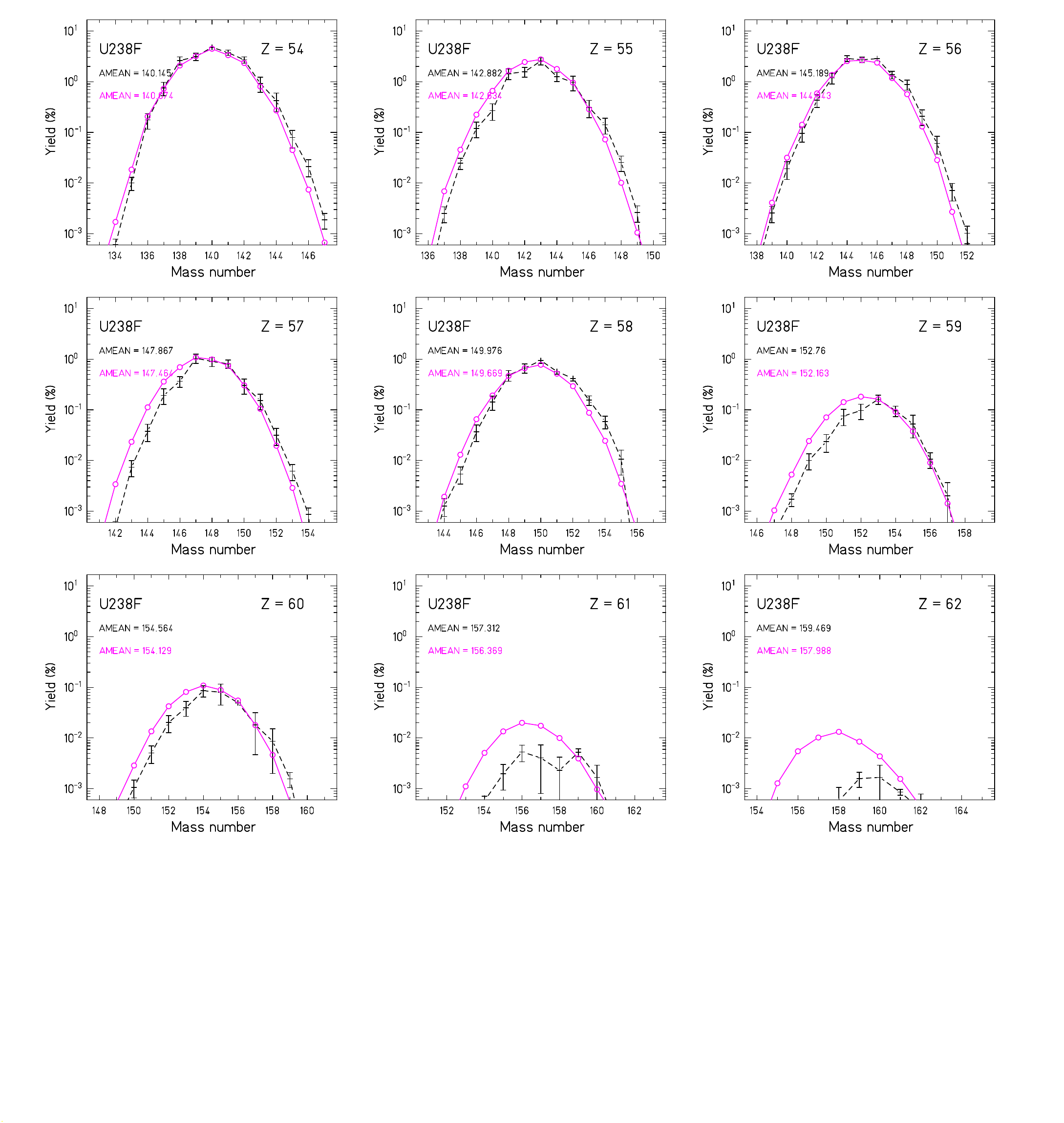}
\caption{Isotopic distributions of $^{238}$U(n$_{\text{fast}}$,f) fission-product yields, comparison of JEFF-3.3 and GEF, logarithmic scale.} 
\label{SUB-ZA_18}       
\end{figure}

In Figs. \ref{SUB-ZA_10} to \ref{SUB-ZA_18} that show the isotopic distributions of the reaction $^{238}$U(n$_{\text{fast}}$,f), one observes about the same features as found for $^{235}$U(n$_{\text{th}}$,f). There is some additional erratic scattering, which may be attributed to the lower quality of the evaluated data for this reaction.
In addition, there are some indications for a slight systematic shift of the isotopic distributions from GEF towards the neutron-rich side in the light group and to the neutron-deficient side in the heavy group. This might indicate an underestimated charge polarization or an overestimated amount of energy sorting at scission.

\subsection*{$^{239}$Pu(n$_{\text{th}}$,f)}
\begin{figure}[h]
\centering
\includegraphics[width=1.0\textwidth]{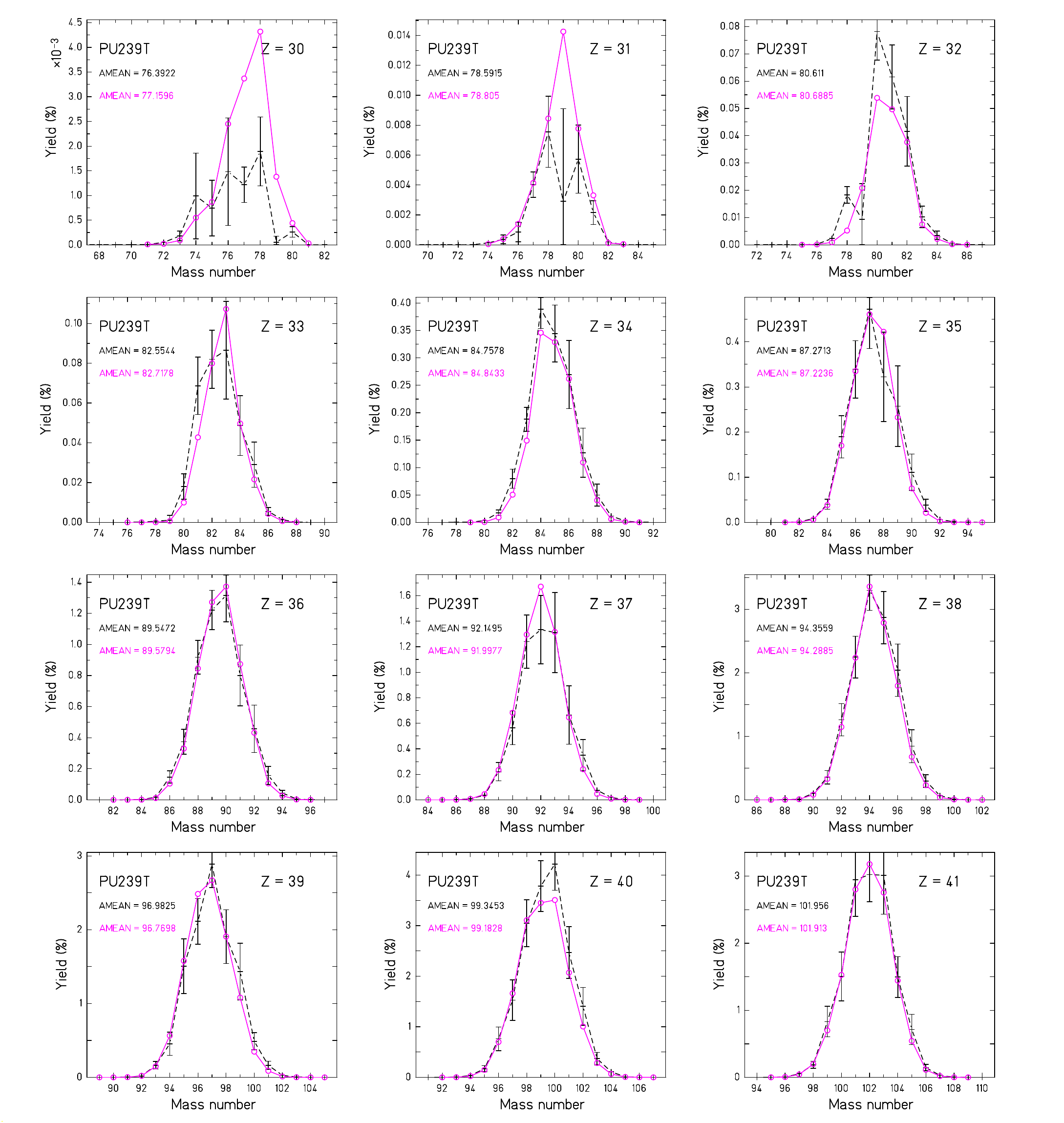}
\caption{Isotopic distributions of $^{239}$Pu(n$_{\text{th}}$,f) fission-product yields, comparison of JEFF-3.3 and GEF, linear scale.} 
\label{SUB-ZA_19}       
\end{figure}

\begin{figure}[h]
\centering
\includegraphics[width=1.0\textwidth]{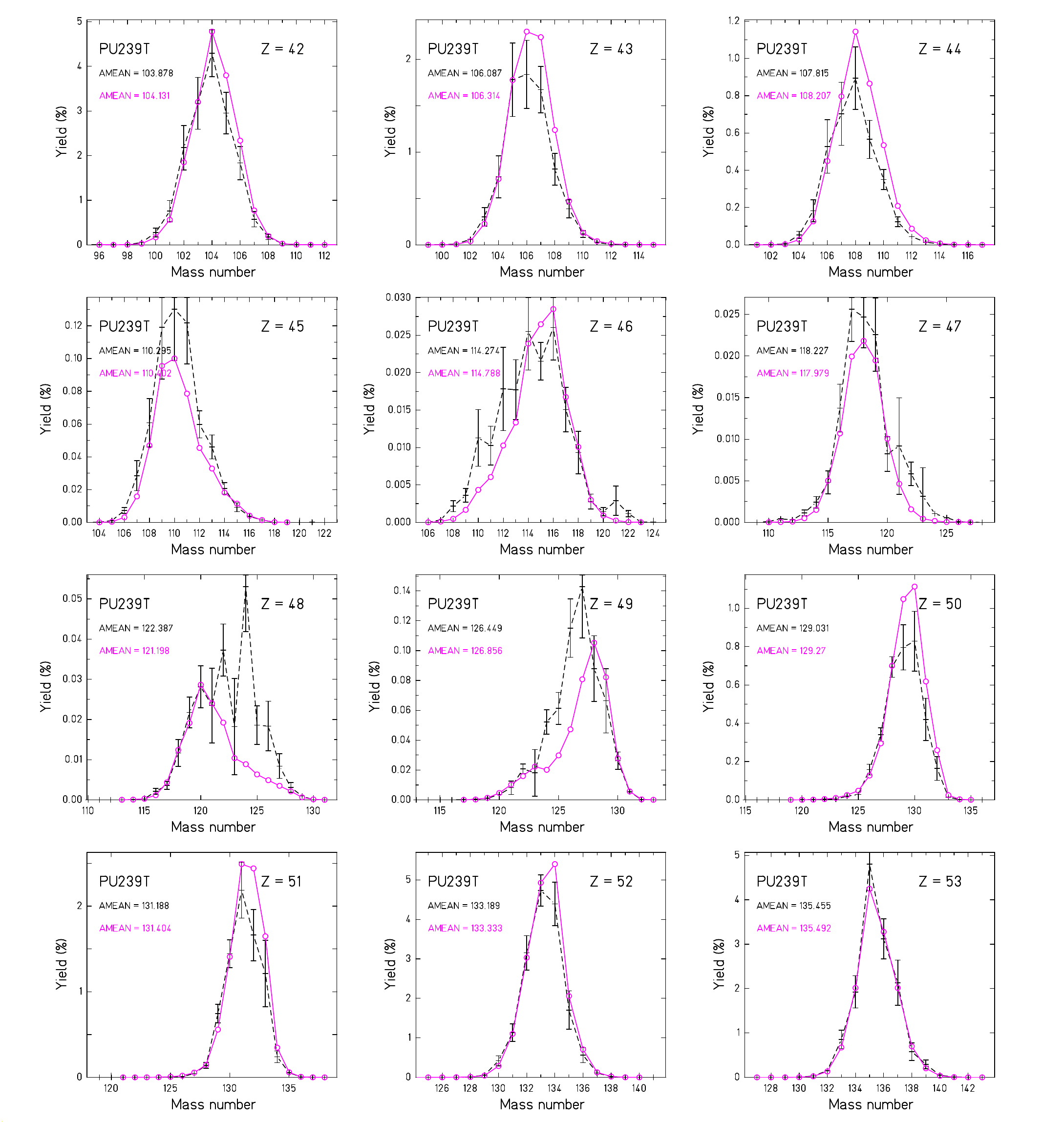}
\caption{Isotopic distributions of $^{239}$Pu(n$_{\text{th}}$,f) fission-product yields, comparison of JEFF-3.3 and GEF, linear scale.} 
\label{SUB-ZA_20}       
\end{figure}

\begin{figure}[h]
\centering
\includegraphics[width=1.0\textwidth]{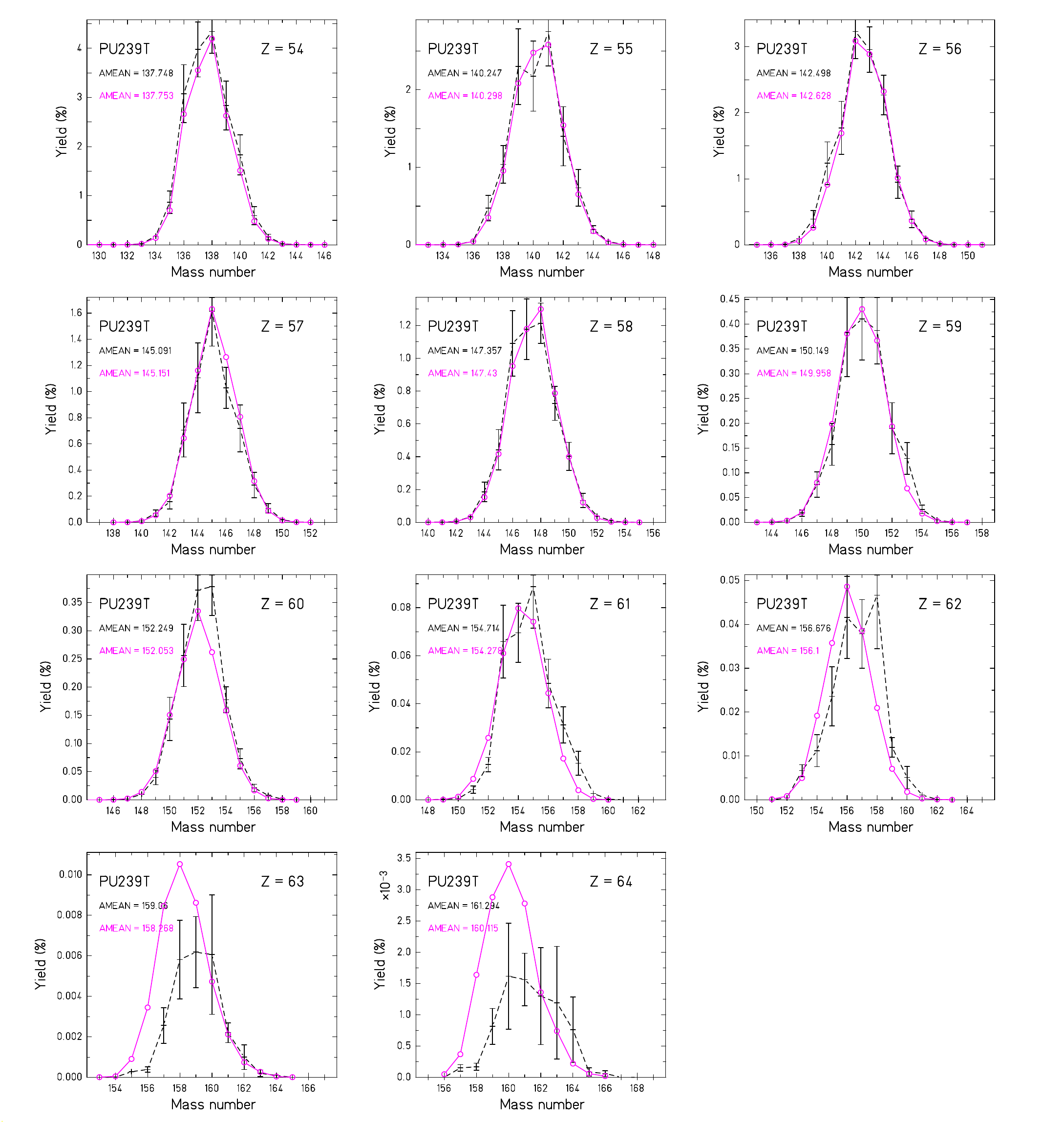}
\caption{Isotopic distributions of $^{239}$Pu(n$_{\text{th}}$,f) fission-product yields, comparison of JEFF-3.3 and GEF, linear scale.} 
\label{SUB-ZA_21}       
\end{figure}

\begin{figure}[h]
\centering
\includegraphics[width=1.0\textwidth]{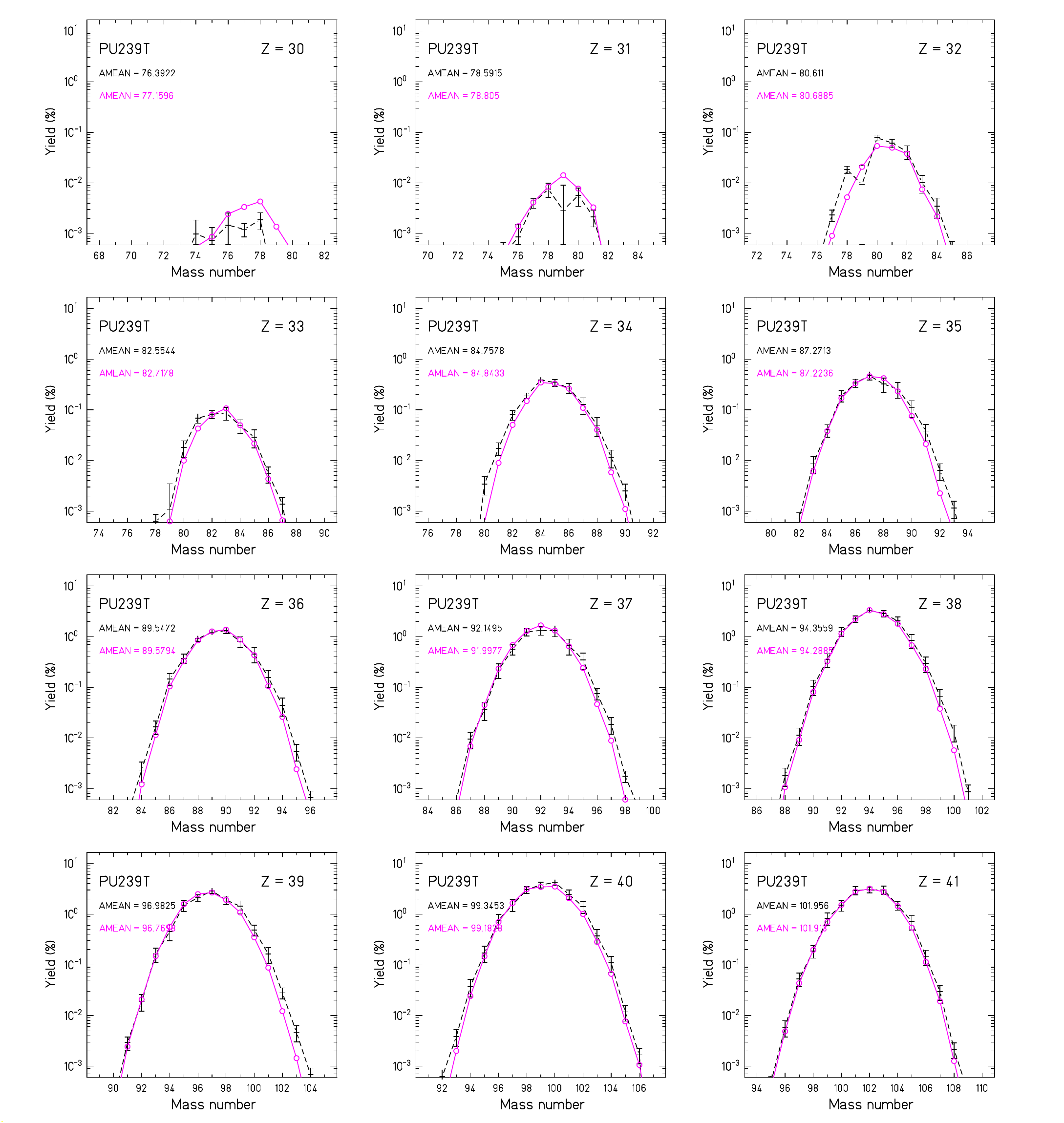}
\caption{Isotopic distributions of $^{239}$Pu(n$_{\text{th}}$,f) fission-product yields, comparison of JEFF-3.3 and GEF, logarithmic scale.} 
\label{SUB-ZA_25}       
\end{figure}

\begin{figure}[h]
\centering
\includegraphics[width=1.0\textwidth]{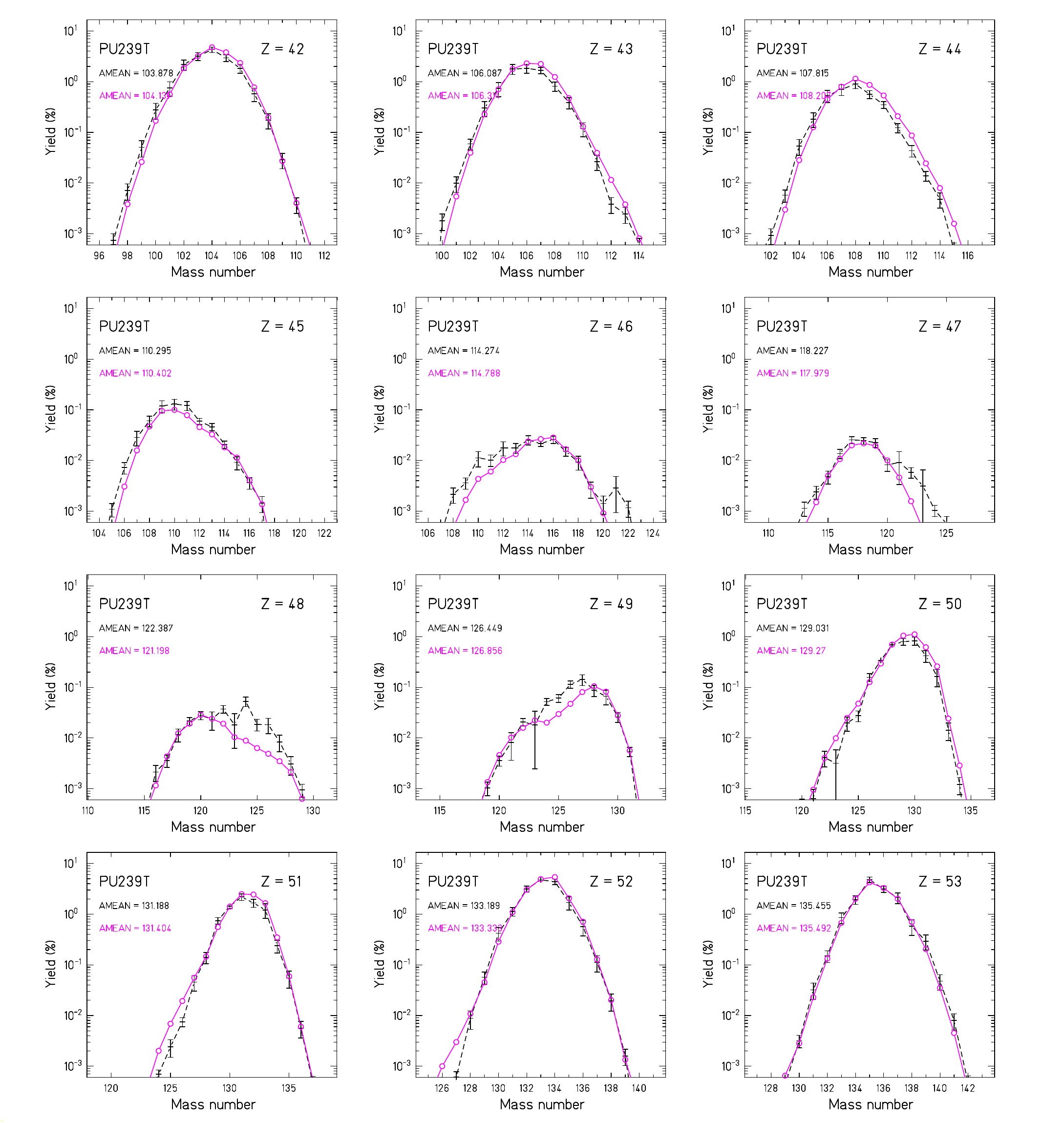}
\caption{Isotopic distributions of $^{239}$Pu(n$_{\text{th}}$,f) fission-product yields, comparison of JEFF-3.3 and GEF, logarithmic scale.} 
\label{SUB-ZA_26}       
\end{figure}

\begin{figure}[h]
\centering
\includegraphics[width=1.0\textwidth]{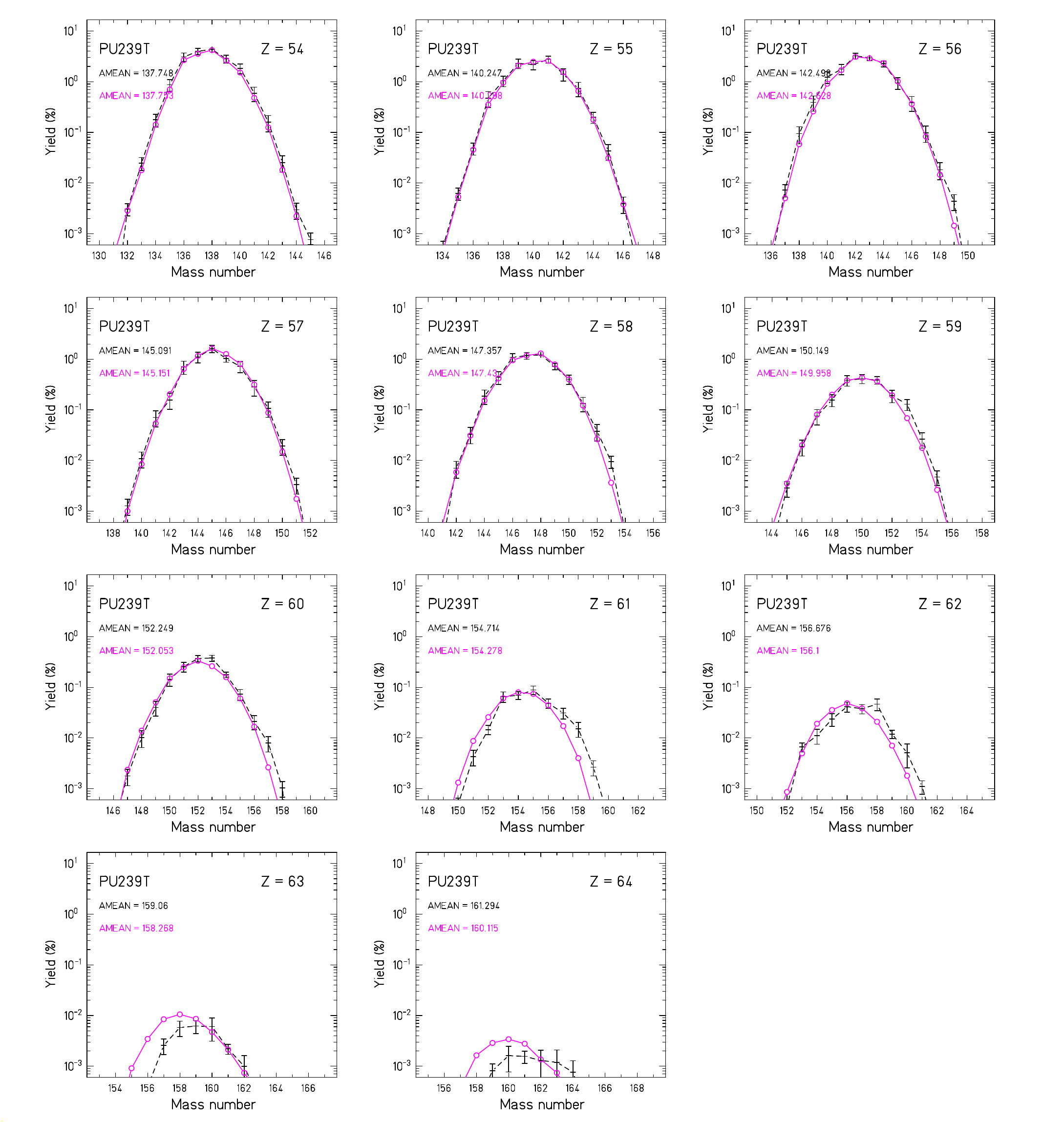}
\caption{Isotopic distributions of $^{239}$Pu(n$_{\text{th}}$,f) fission-product yields, comparison of JEFF-3.3 and GEF, logarithmic scale.} 
\label{SUB-ZA_27}       
\end{figure}

In the isotopic distributions of the reaction $^{239}$Pu(n$_{\text{th}}$,f) shown in Figs. \ref{SUB-ZA_19} to \ref{SUB-ZA_27}, the distributions with peak yields above 1\% are at least fairly well reproduced, except the problems near symmetry.
One observes an increased erratic scattering and larger error bars in the evaluated data than in the uranium cases discussed above. 
Most of the discrepancies between the evaluation and the GEF results are not systematic. 
The problem found for the uranium cases in the asymmetric wings does not appear clearly for $^{239}$Pu(n$_{\text{th}}$,f), except the shift to the neutron-deficient side in the heavy wing. The problem at the transition from the symmetric component to the heavy asymmetric component, here appearing at $Z = 47$ and $Z = 48$, is again clearly visible.

\clearpage
\clearpage

\clearpage

\clearpage
\clearpage

\clearpage

\clearpage
\clearpage

\clearpage

\clearpage
\clearpage

\clearpage

\clearpage
\clearpage

\clearpage

\clearpage
\clearpage

\clearpage

\begin{figure}[h]
\centering
\includegraphics[width=1.0\textwidth]{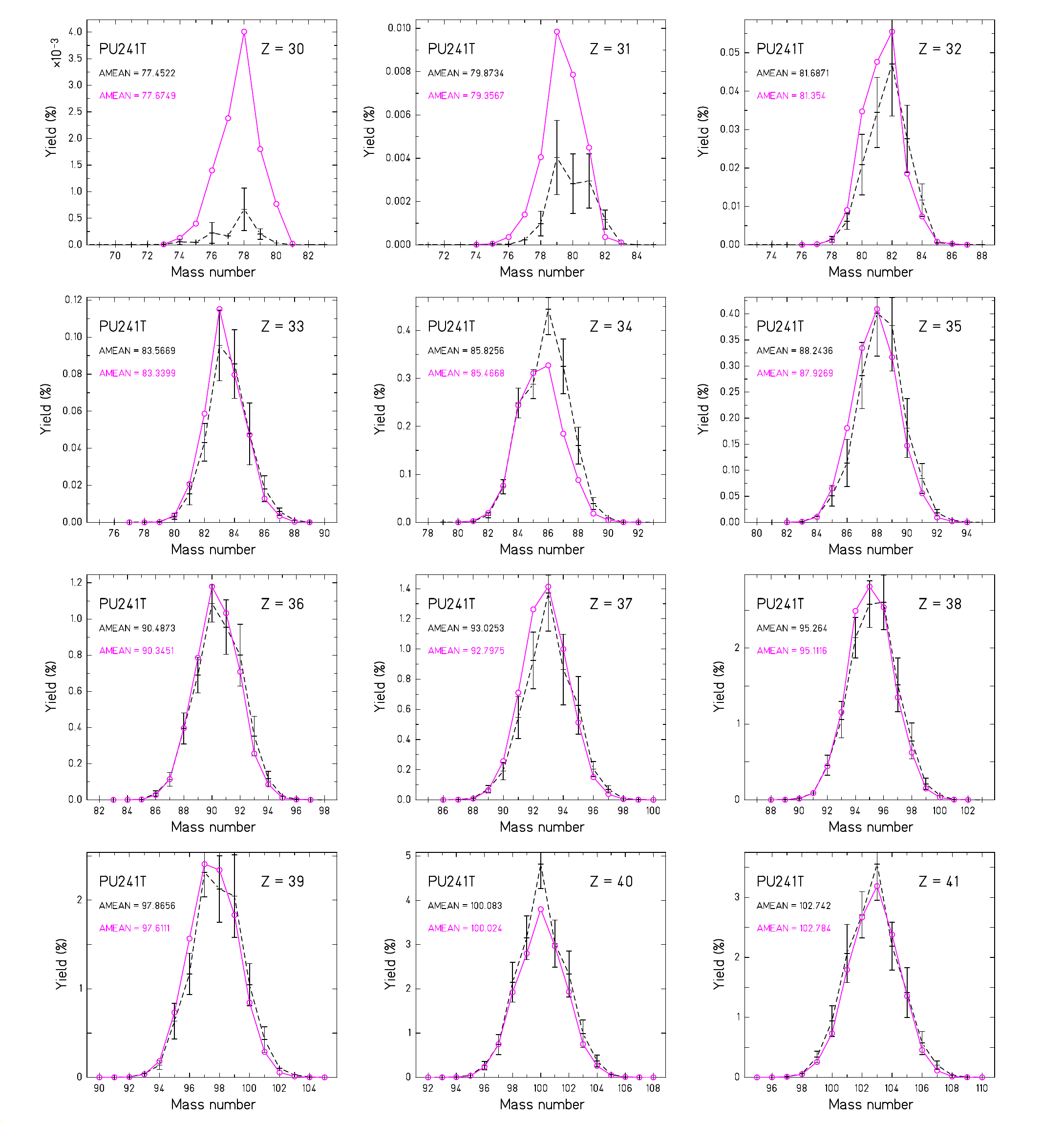}
\caption{Isotopic distributions of $^{241}$Pu(n$_{\text{th}}$,f) fission-product yields, comparison of JEFF-3.3 and GEF, linear scale.} 
\label{SUB-ZA_28}       
\end{figure}
\clearpage
\begin{figure}[h]
\centering
\includegraphics[width=1.0\textwidth]{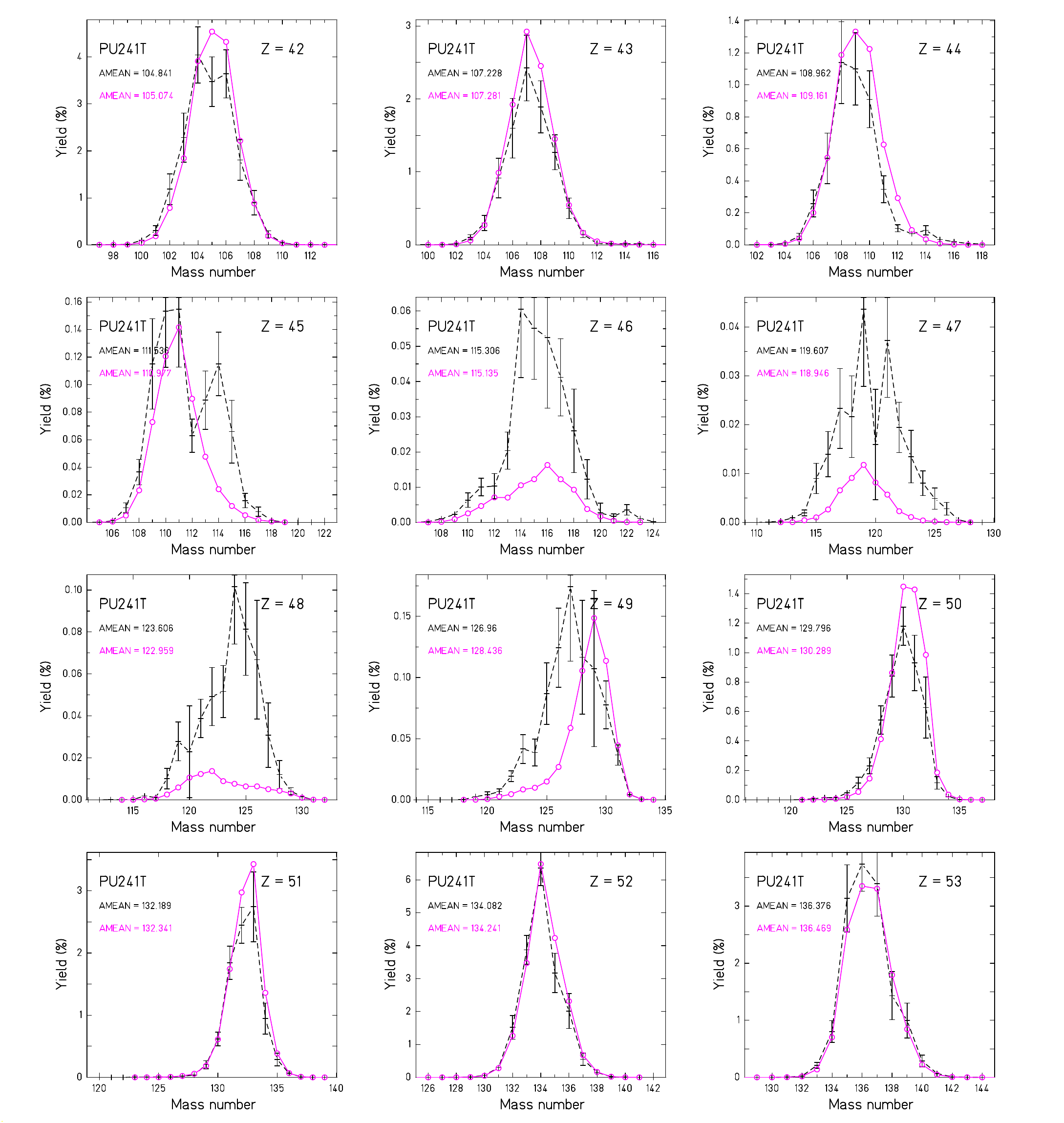}
\caption{Isotopic distributions of $^{241}$Pu(n$_{\text{th}}$,f) fission-product yields, comparison of JEFF-3.3 and GEF, linear scale.} 
\label{SUB-ZA_29}       
\end{figure}
\clearpage
\begin{figure}[h]
\centering
\includegraphics[width=1.0\textwidth]{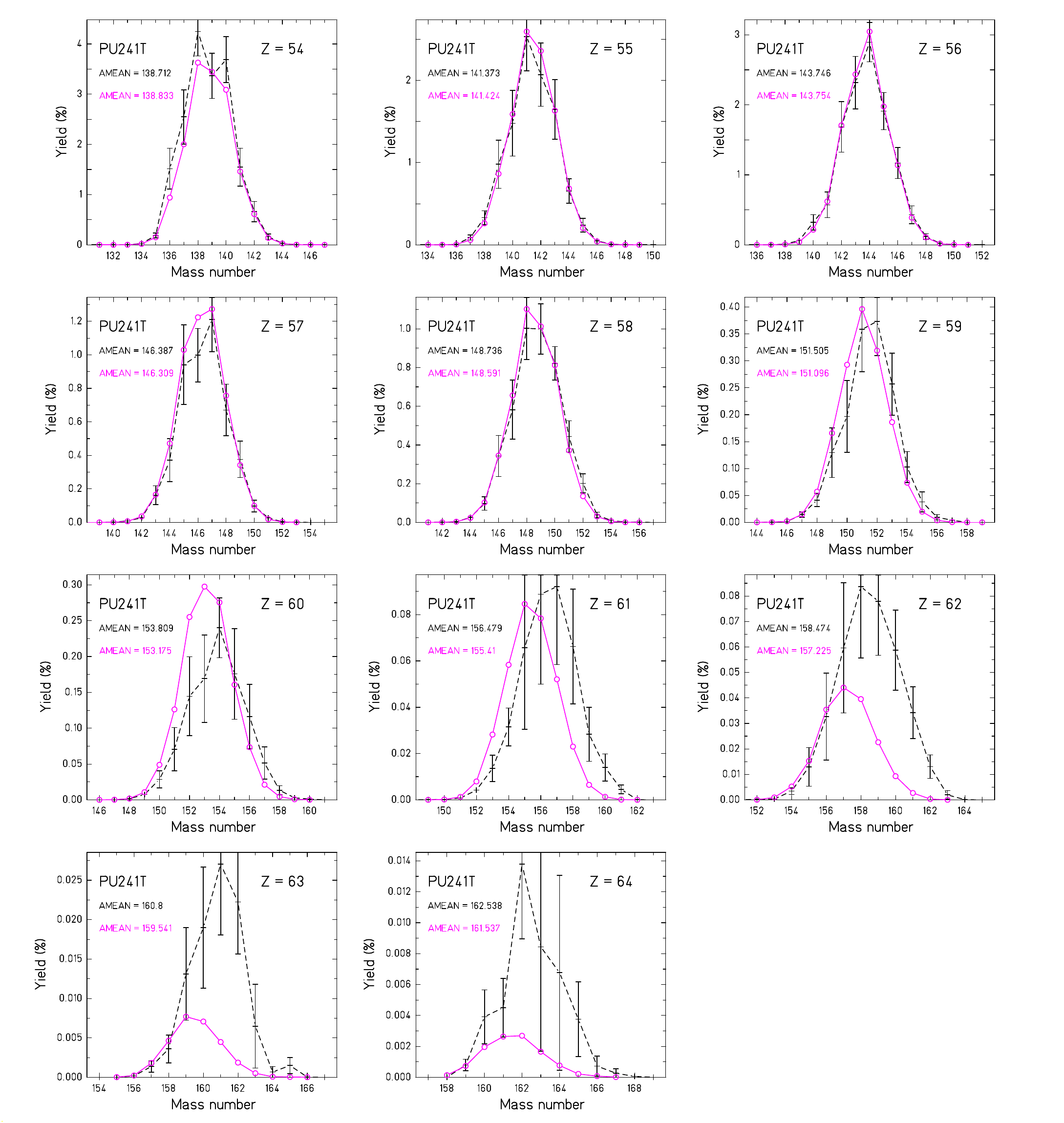}
\caption{Isotopic distributions of $^{241}$Pu(n$_{\text{th}}$,f) fission-product yields, comparison of JEFF-3.3 and GEF, linear scale.} 
\label{SUB-ZA_30}       
\end{figure}

\clearpage

\begin{figure}[h]
\centering
\includegraphics[width=1.0\textwidth]{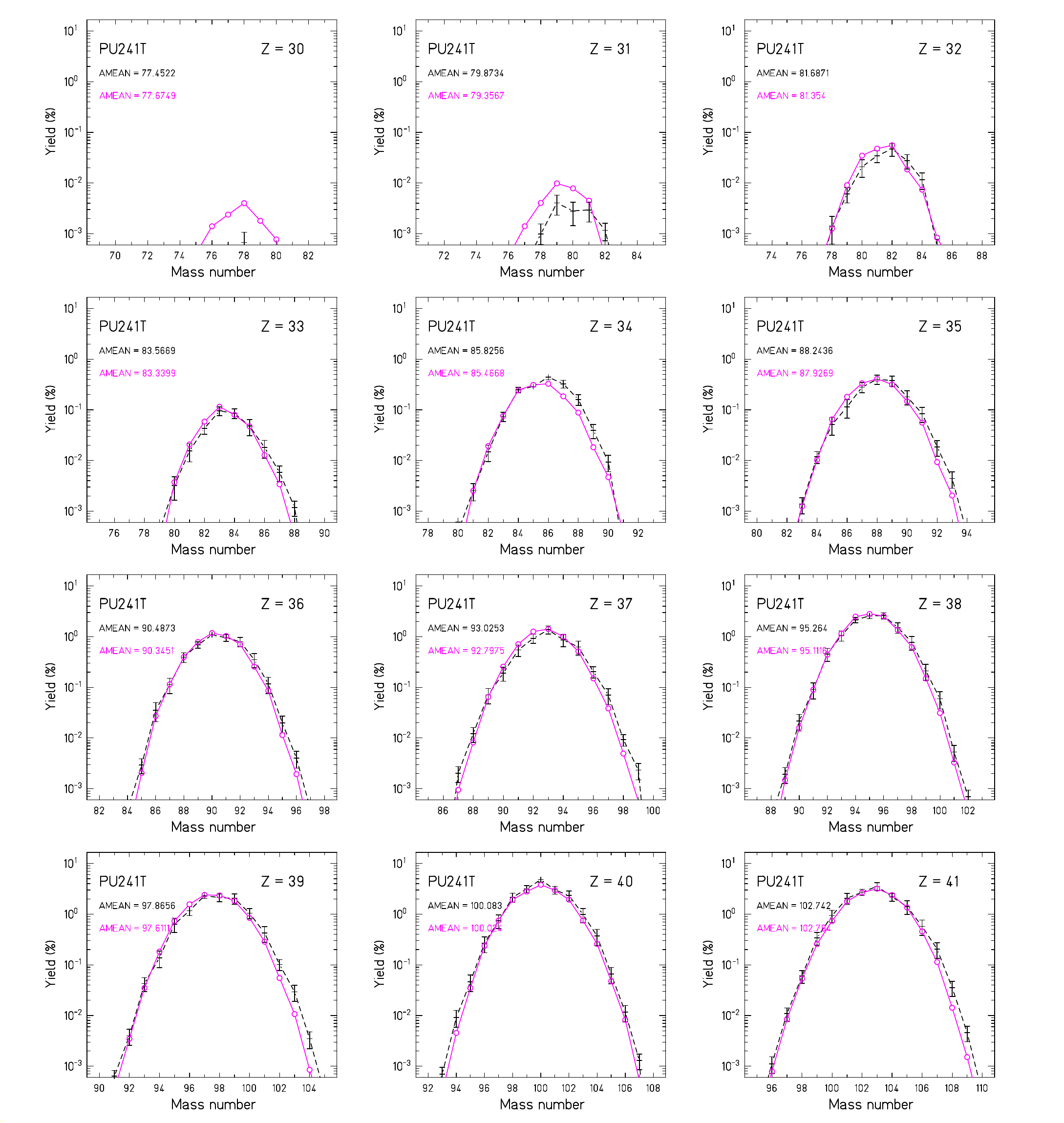}
\caption{Isotopic distributions of $^{241}$Pu(n$_{\text{th}}$,f) fission-product yields, comparison of JEFF-3.3 and GEF, logarithmic scale.} 
\label{SUB-ZA_34}       
\end{figure}
\clearpage
\begin{figure}[h]
\centering
\includegraphics[width=1.0\textwidth]{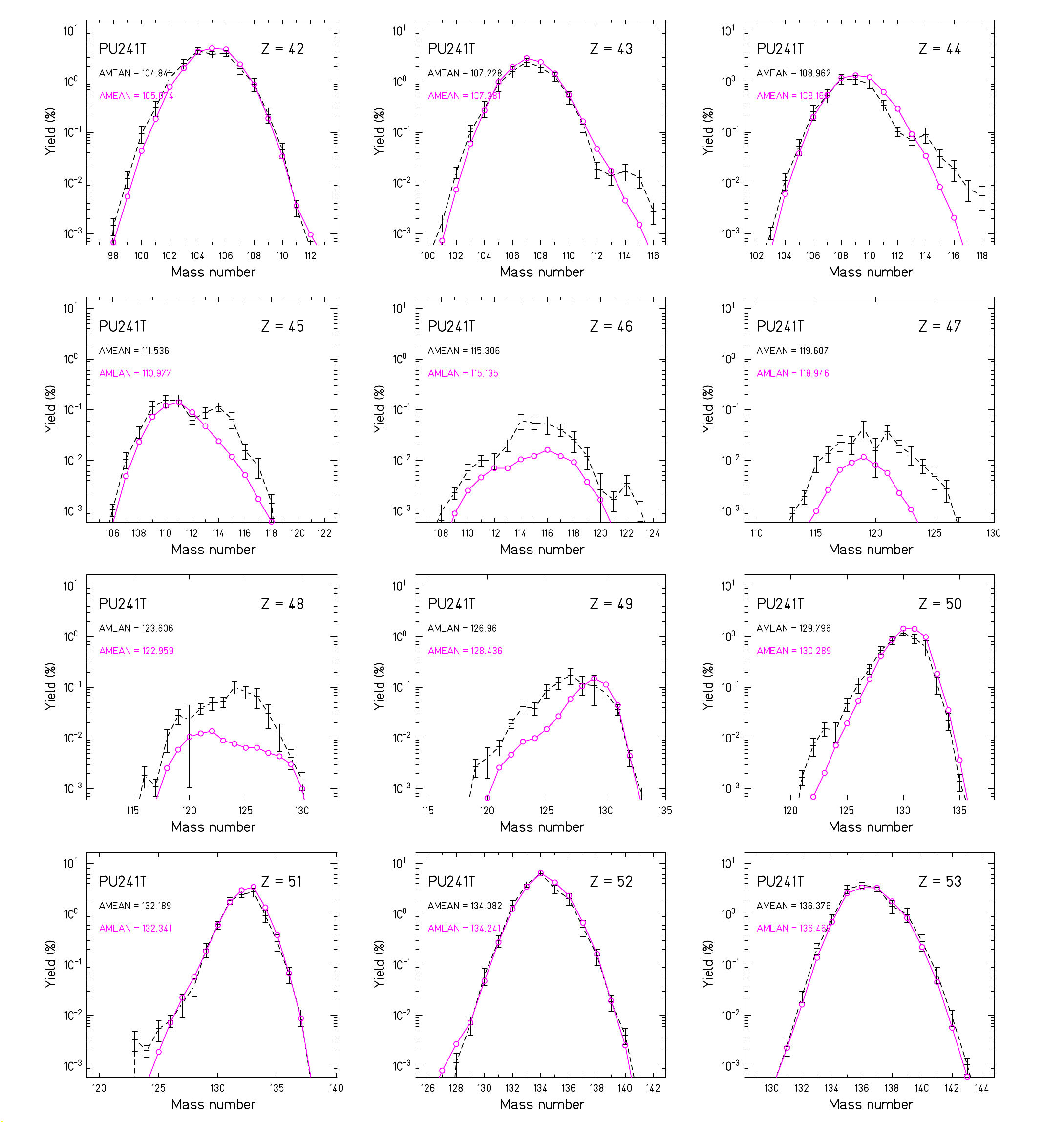}
\caption{Isotopic distributions of $^{241}$Pu(n$_{\text{th}}$,f) fission-product yields, comparison of JEFF-3.3 and GEF, logarithmic scale.} 
\label{SUB-ZA_35}       
\end{figure}
\clearpage
\begin{figure}[h]
\centering
\includegraphics[width=1.0\textwidth]{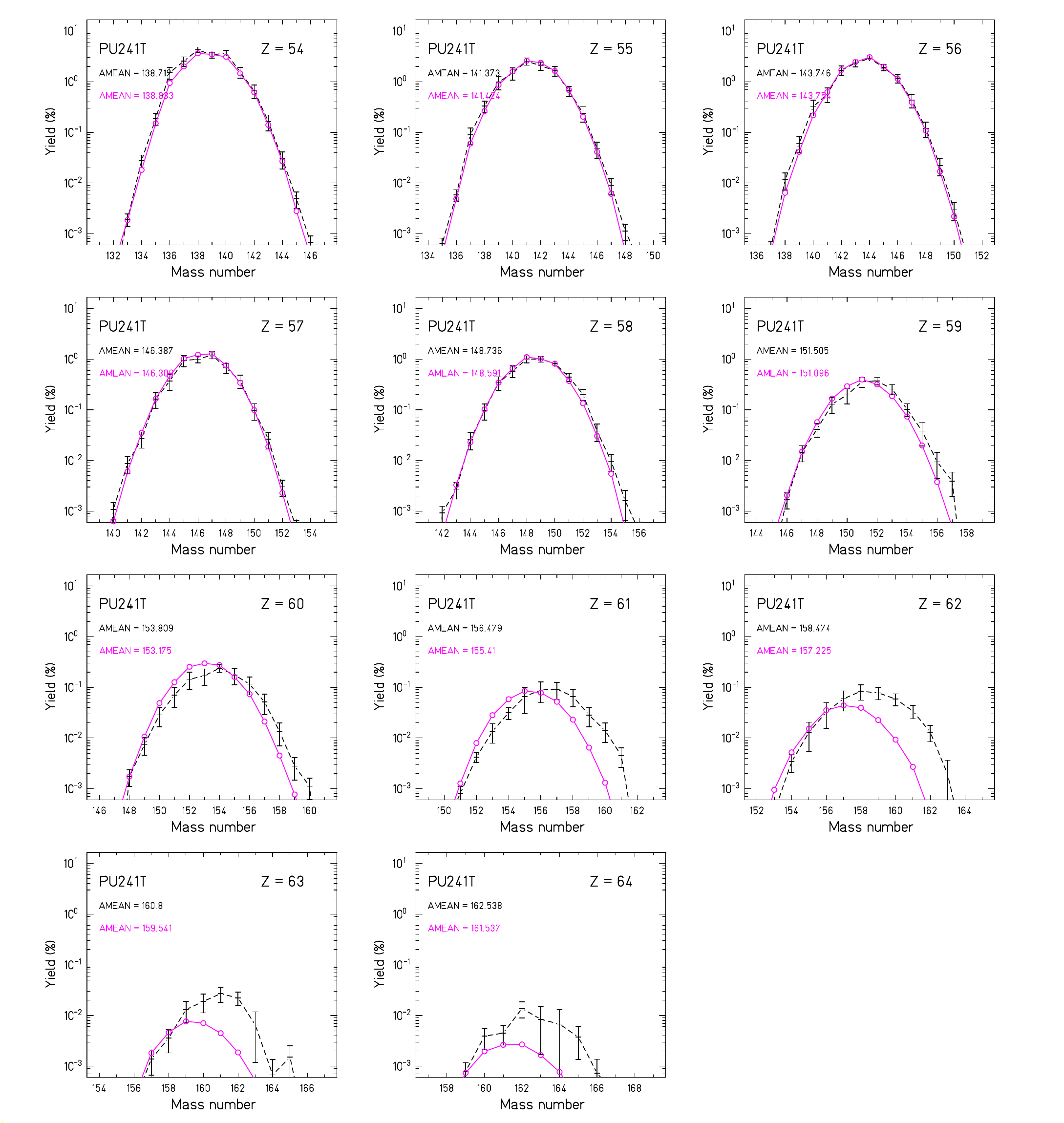}
\caption{Isotopic distributions of $^{241}$Pu(n$_{\text{th}}$,f) fission-product yields, comparison of JEFF-3.3 and GEF, logarithmic scale.} 
\label{SUB-ZA_36}       
\end{figure}

\clearpage

\subsection*{$^{241}$Pu(n$_{\text{th}}$,f)}
The isotopic distributions of $^{241}$Pu(n$_{\text{th}}$,f) fission-product yields in Figs. \ref{SUB-ZA_28} to \ref{SUB-ZA_36} show strong discrepancies between the evaluation and the GEF results. Most of these discrepancies are related to the serious problems found in the mass yields of this system, see Figs. 101 and 102 of the main document. These problems do not allow to make a more detailed discussion of the isotopic distributions.

\subsection*{Summary}
The comparison of the isotopic distributions from the JEFF-3.3 evaluation with the results from the GEF code reveals a rather good, often almost perfect agreement for the yields of the elements with values above 1\%. 
This comparison also reveals the exceptionally good quality of the empirical data for the system $^{235}$U(n$_{\text{th}}$,f) by the small amount of erratic fluctuations in the discrepancies, compared to the other systems. 

Discrepancies between the GEF yields and the values from the JEFF-3.3 evaluation, especially in certain regions of low yields of the fission-product distributions, are found.
Many of these appear to be erratic.
Some of the deviations of the isotopic distributions in the extreme asymmetric wings of the fission-yield distribution and at the transition from the symmetric to the heavy hump of the asymmetric component are common to many fissioning nuclei. 
The problems in the asymmetric wings may be cured by an improved description of the weak S3 fission channel in GEF, while the other ones seem to indicate the presence of a tail in the distribution of the heavy asymmetric component towards symmetry, which is not present in GEF, although the higher yields near the asymmetric peaks are well reproduced.

A detailed analysis with the inclusion of the original experimental yield data would be needed in order to better identify the origin of the observed discrepancies.
At this stage, a reasonable assumption may be that erratic fluctuations in the JEFF yields are indications for problems in the evaluation, whereas
similar discrepancies between evaluated data and GEF results found for several fissioning nuclei hint to underlying problems in GEF. 

As a practical consequence, the comparisons shown in this work allow to apply some empirical corrections to the independent yields from GEF or from the evaluation, resulting in a revised evaluation with an improved quality. 




\begin{thebibliography}{99}
\bibitem{Roskovec18} Bedrich Roskovec, arXiv 1812.03206 (2018).
\bibitem{ReinesCowan}  F. Reines, C.L. Cowan, Jr., Nature {\bf 178} (1956) 446.
\bibitem{DC} Y. Abe  {\it  et al.}, Phys. Rev. Lett.  {\bf 108} (2012) 131801.
\bibitem{DB} F. P. An {\it  et al.}, Phys. Rev. Lett.  {\bf 108} (2012) 171803.
\bibitem{RENO} J. K. Ahn  {\it  et al.}, Phys. Rev. Lett.  {\bf 108} (2012) 191802. 
\bibitem{Mikaelian} L.A. Mikaelian, 1977, Proc. Int. Conf. Neutrino-77, v.2, p.383. 
\bibitem{SchreckU5-1} K.~Schreckenbach {\it et al.}, Phys.\ Lett.\ {\bf 99B} (1981) 251.
\bibitem{SchreckU5-2} K. Schreckenbach {\it et al.}, Phys.\ Lett.\ {\bf 160B} (1985) 325. 
\bibitem{SchreckU5Pu9} F.~von~Feilitzsch, A.~A.~Hahn, K. Schreckenbach, Phys.\ Lett.\ {\bf 118B} (1982) 162.
\bibitem{Hahn} A. A. Hahn {\it et al.}, Phys. Lett. {\bf B 218} (1989) 365.
\bibitem{Haag} N. Haag {\it et al.}, Phys. Rev. Lett.  {\bf 112}, 122501 (2014).
\bibitem{Mueller} T. A. Mueller  {\it et al.}, Phys. Rev.  {\bf C 83} (2011) 054615.
\bibitem{Huber} P. Huber, Phys. Rev.  {\bf C84} (2011) 024617.
\bibitem{Mention} G. Mention {\it et al.}, Phys. Rev.  {\bf D 83} (2011) 073006.
\bibitem{ShapeAnomaly} F. P. An  {\it et al.}, Chinese Phys. {\bf C 41} (2017) 013002.
\bibitem{WhitePaper} K. N. Abazajian {\it  et al.}, http://arxiv.org/abs/1204.5379.
\bibitem{Hayes} A. C. Hayes, J.L. Friar, G.T. Garvey, G.Jungman, G. Jonkmans, Phys. Rev. Lett. {\bf 112} (2014) 202501.
\bibitem{FangBrown} D. L. Fang, B. A. Brown, Phys. Rev. {\bf C 91} (2015) 025503.
\bibitem{Hayen} L. Hayen {\it et al.}, Phys. Rev. {\bf C 99} (2019) 031301.
\bibitem{Fallot} M. Fallot {\it et al.}, { Phys. Rev. Lett.}  {\bf109} (2012) 202504.
\bibitem{Zak} A. A. Zakari-Issoufou {\it et al.}, { Phys. Rev. Lett.}  {\bf 115} (2015) 102503.
\bibitem{Sonzogni} A. A. Sonzogni, T. D. Johnson, E. A. McCutchan, Phys. Rev. {\bf C 91} (2015) 011301(R).
\bibitem{Estienne} M. Estienne {\it  et al.}, Phys. Rev. Lett. {\bf 123} (2019) 022502, http://arxiv.org/abs/1904.09358
\bibitem{Hayes15} A. C. Hayes {\it  et al.}, Phys. Rev. {\bf D 92} (2015) 033015.
\bibitem{Xubo18} Ma Xubo, Yang le, Zhan Liang, An Fengpeng, Cao Jun, arXiv 1807.09265 (2018).
\bibitem{Dwyer} D. A. Dwyer and T. J. Langford, Phys. Rev. Lett. 114, 012502 (2015)
\bibitem{Sonzo2015} A. A. Sonzogni, T. D. Johnson, and E. A. McCutchan, Phys. Rev. C 91, 011301(R) (2015).
\bibitem{Schmidt16} K.-H. Schmidt, B. Jurado, C. Amouroux, C. Schmitt, Nucl. Data Sheets {\bf 131} (2016) 107.
\bibitem{CormonND} S. Cormon {\it et al.}, Nucl. Data Sheets {\bf 120} (2014) 141.
\bibitem{IAEAReport} IAEA Report SG-EQGNRL-RP-0002 (2012) and IAEA Report STR-361, (2009).
\bibitem{Schmidt18} K.-H. Schmidt, B. Jurado, Rep. Progr. Phys. {\bf 81} (2018) 106301.
\bibitem{Schmitt18} Ch. Schmitt, K.-H. Schmidt, B. Jurado, Phys. Rev. {\bf C 98} (2018) 044605.
\bibitem{GEF} http://www.cenbg.in2p3.fr/GEF, \\ http://www.khschmidts-nuclear-web.eu/GEF.html  
\bibitem{Myers96} W. D. Myers, W. J. Swiatecki, Nucl. Phys. {\bf A 601} (1996) 14.
\bibitem{Mosel71} U. Mosel, H. W. Schmitt, Nucl. Phys. {\bf A 165} (1971) 73.
\bibitem{Schmidt08} K.-H. Schmidt, A. Kelic, M. V. Ricciardi, Europh. Lett. {\bf 83} (2008) 32001.
\bibitem{Wilkins76} B. D. Wilkins, E. P. Steinberg, R. R. Chasman, Phys. Rev. {\bf C 14} (1976) 1832.
\bibitem{Nifenecker80} H. Nifenecker, J. Physique Lett. {\bf 41} (1980) 47.
\bibitem{Schmidt12} K.-H. Schmidt, B. Jurado, Phys. Rev. {\bf C 86} (2012) 044322.
\bibitem{Schmidt10} K.-H. Schmidt, B. Jurado, Phys. Rev. Lett. {\bf 104} (2010) 212501.
\bibitem{Schmidt11a} K.-H. Schmidt, B. Jurado, Phys. Rev. {\bf C 83} (2011) 014607.
\bibitem{Schmidt11b} K.-H. Schmidt, B. Jurado, Phys. Rev. {\bf C 83} (2011) 061601.
\bibitem{Jurado15} B. Jurado, K.-H. Schmidt, J. Phys. G: Nucl. Part. Phys. {\bf 42} (2015) 055101.
\bibitem{King} R. W. King, J. F. Perkins, Phys. Rev. {\bf 112} (1958) 963.
\bibitem{Avignonne} F. T. Avignone {\it et al.} , Phys. Rev. {\bf 170} (1968) 931.
\bibitem{Vogel81} P.~Vogel, G. K.~Schenter, F. M.~Mann, R. E.~Schenter, Phys. Rev. {\bf C 24} (1981) 1543.
\bibitem{Tengblad} O. Tengblad  {\it et al.}, Nuclear Physics  {\bf A 503} (1989) 136. G. Rudstam  {\it et al.}, Atomic Data and Nucl. Data Tables  {\bf 45} (1990) 239. 
\bibitem{Hardy77} J. C. Hardy  {\it et al.}, Phys. Lett. {\bf B 71} (1977) 307.
\bibitem{RubioGelletly} B. Rubio, W. Gelletly, Romanian Reports in Physics, {\bf 59} (2007) 635.
\bibitem{PRLAlgora}   A. Algora {\it et al.}, Phys. Rev. Lett. {\bf 105} (2010) 202501.
\bibitem{Valencia} E. Valencia  {\it et al.}, { Phys. Rev.} {\bf C 95} (2017) 024320.
\bibitem{Rice} S. Rice  {\it et al.}, Phys. Rev. {\bf C 96} (2017) 014320.
\bibitem{Guadilla2018} V. Guadilla  {\it et al.}, Phys. Rev. Lett.  {\bf 122} (2019) 042502.
\bibitem{ENDF} 
The values of all fission-fragment yields of ENDF/B-VI.8 to ENDF/B-VIII.0 for spontaneous and neutron-induced fission are identical, with the exception of fission induced by fast and 14-MeV neutrons of $^{239}$Pu, which were revised in ENDF/B-VII.1. ENDF/B-VI.8 yields were taken from Ref.~\cite{England94}, complemented by some yields of $^{241}$Pu(n,f).
\bibitem{England94} T. R. England, B. F. Rider, LA-UR-94-3106 (ENDF-349) (1994) Los Alamos National Laboratory.
\bibitem{Specht74a} H. J. Specht, Physica Scripta {\bf10A} (1974) 21.
\bibitem{Specht74b} H. J. Specht, Rev. Mod. Phys. {\bf46} (1974) 773.
\bibitem{Andreyev18} A. N. Andreyev, K. Nishio, K.-H. Schmidt, Rep. Progr. Phys. {\bf 81} (2018) 016301.
\bibitem{Laurec10} J. Laurec {\it  et al.}, Nucl. Data Sheets {\bf 111} (2010) 2965.
\bibitem{Wahl80} A. C. Wahl, J. Radioanalytical Chem. {\bf 55} (1980) 111.
\bibitem{Wahl02} A. C. Wahl, “Systematics of Fission-Product Yields”, LA-13928, (2002).
\bibitem{Mills95} R. W. Mills, “Fission product yield evaluation”, PhD thesis, university of Birmingham, 1995. 
\bibitem{Moll75} E. Moll {\it  et al.}, Nucl. Instrum. Methods {\bf 123} (1975) 615.
\bibitem{Koester10} U. K\"oster {\it  et al.}, Nucl. Instrum. Methods {\bf A 613} (2010) 363.
\bibitem{Wohlfarth78} H. Wohlfarth {\it  et al.}, Z. Phys. {\bf A 287} (1978) 153.
\bibitem{Martin14} F. Martin {\it  et al.}, Nucl. Data Sheets {\bf 119} (2014) 328.
\bibitem{Schmidt00} K.-H. Schmidt {\it et al.}, Nucl. Phys. {\bf A 665} (2000) 221.
\bibitem{Caamano13} M. Caamano {\it  et al.}, Phys. Rev. {\bf C 88} (2013) 024605.
\bibitem{Boutoux13} G. Boutoux {\it  et al.}, Physics Procedia {\bf 47} (2013) 166.
\bibitem{Chatillon19} A. Chatillon {\it  et al.}, Phys. Rev. {\bf C 99} (2019) 054628.
\bibitem{Bocquet90} J. P. Bocquet {\it  et al.}, Z. Phys. {\bf A 335} (1990) 41.
\bibitem{Quade88} U. Quade {\it et al.}, Nucl. Phys. {\bf A 487} (1988) 1.
\bibitem{Lang80} W. Lang {\it  et al.}, Nucl. Phys. {\bf A 345} (1980) 34.
\bibitem{Sida89} J. L. Sida {\it  et al.}, Nucl. Phys. {\bf A 502} (1989) 233c-242c.
\bibitem{Martinez90} G. Martinez {\it  et al.}, Nucl. Phys. {\bf A 515} (1990) 433.
\bibitem{Tsekhanovich01} I. Tsekhanovich {\it et al.}, Nucl. Phys. {\bf A 688} (2001) 633.
\bibitem{Schmitt84} C. Schmitt {\it et al.} Nucl. Phys. {\bf A 430} (1984) 21.
\bibitem{Djebara89} M. Djebara {\it et al.} Nucl. Phys. {\bf A 496} (1989) 346.  
\bibitem{Hentzschel94} R. Hentzschel {\it et al.} Nucl. Phys. {\bf A 571} (1994) 427.
\bibitem{Kern12} Kilian Kern, Maarten Becker, Cornelis Broeders, PHYSOR Advances in Reactor Physics (2012) 1. 
\bibitem{Noguere16} E. Privas {\it et al.}, EPJ Nuclear Sci. Technol. {\bf 2} (2016) 32.
\bibitem{Fowler74} M. M. Fowler, A. C. Wahl, J. Inorganic Nucl. Chem. {\bf 36} (1974) 1201.
\bibitem{Diiorio77} G. Diiorio, B. W. Wehring, Nucl. Instrum. Methods B {\bf 147} (1977) 487.
\bibitem{Thierens76} H. Thierens {\it et al.}, 
  Nucl. Instrum. Methods {\bf 134} (1976) 299.
\bibitem{Birgersson07} E. Birgersson {\it et al.}, Nucl. Phys. A {\bf 791} (2007) 1.
\bibitem{Report-NDS-676} P. Dimitriou and A. L. Nichols, { IAEA report INDC(NDS)-0676}, Feb. 2015, IAEA, Vienna, Austria.
\bibitem{DayaBay17} F. P. An {\it et al.}, (Daya Bay Collaboration), Phys. Rev. Lett. {\bf 118} (2017) 251801.
\bibitem{Asghar84} M. Asghar, R. W. Hasse, J. Phys. Colloques {\bf 45} (1984) C6-455.
\bibitem{Terrell62} J. Terrell, Phys. Rev. 127 (1962) 880.

\end{thebibliography}
\end{document}